\documentstyle[12pt,twoside,french,epsf]{report}

\def\journal#1, #2, #3, #4 { {\sl #1~}{\bf #2~}(#3) #4 }

\def\mpl{\journal Mod. Phys. Lett., }

\def\prl{\journal Phys. Rev. Lett., }

\def\cmp{\journal Comm. Math. Phys., }

\def\np{\journal Nucl. Phys., }

\def\pl{\journal Phys. Lett., }

\def\annp{\journal Ann. Phys. (N.Y.), }

\catcode`\@=11
\def\marginnote#1{}
\newcount\hour
\newcount\minute
\newtoks\amorpm
\hour=\time\divide\hour by60
\minute=\time{\multiply\hour by60 \global\advance\minute
by-\hour}\edef\standardtime{{\ifnum\hour<12
\global\amorpm={am}%
        \else\global\amorpm={pm}\advance\hour by-12 \fi
        \ifnum\hour=0 \hour=12 \fi
        \number\hour:\ifnum\minute<10
0\fi\number\minute\the\amorpm}}
\edef\militarytime{\number\hour:\ifnum\minute<10
0\fi\number\minute}

\def\draftlabel#1{{\@bsphack\if@filesw {\let\thepage\relax
   \xdef\@gtempa{\write\@auxout{\string
      \newlabel{#1}{{\@currentlabel}{\thepage}}}}}\@gtempa
   \if@nobreak \ifvmode\nobreak\fi\fi\fi\@esphack}
        \gdef\@eqnlabel{#1}}
\def\@eqnlabel{}
\def\@vacuum{}
\def\draftmarginnote#1{\marginpar{\raggedright\scriptsize\tt#1}}
\def\draft{\oddsidemargin -.5truein
        \def\@oddfoot{\sl preliminary draft \hfil
        \rm\thepage\hfil\sl\today\quad\militarytime}
        \let\@evenfoot\@oddfoot \overfullrule 3pt
        \let\label=\draftlabel
        \let\marginnote=\draftmarginnote

\def\@eqnnum{(\theequation)\rlap{\kern\marginparsep\tt\@eqnlabel}%
\global\let\@eqnlabel\@vacuum}  }

\def\numberbysection{\@addtoreset{equation}{section}
        \def\theequation{\thesection.\arabic{equation}}}

\def\underline#1{\relax\ifmmode\@@underline#1\else
 $\@@underline{\hbox{#1}}$\relax\fi}

\catcode`@=12
\relax

\def\fin{\end{document}}
\def\bfg{\begin{figure}}
\def\efg{\end{figure}}
\def\beq{\begin{equation}}
\def\eeq{\end{equation}}
\def\beqa{\begin{eqnarray}}
\def\eeqa{\end{eqnarray}}
 \def\nnn{\nonumber \\}
\def\sqr#1#2{{\vcenter{\vbox{\hrule height.#2pt
\hbox{\vrule width.#2pt height#1pt \kern#1pt
\vrule width.#2pt}
\hrule height.#2pt}}}}

\def\verthat{{\hat |}}

\def\Je{J^e}
\def\Jeb{{\overline J}^e\, \!}

\def\Jne#1 {J_{#1}^e\, \!}
\def\Jneb#1 {{\overline J}_{#1}^e\, \!}
\def\Jnep#1 {J_{#1}'\, \!^e\, \!}
\def\Jnebp#1 {{\overline J}_{#1}'\, \!  \!^e\, \!}

\def\Jehat{{\widehat J}^e}
\def\hhat{{\widehat h}}

\def\Jhat{{\widehat J}}

\def\mhat{{\widehat m}}
\def\nhat{{\widehat n}}

\def\Vhat{{\widehat V}}

\def\gp{g' \, \!}
\def\gb{{\overline g}}
\def\gbp{{\overline g' \,\!}}

\def\psihat{{\widehat \psi}}

\def\psib{{\overline \psi}}

\def\psibhat{{\widehat{\overline \psi}}}

\def\Mhat{{\widehat M}}

\def\chib{{{\overline \chi}}}

\def\Rhat{{\widehat R}}
\def\Vhat{{\widehat V}}

\def\rhat{{\widehat r}}

\def\qhat{{\widehat q}}

\def\varpihat{{\widehat \varpi}}

\def\varpib{{\overline \varpi}}

\def\mhat{{\widehat m}}
\def\nhat{{\widehat n}}
\def\phat{{\widehat p}}

\def\Shat{{\widehat S}}

\def\Pb{{\overline P}}
\def\Jb{{\overline J}}
\def\mb{{\overline  m}}
\def\mub{{\bar \mu}}
\def\zb{{\bar z}}

\def\Vb{{\overline V}}

\def\Vt{{\widetilde V}}

\def\Jge{{\underline J}}

\def\Jgen#1 {  {\underline J_{#1}} }
\def\Jgenp#1 #2 {(J_{#1}+{#2},\Jhat_{#1})}
\def\Jgenm#1 #2 {(J_{#1}-{#2},\Jhat_{#1})}
\def\Jg#1 {J_{#1},\Jhat_{#1}}
\def\Jgp#1 #2 {J_{#1}+{#2},\Jhat_{#1}}
\def\Mgen#1 {{\underline M_{#1}}}

\def\produit#1,#2,#3,#4 {P\Bigl ( [{#1},{#2}]\otimes\{{#3}\},{#4}\Bigr )}
\def\produitscript#1,#2,#3,#4 {P\Bigl (
[{\scriptstyle{#1},{#2}}]\otimes\{{\scriptstyle{#3}}\},{#4}\Bigr )}
\def\pprod#1,#2,#3,#4,#5 {P\Bigl ( [{#1},{#2}]\otimes[{#3},{#4}],{#5}\Bigr )}
\def\pprodscript#1,#2,#3,#4,#5 {P\Bigl (
[{\scriptstyle{#1},{#2}}]\otimes[{\scriptstyle{#3},{#4}}],{#5}\Bigr )}

\def\fusV#1,#2,#3,#4,#5,#6 {f_V(
\Jgen{#1} ,
\Jgen{#2} ,
\Jgen{#3} ,
\Jgen{#4} ,
\Jgen{#5} ,
\Jgen{#6} )}

\def\brdV#1,#2,#3,#4,#5,#6 {b_V(
\Jgen{#1} ,
\Jgen{#2} ,
\Jgen{#3} ,
\Jgen{#4} ,
\Jgen{#5} ,
\Jgen{#6} )}

\def\fusxi#1,#2,#3 {f_\xi (\Jgen{#1} ,
\Mgen{#1} ,
\Jgen{#2} ,
\Mgen{#2} ,
\Jgen{#3} )}

\def\ghat{{\widehat g}}
\def\pghat{{\hat (}}
\def\pdhat{{\hat )}}

\def\gaghat{{\hat {\bigl \{}}}
\def\gadhat{{\hat {\bigr \}}}}

\def\bverthat{{\hat {\bigl |}}}

\def\sixjxi#1,#2,#3,#4,#5,#6 {{\left\{\left . \!\! \,^{#1}_{#2}
\,^{#3}_{#4} \right | \!\, ^{#5}_{#6}\right\}}}
\def\sixje#1,#2,#3,#4,#5,#6 {{\left\{\left\{\left . \!\! \, ^{#1}_{#2}
\, ^{#3}_{#4} \right | \!\, ^{#5}_{#6}\right\}\right\}}}
\def\sixjxihat#1,#2,#3,#4,#5,#6 {{{\gaghat\left . \!\! \, ^{#1}_{#2}
\, ^{#3}_{#4} \right | \!\, ^{#5}_{#6}\gadhat}}}
\def\sixjehat#1,#2,#3,#4,#5,#6 {{\gaghat\gaghat\left . \!\! \, ^{#1}_{#2}
\, ^{#3}_{#4} \right | \!\, ^{#5}_{#6}\gadhat\gadhat}}

\def\Jpe {J^{e+}}
\def\Jpeb {{\overline J}\,^{e+}}
\def\Jme {J^{e-}}
\def\Jmeb {{\overline J}\,^{e-}}
\def\Jpehat {\Jhat^{e+}}
\def\Jmehat {\Jhat^{e-}}
\def\Jehat {{\widehat J^e}}
\def\me {{m^e}}
\def\meb {{{\overline m}^e}}

\def\pb{{\overline p}}

\def\Jhatb{\widehat {\overline J}}
\def\mhatb{\widehat {\overline m}}

\def\sp {{\sigma_+}}
\def\sm {{\sigma_-}}
\def\Tb {{\overline T}}

\def\Rb {{\overline R}}
\def\Deltab {{\overline \Delta}}
\def\Lb {{\overline L}}

\def\im {{\hbox{Im}}}

\def\Jpe {J^{e+}}
\def\Jme {J^{e-}}
\def\Jpehat {\Jhat^{e+}}
\def\Jmehat {\Jhat^{e-}}
\def\Jehat {{\widehat J^e}}
\def\me {{m^e}}

\def\Jhatb{\widehat {\overline J}}
\def\mhatb{\widehat {\overline m}}
\def\vertex#1,#2,{{{\cal V}^{{#1,#2}}}}
\def\vertexc#1,#2,{{{\cal V}^{{#1,#2}}_{\hbox{\scriptsize conj}}}}
\def\lf#1,#2,{{L_{#1,#2}}}
\def\lfc#1,#2,{{L^{\hbox{\scriptsize conj}}_{#1,#2}}}
\def\ns{{\nu}}
\def\nsb{{\bar \nu}}

\def\fb{{\overline f}}
\def\Fb{{\overline F}}
\def\phib{{\overline \phi}}

\def\ib{{\overline i}}
\def\jb{{\overline j}}
\def\kb{{\overline k}}
\def\lb{{\overline l}}

\def\Xt{{\widetilde X}}

\def\Me{M^e}
\def\Mehat{{\widehat M}^e}
\def\Mep{M^{e'}}
\def\Mehatp{{\widehat M}^{e'}}

\begin{document}

\tableofcontents

\newpage

\centerline{\bf\large INTRODUCTION}

\vskip 35mm

La th\'eorie des cordes a connu un regain d'int\'er\^et
dans les ann\'ees 80
\`a la suite des travaux fondateurs
de Polyakov puis de Green et Schwarz.
Elle s'est av\'er\'ee
coh\'erente et renormalisable,
on a pu la traiter comme les th\'eories de jauge habituelles
et sortir des dimensions critiques
\`a condition d'introduire une gravitation quantique
bidimensionnelle.

Une corde \'evoluant dans un espace-temps
de dimension $d$ est
d\'ecrite (en premi\`ere quantification)
par une th\'eorie quantique des champs
bidimensionnelle sur la surface qu'elle balaye
dans son mouvement, dite surface d'univers.
Pour un espace-temps plat,
les coordonn\'ees des points de la corde
sont d'un point de vue bidimensionnel tout simplement $d$ champs libres
(on a un mod\`ele sigma non lin\'eaire dans le cas d'un
espace-temps non trivial).
Il a n\'eanmoins \'et\'e n\'ecessaire d'introduire
dans l'action dite de Polyakov une
m\'etrique bidimensionnelle intrins\`eque.
Bien qu'elle ne soit classiquement qu'un multiplicateur de Lagrange,
lors de la quantification l'anomalie de Weyl la rend dynamique.
C'est l'origine de la gravitation quantique bidimensionnelle.
Les $d$
champs libres coupl\'es \`a la gravit\'e sont habituellement
appel\'es champs de mati\`ere.
Ces th\'eories de cordes ont une grande invariance de jauge
constitu\'ee par les reparam\'etrisations de la surface d'univers.
On la fixe soit dans la jauge conforme,
soit dans celle du c\^one de lumi\`ere.
Nous ferons ici le premier choix.
Comme pour les th\'eories de Yang-Mills,
on a donc \'egalement sur la surface d'univers
des fant\^omes de Faddeev-Popov qui rendent compte du jacobien
provenant de cette fixation de jauge.
Dans la jauge conforme,
c'est le facteur conforme (facteur de proportionnalit\'e de la m\'etrique)
qui devient dynamique.
Son action est celle de Liouville
et c'est le sujet
central de cette th\`ese.

Il reste n\'eanmoins dans cette jauge une sym\'etrie
de r\'esiduelle:
les transformations conformes.
La th\'eorie doit donc \^etre invariante conforme
(en espace-temps de m\'etrique non triviale,
la coh\'erence de la th\'eorie s'exprime
par l'annulation de la fonction beta du mod\`ele sigma).
Cependant, chacun des champs existant sur la surface d'univers
a individuellement une anomalie conforme,
anomalie qui s'annule pour la th\'eorie totale.
On peut voir la dynamique du champ de Liouville comme
une fa\c con pour la th\'eorie de restaurer l'invariance conforme:
les $d$ champs de mati\`ere contribuent de $d$ unit\'es \`a l'anomalie
conforme (ou charge centrale de l'alg\`ebre conforme),
les fant\^omes pour $-26$,
et par cons\'equent le champ de Liouville pour $26-d$
(cas bosonique).

Pour les dimensions critiques ($d=26$ pour les cordes bosoniques,
et $d=10$ pour les supercordes)
pour lesquelles la th\'eorie est exempte d'anomalie de Weyl,
la gravit\'e se d\'ecouple.
Ceci n'est pas \'etonnant puisque dans ce cas les champs de mati\`ere
et les fant\^omes donnent \`a eux seuls une th\'eorie invariante
conforme, de charge centrale totale $c_{\hbox{\scriptsize tot}}=d-26=0$.
N\'eanmoins, nous n'aborderons pas ici le domaine bien connu
des cordes critiques,
et nous nous placerons toujours en dimension
non-critique.

Initialement, le but essentiel de l'\'etude des
cordes non-critiques \'etait de s'affranchir
des dimensions 10 ou 26.
Il est cependant apparu depuis lors de nombreuses mani\`eres
de r\'eduire la dimension de l'espace-temps des cordes critiques,
par compactification ou en consid\'erant des degr\'es de libert\'e internes.
Actuellement l'attention port\'ee aux cordes non-critiques
est davantage motiv\'ee
par l'\'etude de la gravitation quantique bidimensionnelle,
ce qui permet de d\'ecrire des surfaces al\'eatoires
charg\'ees de
champs conformes de mati\`ere.
L'\'etude des th\'eories conformes bidimensionnelles constitue
tout un domaine en soi.
Elles entretiennent des rapports \'etroits avec les mod\`eles
int\'egrables.
Comme nous le verrons dans cette th\`ese,
la th\'eorie de Liouville
s'est ainsi av\'er\'ee int\'egrable quantiquement,
et sa sym\'etrie classique $Sl(2)$ a donn\'e naissance
\`a la sym\'etrie du groupe quantique $U_q(sl(2))$.
Les groupes quantiques jouent \'egalement un r\^ole important
dans de nombreuses th\'eories int\'egrables.
C'est le cas par exemple du mod\`ele de Wess-Zumino-Witten,
qui redonne en principe la th\'eorie de Liouville par r\'eduction
dimensionnelle.
Les groupes quantiques sont d'ailleurs initialement apparus lors
de l'\'etude des mod\`eles int\'egrables et en particulier
de la th\'eorie de Liouville,
avant d'\^etre \'etudi\'es plus syst\'ematiquement.

\vskip 3mm

Ce domaine des cordes non-critiques a \'et\'e tr\`es largement
\'etudi\'e et r\'esolu en dimension $d$ inf\'erieure \`a 1. On
parle souvent un peu abusivement de dimensions non enti\`eres,
en assimilant la dimension $d$ \`a l'anomalie conforme $c$, puisque
les $d$ champs libres de la corde bosonique vivant en dimension
$d$ donnent une charge centrale $c = d$.
Quand on parle de mod\`eles de cordes
en dimensions inf\'erieures \`a 1, il s'agit donc en fait des th\'eories
conformes bidimensionnelles coupl\'ees \`a la gravit\'e \'evoqu\'ees plus haut,
pour une charge centrale $c$ inf\'erieure \`a 1.
Cette th\'eorie conforme
peut \^etre un champ libre avec charge d'arri\`ere-plan (dite
de "background") ou un mod\`ele de physique statistique (d'Ising,
de Potts, O(n)...) \`a un point critique d'une transition de phase
du deuxi\`eme ordre. Les mod\`eles de physique statistique sont
alors d\'ecrits par les mod\`eles minimaux, th\'eories conformes
contenant un nombre fini de familles conformes et
correspondant \`a un ensemble discret de charges centrales particuli\`eres
comprises entre 0 et 1.
Ces mod\`eles
ont pu \^etre trait\'es dans une approche continue, mais un grand
pas en avant a \'et\'e r\'ealis\'e gr\^ace aux succ\`es de
l'approche discr\`ete. En effet, la triangulation de la surface a
permis de ramener le probl\`eme des mod\`eles minimaux coupl\'es
\`a la  gravit\'e \`a des mod\`eles de matrices, qui ont pu \^etre
r\'esolus \`a tous les ordres (c'est-\`a-dire pour tout genre de
surface, ce qui correspond au d\'eveloppement en boucles des diagrammes
de Feynman). L'espoir que la solution soit non-perturbative a
n\'eanmoins \'et\'e d\'e\c cu. L'approche continue a donn\'e des
r\'esultats concordants pour les fonctions \`a trois points et les
exposants critiques.

Au contraire, dans le domaine $d>1$, on a tr\`es peu de r\'esultats. On
l'appelle g\'en\'eralement r\'egime de couplage fort, puisque la
constante de couplage de la th\'eorie de Liouville est alors
sup\'erieure \`a 1/8. La difficult\'e dans cette phase provient de
l'apparition de grandeurs complexes.
Les poids conformes des op\'erateurs
de la th\'eorie, donn\'es par les poids de Kac, deviennent complexes, ce
qui a pour cons\'equence des exposants critiques et des masses carr\'ees
complexes. C'est un des grands m\'erites de la quantification canonique
de la th\'eorie de Liouville introduite par J.-L. Gervais et A. Neveu
au d\'ebut des ann\'ees 80, que d'avoir permis un d\'ebut de r\'esolution
de cette th\'eorie dans le r\'egime de couplage fort. Et
c'est pour le moment \`a peu pr\`es le seul axe de recherche fructueux
dans cette phase.
Il a en effet \'et\'e mis en \'evidence des dimensions
sp\'eciales ($d = 7, 13, 19$ pour la corde bosonique)
pour lesquelles il semblait exister une sous-alg\`ebre ferm\'ee d'op\'erateurs
physiques de poids r\'eels,
ce qui permettrait de r\'esoudre les probl\`emes
pr\'ec\'edemment \'evoqu\'es
(les dimensions correspondantes sont $d = 3, 5, 7$ pour
les supercordes).
C'est le traitement sym\'etrique
des deux charges d'\'ecran $\alpha_+$ et $\alpha_-$ qui semble
\^etre la cl\'e de ces r\'esultats dans le r\'egime de couplage fort.
La fin de la preuve de cette troncature (cas bosonique) et la construction
d'un mod\`ele topologique fortement coupl\'e, pour lequel nous
calculons les fonctions \`a $N$ points, constituent une partie
importante de cette th\`ese.

\vskip 1cm

Le plan de ce m\'emoire est le suivant. Les trois premiers
chapitres ont pour but d'introduire les notions n\'ecessaires \`a la
compr\'ehension de mes travaux originaux dont l'essentiel
est pr\'esent\'e dans les deux derniers chapitres, tous
les d\'etails figurant dans les articles [P1, P2, P3, P4, P6]
joints en annexe. Le compte-rendu [P6] constitue un petit r\'esum\'e
qui a l'avantage d'\^etre en fran\c cais. L'article [P5] qui sera
publi\'e prochainement n'a pas pu \^etre joint au m\'emoire. Un
certain nombre de r\'esultats non publi\'es en refs.[P1, P2, P3, P4]
(et qui le seront pour l'essentiel en [P5]) seront imprim\'es en italique.

Dans le premier chapitre, je pr\'esente la base des th\'eories conformes.
Je me place \`a un niveau \'el\'ementaire, renvoyant le lecteur
int\'eress\'e aux nombreuses revues existant sur le sujet.
Je pr\'esente ensuite dans un deuxi\`eme chapitre la solution classique
de la th\'eorie de Liouville, connue depuis longtemps, ainsi que le
sch\'ema de quantification introduit par J.-L. Gervais et A. Neveu
au d\'ebut des ann\'ees 80. La r\'esolution compl\`ete de l'alg\`ebre des
op\'erateurs de la th\'eorie de Liouville dans le chapitre 4 fera appel
aux \'equations polynomiales qui sont les \'equations de coh\'erence
de cette alg\`ebre. Elles sont d\'emontr\'ees dans
le chapitre 3 o\`u nous \'etendons la discussion de
Moore et Seiberg sur les th\'eories conformes diagonales (et les
\'equations polynomiales) au cas que nous appelons non-diagonal,
ce qui s'av\'erera n\'ecessaire pour le r\'egime de couplage fort.

Les deux derniers chapitres traitent de ma
contribution \`a la r\'esolution de la th\'eorie de Liouville,
en collaboration avec J.-L. Gervais et E. Cremmer pour le chapitre 4
et J.-L. Gervais seul pour le chapitre 5.
Dans le chapitre 4 (articles [P1,P2]) j'expose la r\'esolution
de l'alg\`ebre d'op\'erateurs dans la base
dite des ondes de Bloch puis dans celle du groupe quantique.
Les coefficients de fusion et d'\'echange
sont essentiellement donn\'ees par des symboles
du groupe quantique $U_q(sl(2))$,
ce qui permet de mettre en \'evidence la sym\'etrie $U_q(sl(2))$
de la th\'eorie.
Nous nous contenterons ici d'obtenir cette alg\`ebre,
sans expliciter l'action des g\'en\'erateurs de
$U_q(sl(2))$ (ce qui est n\'eanmoins possible),
puisque la fusion et l'\'echange suffisent \`a prouver
la localit\'e des op\'erateurs ou des corr\'elateurs physiques.
Par rapport aux travaux pr\'ec\'edents de J.-L. Gervais
et ses collaborateurs,
l'alg\`ebre est compl\`etement r\'esolue au sens de Moore
et Seiberg,
alors qu'auparavant la fusion n'\'etait trait\'ee qu'\`a
l'ordre dominant \`a la Wilson, ce qui mettait
l'associativit\'e en d\'efaut.
Nous l'avons \'egalement \'etendue \`a des op\'erateurs
de spins continus
(non n\'ecesssairement demi-entiers),
c'est-\`a-dire des repr\'esentations infinies
du groupe quantique $U_q(sl(2))$.
Nous relions aussi la matrice de passage entre ces deux bases
\`a une autre matrice de passage connue dans la litt\'erature,
qui est donn\'ee par des Clebsch-Gordan.
Notre matrice de passage est la limite de ces Clebsch-Gordan.
La base des ondes de Bloch
est reli\'ee \`a un mod\`ele SOS ou IRF (Interaction
Round a Face) alors
que la base du groupe quantique est reli\'ee \`a
un mod\`ele \`a vertex.
Nous montrons que le mod\`ele \`a vertex est la limite
du mod\`ele IRF lorsque ses spins sur les faces tendent vers l'infini.
Notre g\'en\'eralisation \`a des spins continus
donne donc aussi une g\'en\'eralisation des invariants
de liens correspondant \`a ces mod\`eles \`a vertex et IRF.
Il est  \'egalement propos\'e en [P2] une repr\'esentation tridimensionnelle
de ces invariants
par des tetra\`edres.
Elles est connue pour les mod\`eles IRF (t\'etra\`edres = 6-j)
mais ne semble pas l'\^etre pour les mod\`eles \`a vertex
(t\'etra\`edres = Clebsch-Gordan, matrice $R$ et 6-j).

Le chapitre 5 (articles [P3,P4,P5]) traite du couplage fort,
c'est-\`a-dire de th\'eories en dimensions
$1<d<25$.
Il appara\^ \i t alors en g\'en\'eral des exposants critiques
et des poids complexes qui ne sauraient \^etre physiques.
J.-L. Gervais a cependant montr\'e avec ses collaborateurs
que pour des dimensions sp\'eciales
($d=7,13,19$ dans le cas bosonique)
il existe une sous-alg\`ebre
ferm\'ee d'op\'erateurs de poids r\'eels.
Ce travail \'etait cependant partiel.
Nous montrons dans ce chapitre,
gr\^ace aux r\'esultats du pr\'ec\'edent,
que l'alg\`ebre d'op\'erateurs est bien ferm\'ee \`a tous
les ordres.
Nous \'etendons \'egalement ceci \`a des spins fractionnaires
qui sont n\'ecessaires pour obtenir des fonctions de partition invariantes
modulaires.
Nous proposons ensuite dans la partie 5.2 un nouvel \'eclairage
de la troncature de cette sous-alg\`ebre d'op\'erateurs
de poids r\'eels.
Il appara\^\i t en effet que les dimensions et spins
mis en \'evidence pr\'ec\'edemment sont exactement
ceux qui donnent une d\'eg\'en\'erescence maximale
de la matrice de monodromie,
et nous montrons que cette d\'eg\'en\'erescence
est un cas particuli\`erement favorable \`a l'obtention
d'une troncature de l'alg\`ebre d'op\'erateurs.
N'\'etant pas encore capables
de calculer les amplitudes de v\'eritables th\'eories
de cordes dans ces dimensions sp\'eciales,
nous \'etudions dans la partie 5.3 un mod\`ele simplifi\'e
\`a deux degr\'es de libert\'e,
du m\^eme genre que les mod\`eles minimaux dans
le r\'egime de couplage faible:
la mati\`ere et la gravit\'e
sont engendr\'ees par deux copies de la construction
sp\'eciale pr\'ec\'edente.
Les excitations correspondant
\`a ces deux degr\'es de libert\'e (un pour la mati\`ere, un pour la gravit\'e)
doivent donc ensuite dispara\^ \i tre par cohomologie BRST
et le mod\`ele devenir topologique.
Nous avons pu calculer la susceptibilit\'e de corde
$\gamma_{\hbox{\scriptsize string}}$
et les fonctions de corr\'elation \`a $N$ points de ce mod\`ele
gr\^ace \`a une action effective.
L'exposant critique $\gamma_{\hbox{\scriptsize string}}$ calcul\'e
est la partie r\'eelle de la formule de KPZ
qui est complexe dans la phase de couplage fort.
La physique du couplage fort semble en outre tr\`es diff\'erente
de celle du couplage faible,
avec en particulier un d\'econfinement de la chiralit\'e.

\chapter{INTRODUCTION AUX THEORIES CONFORMES}
\markboth{1. Introduction aux th\'eories conformes}{1. Introduction aux
th\'eories conformes}
\label{p2.1}

Les th\'eories conformes
sont aujourd'hui devenues un des outils fondamentaux
de la th\'eorie des cordes. La sym\'etrie conforme
est en effet la sym\'etrie r\'esiduelle apr\`es fixation
de la jauge conforme
sur la surface d'univers de la corde.
Elles ont \'egalement
une importance fondamentale en physique statistique
car la sym\'etrie conforme est aussi pr\'esente
aux transitions de phase du deuxi\`eme ordre.
Cela fait donc aujourd'hui partie de la culture du
physicien th\'eoricien
et c'est pourquoi
je me contenterai souvent dans ce chapitre de citer
des r\'esultats et de renvoyer le lecteur aux
refs.\cite{BPZ,Pe,LT} ainsi qu'aux nombreuses
autres ref\'erences existantes.

\vskip 2mm

En dimension
$D$ quelconque les transformations conformes sont
les translations, les rotations,
les dilatations et les transformations dites conformes sp\'eciales.
Elles sont donc engendr\'ees par un nombre fini d'entre elles
et le groupe conforme est de dimension finie.
En dimension 2, en revanche, il en appara\^\i t d'autres
et le groupe conforme est alors constitu\'e
des transformations analytiques et anti-analytiques du plan complexe.
Il devient de dimension infinie.
C'est la source de la richesse des th\'eories conformes bidimensionnelles.
Ce nombre infini de sym\'etries a en effet permis
d'en r\'esoudre un grand nombre.
Il semble m\^eme y avoir un rapport profond entre
th\'eories conformes bidimensionnelles et
mod\`eles int\'egrables.
Nous ne nous int\'eresserons ici qu'aux th\'eories bidimensionnelles.

\vskip 2mm

Je me contente ici de
pr\'esenter la base des th\'eories conformes,
et le lecteur un minimum familier de ces th\'eories
pourra passer \`a la suite.
Un des buts de ce chapitre est de montrer
comment les \'equations dites de ``bootstrap''
sont obtenues
comme \'equations de Ward li\'ees au d\'ecouplage
des vecteurs nuls de la th\'eorie conforme.
En effet, la plus simple d'entre elles,
une \'equation hyperg\'eom\'etrique,
est \`a la base de l'approche op\'eratorielle
de la th\'eorie de Liouville.
En 1983 Gervais et Neveu l'on prouv\'ee explicitement
gr\^ace \`a l'expression des op\'erateurs en termes de champs libres,
comme nous le verrons dans le chapitre suivant.
Il n'\'etait donc pas vraiment
n\'ecessaire pour
la r\'esolution de la th\'eorie de Liouville
de pr\'esenter les th\'eories conformes telles
qu'elles ont \'et\'e formalis\'ees dans l'\oe uvre
fondatrice de BPZ \cite{BPZ},
puisque la sym\'etrie conforme appara\^\i t
naturellement dans la th\'eorie de Liouville.
Ceci permet n\'eanmoins de la situer
dans un cadre plus g\'en\'eral.

\section{Changements de coordonn\'ees et tenseur
\'energie - impulsion}

Les transformations conformes sont les changements de coordonn\'ees
qui conservent
les angles,
et qui conservent donc la m\'etrique proportionnelle \`a elle-m\^eme.
Pour un tel changement de coordonn\'ees
$$
\sigma^a\to\xi^a(\sigma)
\qquad\qquad
a=1,2
$$
la m\'etrique se transforme donc (classiquement) par
\beq
g_{ab}\to
{
\partial\xi^c
\over
\partial\sigma^a
}
{
\partial\xi^d
\over
\partial\sigma^b
}
g_{cd}
=
\rho(\sigma)
g_{ab}
\label{g->g'}
\eeq

Les th\'eories classiques invariantes sous les transformations
conformes ont un tenseur energie-impulsion
de trace nulle.
Ceci est tr\`es facile \`a voir:
on sait d\'ej\`a que pour tout
lagrangien ne d\'ependant pas explicitement du point
consid\'er\'e (donc invariant par translation), le
tenseur \'energie-impulsion $T^{ab}$ est conserv\'e:
\beq
\partial_aT^{ab}=0.
\eeq
De plus, pour un lagrangien invariant conforme, donc invariant par dilatation
$x^a\to x^a+\epsilon x^a$, le courant de dilatation
$T^{ab}x_b$ est \'egalement conserv\'e.
Ceci a pour cons\'equence l'annulation de la trace de $T^{ab}$:
\beq
\partial_a(T^{ab}x_b)=(\partial_aT^{ab})x_b+T^{ab}(\partial_a)x_b
=T^a_a=0.
\eeq
En dimension 2 la r\'eciproque est \'egalement vraie:
$T^a_a=0$ et $\partial_aT^{ab}=0$ impliquent la sym\'etrie conforme.
Au niveau quantique, en revanche, une action
classiquement invariante conforme
pourra donner lieu \`a une anomalie conforme,
qu'on appelle donc parfois anomalie de trace.

Nous avons jusqu'ici \'ecrit des formules valables pour
une m\'etrique quelconque,
mais nous nous placerons en fait toujours
dans la jauge conforme o\`u
la m\'etrique de r\'ef\'erence est $\delta_{ab}$
(en euclidien).
Dans un syst\`eme de coordonn\'ees complexes\footnote{
Il faut consid\'erer les deux variables $z$ et $\zb$ comme ind\'ependantes
(donc $\sigma$ et $\tau$ complexes).
On se ram\`ene au cas physique \`a la fin par restriction
\`a $\sigma$ et $\tau$ r\'eels donc $\zb=(z)^*$.
}
$z=\sigma_2+i\sigma_1,\zb=\sigma_2-i\sigma_1$,
la m\'etrique euclidienne s'\'ecrit
$g_{z\zb}=g_{\zb z}\propto 1$, $g_{z\zb}=g_{\zb z}=0$
et les transformations conformes
sont simplement les transformations analytiques et anti-analytiques
\beq
z\to z'(z)
\ ,\
\zb\to \zb'(\zb).
\label{z->z'}
\eeq
Cependant, la surface d\'ecrite par une corde
ferm\'ee dans son mouvement (surface-univers)
a la topologie d'un cylindre.
La coordonn\'ee $\sigma_1=\sigma$ (de genre temps en minkowskien)
est p\'eriodique, la p\'eriode pouvant \^etre ramen\'ee \`a $2\pi$.
C'est pourquoi
on utilise souvent des coordonn\'ees complexes qui sont
les exponentielles de celles d\'efinies pr\'ec\'edemment
($\sigma^1\equiv\sigma$,
$\sigma^2\equiv\tau$):
\beq
z=e^{\tau+i\sigma}
\ ,\
\zb=e^{\tau-i\sigma}
\label{z}
\eeq
ce qui transforme le cylindre en plan.
Dans ces coordonn\'ees du plan complexe,
le temps est radial
et $\sigma$ est naturellement p\'eriodique.
Le pass\'e et le futur infinis correspondent \`a $z=0$ et
$z=\infty$ respectivement.
Pour des th\'eories conformes quantiques
on parlera donc de quantification radiale,
le T-produit \'etant remplac\'e par un ordre radial.
La transformation des coordonn\'ees $z=\tau+i\sigma$, $\zb=\tau-i\sigma$
du cylindre aux coordonn\'ees \ref{z} du plan complexe
est elle-m\^eme conforme, et les coordonn\'ees \ref{z}
sont donc de bonnes coordonn\'ees:
la m\'etrique garde la m\^eme forme et les transformations
conformes sont toujours les transformations
analytiques et anti-analytiques \ref{z->z'}.

Dans ce syst\`eme de coordonn\'ees, la nullit\'e de la trace
de $T$ (utilisant la sym\'etrie de $T$) s'\'ecrit $T_{z\zb}=T_{\zb z}=0$
et la conservation de $T$ prouve que ses autres composantes
$T_{zz}$ et $T_{\zb\zb}$ sont respectivement analytique
et anti-analytique:
$$
\partial_zT_{\zb\zb}=0
\quad \Rightarrow\quad
T_{\zb\zb}(z,\zb)=T_{\zb\zb}(\zb)
\hbox{ not\'e }
\Tb(\zb)
$$
\beq
\partial_\zb T_{zz}=0
\quad \Rightarrow\quad
T_{zz}(z,\zb)=T_{zz}(z)
\hbox{ not\'e }
T(z).
\eeq
$T(z)$ et $\Tb(\zb)$ sont les g\'en\'erateurs des transformations conformes
$z\to z'(z)$ et $\zb\to\zb'(\zb)$ respectivement,
ou
\beq
z\to z+\epsilon(z)
\label{z+eps}
\eeq
pour les transformations infinit\'esimales engendr\'ees par $T(z)$.
Le groupe (des tranformations) conforme(s) est factoris\'e:
c'est le produit tensoriel de sa partie analytique par sa partie
anti-analytique,
c'est pourquoi nous ne consid\'ererons souvent par exemple que la partie
analytique
engendr\'ee par $T(z)$.

Nous pourrions traiter
le cas des th\'eories conformes classiques
mais pr\'ef\'erons passer tout de suite
au cas quantique.
Dans ce cas,
les transformations du tenseur \'energie-impulsion $T$
ne se d\'eduisent pas simplement des lois de transformations tensorielles.
On \'ecrit donc
la forme la plus
g\'en\'erale possible de ses variations sous les
transformations conformes (qu'il engendre lui-m\^eme).
Nous donnons le r\'esultat pour ses modes $\L_n$
qui sont d\'efinis par
\beq
T(z)=
\sum_{n=-\infty}^{\infty}
z^{-n-2}
L_n
\label{TdeLn}
\eeq
chacun des $L_n$ \'etant le g\'en\'erateur
des transformations infinit\'esimales \ref{z+eps} avec
\beq
\epsilon(z)=\epsilon z^{n+1}.
\label{epsn}
\eeq
Les lois de transformation les plus g\'en\'erales
pour ces modes
sont donn\'ees par l'alg\`ebre de Virasoro
\beq
[L_n,L_m]=(n-m)L_{n+m}
+{c\over 12}n(n-1)(n+1)\delta_{n,-m}.
\label{Virasoro}
\eeq
On a une alg\`ebre similaire pour les
modes
$\Lb_n$ de $\Tb$ qui commutent par ailleurs avec les $L_n$.
Le param\`etre $c$ ne peut pas \^etre fix\'e par des consid\'erations
g\'en\'erales.
Il d\'epend de la th\'eorie consid\'er\'ee.
C'est la charge centrale ou anomalie conforme.
Il est en effet imm\'ediat de v\'erifier
que l'alg\`ebre classique des g\'en\'erateurs $z^{n+1}d/dz$
des transformations conformes \ref{epsn}
est donn\'ee par \ref{Virasoro}
avec $c=0$.
L'alg\`ebre de Virasoro \ref{Virasoro} est donc l'extension centrale
de l'alg\`ebre conforme.
La sym\'etrie conforme n'est restaur\'ee que pour
une charge centrale totale nulle.

\vskip 2mm

En th\'eorie des cordes,
l'invariance conforme est ce qu'il reste
de l'invariance par reparam\'etrisation lorsqu'on
s'est plac\'e dans la jauge conforme.
L'invariance par reparam\'etrisation est une invariance de jauge fondamentale
de la th\'eorie et ne doit pas \^etre bris\'ee quantiquement.
Une th\'eorie de corde ne sera donc coh\'erente
que si les contributions \`a la charge centrale de tous
ses degr\'es de libert\'e donnent une charge centrale totale nulle.
Pour une corde bosonique (resp. pour une supercorde)
les fant\^omes de Faddeev-Popov introduits lors de la fixation
de jauge contribuent pour -26 (resp. -15) \`a la charge centrale.
Chaque coordonn\'ee d'espace-temps
correspond
du point de vue bidimensionnel
\`a un (super-)champ (super-)conforme
de charge centrale 1 (resp. 3/2).
Ces seuls degr\'es de libert\'e donnent donc
dans le cas des cordes critiques, d\'efinies
en dimension critique 26 (resp. 10),
une th\'eorie libre d'anomalie conforme.
En dimension non-critique, en revanche,
ces degr\'es de libert\'e n'annulent pas la charge centrale.
Il est donc normal que pour une d\'efinition
coh\'erente de th\'eorie de cordes (i.e. invariante
par reparam\'etrisation et donc invariante conforme)
la m\'etrique
devienne dynamique et contribue pour $C=26-c$
\`a la charge centrale, si $c$ est est la contribution des
autres champs conformes (dits champs de mati\`ere,
par opposition \`a la m\'etrique dont la dynamique
constitue une gravitation quantique bidimensionnelle).

\vskip 2mm

Comme on peut le voir sur
l'\'equation \ref{epsn}, $L_{0}$ est le g\'en\'erateur des dilatations
du plan,
qui ne sont pas autre chose que les translations du temps $\tau$.
Le hamiltonien est donc
\beq
H=L_0+\Lb_0.
\eeq
On remarque \'egalement que $L_{-1}$ est le g\'en\'erateur des translations
alors que
$L_1$ est celui des transformations conformes sp\'eciales.
Ce sont les transformations conformes qui existent en toute dimension.
En dimension 2, ce sont les seules qui soient globalement inversibles.
On peut exhiber la forme globale de ces transformations dites projectives:
\beq
z\to
{az+b
\over
cz+d}
\quad
\hbox{  avec  }
\quad
ad-bc=1
\label{sl2C}.
\eeq
Comme on peut aussi le v\'erifier sur l'alg\`ebre de leurs g\'en\'erateurs
(alg\`ebre de Virasoro \ref{Virasoro} restreinte \`a
$m,n\in\{-1,0,1\}$)
la sous-alg\`ebre des transformations globalement inversibles engendr\'ee
par $L_{-1}$, $L_0$, $L_1$ est un groupe $Sl(2,C)$.
Leur composition et leur inversion se font par la multiplication
habituelle des matrices $\left(^{ab}_{cd}\right)$ de $Sl(2,C)$.

\section{Champs primaires - Etats de plus haut poids - Descendants}

Int\'eressons-nous maintenant aux op\'erateurs de la th\'eorie.
La loi de transformation la plus naturelle et la plus simple
d'un op\'erateur $\phi^i$ sous une transformation conforme
\ref{z->z'} est
\beq
\phi^i(z,\zb)
\to
\left(
{dz'
\over
dz}
\right)^{\Delta_i}
\left(
{d\zb'
\over
d\zb}
\right)^{\Deltab_i}
\phi^i(z',\zb')
\label{trphi}.
\eeq
L'indice (ou le multi-indice) $i$ sert \`a distinguer
les divers champs.
Le champ $\phi^i$ est dit de poids $\Delta_i$ et $\Deltab_i$.
L'\'equation \ref{trphi} est une extension
de la loi de transformation tensorielle
classique.
Les champs qui ont une loi de transformation donn\'ee
par \ref{trphi} sont dits primaires.
Tous les champs ne pourront pas \^etre dans ce cas,
la d\'eriv\'ee d'un champ primaire n'est d\'ej\`a
clairement pas un champ primaire.
Les transformations \ref{epsn} engendr\'ees par les $L_n$
s'\'ecrivent donc pour un tel champ
\beq
[L_n,\phi^i(z)]=
z^n
(z{d\over dz}
+(n+1)\Delta_i)
\phi^i(z)
\label{[Lnphi]}
\eeq
o\`u nous n'avons pas \'ecrit la d\'ependance en $\zb$.
Il y a, comme toujours, une expression similaire pour les $\Lb_n$.

En th\'eorie conforme il y a une correspondance entre les op\'erateurs
et les \'etats.
Les op\'erateurs engendrent les \'etats \`a partir du vide.
Au vu du d\'eveloppement en modes \ref{TdeLn} de $T$,
le vide $|0\! >$ ``entrant'' (correspondant donc \`a $z=0$)
doit \^etre annihil\'e par les modes $L_n$ tels que $-n-2<0$,
sinon $T(z)$ aurait une singularit\'e en $0$:
\beq
L_n|0\! >=0
\hbox{  pour  }
n\ge-1.
\eeq
Le vide $|0\! >$ est donc en particulier invariant sous les transformations
globales de $Sl(2,C)$ engendr\'ees par $L_1,L_0,L_{-1}$.
Les \'etats sont obtenus par action
des champs sur le vide:
\beq
|i\! >
=
\phi^i(0)
|0\! >
\label{etats}.
\eeq
Si le champ $\phi^i$ est primaire,
l'\'etat correspondant est aussi dit primaire
et nous r\'eservons d'ailleurs la notation \ref{etats}
aux \'etats primaires.
On d\'eduit des relations de commutation des champs primaires
que les \'etats primaires v\'erifient
\beq
L_0|i\! >=\Delta_i|i\! >
\hbox{  et  }
L_n|i\! >=0
\hbox{ pour }
n\ge 1.
\label{primLn}
\eeq
Comme les \'etats, ils sont dits de poids $\Delta_i$.
Contrairement au vide $|0\! >$ ils ne sont pas
$Sl(2,C)$ invariants: il y a maintenant un op\'erateur
ins\'er\'e en $z=0$ (i.e. $\tau=-\infty$) qui est
modifi\'e par les transformations $Sl(2,C)$.

La transformation $z\to1/z$ permet de
d\'efinir sym\'etriquement les \'etats sortants
\beq
<\! i|
=
\lim_{z,\zb\to\infty}
<\! 0|\phi^i(z,\zb)
z^{2L_0}\zb^{2\Lb_0}
\eeq
o\`u on a explicit\'e la d\'ependance en $z$ et $\zb$.
Les facteurs $z^{2L_0}$ issus de
la transformation $z\to1/z$ permettent d'avoir une normalisation
non singuli\`ere malgr\'e la limite infinie (cf Eq.\ref{f2p} plus loin)
et l'orthogonalit\'e des \'etats
\beq
<\! i|
j\!>
=
\delta_{i,j}.
\eeq
La m\^eme transformation permet d'obtenir la relation d'orthogonalit\'e
\beq
(L_n)^\dagger=L_{-n}.
\eeq
On en d\'eduit que
sym\'etriquement le vide et les \'etats sortants sont annihil\'es par les
$L_{-n}$:
$$
<\! 0|L_{-n}=0
\quad
\hbox{ pour }
\quad
-n\le 1
$$
\beq
<\! i|L_{-n}=0
\quad
\hbox{ pour }
\quad
-n\le -1.
\label{primL-n}
\eeq
La propri\'et\'e
$<\! 0|L_{-n}=0$ exprime la r\'egularit\'e du tenseur \'energie-impulsion
\`a l'infini.

Nous avons introduit les champs et \'etats primaires
et il est temps maintenant d'examiner quels sont les autres
champs et \'etats
de la th\'eorie, car
ils ne peuvent pas tous \^etre primaires.
On a en effet d\'ej\`a remarqu\'e que la d\'eriv\'ee d'un champ primaire
ne l'\'etait pas.
De fa\c con plus g\'en\'erale que la d\'eriv\'ee, qui correspond \`a
une translation locale engendr\'e par
$L_{-1}$, on peut appliquer \`a un champ $\phi^i$
tous les g\'en\'erateurs $L_n$ des transformations conformes locales.
Ces op\'erateurs doivent n\'ecessairement \^etre inclus dans
l'alg\`ebre car ils apparaissent par exemple dans le d\'eveloppement \`a
courte distance de $T(z)\phi^i(z')$.
Ils engendr\'es \`a partir des champs primaires par
\beq
\phi^{i,(-k_1,...,-k_N)}(z)=
L_{-k_1}(z)...L_{-k_N}(z)\phi^i(z)
\label{phi-k}
\eeq
pour des entiers positifs $k_1,...,k_N$,
d'o\`u leurs noms de descendants.
Comme les $L_n$ qui sont les modes de $T(z)$ d\'evelopp\'e autour
de $0$,
les $L_n(z)$ de \ref{phi-k} sont les modes pour un d\'eveloppement autour de
$z$.
Pour les \'etats (donc $z=0$) on a
\'egalement les descendants ou \'etats secondaires
\beq
|i,(-k_1,...,-k_N)\! >
=
L_{-k_1}...L_{-k_N}
|i\! >.
\eeq
L'\'etat $|i,\{-k_1,...,-k_N\}\! >$ n'est plus annihil\'e par
les $L_n$, $n>0$,
et la condition \ref{primL-n} constitue une
caract\'erisation des \'etats primaires.
Son poids se calcule ais\'ement par \ref{primLn}
et \ref{Virasoro}:
\beq
L_0
|i,(-k_1,...,-k_N)\! >
=(\Delta_i+k_1+...+k_N)
|i,(-k_1,...,-k_N)\! >.
\eeq
Le poids d'un descendant est donc obtenu par accroissement
d'un entier positif \`a partir du poids $\Delta_i$
de l'\'etat primaire dont il est issu.
C'est pourquoi on appelle aussi ces derniers,
``\'etats de plus haut poids'', par r\'ef\'erence \`a une convention de signe
oppos\'ee\footnote{
On a \'evidemment aussi les descendants
$<\! i,(k_1,...,k_N)|
=
<\! i|
L_{k_N}...L_{k_1}$
pour les \'etats sortants.}.

On a ainsi \`a partir de chaque champ primaire toute
une s\'erie de champs dits secondaires ou descendants.
Ils constituent la famille conforme de $\phi^i$.
L'espace de tous les \'etats engendr\'es par action des $L_{-k}$
sur l'\'etat de plus haut poids $|i\! >$
est appel\'e module de Verma
et not\'e ici ${\cal H}_i$.
C'est une repr\'esentation de l'alg\`ebre de Virasoro.
Elle est caract\'eris\'ee par le poids $\Delta_i$.

\section{Propri\'et\'es des fonctions de corr\'elation}

Un des int\'er\^ets de ce ``classement''
des op\'erateurs en repr\'esentations de l'alg\`ebre
de Virasoro,
est qu'\`a partir des fonctions de corr\'elation
des champs primaires
\beq
<\! 0|
\prod_{i=1}^n
\phi^i(z_i)
|0\! >
\label{corr}
\eeq
on peut obtenir toutes celles des descendants $\phi^{i(\{\nu\})}$.
En effet, \'ecrivant les descendants sous la forme \ref{phi-k}
et commutant les $L_n$ jusqu'\`a une extr\'emit\'e
gr\^ace \`a \ref{[Lnphi]},
on obtient finalement la fonction de corr\'elation des descendants
comme le r\'esulat de l'action d'un op\'erateur diff\'erentiel
sur la fonction de corr\'elation des champs primaires.
Cons\'equence (entre autres): des champs primaires
d\'ecoupl\'es (fonction de corr\'elation nulle)
ont des descendants \'egalement d\'ecoupl\'es,
on peut donc parler de r\`egles de s\'election ou
de fusion (cf note \ref{selec} page \pageref{selec})
pour des familles conformes enti\`eres.

Mais on peut dire encore beaucoup plus sur
les fonctions de corr\'elation entre champs primaires.
En effet, les fonctions de corr\'elation doivent \^etre
invariantes par translation, dilatation et transformations conformes
sp\'eciales, les transformations projectives globales de $Sl(2,C)$.
Ceci peut se voir en commutant leurs g\'en\'erateurs
$L_{-1}$, $L_0$ et $L_1$, qui laissent le vide invariant,
avec tous les op\'erateurs\footnote{
Ces op\'erateurs doivent en fait v\'erifier \ref{[Lnphi]} pour
$n=-1,0,1$ uniquement, ce qui est une contrainte moins forte
que pour tout $n$ (champs primaires).
De tels op\'erateurs sont appel\'es quasi-primaires.
Le tenseur \'energie-impulsion $T$ par exemple est quasi-primaire,
mais il n'est pas primaire,
c'est un descendant d'ordre 2 de la famille conforme
de l'identit\'e.
}
de \ref{corr}.
On obtient ainsi trois \'equations diff\'erentielles
des plus simples qui sont les identit\'es de Ward projectives.
Ceci contraint tr\`es fortement la forme des fonctions de
corr\'elation.
La fonction \`a deux points est de la forme
\beq
<\! \phi^1(z_1,\zb_1)
\phi^2(z_2,\zb_2)\! >
\propto
(z_1-z_2)^{-2\Delta_1}
(\zb_1-\zb_2)^{-2\Deltab_1}
\hbox{ si }
\Delta_1=\Delta_2
\hbox{ et }
\Deltab_1=\Deltab_2
\label{f2p}
\eeq
o\`u la constante de proportionnalit\'e peut \^etre normalis\'ee \`a un.
Sinon, pour des poids diff\'erents, elle est nulle.
La fonction \`a trois points est n\'ecessairement de la forme
\beq
<\! \phi^1(z_1,\zb_1)
\phi^2(z_2,\zb_2)
\phi^3(z_3,\zb_3)\! >
=
C_{1,2,3}
\prod_{1\le i<j\le 3}
(z_i-z_j)^{-\Delta_{ij}}
(\zb_i-\zb_j)^{-\Deltab_{ij}}
\label{f3p}
\eeq
o\`u $\Delta_{1,2}=\Delta_1+\Delta_2-\Delta_3$ et de m\^eme pour les autres
$\Delta_{ij}$.
Les constantes $C_{1,2,3}$ qui d\'ependent des trois op\'erateurs
$\phi^1$, $\phi^2$ et
$\phi^3$ ne peuvent pas \^etre
pr\'edites par des consid\'erations g\'en\'erales.
Elles d\'ependent de la th\'eorie consid\'er\'ee et
sont appel\'ees constantes de structure de la th\'eorie.
Elles constituent en fait la seule
information non-triviale contenue
dans les fonctions \`a trois points.
Les fonctions d'ordre sup\'erieur, \`a $N$ points,
sont de la m\^eme fa\c con le produit:

\noindent
- d'un facteur
``trivial'', c'est-\`a-dire connu gr\^ace aux identit\'es de Ward projectives,
ind\'ependamment de la th\'eorie
($(z_i-z_j)^{-\Delta_{ij}}...$ dans le cas $N=3$).

\noindent
- d'un facteur qui d\'epend de la th\'eorie consid\'er\'ee
($C_{1,2,3}$ pour le fonction \`a 3 points)
et ne peut pas \^etre d\'eduit des propri\'et\'es
conformes g\'en\'erales.
C'est une fonction des $N-3$ quotients anharmoniques
$(z_i-z_j)(z_k-z_l)/(z_i-z_l)(z_k-z_j)$
invariants par transformations projectives
(C'est donc bien une constante pour $N=3$).

C'est pourquoi on fixe souvent trois des points des fonctions
de corr\'elation. Ceci permet
de conna\^\i tre la fonction non triviale des $N-3$ quotients anharmoniques
et donc toute la fonction \`a $N$ points.
On peut le voir autrement:
de la fonction de corr\'elation avec trois points fix\'es
(\`a $0$, $1$ et $\infty$ par exemple, comme on le fait souvent)
on peut d\'eduire la fonction \`a $N$ points quelconques
par les trois transformations projectives qui permettent
de transformer $0$, $1$ et $\infty$ en trois autres points, quels qu'ils
soient.
Et la loi de transformation des fonctions de corr\'elation
pour les transformations projectives globales
est connue par \ref{trphi}, comme pour toute transformation conforme locale.

\section{Repr\'esentations r\'eductibles}

La description des \'etats $|i,(-k_1,...,-k_N)\! >$
par le multi-index $(-k_1,...,-k_N)$ n'est pas unique,
on peut en effet commuter les $L_{-k_i}$ gr\^ace \`a \ref{Virasoro}.
On peut imposer par exemple de d\'ecrire les descendants
par des $k_i$ ordonn\'es, ce qui \'elimine les commutations
possible et les ambigu\"\i t\'es.
On a alors, a priori, une description univoque des \'etats du module de Verma.
Et, dans le cas g\'en\'eral, c'est bien vrai.

Cependant, pour certaines valeurs discr\`etes de $\Delta$,
cela cesse de l'\^etre.
Les \'etat $L_{-2}|i\! >$ et $(L_{-1})^2|i\! >$
par exemple
ont tous les deux pour poids $\Delta_i+2$.
Le nombre d'\'etats de poids $\Delta_i+N$ est encore plus grand.
Il est donc l\'egitime de se demander si ces
\'etats ne pourraient pas \^etre li\'es.
Et en effet, pour certaines valeurs du poids,
c'est le cas.

Pour N=2 par exemple,
on montre d'abord que la combinaison lin\'eaire
\beq
|\chi\! >
=
\left(
L_{-2}+
{3\over
2(2\Delta_i+1)}
(L_{-1})^2
\right)
|i\! >
\label{chinul}
\eeq
a les propri\'et\'es d'un champ primaire: on v\'erifie en effet facilement
gr\^ace \`a l'alg\`ebre de Virasoro qu'il est annihil\'e
par les $L_n$
\beq
L_n|\chi\! >=0
\hbox{ pour }
n\ge 1
\label{chiprim}
\eeq
pourvu que $\Delta_i$ v\'erifie une \'equation du deuxi\`eme degr\'e
dont les solutions sont
\beq
\Delta_i=
{1\over16}
\left(
5-c\pm
\sqrt{(c-1)(c-25)}
\right).
\eeq
Mais pour un descendant comme $|\chi\! >$ qui est d\'ej\`a
orthogonal aux \'etats primaires $<\! j|$ (par Eqs.\ref{chinul} et
\ref{primL-n}), le fait d'\^etre aussi un \'etat
primaire (\ref{chiprim}) le rend \'egalement orthogonal
aux descendants $<\! j|L_{k_N}...L_{k_1}$.
L'\'etat $|\chi\! >$ est donc finalement orthogonal
\`a tous les \'etats, y compris \`a lui-m\^eme.
Il est nul au sens faible.
Toutes les fonctions de corr\'elation contenant l'\'etat $|\chi\! >$
ou le champ correspondant (ou leurs descendants) sont nulles.
$|\chi\! >$ se d\'ecouple de la th\'eorie.
On l'appele le champ nul.

Les descendants issus de $|i\! >$ est donc dans ce cas une
repr\'esentation r\'eductible de l'alg\`ebre de Virasoro.
On obtient la repr\'esentation r\'eduite en effectuant
le quotient
par la relation d'annulation de $\chi$:
$L_{-2}|i\! >=$ $3(2\Delta_i+1)/2L_{-1}^2|i\! >$.
Quand dans la suite nous parlerons de modules de Verma,
il s'agira toujours de l'espace r\'eduit dont ont
ainsi \'et\'e enlev\'es les \'etats de norme nulle.

Nous avons exhib\'e un \'etat nul au niveau $N=2$,
mais il y en a beaucoup d'autres
(pour $N=4$ par exemple on peut rechercher
les combinaisons lin\'eaires de $L_{-4}|i\! >$,
$L_{-3}L_{-1}|i\! >$ et $(L_{-2})^2|i\! >$ qui soient orthogonales
\`a tout \'etat).
Kac en a dress\'e la liste compl\`ete (table de Kac)
\cite{K}:
pour les poids de Kac index\'es par deux entiers positifs $(n,m)$
\beq
\Delta(n,m)
=
{c-1\over24}
+{1\over2}
\left(
{n\over2}
\alpha_+
+{m\over2}
\alpha_-
\right)^2
\label{Kac}
\eeq
avec\footnote{
Attention \`a la variabilit\'e des
conventions concernant $\alpha_\pm$:
leur signe est parfois modifi\'e, $\alpha_-$ et $\alpha_+$
sont parfois \'echang\'es, et dans les travaux les plus anciens \cite{BPZ,DF}
il y a un rapport $1/\sqrt{2}$ avec la convention moderne utilis\'ee ici.}
\beq
\alpha_\pm
=
\sqrt{1-c\over12}
\pm
\sqrt{25-c\over12}
\label{alpha}
\eeq
il existe un vecteur null au niveau $N=nm$ donc de poids
$\Delta(n,m)+nm$.
Le cas consid\'er\'e pr\'ec\'edemment en Eq.\ref{chinul}
\'etait donc le premier cas non trivial:
$(1,2)$ ou $(2,1)$.
Le cas $(1,1)$ est trivial,
c'est l'op\'erateur identit\'e, de poids nul.

Un des grands int\'er\^ets de ces repr\'esentations r\'eductibles
r\'eside dans les \'equations diff\'erentielles
qu'on peut \'ecrire pour toute fonction de corr\'elation
contenant un op\'erateur de ce type $(n,m)$.
En effet la m\^eme fonction de corr\'elation
o\`u le champ primaire $(n,m)$ a \'et\'e remplac\'e
par son descendant nul de poids $\Delta(n,m)+nm$
a pour valeur z\'ero.
Or, comme on l'a vu,
cette derni\`ere fonction de corr\'elation est obtenue
\`a partir de la premi\`ere par application
d'un certain op\'erateur diff\'erentiel.
Sa nullit\'e nous donne donc une \'equation lin\'eaire
aux d\'eriv\'ees partielles d'ordre $nm$, appel\'ee \'equation
de ``bootstrap''.

Sa forme est en g\'en\'eral assez compliqu\'ee \`a cause de l'expression
du vecteur nul et du nombre de variables.
Nous donnons ici l'\'equation de bootstrap pour la fonction
\`a quatre points incluant un champ de type $(1,2)$ ou $(2,1)$.
Nous savons que gr\^ace aux trois transformations projectives
la fonction \`a quatre points peut en fait \^etre r\'eduite
\`a une fonction de $4-3=1$ variable.
L'\'equation aux d\'eriv\'ees partielles peut donc dans ce cas
\^etre transform\'ee en
\'equation diff\'erentielle ordinaire:
$$
\left(
{3\over
2(2\Delta_i+1)}
{d^2
\over
dz^2}
+
\sum_{j=1}^3
\left(
{1\over z-z_j}
{d\over dz}
-
{\Delta_j\over
(z-z_j)^2}
\right)
+
\right.
$$
\beq
\left.
\sum_{1\le k<j\le 3}
{\Delta_i+\Delta_{jk}
\over
(z-z_j)(z-z_k)}
\right)
<\! \phi^i(z)
\phi^1(z_1)
\phi^2(z_2)
\phi^3(z_3)
\! >=0
\label{bootstrap}
\eeq
si $\Delta_i=\Delta(1,2)$ ou $\Delta_i=\Delta(2,1)$.
La solution de cette \'equation est proportionnelle \`a une fonction
hyperg\'eom\'etrique $\,_2F_1$ dont la variable est le quotient
anharmonique $(z-z_1)(z_2-z_3)/(z-z_3)(z_2-z_1)$ qui se r\'eduit
\`a $z/z_2$ si on prend $z_1=0$ et $z_3=\infty$.
Nous y reviendrons dans le prochain chapitre sur la th\'eorie de Liouville.

Il convient de remarquer que nous contredisons ici
l'affirmation g\'en\'erale pr\'ec\'edente selon
laquelle on ne pouvait rien dire sur la partie
non-triviale des fonctions \`a $N$ points.
On sait par exemple ici
gr\^ace \`a l'annulation du vecteur nul
que les fonctions \`a quatre points particuli\`eres incluant
un op\'erateur de type $(1,2)$ ou $(2,1)$
sont soit nulles soit proportionnelles
\`a une fonction hyperg\'eom\'etrique que l'on peut d\'eterminer.

L'application de ces \'equations diff\'erentielles aux r\`egles
de fusion \`a l'ordre dominant (\`a la Wilson) permet
\'egalement de montrer que la fusion de deux op\'erateurs de ce genre
donne encore (uniquement) des op\'erateurs de familles d\'eg\'en\'er\'ees.
Mais nous verrons ceci plus en d\'etail dans les chapitres \ref{p3} et \ref{p4}
o\`u nous l'appliquerons \`a la th\'eorie de Liouville.

\section{Mod\`eles minimaux}

Jusqu'ici nous avons consid\'er\'e une
th\'eorie dont la charge centrale $c$ est quelconque.
Nous avons uniquement sp\'ecifi\'e les poids des repr\'esentations
r\'eductibles.
Mais on sait bien que les th\'eories pour lesquelles
$$
-
{\alpha_-
\over
\alpha_+}
={p\over q}
\quad
p,q \in N,
\quad
\hbox {   donc   }
\quad
c=1-{6(p-q)^2\over pq},
$$
ou mod\`eles minimaux,
ont des propri\'et\'es tr\`es sp\'eciales.
On peut en effet prouver que les repr\'esentations d\'eg\'en\'er\'ees
pr\'ec\'edentes contiennent un nombre infini de vecteurs nuls.
Le nombre infini d'\'equations diff\'erentielles qui en d\'ecoulent
permettent de prouver que la fusion n'engendre qu'un nombre
fini de familles conformes dans la th\'eorie:
les op\'erateurs pr\'ec\'edents de type $(n,m)$ pour $0<n<p$ et $0<m<q$.
L'alg\`ebre est tronqu\'ee.

Les mod\`eles minimaux sont de la plus haute importance car ce sont eux
qui d\'ecrivent les mod\`eles de physique statistique \`a leurs
points critiques.
Comme nous l'avons dit en introduction \`a cette th\`ese,
de nombreuses th\'eories conformes ont une structure sous-jacente de groupe
quantique,
et c'est le cas en particulier pour les mod\`eles minimaux
qui sont caract\'eris\'es par une sym\'etrie $U_q(sl(2))$
avec
$q$ racine de l'unit\'e.

De plus les mod\`eles minimaux avec $q=p+1$ sont les seuls
mod\`eles unitaires \`a $c<1$.
On appelle unitaire une th\'eorie qui poss\`ede un espace
de Hilbert de m\'etrique d\'efinie positive.
Ceci impose de n'avoir que des \'etats primaires
de poids positifs (on peut voir par exemple
que l'\'etat $L_{-1}|\Delta>$ a pour norme carr\'ee $2\Delta$).
Or, pour $c<1$ certains poids de Kac \ref{Kac} peuvent \^etre n\'egatifs.
Dans tout le domaine $c<1$,
seuls les mod\`eles minimaux $q=p+1$ ont une alg\`ebre (tronqu\'ee)
d'op\'erateurs et d'\'etats
de poids tous positifs.

Ils sont \'egalement tr\`es importants en th\'eorie des cordes
o\`u on a pu r\'esoudre de tels mod\`eles coupl\'es \`a la gravit\'e
(donc une surface al\'eatoire ou une corde vibrante),
\`a tous les ordres, gr\^ace aux mod\`eles de matrices
en triangulant la surface.

Nous n'en dirons cependant pas davantage sur ce cas et nous travaillerons
\`a charge centrale quelconque.
D'une part, on pourra donner \`a la charge centrale les valeurs
discr\`etes pr\'ec\'edentes et v\'erifier que l'alg\`ebre des op\'erateurs que
nous
calculons dans le chapitre \ref{p4} se tronque alors naturellement
\`a un nombre fini de familles conformes.
D'autre part, un des buts de cette th\`ese \'etant
d'atteindre le r\'egime de couplage fort de la gravit\'e
($c>1$ pour la mati\`ere et donc $C_{\hbox{\scriptsize Liouville}}<25$),
il est souhaitable de conserver $c$ g\'en\'erique.
Dans le r\'egime de couplage fort, les poids de Kac
deviennent complexes et nous d\'emontrerons dans le
dernier chapitre un th\'eor\`eme de troncature donnant une
sous-alg\`ebre ferm\'ee d'op\'erateurs de poids r\'eels.
Comme celle des mod\`eles minimaux,
cette troncature n'a lieu que pour certaines valeurs
discr\`etes de la charge centrale,
et les deux troncatures pourraient \^etre reli\'ees.

\chapter{THEORIE DE LIOUVILLE: INTRODUCTION}

\markboth{2. Th\'eorie de Liouville: introduction}
{2. Th\'eorie de Liouville: introduction}
\label{p3}

L'action de Liouville d\'ecrit la dynamique
du facteur conforme dans les cordes non critiques.
En effet, en dehors des dimensions critiques
($d=26$ pour la corde bosonique
et $d=10$ pour la supercorde),
l'anomalie de Weyl rend dynamique la m\'etrique bidimensionnelle
qui a \'et\'e introduite classiquement
comme un simple multiplicateur de Lagrange.
Il s'agit donc d'une th\'eorie
de gravitation quantique bidimensionnelle.
Le degr\'e de libert\'e de la m\'etrique
qui devient dynamique est,
dans la jauge conforme,
le facteur conforme dont l'action est celle de Liouville.
Il permet de restaurer l'invariance conforme
de la th\'eorie:
dans le cas bosonique
les fant\^omes contribuent pour $-26$
\`a l'anomalie conforme ou charge centrale,
et pour des champs de mati\`ere de charge centrale $c$
le facteur conforme aura alors une charge centrale\footnote{
Nous d\'esignons toujours par
$c$ la charge centrale de la ``mati\`ere''
et par $C$ celle de la gravit\'e} $C=26-c$.
Bien qu'issue de la quantification des champs
de mati\`ere, la th\'eorie de Liouville peut
\^etre trait\'ee classiquement,
ce que nous faisons ici dans une premi\`ere partie.
Nous pr\'esentons ensuite dans une deuxi\`eme partie
le sch\'ema de quantification canonique de Gervais et Neveu.

\vskip 2mm

La solution classique de l'\'equation de Liouville
admet une d\'ecomposition chirale.
L'\'equation Liouville est alors \'equivalente
\`a une \'equation de Schr\"odinger
pour les composantes chirales.
Son potentiel est le tenseur \'energie-impulsion.
La structure de crochets de Poisson canoniques montre
ensuite que les d\'eriv\'ees logarithmiques
des composantes chirales sont des champs libres
(transformation de B\"acklund).

Le sch\'ema de quantification de Gervais et Neveu
consiste \`a effectuer une quantification
canonique sur ces champs libres.
Les composantes chirales de la m\'etrique sont ensuite reconstruits
\`a partir des op\'erateurs quantiques de champs libres.
On choisit leur constante de normalisation de mani\`ere
\`a ce qu'ils soient des champs conformes primaires de
type (2,1) ou (1,2).
L'\'equation diff\'erentielle de ``bootstrap''
(ou \'equation de Ward li\'ee au d\'ecouplage du vecteur nul) qu'ils
v\'erifient alors, appara\^\i t comme
la version quantique de l'\'equation
de Schr\"odinger classique.
Il est finalement possible de construire \`a partir de ces composantes
chirales un champ physique local de Liouville.
De la m\^eme fa\c con que l'\'equation de Schr\"odinger classique
\'etait \'equivalente \`a l'\'equation de Liouville classique,
l'\'equation de bootstrap (quantique) est \'equivalente
\`a une \'equation de Liouville quantique pour ce champ de Liouville local.

Cet op\'erateur local \ref{expphi} n'est cependant valable
que pour le couplage faible.
Dans le r\'egime de couplage fort,
il reste local mais de poids complexe.
Il existe n\'eanmoins dans ce cas des op\'erateurs physiques
locaux de poids r\'eels,
mais uniquement dans les dimensions sp\'eciales
7, 13, 19 comme nous le verrons dans le chapitre \ref{p5}.

On peut citer deux justifications  de ce
sch\'ema de quantification, obtenues
par Gervais et Schnittger.
Comme cela a \'et\'e \'evoqu\'e pr\'ec\'edemment,
ils ont montr\'e que le champ local obtenu
v\'erifiait bien une \'equation de Liouville quantique \cite{GS3}.
Ceci n\'ecessite cependant un travail assez \'elabor\'e,
puisque quantiquement les exponentielles
du champ de Liouville obtenues \`a partir des familles
d\'eg\'en\'er\'ees de BPZ (ce qui correspond \`a des spins demi-entiers
comme nous le verrons) ont un spectre discret de moments possibles.
Pour obtenir une \'equation de Liouville quantique,
il a donc d'abord fallu construire des exponentielles du champ
de Liouville avec moments continus (donc des spins continus)
pour pouvoir obtenir par d\'erivation le champ lui-m\^eme et
l'\'equation quantique qu'il v\'erifie.
Mais il est beaucoup plus facile
(sans g\'en\'eralisation \`a des spins continus)
de prouver que les exponentielles du champ de Liouville
v\'erifient une \'equation de Hirota quantique,
ce qu'ils avaient fait pr\'ec\'edemment en ref.\cite{GS2}.
On sait en effet que, classiquement, l'\'equation de Liouville
\'ecrite en termes du champ de Liouville $\Phi$
est \'equivalente \`a une \'equation de Hirota
qui s'exprime uniquement en termes de de son
exponentielle $e^{-\Phi/2}$.

Il convient aussi de mentionner
deux approches concurrentes de la quantification \cite{BCGT,OW}.
J.-L. Gervais et J. Schnittger \cite{GS2} ont montr\'e qu'elles
\'etaient \'equivalentes \` a celle pr\'esent\'ee ici.
Elles ont conduit \`a un formalisme un peu plus compliqu\'e
et \`a un moindre avancement de la r\'esolution.

Par ailleurs, le lecteur int\'eress\'e par le cas supersym\'etrique
pourra consulter le travail initial de Babelon en ref.\cite{B0}
et les travaux de Gervais et coll. en refs.\cite{BG,GR}.

\vskip 2mm

La premi\`ere partie de ce chapitre traite de la solution classique,
connue depuis longtemps.
Nous mettons en \'evidence sa sym\'etrie $Sl(2)$
qui donnera naissance \`a la sym\'etrie $U_q(sl(2))$
dans le cas quantique (voir la partie 4.2).
La deuxi\`eme partie pr\'esente la quantification canonique
men\'ee \`a bien par J.-L. Gervais et A. Neveu \cite{GN1,GN2,GN3},
travail poursuivi par J.-L. Gervais seul ou
en collaboration avec A. Bilal, E. Cremmer et B. Rostand \cite{BG,GR,G1,G3,CG}.
Nous terminons par le calcul de $\gamma_{\hbox{\scriptsize string}}$ qui
peut  \^etre calcul\'e dans cette approche,
redonnant la valeur de KPZ.

Le chapitre \ref{p4} traitera de la suite de la r\'esolution (cas quantique)
\`a laquelle j'ai personnellement contribu\'e.

\section{Solution classique}
\label{p3.1}

L'action de Liouville s'\'ecrit en m\'etrique
euclidienne
\beq
S_L[\Phi]=
{1\over 16\pi\gamma}
\int d\sigma d\tau
\left(
{1\over 2} (\partial_\tau \Phi)^2 + {1\over 2} (\partial_\sigma \Phi)^2
 +\mu e^{\displaystyle \Phi}
\right)
{}.
\label{Liou0}
\eeq
Nous nous pla\c cons sur une surface cylindrique sur
laquelle $\sigma$ est p\'eriodique de p\'eriode
$2\pi$.
Ce cylindre peut \^etre une poign\'ee d'une surface de genre
quelconque.
Ce choix n'est possible globalement que dans le cas du tore,
mais nous ne nous pr\'eoccuperons le plus
souvent que des propri\'et\'es locales de la th\'eorie.

Pour une corde bosonique en dimension $d$,
il ressort de l'int\'egrale fonctionnelle de Polyakov
une action de Liouville \ref{Liou0} avec constante de couplage
$\gamma=3/(26-d)$.
Nous pr\'ef\'erons renormer le champ $\Phi$
par $\Phi\to 2\sqrt\gamma\Phi$,
ce qui fait appara\^\i tre la constante de couplage $\gamma$
dans l'exponentielle.
Cette normalisation est \'equivalente \`a la premi\`ere,
mais la solution, dans cette normalisation, a l'avantage d'\^etre
la limite classique de la solution quantique que nous
verrons dans la prochaine partie.
De plus, transformant $\Phi$ en $\Phi-$ln$(\mu)/2\sqrt{\gamma}$,
nous nous pla\c cons en $\mu=1$
(nous verrons plus tard comment on peut retrouver
le cas $\mu$ quelconque),
et obtenons
\footnote{
(\ref{Liou4}) n'est pas la m\^eme fonctionnelle $S_L$ de $\Phi$,
puisque c'est pr\'ecis\'ement
la m\^eme fonctionnelle de $2\sqrt\gamma\Phi-$ln$(\mu)$,
mais on garde la m\^eme notation par souci de simplicit\'e.}
\beq
S_L[\Phi]=
{1\over 4\pi}
\int d\sigma d\tau
\left(
{1\over 2} (\partial_\tau \Phi)^2 + {1\over 2} (\partial_\sigma \Phi)^2
 + e^{\displaystyle 2\sqrt\gamma\Phi}
\right)
{}.
\label{Liou4}
\eeq
On en d\'eduit l'\'equation de Liouville
\beq
\partial_\sigma^2\Phi
+
\partial_\tau^2\Phi
=
2\sqrt\gamma
e^{\displaystyle 2\sqrt{\gamma} \Phi}
\label{eqLiou}.
\eeq
Dans le syst\`eme de coordonn\'ees
$$
\sigma_\pm=\sigma
\pm i\tau ,
$$
on a l'\'equation \'equivalente
\beq
4\partial_\sp\partial_\sm\Phi+2\sqrt\gamma
e^{\displaystyle 2\sqrt{\gamma} \Phi}
=0
\label{eqLiou2}.
\eeq
On \'ecrit habituellement sa solution bien connue
sous la forme
\beq
\Phi(\sp,\sm)=
{1\over 2\sqrt\gamma}
\ln
\left(
{2A'(\sp)B'(\sm)
\over
\gamma(A(\sp)-B(\sm))^2}
\right)
\label{solLiou}
\eeq
o\`u $A$ et $B$ sont des fonctions quelconques d'une seule variable,
et o\`u $A'$ et $B'$ d\'esignent leurs d\'eriv\'ees.
Ce n'est cependant pas la forme la plus adapt\'ee \`a
la quantification que nous voulons effectuer prochainement.
Nous la transformons donc en \'ecrivant
$$
f_1(\sp)={1\over\sqrt{A'(\sp)}},
f_2(\sp)={A(\sp)\over\sqrt{A'(\sp)}},
\hbox{ ou r\'eciproquement }
A(\sp)={f_1(\sp)\over f_2(\sp)}
$$
et de m\^eme pour $B$ et $\fb_{1,2}$.
Par cette transformation,
le choix d'une fonction $A$ quelconque est \'equivalent
au choix de deux fonctions $f_1$ et $f_2$ soumises \`a la
seule contrainte d'avoir un Wronskien constant (que nous
normalisons \`a 1):
\beq
f_1f'_2-f_2f'_1=1
\quad
\hbox{ et }
\quad
\fb_1\fb'_2-\fb_2\fb'_1
=1.
\label{wronsk}
\eeq
La solution de l'\'equation de Liouville s'\'ecrit
alors  en termes de ces
fonctions:
\beq
e^{\displaystyle -\sqrt{\gamma} \Phi}
=
i\sqrt\gamma
\left(
f_1(\sp)\fb_2(\sm)
-
f_2(\sp)\fb_1(\sm)
\right)
/2
\label{solLiou2}
\eeq
o\`u les quatre fonctions $f_{1,2},\fb_{1,2}$
sont quelconques, soumises \`a la seule contrainte \ref{wronsk}.

\vskip 2mm

On peut \'egalement obtenir \ref{solLiou2} directement en montrant
que l'\'equation de Liouville \ref{eqLiou2} a pour cons\'equence
les deux \'equations aux d\'eriv\'ees partielles
dans les variables $\sp$ et $\sm$
\beq
\left(
-{\partial^2\over\partial{\sp}^2}+T(\sp)
\right)
e^{\displaystyle -\sqrt{\gamma} \Phi(\sp,\sm)}
=
0
=
\left(
-{\partial^2\over\partial{\sm}^2}+\Tb(\sm)
\right)
e^{\displaystyle -\sqrt{\gamma} \Phi(\sp,\sm)}
\label{Schrod}
\eeq
pour des potentiels $T,\Tb$ quelconques.
On peut, \`a partir de ces \'equations aux
d\'eriv\'ees partielles, \'ecrire deux \'equations de Schr\"odinger
pour des fonctions d'une variable
(de solutions $f_1$ et $f_2$ pour la premi\`ere,
$\fb_{1}$ et $\fb_{2}$ pour la deuxi\`eme)\footnote{
Pour deux fonctions $f_{1}$ et $f_{2}$ quelconques,
il existe toujours une \'equation diff\'erentielle
\`a coefficients non constants dont elles sont solutions.
Les coefficients d\'ependent \'evidemment de $f_{1,2}$,
et celui du terme d'ordre 1 est proportionnel \`a la d\'eriv\'ee
du Wronskien de $f_1$ et $f_2$.
Ici, se donnant deux fonctions $f_{1,2}$ quelconques,
dire qu'il existe un potentiel $T$ pour lequel elles v\'erifient une \'equation
de Schr\"odinger (\ref{Schrod2}),
revient donc uniquement \`a dire qu'elles ont un Wronskien constant
(\ref{wronsk}).
Ceci prouve l'\'equivalence de la solution ainsi obtenue
avec la pr\'ec\'edente.
Et ce potentiel $T$ est fonction de $f_1$ et $f_2$
(c'est tout simplement $f''_1/f_1=f''_2/f_2$).
}
\beq
\left(
-{d^2\over d{\sp}^2}+T(\sp)
\right)
f_{1,2}(\sp)
=
0
=
\left(
-{d^2\over d{\sm}^2}+\Tb(\sm)
\right)
\fb_{1,2}(\sm)
\label{Schrod2}
\eeq
et on conclut que $e^{-\Phi/2}$ est une combinaison lin\'eaire de produits
de $f_{1,2}$ et $\fb_{1,2}$, ce qui donne \ref{solLiou2}.

D'apr\`es la note pr\'ec\'edente on sait que ces potentiels $T(\sp)$ et
$\Tb(\sm)$
sont fonction de $f_{1,2}$ et $\fb_{1,2}$
(et de leurs d\'eriv\'ees), ou du champ $\Phi$,
et le calcul montre qu'ils ne sont pas autre chose que
les deux composantes chirales du tenseur
\'energie-impulsion.
Partant des crochets canoniques entre le champ $\Phi$
et son champ conjugu\'e
on peut ensuite calculer ceux de $T(\sp)$ et $\Tb(\sm)$.
On obtient
\beq
\left\{
T(\sp),T(\sp')
\right\}_{\tau=\tau'}
=
-2\pi\gamma
\delta'''(\sp-\sp')
+4\pi\gamma(\partial_{\sp}-\partial_{\sp'})
\left(
T(\sp)\delta(\sp-\sp')
\right).
%
%
\eeq
En d\'ecomposant le potentiel/tenseur \'energie-impulsion
$T$ p\'eriodique (c'est une fonction du champ
physique $\Phi$ lui-m\^eme p\'eriodique) en s\'erie de Fourier\footnote{
Attention, nous sommes ici sur le cylindre
alors que \ref{TdeLn} dans le chapitre
pr\'ec\'edent a \'et\'e \'ecrit sur le plan complexe.}:
\beq
L_m=
{1\over 4\pi\gamma}
\int_0^{2\pi}
d\sigma T(\sigma) e^{im\sigma}
+{1\over 8 \gamma} \delta_{m,0},
\eeq
on obtient de mani\`ere \'equivalente
\beq
i
\{
L_m,L_n
\}
=
(m-n)
L_{m+n}
+C
{m^3-m\over 12}
\delta_{m,-n}
\ ,\ C={3\over\gamma}.
\eeq
On voit donc appara\^\i tre l'alg\`ebre de Virasoro avec une charge
centrale $C$, d\'ej\`a au niveau classique.

On montre ensuite que les d\'eriv\'ees logarithmiques des champs $f_{1}$
et $f_{2}$ sont des champs libres:
\beq
P^{(1)}(\sp)\equiv f'_1(\sp)/\sqrt{\gamma}f_1(\sp)
\ ,\
P^{(2)}(\sp)\equiv f'_2(\sp)/\sqrt{\gamma}f_2(\sp)
\label{derlog}
\eeq
ont des crochets de Poisson de champs libres:
$$
\{P^{(i)}(\sp),P^{(i)}(\sp')\}_{\tau=\tau'}
=
2\pi
\delta'(\sp-\sp')
\hbox{ pour }i=1 \hbox{ ou } 2.
$$
Les champs $f_{1,2}$,
solutions d'une \'equation diff\'erentielle \`a potentiel p\'eriodique $T$
ont a priori une monodromie non triviale,
qu'on choisit ici de diagonaliser (voir la discussion
de la monodromie dans le chapitre suivant et en particulier la
note \ref{notemonodr} page \pageref{notemonodr}).
Ils sont donc p\'eriodiques \`a une constante multiplicative pr\`es,
et leurs d\'eriv\'ees logarithmiques $P^{(1,2)}$ sont p\'eriodiques.
On peut alors les d\'ecomposer en s\'erie de Fourier
\beq
P^{(i)}(\sigma)=\sum_{n=-\infty}^{\infty} e^{-in\sigma}p^{(i)}_n
\ ,\ i=1,2.
\eeq
On en d\'eduit les crochets de Poisson
\beq
i\{p^{(i)}_m,p^{(i)}_n\}=m\delta_{m,-n}
\ ,\ i=1,2.
\label{comutpn}
\eeq
Les champs $f_{1,2}$ sont soumis \`a une \'equation de Schr\"odinger
\ref{Schrod},
on en d\'eduit que leurs d\'eriv\'ees logarithmiques $P^{(1,2)}$
v\'erifient une \'equation de Ricatti:
\beq
T(\sp)=\gamma \left[ P^{(1)}(\sp)\right] ^2+
\sqrt{\gamma}\left[P^{(1)}(\sp)\right]'
=\gamma\left[ P^{(2)}(\sp)\right] ^2+
\sqrt{\gamma}\left[ P^{(2)}(\sp)\right]'
\label{Ricatti}.
\eeq

\vskip 4mm

Nous devons ici souligner que,
bien que champs libres tous les deux, $P^{(1)}$ et $P^{(2)}$ ne sont pas
ind\'ependants\footnote{
Les crochets de Poisson de $P^{(1)}$ avec $P^{(2)}$
sont compliqu\'es,
contrairement \`a ceux de leurs exponentielles
$f_{1,2}$,
qui ont une expression simple et ferm\'ee
(i.e. en termes de $f_{1}$ et $f_{2}$ elles-m\^eme):
elle est donn\'ee par le deuxi\`eme ordre du d\'eveloppement
classique de la matrice d'\'echange quantique.}.
Il sont li\'ees par l'\'equation de Ricatti \ref{Ricatti}
dont ils sont les deux solutions.
Nous rappelons que cette \'equation de Ricatti provient
de l'\'equation de l'\'equation de Schr\"odinger
qui est elle-m\^eme \'equivalente \`a l'\'equation de Liouville.
L'existence de ces deux champs libres
est une caract\'eristique fondamentale de la th\'eorie.
Ils sont en revanche
ind\'ependants des champs $\Pb^{(1,2)}$ d\'eriv\'es
parall\`element de $\fb_{1,2}$.

\vskip 2mm

L'\'equation de Ricatti \ref{Ricatti} permet de calculer
les coefficients de Fourier de $T$ en fonction des modes de $P^{(1,2)}$:
\beq
L_n=(\sum_m p^{(i)}_m p^{(i)}_{n-m} -inp^{(i)}_n/\sqrt\gamma)/2+\delta_{n,0}/8
\label{Lndepn}
\eeq
pour $i=1$ ou 2, de mani\`ere \'equivalente.

\vskip 2mm

On a maintenant tous les outils pour, remontant aux champs
$f_{1,2}$, montrer que ce sont des champs primaires (classiques)
pour l'alg\`ebre conforme (classique).
De \ref{derlog} on d\'eduit
\beq
f_i(\sp)=
d_i
e^{\sqrt\gamma\left(q_0^{(i)}+p_0^{(i)}\sp
+i\sum_{n\ne 0}e^{-in\sp}p_n^{(i)}/n\right)}
\label{psi}
\eeq
o\`u $d_ie^{\sqrt\gamma q_0^{(i)}}$ est la constante d'int\'egration:
$d_i$ ne d\'epend que de $p_0^{(i)}$,
et $q_0^{(i)}$ est la position du centre de masse de
la corde (classique).
Il n'est donc pas \'etonnant de poser
\beq
\left\{
q_0^{(i)},p_0^{(i)}
\right\}
=1.
\label{q0p0}
\eeq
Ceci donne pr\'ecis\'ement
des champs $f_{1,2}$ primaires,
de poids $-1/2$:
\`a l'aide des \'equations \ref{comutpn}, \ref{Lndepn},
on montre que
\beq
i
\left\{
L_n,
f_i(\sp)
\right\}
=
e^{in\sp}
\left(
-id_\sp+n\Delta
\right)
f_i(\sp)
\ ,\ \Delta=-1/2.
\label{Lnpsi}
\eeq
C'est l'\'equivalent classique de la relation de commutation \ref{[Lnphi]}
qui caract\'erise les champs primaires quantiques.
Le champ $e^{-\Phi/2}$ est donc de poids $(-1/2,-1/2)$.
Pour cette th\'eorie classique on peut en d\'eduire directement
que le terme cosmologique $e^\Phi$ est
de poids $(1,1)$.
Ceci fait de $e^\Phi d\sigma_+ d\sigma_-$
un invariant pour les transformations conformes.
C'est n\'ecessaire pour pouvoir l'int\'egrer
et l'inclure dans une action invariante conforme.
Il sera fondamental au niveau quantique de chercher
\`a obtenir des champs primaires (de poids modifi\'e)
qui permettent de construire un terme cosmologique de poids $(1,1)$,
pour pouvoir l'int\'egrer sur la surface d'univers.

\vskip 5mm

Nous terminons cette partie sur la solution
classique en exhibant sa sym\'etrie $Sl(2)$.
Elle sera transform\'ee au niveau quantique
en une
sym\'etrie $U_q(sl(2))$.
Cette sym\'etrie de groupe quantique sera de la plus
haute importance pour la r\'esolution de la th\'eorie.

L'\'equation de Ricatti \ref{Ricatti} est l'\'equivalent en termes
de $P^{(1,2)}$ de l'\'equation de Schr\"odinger \ref{Schrod} en termes
de $f_{1,2}$.
On passe de l'une \`a l'autre par la transformation \ref{derlog}
de $f_{1,2}$ en $P^{(1,2)}$.
Elles ont toutes les deux la propri\'et\'e tr\`es importante
d'avoir deux solutions ind\'ependantes.
Ces deux solutions $f_1$ et $f_2$ sont ici trait\'ees de fa\c con
sym\'etrique.
C'est un point cl\'e qui distingue cette approche en particulier de celle
du gaz de Coulomb o\`u un des deux champs libres
est privil\'egi\'e (cf \cite{DF} par exemple).
On voit ainsi appara\^\i tre une invariance $Sl(2)$ de la solution
\ref{solLiou2}
qui peut \^etre \'ecrite sch\'ematiquement
$$
e^{\displaystyle -\sqrt{\gamma} \Phi}
\propto
{\scriptstyle (\fb_1\ \fb_2)}
\left(
^{0\ -1}
_{1\ \ 0}
\right)
\left(
_{f_1}
^{f_2}
\right)
$$
et n'est donc pas modifi\'ee par transformation $Sl(2)$
simultan\'ee de $f_{1,2}$ et $\fb_{1,2}$ gr\^ace \`a
l'identit\'e
$$
M^t
\left(
^{0\ -1}
_{1\ \ 0}
\right)
M
=
\left(
^{0\ -1}
_{1\ \ 0}
\right)
\hbox{ pour }
M
\in
Sl(2).
$$
En \'elevant \`a la puissance $2J$ l'\'equation \ref{solLiou2}
on obtient aussi des repr\'esentations de spin $J$ de $Sl(2)$,
combin\'ees en $e^{-2J\sqrt\gamma\Phi}$, singlet de $Sl(2)$:
\beq
e^{\displaystyle -2J\sqrt{\gamma} \Phi}
\propto
\sum_{M=-J}^J (-1)^{J-M}
f_M^{(J)}(\sp) \fb_{M}^{(J)}(\sm)
\label{expphic}
\eeq
avec
\beq
f_M^{(J)}\equiv \sqrt { \textstyle {2J\choose J+M}}
\left (f_1\right )^{J-M} \left(f_2\right )^{J+M},  \quad
\overline f_M^{(J)}\equiv\sqrt { \textstyle {2J\choose J+M}}
\left (\overline f_1\right )^{J+M}
\left(\overline f_2\right )^{J-M}.
\eeq
Ceci peut \^etre \'etendu \`a des spins $J$ n\'egatifs en continuant
les coefficients du binome et en \'etendant cependant la somme
\`a un domaine infini ($M>J$ ou $M<J$).
On peut ainsi \'ecrire le terme cosmologique $e^{2\sqrt\gamma\Phi}$ lui-m\^eme
comme
singlet obtenu \`a partir d'une repr\'esentation (infinie)
de $Sl(2)$.

\vskip 3cm

\section{Quantification}
\label{p3.2}

On proc\`ede \`a la quantification canonique en rempla\c cant
les crochets de Poisson par des commutateurs (fois $-i$).
On le fait au niveau des deux champs libres $P^{(1)}$
et $P^{(2)}$:
\beq
[p^{(i)}_n,p^{(i)}_m]
=n\gamma\delta_{n,-m}
\qquad,\qquad
i=1,2
\eeq
\`a partir desquels nous allons ensuite reconstruire
les op\'erateurs chiraux quantiques
(et de m\^eme
pour l'autre chiralit\'e que nous laissons donc de c\^ot\'e pour le moment).
Il y a un ordre normal par rapport aux modes $p_n^{(1)}$
et un autre par rapport aux $p_n^{(2)}$.
Le lien \ref{Lndepn} entre les $L_n$ et les $p_n^{(i)}$
est modifi\'e par l'ordre normal en:
\beq
L_n=\ :(\sum_m p^{(i)}_m p^{(i)}_{n-m}
-inp^{(i)}_n/\sqrt\gamma):/2+\delta_{n,0}/8
\qquad,\qquad
i=1,2
\label{Lndepnq}
\eeq
ce qui donne une alg\`ebre de Virasoro quantique
avec une charge centrale modifi\'ee
\beq
C=1+3/\gamma.
\eeq
Pour une charge centrale de la mati\`ere \'egale \`a $c$,
l'annulation de la charge centrale totale indique
que la constante de couplage vaut $\gamma=3/(25-d)$.
La quantification du champ de Liouville am\`ene
donc une renormalisation de la constante de couplage
puisqu'on avait classiquement $\gamma=3/(26-d)$.
Ceci sera retrouv\'e plus tard par DDK \cite{Dav,DK}
pour la  quantification par int\'egrale fonctionnelle,
la renormalisation de $\gamma$ apparaissant comme le
r\'esultat du traitement correct de la mesure
d'int\'egration sur $\Phi$.

Il faut ensuite reconstruire \`a partir
de leurs d\'eriv\'ees logarithmiques $P^{(1,2)}$
les composantes chirales
que nous notons $\psi_{1,2}$ dans le cas quantique.
On prend pour d\'efinition de $\psi_{1,2}$ la version
quantique de \ref{psi}:
\beq
\psi_i(\sigma)=
d_i
e^{\sqrt{h/2\pi}q_0^{(i)}}
\> : \!
e^{\sqrt{h/2\pi}(p_0^{(i)}\sigma
+i\sum_{n\ne 0}e^{-in\sigma}p_n^{(i)}/n)}
\!:\
\qquad,\qquad
i=1,2.
\label{psiq}
\eeq
L'int\'er\^et d'introduire la constante devant le champ libre sous la forme
$\sqrt{h/2\pi}$ appara\^\i tra plus tard.
Cette constante \'etait $\sqrt{\gamma}$
au niveau classique.
Au niveau quantique, elle peut \^etre modifi\'ee,
et on suppose donc pour le moment $h$ quelconque,
non reli\'ee \`a la constante de couplage $\gamma$ ou \`a la charge centrale.
Le crochet de $q_0$ avec $p_0$ d\'eduit de \ref{q0p0}
\beq
[q_0^{(i)},p_n^{(i)}]
=
i\delta_{n,0}
\qquad,\qquad
i=1,2
\eeq
fait de $e^{...q_0}$ un op\'erateur d'accroissement de $p_0$.
Ceci permet de prouver pour $\psi_{1,2}$ les relations de commutation
\ref{[Lnphi]} montrant que ce sont
des champs primaires
(quantiques) de poids
\beq
-{1\over 2}(\sqrt{h\over 2\pi\gamma}+{h\over 2\pi}).
\label{wdeh}
\eeq

Ces op\'erateurs engendrent par
d\'eveloppement \`a courte distance
toute une alg\`ebre d'op\'erateurs.
Comme nous l'avons d\'ej\`a signal\'e dans le cas classique,
cette alg\`ebre doit contenir (au moins) un op\'erateur marginal de poids (1,1)
pour qu'il puisse \^etre int\'egr\'e sur la surface-univers
(terme cosmologique).
De plus on ne sait pas, tout au moins dans cette approche,
calculer le d\'eveloppement \`a courte distance
d'op\'erateurs
g\'en\'eriques.
Tout ceci nous conduit \`a sp\'ecifier $h$:
il appara\^\i t que pour $h$ solution de
\beq
{h\over\pi}
+
{\pi\over h}
=
{1\over 2\gamma}
-2
=
{C-13\over 6}
\label{eqh}
\eeq
les champs $\psi_{1,2}$ (ou plus pr\'ecis\'ement
les fonctions \`a quatre points dont un des op\'erateurs
de vertex est un champ $\psi_{1,2}$ avec un tel $h$)
v\'erifient une \'equation lin\'eaire du deuxi\`eme ordre.
Ceci permet de calculer l'alg\`ebre de ces op\'erateurs,
au sens traditionnel
de la th\'eorie quantique des champs (d\'eveloppement \`a courte distance),
et mieux, au sens de Moore et Seiberg que nous verrons
dans le chapitre suivant
(fusion et \'echange, exacts ``\`a toute distance'').
Il sera ensuite possible, chose fondamentale,
de construire un op\'erateur de poids (1,1)
\`a partir de ces $\psi_{1,2}$
(moyennant l'introduction de spins n\'egatifs, cf plus loin).

Ces op\'erateurs particuliers, d\'efinis par un $h$ particulier
ou de mani\`ere \'equivalente par un poids conforme bien particulier,
ne sont en fait pas autre chose que ceux de la table de Kac.
L'\'equation du deuxi\`eme degr\'e \ref{eqh} a deux solutions
\beq
h_\pm=
{\pi\over12}
(C-13\pm\sqrt{(C-25)(C-1)}
=
{\pi\over4\gamma}
(1-4\gamma\pm\sqrt{1-8\gamma})
\label{hpm}
\eeq
qui donnent respectivement pour poids \ref{wdeh}
les poids de Kac $\Delta(1,2)$ et $\Delta(2,1)$,
donc les
op\'erateurs d\'eg\'en\'er\'es\footnote{
Ou de fa\c con plus juste dans une repr\'esentation
d\'eg\'en\'er\'ee de l'alg\`ebre de Virasoro.}
$(1,2)$\footnote{
La notation ne doit pas amener de confusion entre les poids
conformes $(\Delta,\overline \Delta)$ r\'eels
et les entiers $(n,m)$ num\'erotant les repr\'esentations
d\'eg\'en\'er\'ees de l'alg\`ebre conforme dans BPZ.}
et $(2,1)$
dans la classification de BPZ \cite{BPZ},
cf partie \ref{p2.1}.4 page \pageref{chinul}.
Nous les noterons plus souvent
\beq
h\equiv h_-
\quad , \quad
\hhat\equiv h_+
\eeq
et d\'enoterons \'egalement par un chapeau tout
ce qui a trait \`a $\hhat$
(comme les familles d\'eg\'en\'er\'ees $(n,1)$)
et sans chapeau ce qui est li\'e \`a $h$
(op\'erateurs $(1,n)$).
Nous verrons dans le chapitre 4 que la sym\'etrie $Sl(2)$
mise en \'evidence pour la solution classique
est d\'eform\'ee dans le cas quantique
en une double sym\'etrie de groupe quantique
$U_q(sl(2))\odot U_{\qhat}(sl(2))$.
Les param\`etres de d\'eformation de ces groupes quantiques
seront $h$ et $\hhat$ qui sont reli\'es aux param\`etres
\'equivalents $q$ et $\qhat$ par $q=e^{ih}$ et $\qhat=e^{i\hhat}$.
Ils sont reli\'es aux param\`etres $\alpha_\pm$ des poids de Kac
par
\beq
h=
{\alpha_-^2
\over
2}
=
{\alpha_-
\over
\alpha_+}
\qquad,\qquad
\hhat=
{\alpha_+^2
\over
2}
=
{\alpha_+
\over
\alpha_-}
\eeq
et v\'erifient les propri\'et\'es suivantes que nous
utiliserons souvent
\beq
h\hhat=
\pi^2
\
\left(
\hbox{ou }
{\pi\over h}
=
{\hhat\over\pi}
\right)
\qquad
\hbox{et}
\qquad
{h\over\pi}
+
{\hhat\over\pi}
=
{C-13\over 6}
{}.
\eeq

Calculant d'abord les fonctions \`a quatre points contenant
(au moins) un op\'erateur de type $(1,2)$ ou $(2,1)$,
puis le lien entre ces fonctions,
on trouve comme BPZ que le d\'eveloppement \`a courte
distance (et m\^eme la fusion ou l'\'echange) de tels op\'erateurs
de type $(n,m)$ est ferm\'ee.
On peut aussi calculer l'alg\`ebre de ces op\'erateurs avec les op\'erateurs
les plus g\'en\'eraux (hors table de Kac),
nous verrons ceci en d\'etail dans le chapitre 4.
Poursuivant le parall\`ele avec l'approche
de BPZ par les th\'eories conformes,
on voit que l'\'equation diff\'erentielle en question
n'est pas autre chose que l'identit\'e de Ward qui exprime
le d\'ecouplage du vecteur nul de niveau 2 de la repr\'esentation
d\'eg\'en\'er\'ee $(1,2)$ ou $(2,1)$ (cf Eq.\ref{bootstrap} ds la partie 2.1).
On peut \'egalement obtenir les \'equations diff\'erentielles
de degr\'e plus \'elev\'e correpondant aux familles suivantes.

On peut se demander quels sont les int\'er\^ets de cette approche,
par rapport \`a celle purement conforme de BPZ.
Ils sont multiples.
Son premier m\'erite est de fournir une r\'ealisation explicite
des op\'erateurs d\'eg\'en\'er\'es,
de leur alg\`ebre conforme (avec les $L_n$)
et m\^eme de l'\'equation diff\'erentielle.
Tout ceci a en effet \'et\'e v\'erifi\'e de fa\c con
tout \`a fait explicite par J.-L. Gervais et A. Neveu\cite{GN3}\footnote{
Attention au changement de notations.
Avant la ref.\cite{G1} (o\`u on trouvera la correspondance, p.261, note 2),
et en particulier dans \cite{GN1,GN2,GN3},
$\gamma$ est not\'e $\hbar$,
$h_\pm$ correspond \`a $h\eta_\pm^2$...},
en partant de l'expression des
$\psi$ et des $L_n$ en termes des champs libres,
d\`es 1983 parall\`element aux travaux de BPZ.
Son deuxi\`eme m\'erite, et non des moindres,
est de faire appara\^\i tre pour chaque famille de BPZ,
pour la famille $(1,2)$ par exemple,
les deux op\'erateurs de la th\'eorie $\psi_1$ et $\psi_2$
et de les traiter sur un pied d'\'egalit\'e\footnote{
Il faut bien distinguer les trois duplications (ou sym\'etries) successives.
La premi\`ere est la d\'ecomposition chirale on ne peut plus
habituelle entre les modes gauches qui donnent $\psi(\sp)$
et les modes droits qui donnent $\psib(\sm)$.
Nous ne consid\'erons souvent qu'une chiralit\'e,
l'autre \'etant identique.
La deuxi\`eme correspond aux deux charges d'\'ecran
$\alpha_\pm$ de BPZ \cite{BPZ} ou de ref.\cite{DF},
ou de mani\`ere \'equivalente ici aux deux param\`etres
$h_\pm=\pi\alpha_\pm^2/2$.
Elle provient des deux solutions de l'\'equation du deuxi\`eme
degr\'e \ref{eqh} pour $h_\pm$ ou de l'\'equation
$-\alpha(\alpha+\sqrt{(1-C)/3})/2=1$ pour $\alpha_\pm$.
Elle engendre les familles $(1,2)$ et $(2,1)$ de BPZ
puis les suivantes $(n,m)$ par fusion.
La troisi\`eme sym\'etrie discut\'ee ici entre $\psi_1$ et $\psi_2$
est moins connue (car bris\'ee ailleurs)
et n'a rien \`a voir avec la deuxi\`eme:
comme on a des $\psi_1$ et $\psi_2$ pour $h_-$ donc la famille $(1,2)$,
on a des champs similaires not\'es $\psihat_1$ et $\psihat_2$ pour $h_+$
c'est-\`a-dire la famille $(2,1)$
(et des $\psib_1$ et $\psib_2$, $\psibhat_1$
et $\psibhat_2$ pour l'autre chiralit\'e)
{}.
}.
L'origine de cette duplication r\'eside dans les \'equations
dont nous examinons les solutions:
au niveau classique, les \'equations de Schr\"odinger ou Ricatti
avaient deux solutions, d'o\`u les deux champs libres $P^{(1,2)}$
et les deux champs $\psi_{1,2}$.
Au niveau quantique, l'\'equation de Ward, \'equivalent
quantique de l'\'equation de Schr\"odinger,
a aussi un espace des solutions de dimension deux
(les blocs conformes en langage ``conforme'')
qui correspondent chacune \`a l'insertion d'un op\'erateur
diff\'erent (en langage ``op\'erateur de vertex'').
Dans l'autre grande approche qui a men\'e \`a une construction
explicite des op\'erateurs, celle du gaz de Coulomb \cite{DF},
la sym\'etrie entre les deux champs libres est bris\'ee:
on en introduit un seul qui engendre de premiers op\'erateurs
sur lesquels il faut ensuite agir
par des op\'erateurs d'\'ecran (dits de ``screening''
en anglais) pour
obtenir tous ceux de la th\'eorie,
faute de quoi on se limiterait \`a un th\'eorie triviale avec
des fonctions de corr\'elation presque toutes nulles\footnote{
Il n'y a aucune contradiction \`a pouvoir l\`a-bas faire
une construction \`a partir d'un seul champ libre
et ici une construction \'equivalente \`a partir
de deux champs libres,
puisqu'en fait, comme on l'a r\'ep\'et\'e,
ils sont reli\'es
(par \ref{Lndepnq} pour leurs modes au niveau quantique).}.
Ici, en revanche, les op\'erateurs $\psi_1$ et $\psi_2$
engendrent par fusion tous les op\'erateurs de la th\'eorie,
et la sym\'etrie entre eux deux est conserv\'ee en permanence.
Comme nous l'avons d\'ej\`a signal\'e dans le cas classique pour $Sl(2)$,
ceci est \`a l'origine de la sym\'etrie (interne) $U_q(sl(2))$
de la th\'eorie qui s'av\'erera d'une richesse extr\^eme
ainsi que d'une grande aide pour la r\'esoudre.
Cette sym\'etrie n'appara\^\i t que difficilement dans l'approche
traditionnelle du gaz de Coulomb
fond\'ee sur la repr\'esentation int\'egrale des fonctions de Green.
Elle a n\'eanmoins pu
\^etre explicit\'ee en refs.\cite{GS1,GS3}
dans une approche op\'eratorielle du gaz de Coulomb en s'inspirant de la
construction
pr\'esente.

Apr\`es toutes ces consid\'erations g\'en\'erales,
rentrons un peu plus dans les d\'etails en cherchant
toutefois \`a ne pas \^etre trop technique.
La solution classique et la structure de crochets de Poisson
sont \'etudi\'ees en ref.\cite{GN1},
la quantification en ref.\cite{GN2}
et l'\'equation diff\'erentielle est obtenue en ref.\cite{GN3}.
Le lecteur int\'eress\'e pourra s'y r\'ef\'erer,
et nous ne la r\'e\'ecrivons pas car ce n'est pas autre chose que
l'\'equation de bootstrap \ref{bootstrap}.
La (l\'eg\`ere) diff\'erence dans leurs expressions est due \`a
l'utilisation des coordonn\'ees $\sigma_\pm$ sur le cylindre
en ref.\cite{GN3} et ici, et des coordonn\'ees $z,\zb$
du plan complexe dans le premier chapitre.
Nous donnons juste la solution:
$$
< \varpi_1 | \, V^{(1/2)}_{\pm 1/2}(x)\, A_\Delta(x')\,
 | \varpi_4> =
e^{ix'(\varpi_4^2-\varpi_1^2)h/4\pi}e^{i(x'-x)
(-1/2\mp\varpi_1)h/2\pi }
$$
\beq
\times \bigl(1-e^{-i(x-x')}\bigr)^\beta \,
 F(a_\pm ,b_\pm ;c_\pm ;e^{i(x'-x)}),
\label{f4p}
\eeq
$$
a_\pm =\beta -{h\over 2\pi}\pm
h({\varpi_4-\varpi_1 \over 2\pi});\quad
b_\pm =\beta  -{h\over 2\pi}\mp h({\varpi_4+\varpi_1 \over
2\pi});\quad c_\pm =1\mp{h\varpi_1 \over \pi};
$$
\beq
\beta ={1\over 2} (1+h/\pi)\bigl(1-
\sqrt{1-{8h\Delta  \over 2\pi(1+h/\pi)^2}}\bigr);
\label{beta}
\eeq
qui n\'ecessite quelques explications.
Tout d'abord
les op\'erateurs $V^{(1/2)}_{\pm 1/2}$
ne sont pas autre chose que $\psi_1\equiv V^{(1/2)}_{-1/2}$
et $\psi_2\equiv V^{(1/2)}_{1/2}$
renomm\'es pour entrer dans le cadre de l'indexation
par des spins des familles suivantes que nous construirons prochainement.
Ensuite, $F(a,b;c;z)$ est la fonction hypergeometrique standard.
$A_\Delta$ d\'esigne un op\'erateur primaire quelconque
de poids $\Delta$, seules ses propri\'et\'es conformes sont
utilis\'ees, il n'est pas n\'ecessaire de savoir le construire
explicitement en fonction des $p_n$.
Les \'etats $< \varpi_4 |$ et
$| \varpi_1>$ sont les \'etats de plus haut poids
de modules de Verma, habituellement caract\'eris\'es
par leur poids conforme $\Delta$,
mais d\'esign\'es ici de mani\`ere (presque) \'equivalente
par leur mode z\'ero $\varpi$.
$\varpi$ n'est pas autre chose que le moment $p_0$
renorm\'e pour \'eliminer les racines carr\'ees
de la plupart des formules\footnote{
Le choix de $p_0^{(1)}$ dans cette
formule semble introduire une notation asym\'etrique
entre $P^{(1)}$ et $P^{(2)}$ mais \c ca n'est pas vraiment
le cas car la contrainte d'unimodularit\'e de la matrice
de monodromie impose des valeurs propres
$e^{\sqrt{h/2\pi}p_0^{(1)}}$ et $e^{\sqrt{h/2\pi}p_0^{(2)}}$
inverses l'une de l'autre et donc
$p_0^{(1)}=-p_0^{(2)}$.}:
\beq
\varpi=
i\sqrt{2\pi\over h} p_0^{(1)}.
\label{wdep}
\eeq
Le moment d\'etermine bien le plus haut poids du module de Verma,
car de \ref{Lndepnq} appliqu\'ee \`a $n=0$,
on obtient
\beq
L_0={1\over 2}
\sum_n
\ :\!p_n p_{-n} \! :\
+{1\over 8 \gamma}
={1\over 2}
\sum_n
\ :\!p_n p_{-n} \! :\
+{1\over 4}
\left(
{h\over\pi}
+{\hhat\over\pi}
+2\right)
\eeq
qui appliqu\'e \`a un \'etat de plus haut poids
donne
\beq
L_0 |\varpi\! >=
{h\over 4\pi}
\left(
(1+{\pi\over h})^2
-\varpi^2
\right)
|\varpi\! >
\hbox{  c'est-\`a-dire  }
\Delta(\varpi)=
{h\over 4\pi}
\left(
(1+{\pi\over h})^2
-\varpi^2
\right).
\label{deltaw}
\eeq
On caract\'erisera donc souvent les champs primaires
ou les \'etats de plus haut poids par leur moment $\varpi$
plut\^ot que par leur poids correspondant $\Delta=\Delta(\varpi)$.
Aux valeurs discr\`etes des poids de la table de Kac
correspondent des valeurs discr\`etes de $\varpi$
tous les deux \'etiquet\'es par des ``spins'' demi-entiers,
nous allons y venir.
Dans la relation quadratique \ref{deltaw} on remarque
n\'eanmoins que deux valeurs oppos\'ees du moment
donnent le m\^eme poids,
c'est-\`a-dire pour le cas discret, qu'il existe pour le m\^eme poids
de Kac deux \'etats et op\'erateurs diff\'erents.

\vskip 2mm

Nous construisons maintenant les familles d\'eg\'en\'er\'ees
suivantes, traitant dans un premier temps les familles $(1,n)$
qui d\'erivent de $\alpha_-$ ou de $h$
(et non $\alpha_+$ et $\hhat$).
Nous les d\'esignons par un spin $J$ avec $J=2n+1$
car il appara\^\i tra dans le chapitre 4 que les op\'erateurs
$(1,n)$ de la base du groupe quantique sont dans une repr\'esentation
de spin $J$ de celui-ci.
Pour le seul poids $\Delta(1,2)$, nous avons
les deux op\'erateurs primaires $\psi_1$ et $\psi_2$,
not\'es $V^{(1/2)}_{\pm1/2}$,
issus des deux champs libres diff\'erents.
Par fusion nous allons engendrer pour chaque
poids $\Delta(1,n)$ non plus 2 mais $n$ champs primaires
not\'es $V^{(J)}_m$ pour $m=-J,\>-J+1... ,J$ avec $2J+1=n$.
Les poids de Kac de ces op\'erateurs\footnote{
Nous conservons la notation fonctionnelle $\Delta(n,m)$
pour les poids de Kac \ref{Kac}
et adoptons une notation indicielle $\Delta_{J}$ et $\Delta_{J,\Jhat}$
pour les m\^emes
poids fonction des spins \ref{deltaj}, \ref{deltajj}.
On a \'egalement la notation $\Delta(\varpi)$
qui donne le poids d'un \'etat ou d'un op\'erateur
en fonction de mode z\'ero en \ref{deltaw},
cette fois pour des poids et mode-z\'ero quelconques.
}
peuvent s'\'ecrire
en termes de $h$ et du spin $J$:
\beq
\Delta_J=\Delta(1,2J+1)=-J-J(J+1){h\over\pi}
\label{deltaj}.
\eeq
Le nombre\footnote{
Attention,
la notation $m$ ne doit pas faire croire qu'il s'agit
d'un nombre magn\'etique d'une repr\'esentation de $Sl(2)$
ou $U_q(sl(2))$.
Ce sera en revanche le cas pour le nombre $M$ des op\'erateurs
$\xi^{(J)}_M$ de la base covariante du groupe quantique
que nous verrons dans le chapitre 4.}
$m$ est conserv\'e
par fusion car c'est l'accroissement du moment $\varpi$
produit par un op\'erateur $V^{(J)}_m$:
\beq
V^{(J)}_m f(\varpi)
=
f(\varpi+2m)V^{(J)}_m
\label{shiftw}
\eeq
pour toute fonction $f$.
On voit que ceci est vrai
pour $J=1/2$ sur l'expression \ref{psiq} de $V^{(1/2)}_{\pm 1/2}$:
le facteur $e^{...q_0^{(1)}}$
est un op\'erateur d'accroissement de $p_0^{(1)}$ et
de la m\^eme mani\`ere $e^{...q_0^{(2)}}$
est un op\'erateur d'accroissement de $p_0^{(2)}$ qui est l'oppos\'e de
$p_0^{(1)}$
ce qui fait donc varier $p_0^{(1)}$ dans l'autre sens.
La propri\'et\'e d'accroissement de $\varpi$
\'etant additive par fusion,
d\'efinissant le nombre $m$ comme une quantit\'e
\'egalement additive,
\ref{shiftw} est automatiquement v\'erifi\'ee
pour les spins sup\'erieurs.

Examinons la fusion.
Le comportement \`a courte distance ($x'\to x$)
de la fonction \`a quatre points \ref{f4p}
permet de montrer que l'ordre dominant de la fusion
d'un op\'erateur de spin 1/2 et d'un op\'erateur de spin $J$ est
donn\'e par un op\'erateur de spin\footnote{
La singularit\'e indique en fait que la fusion d'un op\'erateur
de poids $(1,2)$ et d'un op\'erateur de poids $\Delta(1,2J+1)$
donne un op\'erateur de poids $\Delta(1,2J+2)$.
}
$J+1/2$.
Le cas le plus simple est celui de la fusion de $V^{(1/2)}_{-1/2}$
(par exemple)
avec lui m\^eme.
On a un seul champ libre $P^{(1)}$ et on obtient
encore des exponentielles du m\^eme champ libre,
$V^{(J)}_{-J}$:
$$
V^{(1/2)}_{-1/2}V^{(J-1/2)}_{-(J-1/2)}\sim
$$
\beq
V^{(J)}_{-J}\propto
e^{2J\sqrt{h/2\pi}q_0^{(1)}}
\> : \!
e^{2J\sqrt{h/2\pi}(p_0^{(1)}\sigma
+i\sum_{n\ne 0}e^{-in\sigma}p_n^{(1)}/n)}
\!:
\label{Vj-j}
\eeq
(on peut d'ailleurs dans ce cas \'egalement consid\'erer $J$ non
demi-entier, cf chapiter 4).
Il en est de m\^eme pour l'autre champ libre $P^{(2)}$
qui engendre les op\'erateurs $V^{(J)}_J$.
Pour un op\'erateur g\'en\'erique,
la conservation du ``nombre d'accroissement'' $m$
nous donne
la fusion \`a l'ordre dominant
\beq
V^{(1/2)}_{\pm 1/2}
V^{(J)}_m
\sim
V^{(J+1/2)}_{m\pm1/2}
\label{fusV}
\eeq
ce qui permet de construire les op\'erateurs $V^{(J)}_m$ par
r\'ecurrence\footnote{
Alors que le signe $\sim$ dans \ref{Vj-j} ne symbolise
que la division par la singularit\'e,
dans \ref{fusV} puis \ref{fusordd} il y a des coefficients
non triviaux (des fonctions gamma)\cite{G1}
}.
Ceci montre que $m$ est effectivement born\'e
par les conditions habituelles des spins
et nombres magn\'etiques $|m|\le J$.
C'est ce qui a initialement conduit \`a la d\'enomination
de spins,
bien que seuls les op\'erateurs $\xi$ de la base du groupe
quantique (cf partie 4.2)
soient r\'eellement dans une repr\'esentation de spin $J$
de $U_q(sl(2))$.

Nous n'avons construit ici que les op\'erateurs de la famille $(1,2J+1)$,
\`a partir du param\`etre $h=h_-$.
On peut faire la m\^eme chose \`a partir de la solution
$\hhat=h_+$ de \ref{eqh}.
Les op\'erateurs ainsi construits de la m\^eme fa\c con
avec le param\`etre $\hhat$,
dans la representation $(2\Jhat+1,1)$, sont not\'es
$\Vhat^{(\Jhat)}_{\mhat}$.
Cette notation qui ne respecte pas la sym\'etrie entre $h_-$ et $h_+$
a n\'eanmoins le m\'erite de la simplicit\'e.
Ensuite, toujours par fusion \`a l'ordre dominant
on les combine avec les pr\'ec\'edents pour obtenir
les op\'erateurs $V^{(J,\Jhat)}_{m,\mhat}$
dans la repr\'esentation d\'eg\'en\'er\'ee $(2\Jhat+1,2J+1)$.
Leurs poids sont ceux de Kac:
$$
\Delta_{J,\Jhat}=\Delta(2\Jhat+1,2J+1)=
{C-1\over 24}-{1\over24}
\left(
(J+\Jhat+1)\sqrt{C-1}
-(J-\Jhat)\sqrt{C-25}
\right)^2
$$
\beq
=
-J-J(J+1){h\over\pi}
-\Jhat-\Jhat(\Jhat+1){\hhat\over\pi}
-2J\Jhat
\label{deltajj}.
\eeq
On effectue cette r\'ecurrence en extrayant la singularit\'e
dominante par
\beqa
V^{(J+1/2,\Jhat)}_{m\pm1/2,\mhat}(z_2)
\propto
\lim_{z_1\to z_2}
(z_1-z_2)^{\left( \Delta_{1/2,0}+\Delta_{J,\Jhat}-
\Delta_{J+1/2,\Jhat}\right)}
V^{(1/2,0)}_{\pm1/2,0}(z_1)
V^{(J,\Jhat)}_{m,\mhat}(z_2)\nnn
V^{(J,\Jhat+1/2)}_{m,\mhat\pm1/2}(z_2)
\propto
\lim_{z_1\to z_2}
(z_1-z_2)^{\left( \Delta_{0,1/2}+\Delta_{J,\Jhat}-
\Delta_{J+1/2,\Jhat}\right)}
V^{(0,1/2)}_{0,\pm1/2}(z_1)
V^{(J,\Jhat)}_{m,\mhat}(z_2).
\label{fusordd}
\eeqa

Nous nous sommes restreints dans ce chapitre,
comme dans l'approche initi\'ee dans les ann\'ees 80,
\`a l'ordre dominant du d\'eveloppement \`a courte distance
qui permet de d\'efinir r\'ecursivement les op\'erateurs
de spins sup\'erieurs comme limite du produit
des pr\'ec\'edents (\ref{fusordd}).
Les ordres suivant du d\'eveloppement \ref{fusV} donnent
les descendants et les op\'erateurs $V^{(J-1/2)}_{m\pm1/2}$ (famille
$(1,2J-1)$),
le nombre $m$ \'etant toujours conserv\'e.
Ceci sera trait\'e en d\'etail dans le chapitre 4
dans le cadre de Moore et Seiberg pr\'esent\'e dans le chapitre 3.

Conform\'ement \`a ce que nous avons vu dans le chapitre
pr\'ec\'edent,
ces op\'erateurs engendrent les \'etats par action sur le vide.
Les op\'erateurs $V^{(J,\Jhat)}_{m,\mhat}$
de type $(2\Jhat+1,2J+1)$ de la classification
de BPZ engendrent les modules de Verma
r\'eductibles correspondants not\'es ${\cal H}_{J,\Jhat}$.
Ce sont en fait les exponentielles de champ libre du type $V^{(J)}_{-J}$
(Eq.\ref{Vj-j})
qui engendrent
les \'etats \`a partir du vide $Sl(2,C)$ invariant:
\beq
V^{(J,\Jhat)}_{-J,-\Jhat}(0)
|\varpi_0\!>
=
|\varpi_{J,\Jhat}\!>
\equiv
|J,\Jhat\!>
\label{etatsL}
\eeq
avec
\beq
\varpi_{J,\Jhat}=
2J+1+(2\Jhat+1){\pi\over h}
\label{wjj}
\eeq
ou (par eq.\ref{wdep})
\beq
\varpihat_{J,\Jhat}\equiv
{h\over\pi}\varpi_{J,\Jhat}=
2\Jhat+1+(2J+1){h\over\pi}
\label{wjjhat}.
\eeq
Ceci est coh\'erent avec l'accroissement produit
par les nombres $m$ et $\mhat$ (qui valent ici $-J$ et $-\Jhat$).
Nous avons \'egalement introduit la notation $|J,\Jhat\!>$
pour les \'etats de ces repr\'esentations r\'eductibles.
On v\'erifie que la formule \ref{deltaw} qui exprime les poids des \'etats
en fonction de $\varpi$ redonne bien les poids de Kac \ref{deltajj}
pour ces \'etats.

Il est apparu n\'ecessaire d'utiliser comme vide sortant
$Sl(2,C)$ invariant celui de moment oppos\'e\footnote{
Nous rappelons que $\Delta(\varpi_0)=0=\Delta(-\varpi_0)$.}
$<\!-\varpi_0|$.
Les \'etats sortants sont donc engendr\'es par l'action
des exponentielles du m\^eme champ libre sur celui-ci:
\beq
<\!-\varpi_0|
V^{(-J-1,-\Jhat-1)}_{J+1,\Jhat+1}(z,\zb)
z^{2L_0}\zb^{2\Lb_0}
\to
<\!\varpi_{J,\Jhat}|
=<\!J,\Jhat|
\eeq
quand $z,\zb\to\infty$.

\vskip 5mm

La derni\`ere \'etape,
apr\`es la construction de ces op\'erateurs chiraux,
consiste \`a construire les op\'erateurs
physiques.
Ils doivent tout d'abord \^etre d\'efinis de fa\c con univoque,
c'est-\`a-dire avec une monodromie triviale.
Ils doivent ensuite, et cette contrainte suffira \`a les
d\'eterminer, \^etre locaux,
c'est-\`a-dire commuter entre eux \`a m\^eme $\tau$ (et $\sigma$ diff\'erent).
C'est l'\'equivalent en termes d'op\'erateurs de
la localit\'e pour les fonctions de corr\'elations physiques
que nous examinerons dans le chapitre suivant.
Ceci n\'ecessite de conna\^\i tre la matrice d'\'echange des op\'erateurs
$V^{(J,\Jhat)}_{m,\mhat}$.
Mais celle-ci n'\'etait pas connue (pour des spins sup\'erieurs \`a 1/2)
avant l'approche plus moderne
\`a la Moore et Seiberg pr\'esent\'ee dans le chapitre suivant.
Ces op\'erateurs physiques ont cependant \'et\'e construits
d\`es 1992 par J.-L. Gervais \cite{G4}.
Leur caract\`ere local a pu \^etre v\'erifi\'e en passant par l'autre
base d'op\'erateurs (celle du groupe quantique)
que nous n'abordons \'egalement que dans le chapitre 4.
Nous esquissons juste le r\'esultat pour conclure cette partie:
\beq
e^{\textstyle -(J\alpha_-+\Jhat\alpha_+)\Phi(\sigma,\tau)}
=
\sum_{m=-J}^J
\sum_{\mhat=-\Jhat}^\Jhat
(...)
V^{(J,\Jhat)}_{m,\mhat}(\sp)
\Vb^{(J,\Jhat)}_{m,\mhat}(\sm)
\label{expphi}
\eeq
o\`u les points symbolisent un certain coefficient d\'ependant
de $J,\Jhat,m,\mhat$ et du moment $\varpi$.
On peut noter la ressemblance avec le cas classique (Eq.\ref{expphic}).
Le terme cosmologique de poids $(1,1)$ est \`a nouveau
obtenu pour des spins $(J,\Jhat)=(0,-1)$ ou $(-1,0)$
(cf Eqs.\ref{deltaj}, \ref{deltajj}),
les op\'erateurs de spins n\'egatifs restant encore \`a d\'efinir
dans le cas quantique.

\vskip 3mm

Examinons la limite classique de \ref{expphi}.
Elle est obtenue pour $\gamma$ tendant vers z\'ero
donc $C$ vers l'infini.
On calcule que dans cette limite $\alpha_-$
se comporte comme $2\sqrt{\gamma}$ alors que
$\alpha_+$ tend vers l'infini.
C'est donc $\alpha_-$ qui a le bon comportement classique:
\beq
e^{\textstyle -J\alpha_-\Phi}
\to
e^{\textstyle -2J\sqrt\gamma\Phi}
\hbox{  (classique)}.
\label{lclass}
\eeq
C'est pour que cette limite donne directement
la solution classique que nous avions choisi de
renormer le $\Phi$ classique par ce facteur
$2\sqrt\gamma$.
Ceci conduit en particulier \`a consid\'erer le
terme cosmologique $e^{\alpha_-\Phi}$
et non $e^{\alpha_+\Phi}$.
Cela ne vaut cependant que pour le couplage faible
($c\le 1$, $C\ge 25$).
La cl\'e du couplage fort
($1<C<25$, loin de la limite classique)
est au contraire de traiter
sym\'etriquement $\alpha_+$ et $\alpha_-$.
Nous y reviendrons dans la derni\`ere partie.

\vskip 5mm

Pour terminer ce chapitre,
nous montrons comment, en s'inspirant de la m\'ethode
de DDK \cite{Dav,DK}, on peut dans cette approche retrouver la formule de KPZ
\cite{KPZ}
pour l'exposant critique $\gamma_{\hbox{\scriptsize string}}$.
Il exprime le comportement de la fonction de partition
(\`a aire fix\'ee) en fonction de l'aire $A$:
\beq
Z(A)\propto A^{\gamma_{\hbox{\scriptsize string}}-3}.
\eeq

Suivant la m\'ethode de DDK \cite{Dav,DK}
nous consid\'erons d'abord une th\'eorie de cordes non-critiques
avec l'action suivante pour le champ de Liouville
\beq
S_{\mu_c}
=
{1\over 8\pi}
\int dz d\zb \sqrt{\ghat}
(\ghat^{ab}\partial_a\Phi \partial_b \Phi
+Q \Rhat^{(2)}\Phi + \mu_c e^{\alpha_-\Phi})
\eeq
avec
$$
Q=\sqrt{C-1\over 3}.
$$
Les op\'erateurs de vertex de mati\`ere sont habill\'es
par des exponentielles de Liouville
$e^{-\beta\Phi}$ (avec $\beta=J\alpha_-+\Jhat\alpha_+$
dans nos notations)
de poids $-\beta(\beta+Q)/2$
de mani\`ere \`a obtenir des op\'erateurs de poids (1,1).
Suivant l'argument de DDK, il est facile de voir qu'un shift
de $\Phi\to\Phi-ln(\mu_c)/\alpha_-$
transforme les corr\'elateurs
\beq
\left <
\prod_i
e^{\displaystyle -\beta_i\Phi}
\right >_{\mu_c}
=
\int
{\cal D}\Phi
e^{-S_{\mu_c}}
\prod_i
e^{\displaystyle -\beta_i\Phi}
\eeq
par
\beq
\left <
\prod_i
e^{\displaystyle -\beta_i\Phi}
\right >_{\mu_c}
=
\left <
\prod_i
e^{\displaystyle -\beta_i\Phi}
\right >_1
\mu_c^{
\left[
\sum_i \beta_i +Q(1-h)
\right ]
/\alpha_-
}
\label{scalDDK}
\eeq
o\`u le dernier terme $\propto Q(1-h)$ provient de l'utilisation du
th\'eor\`eme
de Gauss-Bonnet sur une surface de genre $h$.

On peut obtenir le m\^eme r\'esultat dans l'approche
op\'eratorielle pr\'esent\'ee ici.
Il s'agit de repasser du cas particulier $\mu=1$
auquel nous nous sommes restreints en \ref{Liou4}
au cas d'une constante cosmologique $\mu_c$ quelconque
(action \ref{Liou0}).
La loi d'\'echelle la plus g\'en\'erale
pour les composantes chirales $V$ qui conserve
le caract\`ere local de l'op\'erateur de Liouville \ref{expphi}
est la suivante
\beqa
\left.
V^{(J,\Jhat)}_{m,\mhat}
\right | _{(\mu)}
&=&
\mu^{J+\Jhat(\pi/h)}
\mu^{-\varpi/2}
\left.
V^{(J,\Jhat)}_{m,\mhat}
\right | _1
\mu^{\varpi/2} \nnn
\left.
\Vb^{(\Jb,\Jhatb)}_{\mb,\mhatb}
\right | _{(\mub)}
&=&
\mub^{\Jb+\Jhatb(\pi/h)}
\mub^{-\varpib/2}
\left.
\Vb^{(\Jb,\Jhatb)}_{\mb,\mhatb}
\right | _1
\mub^{\varpib/2}
\label{scalV1}
\eeqa
ce qui donne pour le champ de Liouville
\ref{expphi}
\beq
\left .
e^{\textstyle -(J\alpha_-+\Jhat\alpha_+)\Phi(\sigma,\tau)}
\right | _{\mu_c}
=
\mu_c^{J+\Jhat(\pi/h)}
\mu_c^{-\varpi/2}
\left.
e^{\textstyle -(J\alpha_-+\Jhat\alpha_+)\Phi(\sigma,\tau)}
\right | _1
\mu_c^{\varpi/2}
\label{scalphi}
\eeq
en tenant compte de l'\'egalit\'e des spins et moments
gauches et droits, et en d\'efinissant $\mu_c=\mu\mub$.
Ces lois d'\'echelle \ref{scalV1} seront
r\'eutilis\'ees dans le dernier chapitre
pour donner le comportement du nouvel op\'erateur
physique de Liouville.

Appliquons les aux corr\'elateurs
\beq
<\!\varpi_0|
\prod_i
e^{\textstyle -(J_i\alpha_-+\Jhat_i\alpha_+)\Phi(\sigma,\tau)}
|\varpi_0\!>
_{\mu_c}.
\eeq
Chaque op\'erateur de vertex donne bien la m\^eme contribution
que dans la formule \ref{scalDDK}:
$$
\mu_c^{J_i+\Jhat_i(\pi/h)}
=
\mu_c^{(J_i\alpha_-+\Jhat_i\alpha_+)/\alpha_-}.
$$
L'action des op\'erateurs $\mu_c^{\varpi/2}$
sur le vide entrant $|\varpi_0\!>=|1+\pi/h\!>$ et de $\mu_c^{-\varpi/2}$
sur le vide sortant $<\!-\varpi_0|$ contribue pour
$$
\mu_c^{1+\pi/h}
=
\mu_c^{(\alpha_-+\alpha_+)/\alpha_-}
=
\mu_c^{Q/\alpha_-}
$$
ce qui redonne bien au total l'expression \ref{scalDDK}
de DDK pour le genre 0.
Suivant ensuite le m\^eme raisonnement que DDK,
ceci permet de retrouver
la valeur de $\gamma_{\hbox{\scriptsize string}}$ de KPZ (genre 0).
Nous la donnons en genre quelconque
\beq
\gamma_{\hbox{\scriptsize string}}=
(1-h)
{1\over 12}
\left(
d-25
-\sqrt{(25-d)(1-d)}
\right)
+2
\label{gstring}.
\eeq

Ceci a \'et\'e obtenu en ref.\cite{KPZ} (KPZ) en genre $h=0$
dans la jauge du c\^one de lumi\`ere
avant ce r\'esultat en genre quelconque obtenu dans
la jauge conforme en refs.\cite{Dav,DK}.
C'est pourquoi on appelle g\'en\'eralement cette formule (pour $h=0$)
la formule de KPZ.
Elle est en accord avec les r\'esultats de l'approche
discr\'etis\'ee par mod\`eles de matrices.
Ces derniers ont permis de traiter \`a tous les ordres les mod\`eles
minimaux coupl\'es \`a la gravit\'e,
ce qui correspond \`a des charges centrales pour la mati\`ere
$c=1-6(p-q)^2/pq$ pour deux entiers $p$ et $q$ positifs premiers
entre eux,
donc une charge centrale inf\'erieure \`a 1,
dans le r\'egime de couplage faible.
On peut d'ailleurs \'egalement retrouver dans l'approche
op\'eratorielle pr\'esent\'ee ici
le r\'esultat des mod\`eles de matrice pour les fonctions \`a
trois points des mod\`eles minimaux coupl\'es \`a
la gravit\'e (qui vaut essentiellement 1).
Ceci a \'et\'e fait en ref.\cite{G4} en consid\'erant
deux copies de la th\'eorie discut\'ee ici,
la premi\`ere tenant lieu de mati\`ere
et l'autre constituant l'habillage par la gravit\'e
(un r\'esultat concordant a \'egalement \'et\'e obtenu
ant\'erieurement dans l'approche du gaz de Coulomb en ref.\cite{D}).

Comme annonc\'e pr\'ec\'edemment,
dans le r\'egime de couplage fort ($c>1$)
le r\'esultat \ref{gstring} n'est pas acceptable.
Une fonction de partition r\'eelle ne peut pas avoir
une loi d'\'echelle d'exposant complexe.
Il n'y a pas non plus de version discr\'etis\'ee
de mod\`ele fortement coupl\'e qui permette de calculer
$\gamma_{\hbox{\scriptsize string}}$.

C'est pourquoi nous proposons dans le dernier chapitre
une autre approche du r\'egime de couplage fort.
Dans cette phase $c>1$ les op\'erateurs physiques
sont diff\'erents de ceux du couplage faible et de poids r\'eels.
Ceci a en particulier pour cons\'equence
que le nouveau terme cosmologique
a un poids r\'eel et une loi de changement d'\'echelle
r\'eelle.
La nouvelle valeur de $\gamma_{\hbox{\scriptsize string}}$ obtenue
est la partie r\'eelle de \ref{gstring}.

\chapter{LES EQUATIONS POLYNOMIALES}
\markboth{3. Les \'equations polynomiales}
{3. Les \'equations polynomiales}

Je pr\'esente ici une partie des travaux, dus essentiellement \`a
Moore et Seiberg, qui ont men\'e aux \'equations dites polynomiales.
Ce sont les \'equations de coh\'erence de l'alg\`ebre
d'op\'erateurs des th\'eories conformes.
Leur but \'etait de classifier les th\'eories
conformes,
le mien est uniquement de fournir les outils
n\'ecessaires \`a la r\'esolution de l'alg\`ebre d'op\'erateurs
de la th\'eorie de Liouville dans le chapitre \ref{p4}.
Je me contenterai donc de l'essentiel, me limitant en particulier au
cas de la topologie sph\'erique.
Le lecteur pourra se r\'ef\'erer \`a
\cite{MS1,MS2} ou au cours \cite{MS3}
o\`u est trait\'e en particulier le cas des genres sup\'erieurs.

Comme la th\'eorie de Liouville,
les th\'eories conformes qui
admettent une d\'ecomposition chirale
ont des fonctions de corr\'elation
chirales ou blocs conformes de monodromie
non triviale.
Ces derniers
comportent donc des coupures.
Les fonctions de corr\'elation physiques
doivent au contraire \^etre locales ou mono-valu\'ees.
Lors de l'\'etude d'une th\'eorie particuli\`ere,
on cherche donc g\'en\'eralement \`a construire des fonctions
de corr\'elations physiques locales
\`a partir des blocs conformes
qui ne le sont pas.
Ici, au contraire,
nous supposerons qu'il est possible de construire
des corr\'elateurs physiques locaux \`a partir des blocs
conformes,
comme cela doit \^etre le cas pour une th\'eorie conforme g\'en\'erique,
et nous en examinerons les cons\'equences sur
l'alg\`ebre des op\'erateurs chiraux.
Nous ne cherchons pas ici \`a calculer les blocs
conformes, puisqu'ils d\'ependent de la th\'eorie envisag\'ee.
Mais nous montrons que l'hypoth\`ese de localit\'e
des corr\'elateurs physiques permet d'obtenir
la forme g\'en\'erale de leurs transformations
de dualit\'e (\'echange et fusion)
ainsi que les contraintes de coh\'erence
de ces transformations (\'equations polynomiales).

Nous introduisons dans la premi\`ere partie
les notations des op\'erateurs chiraux
et les propri\'et\'es \'el\'ementaires des blocs conformes.
Nous discutons en particulier leurs propri\'et\'es de monodromie.
Nous montrons ensuite dans une deuxi\`eme partie
que la localit\'e des fonctions de corr\'elation physiques
est \'equivalente aux
propri\'et\'es de dualit\'e
des blocs conformes,
qui s'expriment sous forme d'une alg\`ebre
de fusion et d'\'echange
pour les op\'erateurs chiraux.
Nous \'etendons d'ailleurs dans cette partie
la discussion des th\'eories diagonales par Moore et Seiberg
au cas des th\'eories non-diagonales qui peuvent \^etre
envisag\'ees lorsque la matrice de monodromie est d\'eg\'en\'er\'ee.
Ceci s'appliquera en effet aux constructions du dernier
chapitre dans le r\'egime de couplage fort.
L'existence de transformations \'equivalentes
(i.e. entre m\^emes blocs conformes initiaux et finaux)
donne des \'equations de coh\'erence sur
les coefficients de fusion et d'\'echange,
appel\'ees \'equations polynomiales,
que nous d\'eterminons dans une troisi\`eme et derni\`ere partie.

Ces \'equations polynomiales contraignent \'enorm\'ement
l'alg\`ebre de fusion et d'\'e\-change des op\'erateurs chiraux.
Dans le prochain chapitre, nous pourrons
calculer explicitement les blocs conformes
et les coefficients de fusion et d'\'echange les plus simples
de la th\'eorie de Liouville,
et ce sont ces \'equations polynomiales qui nous permettront ensuite de
d\'eterminer
l'alg\`ebre de fusion et d'\'e\-chan\-ge
de tous les op\'erateurs.

\section{Blocs conformes et monodromie}
\label{3.1}

Nous consid\'erons ici une th\'eorie conforme rationnelle\footnote{
\label{ration}
Une th\'eorie (conforme) est souvent dite rationnelle
si elle comporte un nombre fini de familles conformes,
ou de repr\'esentations de l'alg\`ebre de Virasoro.
C'est le cas par exemple des mod\`eles minimaux.
Ce n'est pas le cas ici, comme chez
Moore et Seiberg.
Nous appelons une th\'eorie rationnelle si l'espace de ses blocs
conformes est de dimension finie
($p$ prend un nombre fini de valeurs dans \ref{ijklp}),
ce qui donne des matrices d'\'echange et de fusion de dimension finie.
Nous appliquerons \'evidemment toujours ces r\'esultats
dans ce cadre (chapitre \ref{p4}).
On pourra en effet v\'erifier que m\^eme lorsqu'on
g\'en\'eralisera notre alg\`ebre d'op\'erateurs
\`a des spins continus (elle pourra donc
m\^eme contenir une ``infinit\'e continue''
de familles),
les r\`egles de s\'election
garantiront toujours que les blocs conformes restent de dimension finie.
},
dont les op\'erateurs
physiques $\Phi^i(z,\zb)$ ont une d\'ecomposition chirale
en termes d'op\'erateurs\footnote{
Conform\'ement \`a la partie pr\'ec\'edente,
nous conservons dans cette partie la notation
$\phi$ minuscule pour les op\'erateurs conformes chiraux.
Les op\'erateurs physiques locaux sont not\'es $\Phi$
majuscule.
Ces notations sont contraires \`a celles de certains auteurs,
en particulier
de Moore et Seiberg,
mais nous les conservons ici par souci de coh\'erence \`a
l'int\'erieur de cette th\`ese.}
$\phi^i(z)$ et $\phib(\zb)$.
Au lieu de noter $\phi^i$ les op\'erateurs chiraux g\'en\'eriques
comme dans le premier chapitre,
nous consid\'erons ici une restriction de ces op\'erateurs,
not\'ee $\phi_{i,k}^{j}$.
Outre le(s) nombre(s) quantique(s) $j$ qui caract\'erise(nt)
l'op\'erateur lui-m\^eme,
on pr\'ecise par $k$ et $i$
les modules de Verma auquel cet op\'erateur
s'applique et celui dans lequel il arrive.
On a de fa\c con plus g\'en\'erale pour un descendant\footnote{
\label{asym}
Attention aux notations partiellement asym\'etriques:
la notation $\phi_{i,k}^{j}(z)$ ou $\phi_{i,k}^{j,\{\nu\}}(z)$
indique que l'op\'erateur $\phi$ est un champ primaire ou un descendant
de ${\cal H}_j$ respectivement.
Au contraire les indices $i$ et $k$ indiquent simplement
que ces op\'erateurs s'appliquent \`a des \'etats du module
de Verma ${\cal H}_k$,
qu'il s'agisse de l'\'etat de plus haut poids $|k\!>$
ou d'un descendant $|k,\{\nu\}\!>$,
et sont \`a valeurs dans ${\cal H}_i$.
}
\beq
\phi_{i,k}^{j,\{\nu\}}(z):
{\cal H}_k\to{\cal H}_i.
\label{phidesc}
\eeq
Les op\'erateurs consid\'er\'es ici peuvent donc \^etre obtenus par
restriction de ceux du premier chapitre:
\beq
\phi_{i,k}^{j,\{\nu\}}(z)
=
{\cal P}_{{\cal H}_i}
\phi^{j,\{\nu\}}(z)
{\cal P}_{{\cal H}_k}
\label{restric}
\eeq
o\`u ${\cal P}_{{\cal H}_i}$ est le projecteur sur le module de Verma ${{\cal
H}_i}$.
Dans une base orthonorm\'ee de ${\cal H}_i$, d\'ecrite
par le multi-index $\{\nu\}$,
ce projecteur s'\'ecrit
\beq
{\cal P}_{{\cal H}_i}
=
\sum_{\{\nu\}}
|i,\{\nu\}\!>
<\!i,\{\nu\}|
\label{proj}.
\eeq
L'espace de Hilbert total (pour une chiralit\'e) est constitu\'e
de la somme directe des modules de Verma et on a donc
\'egalement la relation de fermeture
\beq
1=\hbox{Id}=
\sum_i
{\cal P}_{{\cal H}_i}
=
\sum_{i,\{\nu\}}
|i,\{\nu\}\!>
<\!i,\{\nu\}|
\label{ferm}.
\eeq
La correspondance entre les op\'erateurs et les \'etats
est donc dans ces notations
\beq
\phi_{i,0}^{i,\{\nu\}}(0)|0\!>
=
|i,\{\nu\}\!>.
\eeq
Nous verrons dans le prochain chapitre comment les op\'erateurs
de la th\'eorie de Liouville du chapitre pr\'ec\'edent
doivent de la m\^eme mani\`ere \^etre restreints
pour rentrer exactement dans le cadre de ce formalisme.

\vskip 2mm

On peut int\'egrer les transformations infinit\'esimales \ref{[Lnphi]}
dues \`a l'action de $L_0$ sur un champ primaire de poids $\Delta_j$
pour obtenir
\beq
e^{-\alpha L_0}
\phi_{ik}^j(z)
e^{\alpha\Delta_j}
e^{\alpha L_0}
=
\phi_{ik}^j(e^\alpha z)
\label{tourner}.
\eeq
Ceci permet de d\'eterminer la monodromie des op\'erateurs
$\phi_{ik}^j(z)$ autour de l'origine.
Appliquant \ref{tourner} \`a $\alpha=2i\pi\epsilon$
(o\`u $\epsilon=\pm1$) l'op\'erateur effectue en
effet un tour autour de $0$.
Ce qui est particulier \`a cette base d'op\'erateurs,
c'est que les poids des \'etats entrants et sortants sont connus
\`a un entier pr\`es (pour un descendant $|k,\{\nu\}\!>$
d'ordre $N$ on a $\Delta_{k,\{\nu\}}=
\Delta_{k}+N$)
qui ne contribue pas pour $\alpha=2i\pi\epsilon$.
On connait donc la monodromie de ces op\'erateurs autour de $0$
\beq
\phi_{ik}^j(z)
\to
\phi_{ik}^j(e^{2i\pi\epsilon}z)
=
\phi_{ik}^j(z)
e^{2i\pi\epsilon(\Delta_i+\Delta_j-\Delta_k)}
\label{monodr3}.
\eeq
Cet op\'erateur donne la fonction \`a trois points suivante
\beq
\epsffile{ijk.eps}
=
<\!i,\{\nu_i\}|\phi_{i,k}^j(z)|k,\{\nu_k\}\!>
=
z^{\Delta_j+\Delta_{k,\{\nu_k\}}-\Delta_{i,\{\nu_i\}}}
\!
<\!i,\{\nu_i\}|\phi_{i,k}^j(1)|k,\{\nu_k\}\!>
\label{f3pz}
\eeq
dont la d\'ependance en $z$ est obtenue par \ref{tourner},
pour des descendants $<\!i,\{\nu_i\}|$ et $|k,\{\nu_k\}\!>$ quelconques.
La monodromie \ref{monodr3}  de $\phi_{i,k}^j(z)$ autour de 0
donne alors celle du bloc conforme \`a trois points\footnote{
On obtient en fait \ref{monodr3p} uniquement pour une patte sup\'erieure
$i$ qui d\'esigne un champ primaire.
Ceci est facilement \'etendu aux descendants $i,\{\nu_i\}$ par action
des op\'erateurs diff\'erentiels appropri\'es sur les deux membres
de \ref{monodr3p}.}
\beq
\epsffile{ijk.eps}
\to
e^{2i\pi\epsilon(\Delta_i+\Delta_j-\Delta_k)}
\epsffile{ijk.eps}
\label{monodr3p}
\eeq
valable pour des champs primaires ou des descendants.
Les fonctions \`a trois points comprennent donc une coupure issue de 0,
que l'on place g\'en\'eralement sur le demi-axe r\'eel n\'egatif.

\vskip 2mm

Les blocs conformes \`a quatre points s'\'ecrivent
\beq
{\cal F}_p^{ijkl}(z_2,z_3)
\equiv
<\!i|\phi_{ip}^j(z_2)\phi_{pl}^k(z_3)|l\!>
=
\epsffile{ijklp.eps}
\label{ijklp}.
\eeq
Nous pouvons les d\'ecomposer
en somme de fonctions \`a trois points gr\^ace \`a
l'expression \ref{proj} du projecteur sur ${\cal H}_p$.
Ceci donne une somme sur les descendants $p,\{\nu\}$
qui s'exprime, d'apr\`es ce que nous venons de voir (\ref{f3pz}),
comme une s\'erie enti\`ere
de terme g\'en\'eral
$\propto (z_3/z_2)^{\Delta_{p,\{\nu\}}}=$ $(z_3/z_2)^{\Delta_{p}+N}$
pour $N$ entier positif.
La s\'erie est donc convergente pour $z_3/z_2<1$.
On retrouve ainsi l'ordre radial:
les blocs conformes \ref{ijklp} sont d\'efinis
dans le domaine $z_3/z_2<1$
(on a cependant toujours la coupure pour $z_3$ r\'eel
n\'egatif isssue des fonctions \`a trois points).

\begin{figure}
\epsffile{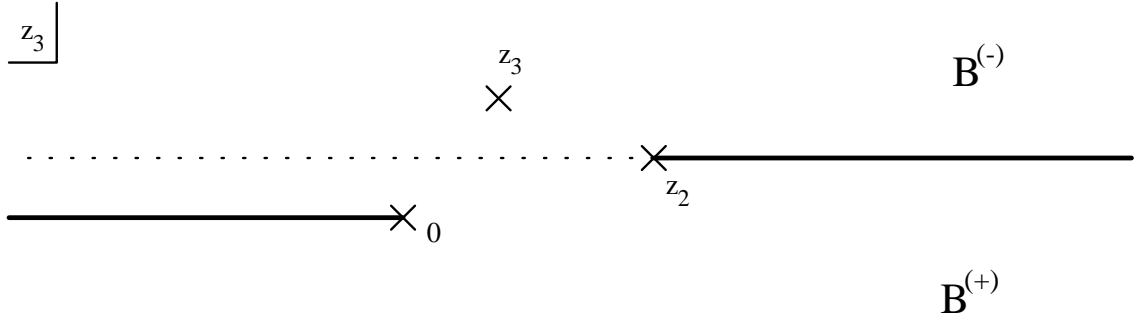}
\caption{Coupures de la fonction \`a quatre points
$<\! i|$
$\phi_{ip}^j(z_2)$ $\phi_{pl}^k(z_3)$
$|l\!>$.
Les pointill\'es d\'elimitent les zones d'application de
$B^{(-)}$ et $B^{(+)}$ (cf partie 3.2).}
\label{monodrf}
\end{figure}

Examinons les monodromies des fonctions \`a quatre points \ref{ijklp}.
Fixant $z_2=1$ par les transformations projectives,
les \'equation de bootstrap qui permettent de calculer
ces blocs conformes sont des \'equations diff\'erentielles
lin\'eaires comportant des p\^oles en 0, 1 et $\infty$.
Leurs solutions, les blocs conformes,
ont donc une monodromie non-triviale autour
de ces p\^oles\footnote{
\label{notemonodr}Chaque solution, c'est-\`a-dire chaque bloc conforme
(pour chaque valeur de $p$ dans \ref{ijklp}),
ne peut \^etre d\'efinie sur tout
le plan mais doit inclure une coupure issue du p\^ole.
Au niveau de cette coupure elle ne se raccorde pas continument \`a
elle-m\^eme mais \`a une combinaison lin\'eaire de solutions
de l'\'equation diff\'erentielle.
Ces combinaisons lin\'eaires pour toutes les solutions
donnent la matrice de monodromie,
matrice unitaire de l'espace des solutions dans lui-m\^eme.}.
Contrairement au cas de la fonction \`a trois points,
parmi ces monodromies, il y en plusieurs
(deux en l'occurence) ind\'ependantes,
et elles ne peuvent \^etre diagonalis\'ees simultan\'ement.
Ainsi,
le choix qui a \'et\'e fait en \ref{ijklp}
consiste \`a diagonaliser la monodromie (de $z_3$) autour de 0.
Ceci se d\'eduit imm\'ediatement de la monodromie \ref{monodr3}
des op\'erateurs (ou de la d\'ecomposition des blocs
conformes \`a quatre points en blocs \`a trois points).
Le domaine de d\'efinition des blocs conformes \ref{ijklp}
peut \^etre \'etendu analytiquement en dehors du domaine
$z_3/z_2<1$
mais cette monodromie non-triviale de $z_3$ autour de $z_2$
montre qu'il doit y avoir une coupure issue de $z_2$.
Les coupures dues au monodromies de $z_3$ autour de 0
(diagonale) et autour de $z_2$ (non-diagonale)
sont repr\'esent\'ees en figure \ref{monodrf}.

\begin{figure}
\epsffile{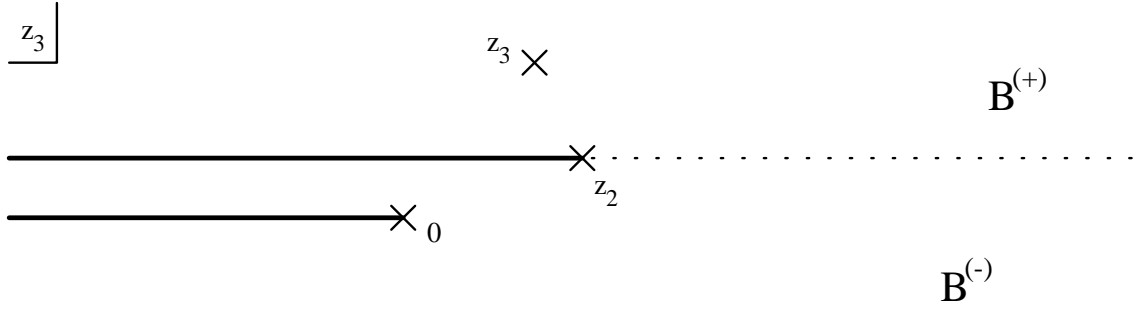}
\caption{Coupures dans l'autre base
$<\!i|$
$\phi_{ip}^k(z_3)$ $\phi_{pl}^j(z_2)$
$|l\!>$, plan $z_3$.}
\label{monodrf3}
\end{figure}

Il existe \'egalement des bases\footnote{
Bien que cela puisse \^etre v\'erifi\'e dans la pratique
au cas par cas (par les \'equations de bootstrap par exemple),
dans la logique de cet expos\'e nous n'avons pas encore prouv\'e
que les blocs
conformes \ref{ikjlq} constituent une autre base
de l'espace vectoriel des blocs conformes \ref{ijklp}.
Pour le cas g\'en\'eral que nous examinons ici,
nous le d\'eduirons de l'hypoth\`ese de localit\'e
des corr\'elateurs physiques dans la prochaine partie.}
qui diagonalisent les autres monodromies.
Ainsi
les blocs conformes ${\cal F}_q^{ikjl}(z_3,z_2)$ toujours donn\'es
par la m\^eme d\'efinition
\beq
{\cal F}_q^{ikjl}(z_3,z_2)=
<\!i|\phi_{iq}^k(z_3)\phi_{ql}^j(z_2)|l\!>=
\epsffile{ikjlq.eps}
\label{ikjlq}
\eeq
diagonalisent de la m\^eme mani\`ere la monodromie de
$z_3$ autour de l'infini,
ou de mani\`ere \'equivalente autour de 0 et $z_2$
simultan\'ement,
par rotation de l'op\'erateur $\phi^k(z_3)$.
Le domaine de converge est ici $z_3/z_2>1$,
et la coupure est donc plac\'ee en $z_3-z_2\in R_-$.
Les coupures de ces blocs sont repr\'esent\'es
en figure \ref{monodrf3}.
Comparant avec les coupures dans la base \ref{ijklp} figure \ref{monodrf}
qui diagonalise la monodromie autour de 0,
on voit qu'il y a eu rotation de la coupure issue de $z_2$.

Il y a \'egalement une troisi\`eme base qui diagonalise
la monodromie autour de $z_2$ (ou 1),
mais nous y reviendrons plus tard.

\section{Localit\'e et dualit\'e}
\label{3.2}

Ayant d\'etermin\'e les propri\'et\'es les plus \'el\'ementaires
des blocs conformes,
nous sommes pr\^ets \`a examiner la localit\'e
des fonctions de corr\'elation physiques.
Nous sortons ici du cadre strict de la discussion de Moore
et Seiberg \cite{MS1,MS2,MS3}.
Alors qu'ils ne discutent que les cas des th\'eories
dites diagonales,
nous envisageons ici \'egalement le cas o\`u la matrice
de monodromie est d\'eg\'en\'er\'ee,
ce qui permet, comme nous le verrons, de d\'efinir des th\'eories
non-diagonales.
Ce sera le cas pour la construction effectu\'ee dans le dernier
chapitre dans le r\'egime de couplage fort.
Nous montrerons l\`a-bas l'importance de la
d\'eg\'en\'erescence de la matrice de monodromie.
C'est en effet dans les cas o\`u cette
d\'eg\'en\'erescence est maximale,
cas qui correspondent aux dimensions sp\'eciales et aux
moments particuliers mis en \'evidence par Gervais et ses
collaborateurs depuis 1985 \cite{GN5,BG,GR}[P3,P4],
qu'il est possible d'obtenir une troncature
de l'alg\`ebre d'op\'erateurs \`a une sous-alg\`ebre
ferm\'ee d'op\'erateurs de poids r\'eels.

\vskip 2mm

Un correlateur physique \`a quatre points
dont les pattes externes ont pour
nom\-bres quantiques $i,j,k,l$ peut
s'\'ecrire
\beq
<\!i|
\Phi^j(z_2,\zb_2)
\Phi^k(z_3,\zb_3)
|l\!>
=
\sum_{p,\pb}
{\cal \Fb}_{\pb}^{ijkl}(\zb_2,\zb_3)
X_{\pb,p}
{\cal F}_p^{ijkl}(z_2,z_3)
\label{f4pphysnd}
\eeq
o\`u les nombres quantiques internes $p$ et $\pb$ ne
sont a priori pas corr\'el\'es.
Ces fonctions de corr\'elation physiques
doivent \^etre mono-valu\'ees et donc
\^etre invariantes lorsqu'un op\'erateur
tourne autour d'un autre.
Ce sont des contraintes tr\`es fortes
qui permettent g\'en\'eralement de d\'eterminer la matrice $X_{\pb,p}$.

Les blocs \ref{ijklp} \`a partir desquels
est \'ecrit le correlateur physique \ref{f4pdiag}
diagonalisent la monodromie autour de l'origine,
c'est donc la premi\`ere que nous examinons.
Si on note ($\epsilon=\pm 1$)
\beq
\lambda_p=
e^{2i\pi\epsilon(\Delta_k+\Delta_l-\Delta_p)}
\label{vpmonodr}
\eeq
la valeur propre du bloc ${\cal F}_p^{ijkl}$ pour la monodromie
(de $z_3$) autour de 0,
le corr\'elateur physique est transform\'e en
\beq
\sum_{p,\pb}
{\cal \Fb}_{\pb}^{ijkl}(\zb_2,\zb_3)
\lambda_{\pb}^*
X_{\pb,p}
\lambda_p
{\cal F}_p^{ijkl}(z_2,z_3)
\label{f4pmon}
\eeq
par cette monodromie.
Pour qu'il soit inchang\'e, c'est-\`a-dire local,
la matrice $X_{\pb,p}$ doit \^etre stable par transformation
par la matrice Diag($\lambda_p$) diagonale unitaire.
On en d\'eduit que la matrice $X_{\pb,p}$ doit \^etre diagonale par blocs,
chaque bloc correspondant \`a un sous espace propre de
Diag($\lambda_p$).
D'apr\`es \ref{vpmonodr},
deux blocs ${\cal F}_p^{ijkl}$ sont dans le m\^eme espace propre
si les poids $\Delta_p$ correpondants diff\`erent d'un entier.
Dans la plupart des cas,
le spectre des poids de la th\'eorie est tel que
ceci ne se produit pas.
Les espaces propres sont de dimension 1 et la matrice $X_{\pb,p}$
doit alors \^etre diagonale pour que le corr\'elateur
physique soit local.
La th\'eorie est dite diagonale et les fonction de corr\'elation
physiques
s'\'ecrivent\footnote{
Si ${\cal \Fb}={\cal F}^*$.}
\beq
<\!i|
\Phi^j(z_2,\zb_2)
\Phi^k(z_3,\zb_3)
|l\!>
=
\sum_p
d_p
|{\cal F}_p^{ijkl}(z_2,z_3)|^2
\label{f4pdiag}
\eeq
avec $X_{\pb,p}=$Diag($d_p$).

Ce n'est cependant pas toujours le cas.
Dans le cas o\`u les valeurs propres
sont d\'eg\'en\'er\'ees, la matrice $X_{\pb,p}$, qui doit \^etre
hermitienne, peut n\'eanmoins \^etre diagonalis\'ee,
bloc par bloc,
c'est-\`a-dire en conservant une monodromie autour de l'origine
diagonale\footnote{
Il n'est pas \'etonnant que $X_{\pb,p}$ et la matrice de monodromie
puissent \^etre diagonalis\'ees simultan\'ement,
puisque la localit\'e affirme pr\'ecis\'ement
que ces deux matrices commutent.}.
La nouvelle base obtenue,
diagonalise simultan\'ement $X_{\pb,p}$ et la monodromie autour de l'origine,
mais en contrepartie, les \'etats interm\'ediaires
des blocs conformes (\'etiquet\'es par $p$ en \ref{ijklp})
de cette base n'appartiennent pas \`a un module de Verma
d\'etermin\'e\footnote{Ce sont des combinaisons lin\'eaires
d'\'etats dont les poids diff\`erent par des entiers.}
et en particulier les poids droits et gauches peuvent
\^etre diff\'erents.
Il semble d'ailleurs coh\'erent de laisser
alors la m\^eme libert\'e
aux pattes externes,
c'est-\`a-dire de consid\'erer une d\'ecomposition chirale du type
$\Phi^{i,\ib}(z,\zb)=\phi^i(z)\phi^\ib(\zb)$
et des blocs gauches et droits ${\cal F}_p^{ijkl}$ et ${\cal
\Fb}_\pb^{\ib\jb\kb\lb}$
de nombres quantiques gauches et droits d\'ecorr\'el\'es.

C'est ce que nous ferons
dans le dernier
chapitre.
Nous verrons en effet
que les dimensions sp\'eciales,
pour lesquelles nous prouverons le d\'ecouplage d'une sous-alg\`ebre
d'op\'erateurs
de poids r\'eels,
correspondent exactement aux dimensions
o\`u les valeurs propres \ref{vpmonodr} sont d\'eg\'en\'er\'ees.
Ceci
fait appara\^\i tre dans ces dimensions
sp\'eciales un espace propre important
correspondant en fait \`a l'espace engendr\'e par tous les blocs
de poids $\Delta_p$ r\'eels.
Ceci semble avoir une r\^ole capital
dans le d\'ecouplage de cette sous-alg\`ebre physique
de poids r\'eels
et sera discut\'e dans la partie \ref{p5.1b}.

Nous pouvons \'egalement remarquer que les espaces propres
de la monodromie autour de l'origine sont aussi
de dimensions sup\'erieures \`a 1 dans le cas des mod\`eles
minimaux.
C'est assez trivial dans ce cas puisque
ceci provient de l'\'egalit\'e des poids de Kac
$\Delta(n,m)=\Delta(n+p,m+q)$
pour un mod\`ele minimal $(p,q)$.
Cependant, les \'etats $(n+p,m+q)$ se d\'ecouplent
et la th\'eorie est diagonale pour les \'etats qui restent
(ce d\'ecouplage n'a n\'eanmoins plus lieu lorsque ces
mod\`eles minimaux sont habill\'es par la gravit\'e).
Bien que les troncatures des l'alg\`ebres d'op\'erateurs
soient diff\'erentes dans le cas de mod\`eles minimaux
et du couplage fort,
il est remarquable qu'elles aient lieu dans les
deux
cas pour les dimensions o\`u les valeurs
propres de la monodromie sont d\'eg\'en\'er\'ees
(voir discussion dans la partie \ref{p5.1b}).

\vskip 5mm

Nous choisirons dans la suite de ce chapitre de toujours diagonaliser
la matrice $X_{\pb,p}$.
Ceci ne rend pas la th\'eorie diagonale dans le sens habituel du terme
puisque les corr\'elateurs physiques ne s'\'ecrivent alors de mani\`ere
diagonale
que dans une base de blocs conformes qui n'ont pas d'\'etat
interm\'ediaire de poids $\Delta_p$ d\'etermin\'e.
Ce sont des combinaisons lin\'eaires d'\'etats de poids $\Delta_p$
variant par quantit\'es enti\`eres.
Cependant, dans la discussion qui suit,
nous n'utiliserons pas la valeur de ces poids mais uniquement
les valeurs propres $\propto e^{2i\pi\epsilon\Delta_p}$
(Eq.\ref{vpmonodr}) qui,
elles, sont parfaitement d\'etermin\'ees dans cette base.
Nous emploierons donc les notations de Moore et Seiberg,
et il faudra comprendre dans le cas d'une matrice de monodromie
d\'eg\'en\'er\'ee que,
partout (Eqs.\ref{symf3p}, \ref{fusbrd}, \ref{hexa}...),
les poids $\Delta_p$ d\'esignent un quelconque des poids des
\'etats interm\'ediaires de la nouvelle base,
qui sont tous \'egaux modulo un\footnote{
Il y a en toute rigueur dans ces formules,
qui contiennent en fait la racine carr\'ee $\propto e^{i\pi\epsilon\Delta_p}$
de ces valeurs propres,
une ambigu\"\i t\'e de signe $e^{i\pi\epsilon N}$
pour un entier $N$.
Mais elle peut \^etre r\'esolue,
puisque dans le chapitre suivant nous d\'eterminons
l'alg\`ebre de la th\'eorie de Liouville pour une charge
centrale g\'en\'erique,
pour laquelle la th\'eorie est diagonale.
Nous l'appliquons ensuite aux dimensions sp\'eciales du couplage fort
o\`u les corr\'elateurs physiques sont construits
diff\'eremment,
de fa\c con non diagonale,
mais pour lesquelles l'alg\`ebre d'op\'erateurs chiraux
obtenue dans le cas g\'en\'erique reste valable.
}.

\vskip 5mm

Nous venons de voir que les fonctions de corr\'elation physiques
\ref{f4pdiag} n'ont pas de coupures pour la monodromie
de $z_3$ autour de l'origine (i.e. quand l'op\'erateur
$\Phi^k(z_3)$ tourne autour de $\Phi^l(0)$).
Il reste \`a examiner les autres monodromies.

Nous avons \'ecrit les fonctions de corr\'elation physiques
sous la forme \ref{f4pdiag} en termes des blocs conformes \ref{ijklp}
dans leur domaine d'analyticit\'e $z_3/z_2<1$.
Dans le domaine $z_3/z_2>1$ elles s'\'ecrivent
plus naturellement en termes des blocs conformes \ref{ikjlq}.
Cependant la domaine de d\'efinition
de ces blocs a pu \^etre \'etendu par continuation
analytique aux plans coup\'es figures \ref{monodrf} et \ref{monodrf3}.
Nous avons donc deux expressions
de la m\^eme fonction de corr\'elation.
La dualit\'e\footnote{
Il s'agt ici de la dualit\'e $s\to u$,
affirmant que la somme sur les voies $s$ et $u$
sont \'equivalentes.
Nous verrons plus loin la dualit\'e $s\to t$.}
qui va assurer la localit\'e des corr\'elateurs
physiques affirme que leurs deux expressions
en fonctions de ces deux types de blocs conformes
sont \'equivalentes:
\beq
<\!i|
\Phi^j(z_2,\zb_2)
\Phi^k(z_3,\zb_3)
|l\!>
=
\sum_p
d_p
|{\cal F}_p^{ijkl}(z_2,z_3)|^2
=
\sum_q
d_q
|{\cal F}_q^{ikjl}(z_3,z_2)|^2.
\label{dualite}
\eeq
De la deuxi\`eme expression de \ref{dualite}
on d\'eduit en effet comme pr\'ec\'edemment
que le corr\'elateur physique est invariant
lorsque $z_3$ tourne autour de $z_2$ et 0 en m\^eme temps,
ou de mani\`ere \'equivalente autour de l'infini.

Pour des sommes finies\footnote{
Nous v\'erifierons quand nous appliquerons ceci \`a
la th\'eorie de Liouville que, malgr\'e nos repr\'esentations
infinies en nombre infini,
nos fonctions de corr\'elation physiques sont des sommes
d'un nombre fini de termes.
}
de fonctions analytiques,
l'\'egalit\'e des deux sommes dans
\ref{dualite} prouve que les deux ensembles de fonctions
analytiques sont en m\^eme nombre et qu'elles
sont \'equivalentes \`a un changement de base pr\`es\footnote{
\label{orthon}
Pour les fonctions
$(\sqrt{d_p}{\cal F}_p^{ijkl}(z_2,z_3))_p$ et
$(\sqrt{d_q}{\cal F}_q^{ikjl}(z_3,z_2))_q$
il s'agit d'un changement de base orthonormal,
ce qui garantit \ref{dualite}.
Pour les fonctions
$({\cal F}_p^{ijkl}(z_2,z_3))_p$ et
$({\cal F}_q^{ikjl}(z_3,z_2))_q$
la matrice de passage $B^{(\pm)}$ (cf plus loin)
ne le sera pas n\'ecessairement si les coefficients $d_p$
sont non triviaux.
}.
Ceci est vrai ind\'ependamment sur chaque connexe du recouvrement
des plans coup\'es.
Ce sont deux demi-plans, un pour chaque signe de Im$(z_2-z_3)$.
Notant $B^{(\pm)}$ les matrices de passage d'une base \`a l'autre
sur chaque connexe (cf figs.\ref{monodrf}, \ref{monodrf3}), ceci signifie
\beq
<\!i|
\phi_{ip}^j(z_2)\phi_{pl}^k(z_3)
|l\!>
=
\sum_q
B^{(\pm)}_{pq}\left [
_{i\ l}^{j\ k}
\right ]
<\!i|
\phi_{iq}^k(z_3)\phi_{ql}^j(z_2)
|l\!>.
\label{brdf4p}
\eeq
Ceci vaut \'egalement pour des \'etats $<\!i|$ et $|l\!>$
remplac\'es par leurs descendants
(par action des op\'erateurs diff\'erentiels $L_n$ appropri\'es
sur \ref{brdf4p})
puisque les \'el\'ements de la matrice $B^{(\pm)}$ ne dependent pas de $z$.
L'\'egalit\'e vaut donc pour les op\'erateurs (au sens faible):
\beq
\phi_{ip}^j(z_2)\phi_{pl}^k(z_3)=
\sum_q
B^{(\pm)}_{pq}\left [
_{i\ l}^{j\ k}
\right ]
\phi_{iq}^k(z_3)\phi_{ql}^j(z_2)
\label{brd}.
\eeq
Graphiquement cela donne
\beq
\epsffile{ijklp.eps}
\ =\
\sum_q
B^{(\pm)}_{pq}\left [
_{i\ l}^{j\ k}
\right ]
\
\epsffile{ikjlq.eps}.
\label{brdg}
\eeq
Les matrices $B^{(\pm)}$ sont donc les matrices de passage
de la base $s$ (\ref{ijklp}) qui diagonalise la monodromie (de $z_3$)
autour de 0,
\`a la base $u$ (\ref{ikjlq}) qui diagonalise la monodromie
autour de l'infini.
Les matrices $B^{(+)}$ et $B^{(-)}$ sont les matrices de passage
dans un sens et dans l'autre, elles sont donc
inverses l'une de l'autre.
Pour les op\'erateurs $\phi^j$ et $\phi^k$ il s'agit d'une op\'eration
d'\'echange d'o\`u la notation conventielle
de la matrice $B$ pour ``braiding'' en anglais.

Examinons plus pr\'ecis\'ement ce qui se passe lors de deux \'echanges
successifs.
Si, par exemple Im$(z_2-z_3)$ est n\'egatif,
on utilise $B^{(-)}$ pour le premier \'echange
(cf fig.\ref{monodrf}).
Pour revenir \`a l'\'etat initial
le deuxi\`eme \'echange doit alors se faire par la matrice $B^{(+)}$,
comme il a \'et\'e indiqu\'e sur la figure \ref{monodrf3}.
La relation d'inverse entre $B^{(-)}$ et $B^{(+)}$
s'\'ecrit donc plus pr\'ecis\'ement
\beq
\sum_q
B^{(-)}_{pq}\left [
_{i\ l}^{j\ k}
\right ]
B^{(+)}_{qp'}\left [
_{i\ l}^{k\ j}
\right ]
=
\delta_{pp'}.
\label{orthogB}
\eeq
On aurait pu, avant le deuxi\`eme \'echange,
d\'eplacer $z_3$ et le faire descendre plus bas que $z_2$
en passant par sa droite
sur la figure \ref{monodrf3}, en ne traversant que la ligne pointill\'ee.
Dans cette base, on reste dans le domaine d'analyticit\'e,
il ne se passe donc rien, si ce n'est que l'\'echange suivant
s'exprime \`a nouveau par la matrice $B^{(-)}$.
Mais apr\`es ce deuxi\`eme \'echange,
$z_3$ se retrouve dans la base initiale (figure \ref{monodrf}) en dessous
de la coupure en \'etant pass\'e \`a droite de $z_2$,
et il peut alors terminer son tour ``gratuitement'' par la gauche de $z_2$.
Dans ce cas, on se retrouve donc dans la base initiale mais
apr\`es un tour de $z_3$ autour $z_2$.
On en conclut que
le carr\'e de la matrice $B$ est la matrice de monodromie
(qui est tout \`a fait non-triviale,
contrairement \`a $B^{(-)}B^{(+)}$):
\beq
\sum_q
B^{(-)}_{pq}\left [
_{i\ l}^{j\ k}
\right ]
B^{(-)}_{qp'}\left [
_{i\ l}^{k\ j}
\right ]
=
(\hbox{Matrice de monodromie})_{pp'}.
\label{monodrB}
\eeq
Autrement dit, la matrice d'\'echange est la racine carr\'ee
de la matrice de monodromie.

\begin{figure}
\epsffile{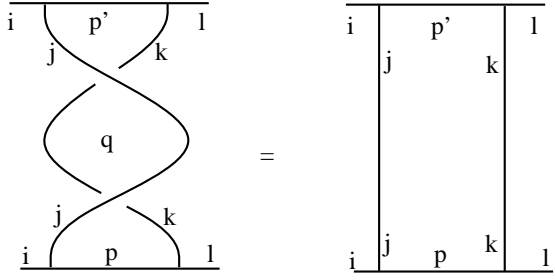}
\caption{$B^{(-)}$ et $B^{(+)}$ sont inverse l'une de l'autre,
la somme sur $q$, nombre quantique interne au
diagramme, est implicite.}
\label{orthogG}
\end{figure}

\begin{figure}
\epsffile{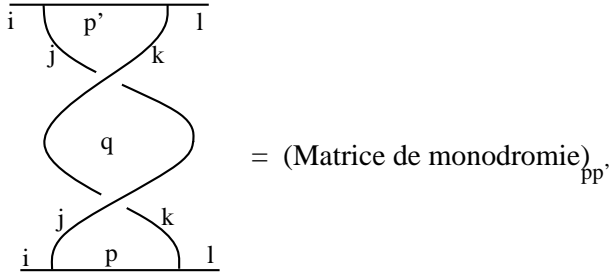}
\caption{L'\'echange est la ``racine carr\'ee'' de la monodromie.}
\label{monodrG}
\end{figure}

Une autre repr\'esentation graphique de l'\'echange,
diff\'erente de \ref{brdg},
illustre bien ce que nous venons de voir.
L'op\'eration \'el\'ementaire d'\'echange est repr\'esent\'ee par
un croisement de lignes (de bas en haut)
\beq
B_{pq}\left [
_{i\ l}^{j\ k}
\right ]
\ \to\
\epsffile{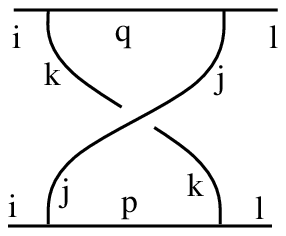}
\label{brdG}
\eeq
o\`u une ligne donn\'ee passe devant ou derri\`ere
selon qu'il s'agit de $B^{(-)}$ ou $B^{(+)}$.
Cette repr\'esentation graphique montre que les coefficients
d'\'echange de l'alg\`ebre d'une th\'eorie conforme coh\'erente
permettent de construire des invariants de liens
(cf refs.\cite{KR}, [P2]).
Combinant deux \'echanges successifs on obtient pour $B^{(-)}B^{(+)}$
la figure \ref{orthogG} qui repr\'esente l'\'equation \ref{orthogB}
et pour $B^{(-)}B^{(-)}$
la figure \ref{monodrG} qui repr\'esente l'\'equation \ref{monodrB}.
Leur interpr\'etation en termes de liens et de n\oe uds est limpide.

On peut appliquer ceci \`a la fonction
\`a trois points $<\!i|\phi_{ik}^j(z)|k\!>$.
On conna\^\i t la monodromie de cette
fonction lorsque $\phi_{ik}^j(z)$ tourne autour de $0$.
Elle acquiert une simple phase (Eq.\ref{monodr3}).
Mais on peut comme pour la fonction \`a quatre points
effectuer un simple \'echange de $\phi^j(z_2)$
et $\phi^k(z_3)$.
On a \`a nouveau le m\^eme genre de coupures selon
la base choisie, et si l'on effectue deux fois de suite
le m\^eme \'echange on obtient la monodromie au lieu
de l'identit\'e.
L'\'echange est donc \`a nouveau donn\'e par la racine carr\'ee
de la monodromie qui, cette fois, est connue (Eq.\ref{monodr3}).
On a par cons\'equent
\beq
\epsffile{ijk.eps}
\!
=
<i|\phi_{i,k}^j(1)|k>
=
\!
e^{i\pi\epsilon(\Delta_j+\Delta_k-\Delta_i)}
\!\!
<i|\phi_{i,j}^k(1)|k>
=
\!
e^{i\pi\epsilon(\Delta_j+\Delta_k-\Delta_i)}
\!
\epsffile{ikj.eps}
\label{symf3p}
\eeq
Ceci peut \^etre vu comme un cas particulier de l'\'echange
\ref{brdg} aver $l=0$, $p=k$ et $q=j$.
Nous apprenons donc que la fonction \`a trois points,
ou de mani\`ere \'equivalente le vertex \'el\'ementaire,
est sym\'etrique quant \`a ses pattes droite et sup\'erieure
(\`a un facteur pr\`es).
Ceci s'av\'erera de la plus grande importance.

\vskip 5mm

Introduisons maintenant l'autre transformation de dualit\'e, la fusion.
Alors que l'\'echange est la dualit\'e $s\to u$,
la fusion n'est autre que la dualit\'e $s\to t$:
\beq
\epsffile{ijklp.eps}=
\sum_q
F_{pq}\left [
_{i\ l}^{j\ k}
\right ]
\epsffile{ijklq.eps}.
\label{fusg}
\eeq
L'expression de la fusion en termes d'op\'erateurs est cependant
l\'eg\`erement diff\'erente de celle de l'\'echange.
Ceci est d\^u \`a l'asym\'etrie de notations
d\'ej\`a remarqu\'ee en note \ref{asym} page \pageref{asym}.
En effet, la ligne interne \'etiquet\'ee par $q$ suppose
que l'on somme sur tous les \'etats $|q,\{\nu\}\!>$ du module de Verma ${\cal
H}_q$,
c'est-\`a-dire les descendants.
Cette somme est implicite sur les ``pattes'' gauches et droites
des operateurs de vertex (\'etats entrants et sortants),
mais pour la patte sup\'erieure, il n'en est rien
car elle d\'esigne l'op\'erateur lui-m\^eme et il faut
expliciter la somme sur les descendants.
La fusion s'\'ecrit donc en explicitant
cette somme:
\beq
\phi_{ip}^j(z_2)\phi_{pl}^k(z_3)=
\sum_q
F_{pq}\left [
_{i\ l}^{j\ k}
\right ]
\sum_{\{\nu\}\in {\cal H}_q}
\phi_{il}^{q,\{\nu\}}(z_2)
<q,\{\nu\}|\phi_{qk}^j(z_2-z_3)|k>
\label{fus}.
\eeq
On pourrait la rendre plus sym\'etrique en explicitant la somme
sur les descendants de ${\cal H}_p$ du membre de gauche
gr\^ace \`a \ref{proj}, \ref{ferm}.
Le membre de gauche peut ainsi s'\'ecrire
$$
\sum_{\{\nu\}\in {\cal H}_p}
\phi_{ip}^j(z_2)
|p,\{\nu\}><p,\{\nu\}|
\phi_{pl}^k(z_3).
$$
Cette base de blocs conformes\footnote{
Nous les notons ${\cal F'}$
contrairement aux blocs \ref{ijklp}
et \ref{ikjlq} not\'es tous les deux ${\cal F}$
puisqu'il s'agissait de la m\^eme fonction
avec indices $k,l$ et arguments $z_2,z_3$
\'echang\'es.
Cette diff\'erence est encore due au
traitement (en apparence)
diff\'erent de la patte sup\'erieure.}
\beq
{\cal F'}_q^{iljk}(z_2,z_3)
=
<\!i|
\phi_{il}^{q,\{\nu\}}(z_2)
|l\!>
<q,\{\nu\}|\phi_{qk}^j(z_2-z_3)|k>
=
\epsffile{ijklq.eps}.
\label{iljkq}
\eeq
est celle qui diagonalise
la derni\`ere monodromie,
celle de $z_3$ autour de $z_2$.
On peut le voir
en faisant effectuer
une rotation $z_2-z_3\to e^{2i\pi}(z_2-z_3)$
\`a l'op\'erateur $\phi_{qk}^j(z_2-z_3)$.
Les fonctions de corr\'elation physiques
s'expriment \'egalement
\`a partir de cette base de blocs conformes par
\beq
<\!i|
\Phi^j(z_2,\zb_2)
\Phi^k(z_3,\zb_3)
|l\!>
=
\sum_q
d_q
|{\cal F'}_p^{iljk}(z_2,z_3)|^2.
\eeq
Et \`a nouveau, la localit\'e de ces corr\'elateurs
physiques pour la monodromie de $z_3$ autour de $z_2$
est \'equivalente \`a la transformation de dualit\'e\footnote{
Comme dans la note \ref{orthon} page \pageref{orthon},
nous notons que ce changement de base doit \^etre
orthonormal pour les blocs renorm\'es
$(\sqrt{d_q}{\cal F'}_q^{ilkj}(z_2,z_3))_q$.
}
\ref{fusg}.

Le terme de fusion n'a pas \'et\'e utilis\'e par
hasard pour d\'esigner cette dualit\'e $s\to t$.
On peut en effet consid\'erer
cette fusion pour $z_2\to z_3$ comme un d\'eveloppement de produit
d'op\'erateurs \`a courte distance.
De mani\`ere standard en th\'eorie des champs,
les op\'erateurs $\phi^i$ admettent un
d\'eveloppement \`a courte distance du type
\beq
\phi^i(z_1)
\phi^j(z_2)
=
\sum_{k,\{\nu\}}
C^{ij}_{k,\{\nu\}}(z_1-z_2)
\phi^{k,\{\nu\}}(z_1)
\label{dominant}
\eeq
les $\phi^{k,\{\nu\}}$ constituant ici un ensemble
complet d'op\'erateurs.
Mais, en th\'eorie confor\-me, la fusion en dit beaucoup plus.
D'abord ce n'est pas qu'un d\'eveloppement \`a courte distance
c'est une \'egalit\'e vraie en tous points $z_i$.
Ensuite, on sait beaucoup de choses sur les coefficients $C^{ij}_{k,\{\nu\}}$
du d\'eveloppement:
non seulement on sait depuis BPZ \cite{BPZ} qu'ils
sont factoris\'es en (d\'ependance en les plus haut poids) $\times$
(d\'ependance en les descendants),
mais on connait de plus la forme de la d\'ependance
en les descendants qui est donn\'ee par
l'\'el\'ement de matrice du quatri\`eme op\'erateur
de l'\'equation \ref{fus}.
C'est cette forme particuli\`ere de la d\'ependance en les descendants
qui fait le lien entre le d\'eveloppement ``traditionnel'' et
l'interpr\'etation de la fusion
en termes de dualit\'e.

La fusion qui est donc en th\'eorie conforme une vraie transformation
de dualit\'e peut \'egalement,
gr\^ace \`a la sym\'etrie
du vertex \'el\'ementaire Eq.\ref{symf3p},
\^etre
reli\'ee \`a l'autre transformation de dualit\'e
qu'est l'\'echange.
La figure \ref{fusbrdg} montre en effet que la fusion
peut \^etre r\'ealis\'ee par deux op\'erations de sym\'etrie
du vertex \`a trois pattes ainsi qu'une op\'eration d'\'echange.
\bfg
\epsffile{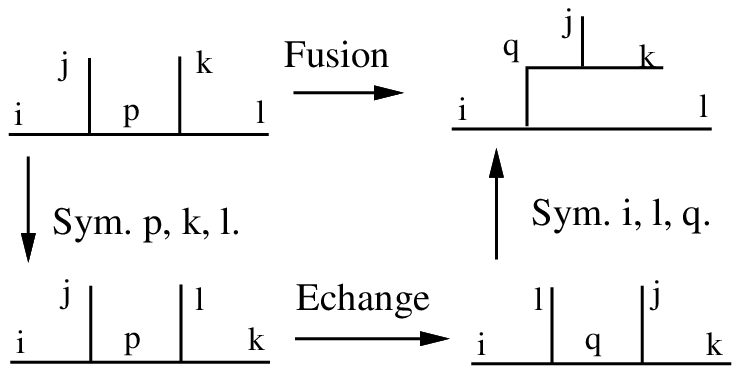}
\caption{La fusion peut \^etre reli\'ee \`a l'\'echange}
\label{fusbrdg}
\efg
Utilisant l'\'equation \ref{symf3p}, le r\'esultat est imm\'ediat:
\beq
F_{pq}\left [
_{i\ l}^{j\ k}
\right ]
=
e^{i\pi\epsilon
(\Delta_p+\Delta_q-\Delta_i-\Delta_k)}
B^{(\epsilon)}_{pq}\left [
_{i\ k}^{j\ l}
\right ]
\label{fusbrd}.
\eeq
Cette relation est d\'ej\`a un exemple, des plus simples,
d'\'equation polynomiale.
Avant de les \'etudier plus syst\'ematiqument
dans la partie suivante,
nous terminons cette partie par deux remarques.

\vskip 2mm

Nous avons discut\'e ici les changements
de bases entre celles des blocs conformes qui diagonalisent
les monodromies autour de 0, 1 et $\infty$.
Nous avons observ\'e que les coupures de ces blocs
\'etaient diff\'erentes selon les bases.
Mais on peut envisager d'autres changements de base,
qui ne diagonalisent aucune monodromie.
C'est ce qui est fait en partie \ref{p4.2} (o\`u on effectue
de mani\`ere \'equivalente
un changement de base sur les op\'erateurs plut\^ot que sur les
blocs conformes).
Les nouveaux op\'erateurs, qui n'ont
pas une monodromie d\'etermin\'ee,
ont en revanche l'avantage
d'\^etre dans des repr\'esentations du groupe quantique
$U_q(sl(2))$.
Ce changement de base est d'un genre plus simple
que ceux que nous venons de voir ici
puisqu'il ne modifie pas la place des coupures.
Et de la m\^eme mani\`ere ces nouveaux op\'erateurs
ont aussi un \'echange (et une fusion) qui relient
les bases avec coupures diff\'erentes.

\vskip 2mm

Dans l'approche de la th\'eorie de Liouville
pr\'esent\'ee ici,
nous parlerons plus souvent (et de mani\`ere \'equivalente)
d'op\'erateurs physiques locaux que de corr\'elateurs physiques locaux.
C'est l'objet de la deuxi\`eme remarque.
Il faudra alors de mani\`ere \'equivalente v\'erifier
que ces op\'erateurs physiques:\\
- sont inchang\'es par monodromie
autour de 0\\
- ont
un \'echange et une fusion
qui ne s'expriment qu'en termes des m\^emes op\'erateurs
ce qui garantira de m\^eme que leurs fonctions
de corr\'elations sont aussi invariantes par monodromie
autour de 1 et de l'infini.

Ceci pourra \^etre fait gr\^ace aux r\'esultats
du chapitre suivant pour l'exponentielle de Liouville \ref{expphi}
du couplage faible introduit dans le chapitre pr\'ec\'edent,
et sera \'egalement fait pour les op\'erateurs
du couplage fort dans le dernier chapitre.

\section{Les \'equations polynomiales}
\label{3.3}

Comme nous l'avons dit en introduction,
les \'equations polynomiales
sont les \'equations de coh\'erence de l'alg\`ebre
qui comprend l'\'echange (Eq.\ref{brdg}),
la fusion (Eq.\ref{fusg}) et la sym\'etrie (Eq.\ref{symf3p}),
ainsi que les transformations modulaires
en genre sup\'erieur ou \'egal \`a un, ce dont nous ne parlerons pas ici.
En effet, vu le nombre important de transformations existantes,
on peut souvent faire la m\^eme transformation
(i.e. passer d'une base de blocs conformes \`a une autre)
de plusieurs fa\c cons diff\'erentes.
La coh\'erence implique donc qu'elles soient \'equivalentes.
La figure \ref{fusbrdg}
en est un exemple.
L'\'equivalence des deux ``voies'' de transformations repr\'esent\'ees
a pour cons\'equence l'\'equation \ref{fusbrd}.
Nous avons aussi vu un autre exemple simple, celui des deux \'echanges
successif $B^{(-)}B^{(+)}$ (Eq.\ref{orthogB}).
Sa repr\'esentation en termes de n\oe uds (fig.\ref{orthogG})
est bien adapt\'ee au cas de l'\'echange.

Appliquons le maintenant \`a la fonction \`a cinq points.
On peut inverser l'ordre des trois op\'erateurs de vertex
de deux fa\c cons diff\'erentes,
en commen\c cant par \'echanger les deux de droite ou
les deux de gauche, comme illustr\'e sur la figure \ref{YBG}.
\bfg
\epsffile{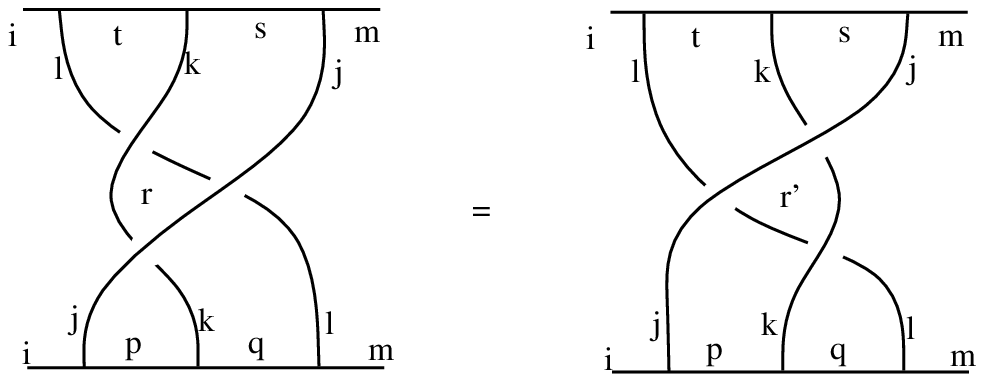}
\caption{Relation de Yang-Baxter}
\label{YBG}
\efg
Ceci, par coh\'erence, ne donne pas autre chose que
la c\'el\`ebre \'equation de Yang-Baxter:
\beq
\sum_r
B^{(\epsilon)}_{pr}\left [
_{i\ q}^{j\ k}
\right ]
B^{(\epsilon)}_{qs}\left [
_{r\ m}^{j\ l}
\right ]
B^{(\epsilon)}_{rt}\left [
_{i\ s}^{k\ l}
\right ]
=\sum_{r'}
B^{(\epsilon)}_{qr'}\left [
_{p\ m}^{k\ l}
\right ]
B^{(\epsilon)}_{pt}\left [
_{i\ r'}^{j\ l}
\right ]
B^{(\epsilon)}_{r's}\left [
_{t\ m}^{j\ k}
\right ]
\label{YB}
\eeq
qui est donc aussi une des \'equations polynomiales.

Combinant fusion et \'echange, on peut obtenir
de tr\`es nombreuses relations.
Moore et Seiberg ont montr\'e
qu'en genre 0 toutes les \'equations polynomiales
pouvaient \^etre obtenues \`a partir
de la relation pentagonale et des deux
relations hexagonales que nous pr\'esentons maintenant.
Ce sont des relations impliquant uniquement la matrice de fusion.
Les \'equations contenant aussi la matrice d'\'echange
sont obtenues gr\^ace \`a la relation \ref{fusbrd}.
L'\'equation pentagonale est l'expression de l'associativit\'e
de la fusion.
La figure \ref{pentag} montre comment on peut fusionner les trois op\'erateurs
de vertex de la fonction \`a cinq points en commen\c cant par la
gauche ou la droite et obtenir le m\^eme r\'esultat.
\bfg
\epsffile{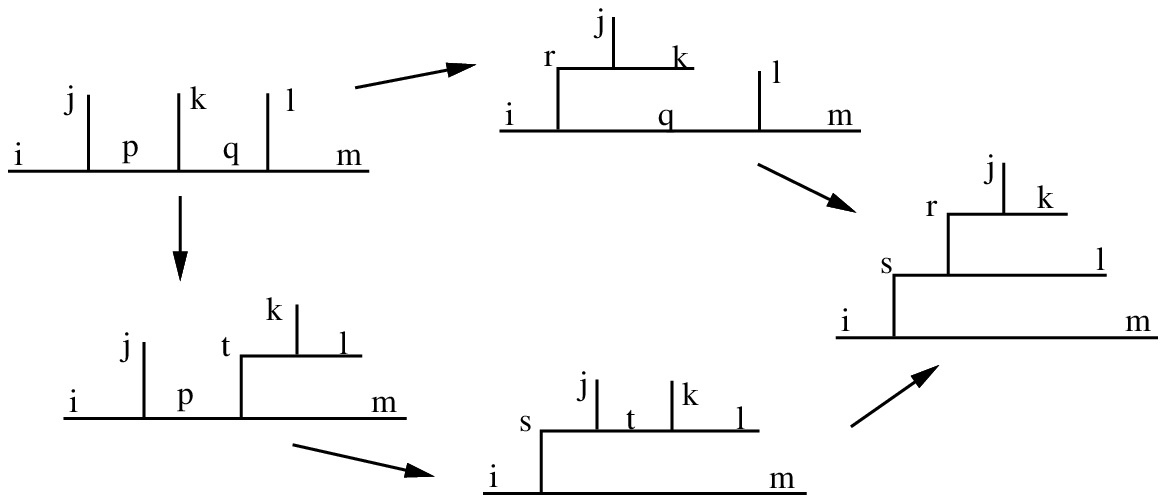}
\caption{Equation pentagonale (associativit\'e de la fusion)}
\label{pentag}
\efg
Contrairement \`a l'associativit\'e usuelle et
comme on le voit sur la figure,
il faut effectuer cinq op\'erations de fusion et non quatre.
En effet, la comparaison de deux fusions
en commen\c cant par la
gauche \`a deux fusions par la droite, donne
$$
\phi_{ip}^j\phi_{pq}^k\phi_{qm}^l=
\sum_{s,r}
F_{pr}\left [
_{i\ q}^{j\ k}
\right ]
F_{qs}\left [
_{i\ m}^{r\ l}
\right ]
\!\!\!\!\!
\sum_{\{\nu_r\}\in {\cal H}_r, \{\nu_s\}\in {\cal H}_s}
\!\!\!\!\!\!\!\!\!\!
\phi_{im}^{s,\{\nu_s\}}
\!
<\!s,\{\nu_s\}|\phi_{sl}^{r,\{\nu_r\}}|l\!>
<\!r,\{\nu_r\}|\phi_{rk}^j|l\!>=
$$
\beq
\sum_{t,s}
F_{qt}\left [
_{p\ m}^{k\ l}
\right ]
F_{ps}\left [
_{i\ m}^{j\ t}
\right ]
\sum_{\{\nu_t\}\in {\cal H}_t, \{\nu_s\}\in {\cal H}_s}
\!\!\!\!\!
\phi_{im}^{s,\{\nu_s\}}
\!
<\!s,\{\nu_s\}|\phi_{st}^j|t,\{\nu_t\}\!>
<\!t,\{\nu_t\}|\phi_{tl}^k|l\!>.
\label{fuspart}
\eeq
Les \'el\'ements de matrice
des deux membres sont diff\'erents,
on ne peut donc rien conclure sur cette expression.
Comme indiqu\'e par la figure \ref{pentag} il faut
effectuer une cinqui\`eme fusion
(d'o\`u le nom d'\'equation pentagonale)
sur les deux op\'erateurs dont les \'el\'ements de matrice
sont dans le membre de droite (apr\`es utilisation
de la relation de fermeture \ref{ferm}).
Ensuite, les \'el\'ements de matrice des deux membres sont identiques,
autrement dit on est dans la m\^eme base des blocs conformes \`a cinq points,
et on peut projeter sur l'\'etat  final de la figure \ref{pentag}
de nombres quantiques $i,j,k,l,m,r,s$ pour obtenir
\beq
F_{pr}\left [
_{i\ q}^{j\ k}
\right ]
F_{qs}\left [
_{i\ m}^{r\ l}
\right ]
=
\sum_{t}
F_{qt}\left [
_{p\ m}^{k\ l}
\right ]
F_{ps}\left [
_{i\ m}^{j\ t}
\right ]
F_{tr}\left [
_{s\ l}^{j\ k}
\right ].
\label{penta}
\eeq

Si, au lieu d'utiliser la forme \ref{fus} de la fusion,
on veut examiner l'associativit\'e de la fusion
par son expression traditionnelle \ref{dominant}
on a deux solutions:
soit on ne consid\`ere que l'ordre dominant,
c'est-\`a-dire uniquement l'op\'erateur primaire $\phi^s$
de poids $\Delta_s$ minimal,
et on n'a plus alors aucune somme dans \ref{fuspart}
et le r\'esultat est assez trivial.
Soit on veut consid\'erer les ordres suivants,
mais alors on doit n\'ecessairement inclure
la multitude des descendants, puisque
la d\'ependance en $z$ du coefficient de fusion
$C^{ij}_{k,\{\nu\}}(z_1-z_2)$
est $(z_1-z_2)^{(\Delta_k+N)-\Delta_i-\Delta_j}$
pour un descendant d'ordre $N$ qui peut
donc tr\`es bien dominer un champ primaire
de poids $\Delta_{k'}$.
Il est fort probable que sans \'ecrire
la d\'ependance dans les descendants sous forme
d'un \'el\'ement de matrice comme en \ref{fus},
on ne puisse pas aller plus loin.

De la m\^eme mani\`ere, les deux \'equations hexagonales
qui avec l'\'equation pentagonale permettent d'engendrer
toutes les \'equations polynomiales,
peuvent \^etre d\'eduites de la figure \ref{hexag}:
\bfg
\epsffile{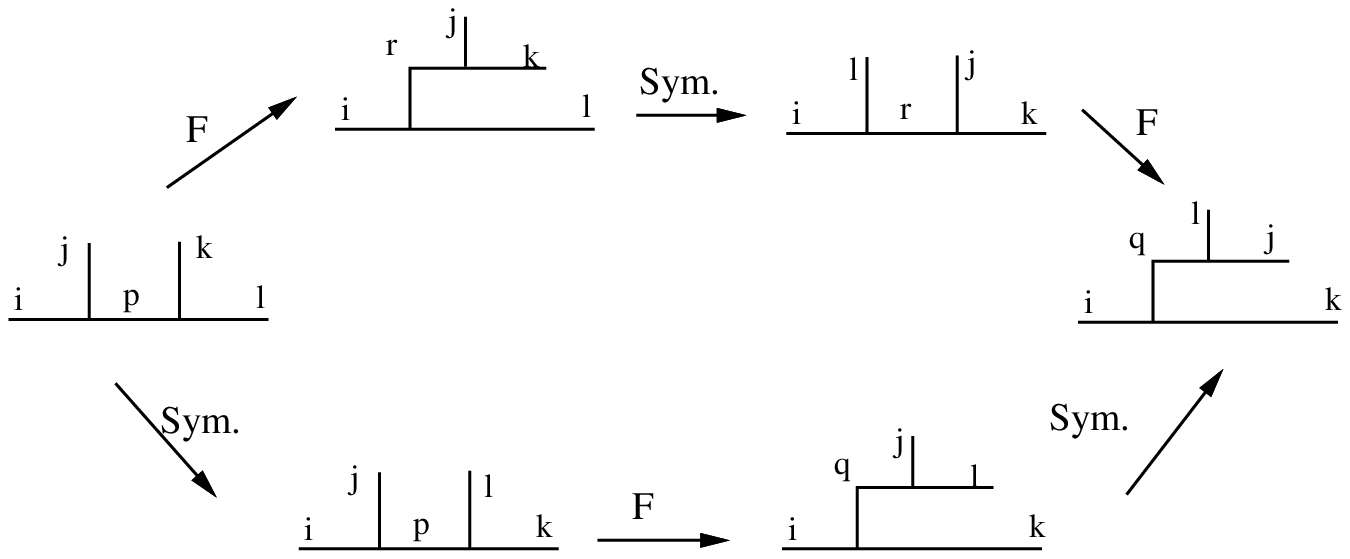}
\caption{Equation hexagonale}
\label{hexag}
\efg
\beq
F_{pq}\left [
_{i\ k}^{j\ l}
\right ]
e^{i\pi\epsilon(\Delta_i+\Delta_j+\Delta_k+\Delta_l-\Delta_p-\Delta_q)}
=
\sum_{r}
F_{pr}\left [
_{i\ l}^{j\ k}
\right ]
F_{rq}\left [
_{i\ k}^{l\ j}
\right ]
e^{i\pi\epsilon\Delta_r}.
\label{hexa}
\eeq

\chapter{THEORIE DE LIOUVILLE:
L'ALGEBRE D'OPERATEURS}
\label{p4}

Je pr\'esente dans ce chapitre,
ainsi que le suivant,
mes contributions \`a la r\'esolution
de la th\'eorie de Liouville.
Elles constituent le corps de cette th\`ese.
Elles sont d\'ecrites en d\'etail dans
les articles [P1] (partie \ref{p4.1})
et [P2] (partie \ref{p4.2})
par E. Cremmer, J.-L. Gervais et moi-m\^eme
pour ce qui est de l'alg\`ebre d'op\'erateurs,
et [P3,P4,P5] (chapitre suivant)
par J.-L. Gervais et moi-m\^eme
pour ce qui est du couplage fort.
Ils sont joints en annexe \`a cette th\`ese
(sauf pour la derni\`ere publication [P5] encore \`a para\^\i tre).
J'y ai \'egalement ajout\'e [P6], ma contribution
aux Journ\'ees Jeunes Chercheurs 93 qui constitue
un r\'esum\'e de la situation, tr\`es succint mais en fran\c cais.
Le lecteur int\'eress\'e trouvera
dans les refs.[P1,P2,P3,P4] jointes en annexe
les consid\'erations
les plus techniques sur lesquelles je ne m'\'etendrai pas ici.

\vskip 2mm

Il est connu depuis longtemps que,
pour l'approche op\'eratorielle de la th\'eorie de Liouville
et plus g\'en\'eralement de certaines th\'eories conformes,
il existe deux bases d'op\'erateurs
ou de blocs conformes
particuli\`erement int\'eressantes.
Les blocs conformes sont en effet l'espace vectoriel
des solutions d'une \'equation diff\'erentielle,
l'\'equation de bootstrap, dont on peut choisir diverses bases.
L'\'equivalent en termes d'op\'erateurs
est l'existence de plusieurs familles conformes de m\^eme poids
$\Delta_{J,\Jhat}$
dont on peut consid\'erer des combinaisons lin\'eaires.

La premi\`ere possibilit\'e consiste \`a
diagonaliser la matrice de monodromie
des op\'erateurs (autour de 0).
Ceci est \'equivalent \`a un choix de base des blocs conformes
dont toutes les ``pattes internes'' (cf \ref{ijklp})
sont des \'etats d'un module de Verma d\'etermin\'e.
C'est le choix fait dans le chapitre 2
pour les op\'erateurs $V^{(J,\Jhat)}_{m,\mhat}$
ainsi que dans le formalisme de Moore et
Seiberg dans le chapitre pr\'ec\'edent.
Ce choix a \'egalement l'avantage de donner des fonctions \`a trois points
qui ont une sym\'etrie des plus simples (Eq.\ref{symf3p}).
En th\'eorie de Liouville,
cette base est g\'en\'eralement appel\'e
base des ondes de Bloch.
Nous d\'eterminons l'alg\`ebre de fusion et d'\'echange de ces
op\'erateurs dans la premi\`ere partie de ce chapitre.
Ses coefficients sont essentiellement donn\'es par des
coefficients de 6-j du groupe quantique.

La deuxi\`eme base digne d'int\'er\^et est celle
du groupe quantique.
Nous d\'eterminons la fusion et l'\'echange de ces op\'erateurs
not\'es
$\xi$ dans la deuxi\`eme partie de ce chapitre.
Ils peuvent \^etre obtenus en demandant que leurs
coefficients de fusion et d'\'echange
ne d\'ependent pas de du module de Verma sur lequel ils
s'appliquent,
ou de mani\`ere \'equivalente du mode-z\'ero $\varpi$.
Il est remarquable que cette seule condition,
qui d\'etermine les op\'erateurs $\xi$ (\`a une normalisation pr\`es),
suffise \`a donner pour ces op\'erateurs des
coefficients de fusion et d'\'echange
qui sont respectivement les Clebsch-Gordan
et la matrice $R$ du groupe quantique.
Ceci met donc en \'evidence la sym\'etrie interne
de la th\'eorie: celle du groupe quantique $U_q(sl(2))$.
C'est clairement l'int\'er\^et majeur de cette base.
L'action de $U_q(sl(2))$ sur ces op\'erateurs n'est
cependant pas explicitement r\'ealis\'ee ici\footnote{
Ceci a \'et\'e fait r\'ecemment par E. Cremmer, J.-L. Gervais
et J. Schnittger et sera publi\'e prochainement.
L'action des g\'en\'erateurs $J_\pm,J_3$
de $U_q(sl(2))$ est r\'ealis\'e par la commutation avec
les op\'erateurs $\xi^{(1/2)}_{\pm1/2}$.
Ceci leur a \'egalement permis de d\'eterminer l'action de $U_q(sl(2))$
sur les op\'erateurs $V$.},
et nous nous int\'eresserons en fait davantage aux
transformations d'\'echange et de fusion
qui permettent d'examiner la localit\'e des corr\'elateurs physiques.
En revanche, le formalisme de Moore et Seiberg
ne s'applique pas directement \`a cette base
(puisqu'elle ne diagonalise aucune monodromie)
et en particulier la fusion de ces op\'erateurs
contient un \'el\'ement de matrice des op\'erateurs $V$.
Alors que les op\'erateurs $V$ de la base de Bloch
sont reli\'es \`a un mod\`ele IRF (Interaction
Round a Face),
les op\'erateurs $\xi$ du groupe quantique
sont reli\'es \`a un mod\`ele \`a vertex.

La base des ondes de Bloch, plus naturelle \`a priori,
a pr\'ec\'ed\'e celle du groupe quantique.
La matrice de changement de base a \'et\'e d\'etermin\'ee
par Babelon \cite{B1} pour le spin $1/2$
puis par Gervais \cite{G1} pour tous spins,
Babelon en donnant finalement une forme universelle \cite{B2}.
Babelon a \'egalement d\'etermin\'e en ref.\cite{B2}
la forme universelle de la matrice d'\'echange des op\'erateurs $V$,
sans remarquer qu'elle \'etait donn\'ee par des 6-j
(\`a des normalisations pr\`es).

\vskip 2mm

Pr\'ecisons ce qu'il y a de nouveau dans les travaux pr\'esent\'es ici.

L'\'echange des op\'erateurs $V$ de type IRF \'etait d\'ej\`a
connu par la formule
universelle de Babelon \cite{B2}.
Cependant, le rapport entre les coefficients d'\'echange
des $V$ et les 6-j de $U_q(sl(2))$
n'\'etait pas \'etabli\footnote{
Il avait \'et\'e \'etabli en ref.\cite{FFK}
dans l'approche du gaz de Coulomb un lien entre
ces coefficients d'\'echange et des poids de Boltzman de
mod\`eles IRF, eux-m\^emes en rapport avec des 6-j.}.
La fusion, en revanche, n'\'etait utilis\'ee
qu'\`a l'ordre dominant pour engendrer des op\'erateurs de spins sup\'erieurs.
Outre les coefficients d'\'echange d\'ej\`a plus ou moins connus,
nous d\'eterminons donc \'egalement la forme de la fusion,
prouvant pour ce cas particulier la forme g\'en\'erale \ref{fus} de
la fusion \`a la Moore et Seiberg, ainsi que l'expression des coefficients de
fusion
des op\'erateurs de type IRF.
Mais surtout la nouveaut\'e importante
est l'extension de ces r\'esultats \`a des spins continus,
c'est-\`a-dire \`a des repr\'esentations non r\'eductibles
de l'alg\`ebre de Virasoro, hors de la table de Kac.
Cette extension s'est av\'er\'ee n\'ecessaire pour
pouvoir traiter le cas des
spins fractionnaires (non demi-entiers) qu'il faut absolument
inclure
dans le r\'egime de couplage fort,
comme nous le verrons dans le prochain chapitre.
Ceci nous a amen\'es en [P3] \`a g\'en\'eraliser les 6-j
\`a des spins continus et \`a prouver les \'equations
de Bidenharn-Elliot, de Racah et d'orthogonalit\'e
correspondantes.
Une caract\'eristique de cette extension \`a des spins
continus est que la double structure de groupe quantique
$U_q(sl(2))\odot U_\qhat (sl(2))$ n'est plus
dans ce cas une simple structure de (quasi) produit tensoriel.
Les quantit\'es correspondant aux param\`etres de d\'eformation
$h$ et $\hhat$ (comme $J$ et $\Jhat$) ne peuvent plus \^etre s\'epar\'ees.

Pour les op\'erateurs $\xi$ du groupe quantique,
on calcule la fusion \`a tous les ordres.
Alors que l'\'echange est donn\'e par
la matrice $R$ universelle,
les coefficients de fusion sont des Clebsch-Gordan de $U_q(sl(2))$.
Nous faisons \'egalement ceci pour des spins continus
qui donnent des repr\'esentations infinies de $U_q(sl(2))$.
Appliquant correctement le changement de base des $V$ aux $\xi$,
l'expression de la fusion des $V$ \`a la Moore et Seiberg
montre que la fusion des $\xi$ contient un \'el\'ement
de matrice des $V$,
qui peut \^etre vu comme exprimant la d\'ependance
en $z$ pour les descendants obtenus par fusion.
Ceci est absolument n\'ecessaire
pour que la fusion
des $\xi$ soit associative.
En effet, du point de vue du groupe quantique,
la fusion des $\xi$ (qui sont dans des repr\'esentations
du groupe quantique)
correspond \`a la d\'ecomposition en repr\'esentations irr\'eductibles
du produit tensoriel de repr\'esentations,
qui s'exprime par des Clebsch-Gordan.
Or, comme chacun le sait,
effectuer cette d\'ecomposition pour le produit
tensoriel de trois repr\'esentations en commen\c cant
par la droite ou par la gauche ne donne pas le m\^eme
r\'esultat, ce qui mettrait en d\'efaut une associativit\'e
na\"\i ve de la fusion des $\xi$.
Le lien entre ces deux d\'ecompositions est fait
par des 6-j, et ce sera pr\'ecis\'ement le r\^ole des \'el\'ements
de matrice des $V$, dont la fusion est donn\'ee par des 6-j,
que d'assurer l'associativit\'e de la fusion des $\xi$.

Avant le changement de base
de Babelon et Gervais,
une autre mani\`ere de passer d'un mod\`ele IRF
\`a un mod\`ele \`a vertex avait \'et\'e propos\'e par
Pasquier (voir ref.\cite{P} puis \cite{MR}).
Partant d'un mod\`ele IRF (ou SOS, ``Solid On Solid''),
dont l'alg\`ebre est donn\'ee par des 6-j,
on peut en effet construire,
par le biais de Clebsch-Gordan,
des op\'erateurs d'un mod\`ele \`a vertex.
Il ne s'agit pas \`a proprement parler
d'un changement de base
puisque les op\'erateurs de type vertex vivent dans un espace
diff\'erent, \'etendu par rapport \`a l'espace de Hilbert initial.
Ici aussi,
l'\'echange des op\'erateurs de type mod\`ele \`a vertex
est donn\'e par la matrice $R$
alors que celui des op\'erateurs de type IRF \'etait
donn\'e par des 6-j.
La coexistence de ces deux transformations est longtemps
rest\'ee une \'enigme, le lien entre elles \'etant inconnu.
Nous avons montr\'e en [P2] que la matrice de changement
de base de Babelon et Gervais est la limite
des Clebsch-Gordan de ref.\cite{P,MR}
lorsque le nombre quantique suppl\'ementaire
introduit en \cite{P,MR} tend vers l'infini.
J'explique ici plus pr\'ecis\'ement qu'en [P2] comment cette limite
rend triviale l'extension de l'espace de Hilbert,
permettant d'identifier compl\`etement cette
limite \`a la construction pr\'ec\'edente.

Ceci nous a donn\'e l'id\'ee d'\'etudier une autre limite infinie:
dans celle-ci,
le mod\`ele IRF correspondant aux $V$
a pour limite le mod\`ele \`a vertex li\'e aux $\xi$
(quand les spins sur les faces
tendent vers l'infini\footnote{
Ce lien entre mod\`eles IRF et \`a vertex a \'et\'e mis en \'evidence par
Witten \cite{W} dans un cadre plus g\'en\'eral.}).
Les 6-j, poids de Boltzman du mod\`ele IRF,
ont pour limite
des \'el\'ements de matrice $R$
ou des Clebsch-Gordan.
Ceci n'a \'et\'e fait en [P2] que pour des spins
demi-entiers\footnote{
De plus, il faut donner une partie imaginaire
aux spins qui tendent vers l'infini, ce qui exige
une extension des 6-j \`a des spins non demi-entiers.
Cette extension avait \'et\'e faite de mani\`ere arbitraire
en [P2], en particulier sans v\'erifier que ces 6-j
g\'en\'eralis\'es satisfaisaient toujours aux \'equations
de Bidenharn-Elliot, de Racah et d'orthogonalit\'e.
C'est pourquoi nous r\'eexaminons ici cette limite
en utilisant les 6-j g\'en\'eralis\'es en [P3],
pour lesquels les \'equations pr\'ec\'edentes ont \'et\'e prouv\'ees.}.
{\sl
La partie qui traite de ces limites n'a donc pas \'et\'e
publi\'ee ailleurs et sera imprim\'ee en italique
de mani\`ere \`a ressortir.}
Cette limite permet ensuite de d\'emontrer \`a partir
des \'equations v\'erifi\'ees par les 6-j
toutes les \'equations impliquant des Clebsch-Gordan
et des \'el\'ements de matrice $R$ pour des spins continus.

Nous avons \'egalement
propos\'e en ref.[P2] une repr\'esentation
g\'eom\'etrique tridimensionnelle
de l'alg\`ebre des $\xi$ en termes de t\'etra\`edres.
Une telle repr\'esentation du mod\`ele IRF,
pour lequel les t\'etra\`edres sont des 6-j
dont les six spins sont sur les ar\^etes,
\'etait d\'ej\`a connue.
C'est ce qui a permis de formuler la gravit\'e simplicielle
tridimensionnelle,
en pavant l'espace de t\'etra\`edres \cite{DisGr}.
Notre repr\'esentation en est une g\'en\'eralisation,
certains t\'etra\`edres \'etant des Clebsch-Gordan
ou des \'el\'ements de matrice $R$
dont les nombres quantiques magn\'etiques $M$
sont figur\'es sur les faces.
Nous ne reprendrons pas cette repr\'esentation
dans le corps de cette th\`ese, par souci de bri\`evet\'e.

\section{La base des ondes de Bloch}
\label{p4.1}

Nous d\'eterminons ici compl\`etement l'alg\`ebre
des op\'erateurs $V$ de la base qui diagonalise la monodromie,
celle des ondes de Bloch.
C'\'etait \'egalement le choix fait par
Moore et Seiberg dans le chapitre pr\'ec\'edent,
et leur formalisme s'appliquera donc directement.
Les coefficients de cette alg\`ebre d'op\'erateurs peuvent \^etre
interpr\'et\'es comme les poids de Boltzmann d'un
mod\`ele IRF ou SOS.

Par souci de lisibilit\'e,
nous \'enon\c cons les r\'esultats dans une premi\`ere
sous-partie et esquissons la m\'ethode de r\'esolution
(par transformation des fonctions \`a quatre points
calculables par \'equation de boostrap)
dans une deuxi\`eme sous-partie,
qui peut \^etre consid\'er\'ee comme un guide
de lecture des articles joints en annexe.
Consacrer une partie enti\`ere \`a
l'expos\'e des r\'esultats nous a paru n\'ecessaire
pour expliquer correctement
la g\'en\'eralisation des spins
demi-entiers aux spins continus.

\subsection{Les r\'esultats}

\subsubsection{Le cas standard: les spins demi-entiers}

Commen\c cons par les op\'erateurs standards
$V^{(J,\Jhat)}_{m,\mhat}$
introduits
pr\'ec\'edemment, de spins $J$ et $\Jhat$
demi-entiers.
Nous rappelons qu'ils engendrent des modules de Verma
r\'eductibles correspondant aux cas $(2\Jhat+1,2J+1)$ de la classification
de BPZ que nous notons ${\cal H}_{J,\Jhat}$ (Eq.\ref{etatsL}).

Les indices $i,j,k,l,p...$ g\'en\'eriques de Moore et Seiberg
sont donc dans le cas pr\'esent chacun un couple de spins\footnote{
Il a \'et\'e introduit en [P1] une notation $\Jge\equiv(J,\Jhat)$
que nous n'utiliserons pas ici afin de ne pas multiplier les notations.}
$(J_i,\Jhat_i)$.
Nous rappelons que
les op\'erateurs du chapitre pr\'ec\'edent $\phi_{ik}^j$, pour lesquels
les nombres quantiques des trois pattes sont pr\'ecis\'es,
sont des restrictions des op\'erateurs de Liouville du chapitre 2.
Utilisant pour la derni\`ere fois la notation $\phi^._{..}$
du chapitre pr\'ec\'edent, on a la correspondance:
\beq
\phi_{(J_{12},\Jhat_{12}),(J_2,\Jhat_2)}^{(J_1,\Jhat_1),\{\nu\}}
=
{\cal P}_{J_{12},\Jhat_{12}}
V^{(J_1,\Jhat_1),\{\nu\}}_{J_2-J_{12},\Jhat_2-\Jhat_{12}}
{\cal P}_{J_{2},\Jhat_{2}}
\eeq
o\`u ${\cal P}_{J_{12},\Jhat_{12}}$ est le projecteur
sur le module de Verma ${\cal H}_{J_{12},\Jhat_{12}}$.

Un bloc conforme \`a quatre points est donc repr\'esent\'e
graphiquement par
\beq
\epsffile{f4p.eps}
=\
<\!J,\Jhat|
V^{(J_1,\Jhat_1)}_{J_{23}-J,\Jhat_{23}-\Jhat}
\,
V^{(J_2,\Jhat_2)}_{J_3-J_{23},\Jhat_3-\Jhat_{23}}
|J_3,\Jhat_3\!>
\label{f4pg}.
\eeq
Les points introduits sur les vertex indiquent que
les r\`egles de s\'election\footnote{
Etant\label{selec} donn\'e trois repr\'esentations de l'alg\`ebre
de Virasoro,
dans une th\'eorie,
il existe ou il n'exite pas
de vertex interpolant (``intertwinning'')
entre elles (comme \ref{ThreeCond},
\ref{ThreeCondEff}, \ref{OneCondEff}...).
C'est ce que nous
appelons ici r\`egles de s\'election.
Elles sont
appel\'ees ailleurs r\`egles de fusion,
en particulier par Verlinde\cite{V}
o\`u elles sont repr\'esent\'ees par les
entiers naturels $N_{ij}^k$ qui valent
0 pour les transitions interdites.
Nous n'utilisons pas ici le terme de r\`egles de fusion
pour ne pas risquer de cr\'eer de confusion
avec la fusion au sens plein du terme (Eqs.\ref{fus}, \ref{fusgjj}...).}
pour des spins demi-entiers sont les in\'egalit\'es
triangulaires standard\footnote{
${\cal N}$ d\'esigne l'ensemble des entiers naturels.}
de $Sl(2)$:
\beq
\epsffile{threecon.eps}
\to
\left\{
\begin{array}{ccc}
J_1+J_2-J_{12}  \in {\cal N},\\
J_{12}+J_2-J_1  \in {\cal N},\\
J_1+J_{12}-J_2  \in {\cal N},\\
\Jhat_1+\Jhat_{12}-\Jhat_{2}  \in {\cal N}\\
\Jhat_1+\Jhat_2-\Jhat_{12}  \in {\cal N}\\
\Jhat_{12}+\Jhat_2-\Jhat_{1}  \in {\cal N}\\
\end{array}
\right\}
\ \Rightarrow\
\left\{
\begin{array}{ccc}
2J_1,2J_2,2J_{12} \in {\cal N}\\
2\Jhat_1,2\Jhat_2,2\Jhat_{12} \in {\cal N}
\end{array}
\right.
\label{ThreeCond}.
\eeq
Nous reviendrons sur la signification de ces points
lors de la g\'en\'eralisation \`a des spins continus.
Nous en reparlerons dans la prochaine sous-partie,
mais nous pr\'ecisons d\'ej\`a ici que les r\`egles de s\'election
\ref{ThreeCond}
ne sont pas du tout {\sl ad hoc} mais proviennent du calcul.

Nous donnons maintenant l'expression de la fusion et de l'\'echange
de ces op\'erateurs.
Conform\'ement au sch\'ema g\'en\'eral de Moore et Seiberg,
la fusion s'\'ecrit
\beq
\epsffile{f4p.eps}
=
\sum_{J_{12},\Jhat_{12}}
F_{({J_{23},\Jhat_{23})},{(J_{12},\Jhat_{12})}}\!\!\left[^{(J_1,\Jhat_1)}_{(J,\Jhat)}
\,^{(J_2,\Jhat_2)}_{(J_3,\Jhat_3)}\right]
\epsffile{f4pfus.eps}
\label{fusgjj}.
\eeq
Nous avons calcul\'e en [P1] la valeur du coefficient
de fusion:
$$
F_{(J_{23},\Jhat_{23}),(J_{12},\Jhat_{12})}\!\!\left[^{(J_1,\Jhat_1)}_{(J,\Jhat)}
\,^{(J_2,\Jhat_2)}_{(J_3,\Jhat_3)}\right]
=
$$
\beq
e^{i\pi f_V(J_i)}
{g_{(J_1,\Jhat_1),(J_2,\Jhat_2)}^{(J_{12},\Jhat_{12})}\
g_{(J_{12},\Jhat_{12}),(J_3,\Jhat_3)}^{(J,\Jhat)}
\over
g_{(J_2,\Jhat_2),(J_3,\Jhat_3)}^{(J_{23},\Jhat_{23})}\
g_{(J_1,\Jhat_1),(J_{23},\Jhat_{23})}^{(J,\Jhat)}
}
\left\{ ^{J_1}_{J_3}\,^{J_2}_{J}
\right. \left |^{J_{12}}_{J_{23}}\right\}
\gaghat
\,^{\Jhat_1}_{\Jhat_3}\,^{\Jhat_2}_{\Jhat}
\bigr. \bverthat ^{\Jhat_{12}}_{\Jhat_{23}}\gadhat,
\label{Fjj}
\eeq
o\`u les 6-j sont des 6-j d\'eform\'es
(du groupe quantique) avec param\`etre de d\'eformation $h$
pour le premier et $\hhat$ pour le deuxi\`eme figur\'e avec des chapeaux.
Il s'agit des 6-j de normalisation non sym\'etrique
(pas ceux de Racah-Wigner),
comme indiqu\'e par le trait vertical qui isole les spins
$J_{12}$ et $J_{23}$
(cf \cite{KR,HHM}).
Leur expression est donn\'ee plus loin en Eq.\ref{6j}
dans le cas de spins continus (cf aussi [P3] Eqs.3.5, 3.25).
Les coefficients $g$ ont pour valeur
\beq
g_{(J_1,\Jhat_1),(J_2,\Jhat_2)}^{(J_{12},\Jhat_{12})}= (-1)^{p\phat}
(i/2)^{p+\phat}{H_{p\phat}(\varpi_{J_1,\Jhat_1})H_{p\phat}(\varpi_{J_2,\Jhat_2})
H_{p\phat}(-\varpi_{J_{12},\Jhat_{12}})
\over{H_{p\phat}(\varpi_{p/2,\phat /2})}}
\label{gjj}
\eeq
avec $p\equiv J_1+J_2-J_{12}$, $\phat\equiv\Jhat_1+\Jhat_2-\Jhat_{12}$ et
\beq
H_{p\phat}(\varpi) =
{
\prod_{r=1}^p \sqrt{F(\varpihat -r h/\pi )}
\prod_{\rhat =1}^{\phat}
\sqrt{F(\varpi - \rhat \pi /h )}
\over
\prod_{r=1}^p \prod_{\rhat =1 }^{\phat }
\left ( \varpi \sqrt{h/\pi}
-r \sqrt{h/\pi} -\rhat \sqrt{\pi /h}\right ) }
\ ,\
F(z)\equiv{\Gamma(z)\over\Gamma(1-z)}
\label{H}.
\eeq
Il y a \'egalement une expression plus
\'el\'egante de $H$ en termes de chemins
(voir ref.[P1] Eq.4.26 ou ref.[P3] Eq.4.20).
La phase $f_V$ (donn\'ee en [P1] Eq.4.8
et avec correction en [P3] Eq.4.21)
est un demi-entier et donne donc juste une puissance de $i$.

L'\'echange est donn\'e par
\beq
\epsffile{f4p.eps}
=
\sum_{J_{13},\Jhat_{13}}
B^{(\pm)}_{({J_{23},\Jhat_{23})},{(J_{13},\Jhat_{13})}}
\!\!\left[^{(J_1,\Jhat_1)}_{(J,\Jhat)}
\,^{(J_2,\Jhat_2)}_{(J_3,\Jhat_3)}\right]
\epsffile{f4pbrd.eps}
\label{brdgjj}
\eeq
avec
$$
B^{(\pm)}_{({J_{23},\Jhat_{23})},{(J_{13},\Jhat_{13})}}
\!\!\left[^{(J_1,\Jhat_1)}_{(J,\Jhat)}
\,^{(J_2,\Jhat_2)}_{(J_3,\Jhat_3)}\right]
=
e^{i\pi f_B(J_i)}
e^{\pm i\pi (\Delta_{J,\Jhat}+\Delta_{J_3,\Jhat_3}
-\Delta_{J_{23},\Jhat_{23}}-\Delta_{J_{13},\Jhat_{13}})}
$$
\beq
{g_{(J_1,\Jhat_1),(J_3,\Jhat_3)}^{(J_{13},\Jhat_{13})}
g_{(J_2,\Jhat_2),(J_{13},\Jhat_{13})}^{(J,\Jhat)} \over
g_{(J_2,\Jhat_2),(J_3,\Jhat_3)}^{(J_{23},\Jhat_{23})}
g_{(J_1,\Jhat_1),(J_{23},\Jhat_{23})}^{(J,\Jhat)}}
\left\{
^{J_1}_{J_2}\,^{J_3}_{J_{123}}
\right. \left |^{J_{13}}_{J_{23}}\right\}
\gaghat
\,^{\Jhat_1}_{\Jhat_2}\,^{\Jhat_3}_{\Jhat_{123}}
\bigr. \bverthat\, ^{\Jhat_{13}}_{\Jhat_{23}}\gadhat
\label{Bjj}
\eeq
o\`u $f_B$  peut par exemple \^etre d\'eduit de $f_V$ par sym\'etrie
de la fonction \`a trois points.

Que ce soit pour l'\'echange
ou la fusion, les coefficients $g$ peuvent \^etre
interpr\'et\'ees comme des normalisations des op\'erateurs
de vertex ce qui permet de d\'efinir
des op\'erateurs renorm\'es $\Vt$
\beq
{\cal P}_{J_{12},J_{12}}
\Vt^{(J_1,J_1)}(z)
{\cal P}_{J_2,J_2}
=
g_{(J_1,\Jhat_1),(J_2,\Jhat_2)}^{(J_{12},\Jhat_{12})}
{\cal P}_{J_{12},J_{12}}
V^{(J_1,J_1)}(z)
{\cal P}_{J_2,J_2}
\label{Vtilde}
\eeq
dont l'alg\`ebre est donn\'ee uniquement par des 6-j
(et les exponentielles des poids conformes pour l'\'echange,
comme il se doit).
Les \'equations d'orthogonalit\'e, de
Racah et de Bidenharn-Elliot pour les 6-j q-d\'eform\'es\cite{KR}
permettent de prouver que les coefficients de fusion et d'\'echange
des $\Vt$ v\'erifient les \'equations polynomiales.
Dans le cas des op\'erateurs $V$, les normalisations $g$
s'annulent deux \`a deux et disparaissent des \'equations polynomiales.
Il faut \'egalement v\'erifier que les phases $f_V$ et $f_B$
s'annulent bien.
Ceci est fait en [P1] (avec correction en [P3])
et nous ne nous \'etendons pas davantage sur ces d\'etails ici,
d'autant que ces phases disparaissent tr\`es \'el\'egamment
dans le cas de spins continus, auquel nous passons maintenant.

\subsubsection{Les spins continus}

L'introduction de spins non demi-entiers,
c'est-\`a-dire de repr\'esentations irr\'eductibles de l'alg\`ebre
de Virasoro,
est apparue n\'ecessaire pour obtenir une fonction
de partition invariante modulaire dans le r\'egime de couplage fort,
que nous abordons dans le prochain chapitre.
Il s'agit donc ici de se donner les outils techniques
n\'ecessaires \`a la r\'esolution de la th\'eorie physique
dans le r\'egime de couplage fort
au chapitre \ref{p5}.

Le lecteur attentif a s\^urement d\'ej\`a remarqu\'e
que les blocs conformes consid\'er\'es
initialement par Gervais et Neveu
(chapitre 2) \'etaient plus g\'en\'eraux
que ceux dont nous venons de d\'eterminer les matrices
$B$ et $F$.
Les moments $\varpi$ des \'etats pouvaient en effet
\^etre quelconques,
alors que nous les avons ici restreints
aux repr\'esentations r\'eductibles de moments $\varpi=\varpi_{J,\Jhat}$.
Mais la correspondance entre \'etats et op\'erateurs exigeait
de consid\'erer \'egalement des op\'erateurs de spins quelconques.

Il est facile pour commencer de g\'en\'eraliser
\`a des spins continus, not\'es $\Je$, l'exponentielle
de champ libre \ref{Vj-j}
\beq
V^{(\Je)}_{-\Je}\propto
e^{2\Je\sqrt{h/2\pi}q_0^{(1)}}
\> : \!
e^{2\Je\sqrt{h/2\pi}(p_0^{(1)}\sigma
+i\sum_{n\ne 0}e^{-in\sigma}p_n^{(1)}/n)}
\!:
\hbox{ pour }
\Je\in R
\hbox{ (ou m\^eme }
C).
\label{VJe-Je}
\eeq
Ce sont les op\'erateurs\footnote{
On peut v\'erifier que la limite \ref{op->et}
est nulle pour les op\'erateurs $V^{(\Je)}_{\Je}$
construits \`a partir de l'autre champ libre $P^{(2)}$
appliqu\'es au vide $|\varpi_0\!>$.
Ces op\'erateurs peuvent \'egalement engendrer les \'etats
mais \`a partir de l'autre vide
$Sl(2,C)$ invariant $|-\varpi_0\!>=|-1-\pi/h\!>$.
Ceci nous m\`enerait ensuite \`a des r\`egles de s\'election TI1
(Eq.\ref{OneCondEff}) sym\'etris\'ees:
$\Je_1+\Je_{12}-\Je_{2} \in {\cal N}\!\!+\!{\pi\over h}{\cal N}$.}
qui engendrent
tous les \'etats de plus haut poids \`a partir du
vide
$Sl(2,C)$ invariant $|\varpi_0\!>=|1+\pi/h\!>$:
\beq
V^{(\Je)}_{-\Je}(z)
|\varpi_0\!>
\to
|\varpi_{\Je}\!>
\hbox{  quand  }
z\to 0
\label{op->et}
\eeq
o\`u on a g\'en\'eralis\'e la notation
\beq
\varpi_{\Je}=\varpi_0+2\Je \hbox{ pour } \Je\in R \ (\hbox{ou } \Je\in C).
\label{varpieff}
\eeq
Mais nous arr\^eterons l\`a les consid\'erations
les plus techniques que nous renvoyons \`a
la sous-partie suivante.
Nous y examinons les possibilit\'es de g\'en\'eralisation
dans l'approche de Gervais et Neveu ainsi que la
compl\'ementarit\'e de celle du gaz de Coulomb.
Nous nous contentons ici d'expliquer les r\'esultats.

\vskip 2mm

Revenons un instant au cas des spins demi-entiers.
Un \'etat (ou un op\'erateur) de spins $J$ et $\Jhat$
peut de mani\`ere \'equivalente \^etre caract\'eris\'e par
le mode-z\'ero $\varpi=$ $\varpi_{J,\Jhat}\equiv$ $\varpi_0+2J+2\Jhat\pi/h$.
Son poids, en particulier, n'est fonction que de $\varpi_{J,\Jhat}$.
On peut \'ecrire $\varpi_{J,\Jhat}=\varpi_0+2\Je$
pour un spin effectif\footnote{
La notation $\Je=J+\Jhat\pi/h$ est asym\'etrique entre les quantit\'es chapeau
et non-chapeau, mais elle a sa contre-partie sym\'etris\'ee
$\Jehat\equiv h/\pi\Je=\Jhat+J\pi/\hhat=\Jhat+J h/\pi$.}
$\Je\equiv J+\Jhat\pi/h$ qui est limit\'e dans le cas
demi-entier
\`a des valeurs discr\`etes.
L'\'etat ou op\'erateur en question est compl\`etement d\'etermin\'e
par la donn\'ee de son spin effectif.
On v\'erifiera en particulier dans la sous-partie suivante
que les coefficients de fusion ou d'\'echange
sont des fonctions analytiques des spins effectifs.
Et la chose \'etonnante est plut\^ot le fait qu'ils
aient pu se factoriser en produits d'un 6-j fonction
des $J_i$ par un 6-j fonction
des $\Jhat_i$.
Ceci est d\^u aux propri\'et\'es de la fonction gamma pour
des variations enti\`eres de son argument.
L'extension \`a des spins continus consiste donc logiquement
\`a consid\'erer des spins effectifs $\Je_i$ quelconques,
comme nous l'avons d\'ej\`a  fait dans le cas particulier \ref{VJe-Je}.
Ceci n'a rien d'\'etonnant du point de vue du gaz de Coulomb\cite{GS1,GS3}.
La correspondance entre les deux approches montre en effet
que nos op\'erateurs
peuvent s'\'ecrire
\beq
V^{(J,\Jhat)}_{m,\mhat} \sim
V^{(J+\Jhat \pi/h)}_{-J-\Jhat\pi/h }
S^{J+m} \Shat^{\Jhat+\mhat}
=
V^{(\Je)}_{-\Je }
S^{J+m} \Shat^{\Jhat+\mhat}.
\label{opCoul}
\eeq
L'op\'erateur $V$ du membre de droite est un op\'erateur
du type \ref{Vj-j} ou \ref{VJe-Je} donc une exponentielle de champ libre,
aussi bien dans l'approche du gaz de Coulomb que dans la notre.
Le spin effectif $\Je\equiv J+\Jhat\pi/h$ peut donc \^etre quelconque.
Les op\'erateurs $S$ et $\Shat$, en revanche,
sont des op\'erateurs int\'egraux,
dits op\'erateurs d'\'ecran\footnote{
Op\'erateurs de ``screening'' en anglais.},
dont on ne sait prendre que des puissance enti\`eres.
Ceci nous donne d\'ej\`a une indication
sur les g\'en\'eralisations possibles.

Conform\'ement \`a cela, il est apparu que la chose la plus importante
pour l'alg\`ebre d'op\'erateurs n'\'etait pas tellement
le caract\`ere demi-entier des spins mais l'expression
des r\`egles
de s\'election en termes de contraintes sur les spins effectifs.
Expliquons nous.
Les r\`egles de s\'elections standard \ref{ThreeCond} de $Sl(2)$
peuvent \^etre \'ecrites en termes de spins effectifs
de la fa\c con suivante\footnote{
${\cal N}\!\!+\!(\pi/h){\cal N}\equiv\{p+(\pi/h)\phat\>;\> p,\phat\in {\cal
N}\}$.}
\beq
\epsffile{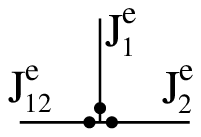}
\to
\left\{
\begin{array}{ccc}
\Je_1+\Je_2-\Je_{12}  \in {\cal N}\!\!+\!\!(\pi/h){\cal N}\\
\Je_{12}+\Je_2-\Je_1  \in {\cal N}\!\!+\!(\pi/h){\cal N}\\
\Je_1+\Je_{12}-\Je_2  \in {\cal N}\!\!+\!(\pi/h){\cal N}\\
\end{array}
\right\}
\ \Rightarrow\
2\Je_1,2\Je_2,2\Je_{12} \in {\cal N}\!\!+\!{\pi\over h}{\cal N}
\label{ThreeCondEff}.
\eeq
Le caract\`ere demi-entier
(non n\'egatif)
des spins $J_i$ et $\Jhat_i$
appara\^\i t alors
uniquement comme une cons\'equence des trois contraintes
$\Je_1+\Je_2-\Je_{12} \in {\cal N}\!\!+\!(\pi/h){\cal N}$...etc.
On appelle ces r\`egles de s\'election TI3
puisque ce sont les trois in\'egalit\'es
standard qui constituent les in\'egalit\'es triangulaires.
Et la bonne mani\`ere de les
g\'en\'eraliser n'est pas d'enlever des contraintes sur les spins
mais sur les combinaisons lin\'eaires de spins
$\Je_1+\Je_2-\Je_{12}$...etc.
Enlevant deux de ces trois contraintes,
on obtient
\beq
{\cal P}_{\Je_{12}}
V^{(\Je_1)}_{\Je_2-\Je_{12}}
{\cal P}_{\Je_{2}}
=
\epsffile{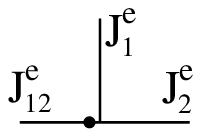}
\ \to\
\Je_1+\Je_2-\Je_{12} \in {\cal N}\!\!+\!{\pi\over h}{\cal N}
\label{OneCondEff}
\eeq
o\`u de la m\^eme mani\`ere que les spins effectifs,
$\Je_2-\Je_{12}$ est un nombre de shift effectif
que l'on note g\'en\'eriquement $\me$.
Ce n'est pas autre chose que la contrainte issue de l'approche
du gaz de coulomb (Eq.\ref{opCoul}):
les entiers $p_{12}$ et $\phat_{12}$
d\'efinis par $\Je_1+\Je_2-\Je_{12}=p_{12}+(\pi/h)\phat_{12}$
sont les nombres de charges d'\'ecran $S$ et $\Shat$
attach\'ees \`a cet op\'erateur.
Les spins effectifs peuvent alors \^etre quelconques
pourvu qu'ils v\'erifient la contrainte \ref{OneCondEff}
que l'on peut aussi \'ecrire $\Je_1+\me_1\in {\cal N}\!\!+\!{\pi\over h}{\cal
N}$.
On appelle TI1 les r\`egles de s\'election \ref{OneCondEff}.

Le vertex \ref{OneCondEff} illustre bien la r\`egle des points
dessin\'es \`a la base des vertex.
Le point situ\'e \`a la base de la patte $\Je_{12}$
indique que la contrainte $\Je_1+\Je_2-\Je_{12} \in {\cal N}\!\!+\!(\pi/h){\cal
N}$
est maintenue.
Dans un vertex du type \ref{ThreeCondEff},
le point sur la patte $\Je_1$ indique de la m\^eme
mani\`ere qu'on impose la condition $\Je_{12}+\Je_2-\Je_1 \in {\cal N}\!\!+\!
(\pi/h){\cal N}$,
et de m\^eme $\Je_{1}+\Je_{12}-\Je_2 \in {\cal N}\!\!+\!(\pi/h){\cal N}$ pour
le point sur la patte $\Je_2$.

Les op\'erateurs de l'alg\`ebre la plus g\'en\'erale
seront tous du type \ref{opCoul} ou \ref{OneCondEff}.
Comme on l'a dit dans la note pr\'ec\'edente,
on aurait aussi bien pu consid\'erer sym\'etriquement
des op\'erateurs avec un point sur la patte droite,
ce qui reviendrait \`a construire une repr\'esentation
de gaz de Coulomb \`a partir de l'autre champ libre $P^{(2)}$
(cf Eqs.\ref{Vj-j}, \ref{VJe-Je}).

Il existe \'egalement un cas interm\'ediaire entre TI3 et TI1.
Ce sont des op\'erateurs dits TI2 qui supportent
les deux contraintes suivantes
\beq
\epsffile{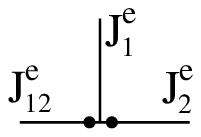}
\to
\left\{
\begin{array}{ccc}
\Je_1+\Je_2-\Je_{12}  \in {\cal N}\!\!+\!(\pi/h){\cal N}\\
\Je_1+\Je_{12}-\Je_2  \in {\cal N}\!\!+\!(\pi/h){\cal N}\\
\end{array}
\right\}
\ \Rightarrow\
2\Je_1 \in {\cal N}\!\!+\!{\pi\over h}{\cal N}
\label{TwoCondEff}.
\eeq
On voit donc que ces r\`egles de s\'election TI2
imposent aux spins $J_1$ et $\Jhat_1$ de l'op\'erateur lui-m\^eme
d'\^etre des demi-entiers avec les conditions standard $m_1=-J_1, ... ,J_1$,
et $\mhat_1=-\Jhat_1, ... ,\Jhat_1$
($m_1$ et $\mhat_1$ sont d\'efinis par $\Je_2-\Je_{12}=m_1+(\pi/h)\mhat_1$).
Les spins effectifs $\Je_{12}$ et $\Je_2$ des \'etats,
ne sont pas contraints \`a \^etre des demi-entiers.
On retrouve donc en fait les op\'erateurs consid\'er\'es
initialement par Gervais et Neveu (chapitre 2).
Ce cas interm\'ediaire engendre aussi une alg\`ebre
coh\'erente, mais nous n'en parlerons pas davantage
puisque nous traitons le cas plus g\'en\'eral TI1.

Une fonction \`a quatre points de ces op\'erateurs de type TI1
s'\'ecrit donc
\beq
\epsffile{f4peff.eps}
\label{f4pggen}.
\eeq
Une diff\'erence essentielle avec la fonction \`a quatre points
\ref{f4pg} r\'eside dans le domaine autoris\'e
pour $\Je_{23}$.
Le spin effectif $\Je_{23}$ peut maintenant prendre
les valeurs autoris\'ees par les r\`egles de s\'election TI1
(\ref{OneCondEff}):
$\Je_2+\Je_3-\Je_{23}=p_{2,3}+\phat_{2,3}\pi/h$
pour deux entiers naturels $p_{2,3}$ et $\phat_{2,3}$ qui sont les nombres
de charges d'\'ecran associ\'ees \`a l'op\'erateur de droite.
De la m\^eme mani\`ere pour l'op\'erateur
de gauche, on a $\Je_1+\Je_{23}-\Je_{123}=
p_{1,23}+\phat_{1,23}\pi/h$ avec\footnote{
Ceci impose clairement des contraintes sur les spins externes:
$\Je_1+\Je_2+\Je_3-\Je_{123}=$ $p+(\pi/h)\phat$ pour $p,\phat\in {\cal N}$.
Ce n'est pas autre chose que la condition de neutralit\'e
d'un syst\`eme de gaz de Coulomb avec $p$ et $\phat$
charges d'\'ecran.
Ces spins g\'en\'eralis\'es sont de toutes fa\c con tr\`es
naturels du point de vue de l'approche int\'egrale du gaz de Coulomb
et la vraie nouveaut\'e est d'avoir pu obtenir
la m\^eme g\'en\'eralisation dans l'approche op\'eratorielle
qui, elle, permet d'obtenir l'alg\`ebre de fusion et d'\'echange.
}
$p_{1,23},\phat_{1,23}\in {\cal N}$.
Le spin effectif $\Je_{23}$ est born\'e inf\'erieurement et sup\'erieurement,
le nombre de ses valeurs autoris\'ees est donc fini.
Un caract\'eristique importante de ces r\`egles de s\'election
est donc que {\sl l'espace des blocs conformes
est de dimension finie}
( pour les fonctions \`a 4 points,
mais aussi d'ordre plus \'elev\'e).
Ceci avait \'et\'e annonc\'e en d\'ebut du chapitre pr\'ec\'edent
sur les \'equations polynomiales de Moore et Seiberg.
C'\'etait une condition n\'ecessaire pour pouvoir
appliquer leurs r\'esultats sur des matrices d'\'echange et de
fusion de dimension finie.

La fusion de ces op\'erateurs de type TI1
s'\'ecrit donc
\beq
\epsffile{f4peff.eps}
=
\sum_{\Je_{12}}
F_{{\Je_{23}},{\Je_{12}}}\!\!\left[^{\Je_1}_{\Je_{123}}
\,^{\Je_2}_{\Je_3}\right]
\epsffile{f4pfusef.eps}
\label{fusggen}.
\eeq
Le domaine autoris\'e pour $\Je_{12}$
est encore obtenu par application des r\`egles de s\'election
TI1 au membre de droite.
La somme sur $\Je_{12}$ est donc encore finie,
le nombre de termes \'etant donn\'e par la dimension de ces blocs conformes.

L'expression du coefficient de fusion n\'ecessite une g\'en\'eralisation
des 6-j.
Il est effectivement possible de g\'en\'eraliser
les 6-j \`a des spins continus.
Ceci est fait en ref.[P3] o\`u nous prouvons
\'egalement les \'equations de Bidenharn-Elliot,
de Racah et d'orthogonalit\'e correspondante
(l'orthogonalit\'e de polynomes reli\'es
\`a ces 6-j avait d\'ej\`a \'et\'e prouv\'ee
par des math\'ematiciens \cite{AW}).
Nous \'ecrivons ces identit\'es et indiquons
la m\'ethode de d\'emonstration dans la sous-partie suivante.
{\sl Le fait remarquable est que cette g\'en\'eralisation des 6-j
correspond exactement \`a celle qui est n\'ecessaire
pour des conditions de type TI1:}
\beq
\left\{
\begin{array}{ccc}
j_1 & j_2 \\
j_3 & j \\
\end{array}
\right .
\left |
\begin{array}{ccc}
 j_{12} \\
 j_{23} \\
\end{array}
\right\}
\hbox{ peut \^etre d\'efini pour }
\left\{
\begin{array}{ccc}
j_1+j_2-j_{12}  \in {\cal N}\\
j_{12}+j_3-j  \in {\cal N}\\
j_1+j_{23}-j  \in {\cal N}\\
j_2+j_3-j_{23}  \in {\cal N}\\
\end{array}
\right .
\label{sixjgen}
\eeq
ce qui permet aux spins $j_i$ de ne pas \^etre demi-entiers.
Chacune de ces quatre conditions
est issue de la condition de type \ref{OneCondEff}
pour les quatre vertex impliqu\'es dans une fusion ou un \'echange.
Le lien entre une condition du type \ref{sixjgen} et les quatre
conditions de type \ref{OneCondEff} fait appel \`a des 6-j dits effectifs.
L'id\'ee est que d'un vertex de type \ref{OneCondEff} avec une condition
$j_1+j_2-j_{12}=p+(\pi/h)\phat\in {\cal N}\!\!+\!(\pi/h){\cal N}$,
il ressort deux entiers positifs $p$ et $\phat$
(le nombre de charges d'\'ecran de type $S$ et $\Shat$ respectivement)
qui, l'un, donnera une condition pour le 6-j ``non-chapeau'',
et l'autre pour le 6-j ``chapeau'',
puisque le coefficient de fusion contient ces deux 6-j.
Ceci nous am\`ene \`a d\'efinir ces 6-j effectifs,
d\'esign\'es par des doubles accolades, par\footnote{
La prescription ci-dessus peut amener
\`a plusieurs d\'efinitions du genre de \ref{sixjeff}
mais elles sont \'equivalentes (cf [P3] Eqs.4.16, 4.17)
}
\beq
\left\{
\left\{ ^{\Je_1}_{\Je_3}\,^{\Je_2}_{\Je_{123}}
\right. \left |^{\Je_{12}}_{\Je_{23}}\right\}\right\}
\equiv
\left\{ ^{\Je_1}_{\Je_3}\,^{\Je_2}_{\Je_{123}+(\phat_{2,3}+\phat_{1,23})\pi/h}
\right. \left |^{\Je_{12}+\phat_{1,2}\pi/h}_
{\Je_{23}+\phat_{2,3}\pi/h}\right\}
\label{sixjeff}
\eeq
o\`u comme pr\'ec\'edemment les nombres de charges d'\'ecran $p$
et $\phat$ sont
d\'efinis par $\Je_i+\Je_k-\Je_{ik}\equiv$
$p_{i,k}+(\pi/h)\phat_{i,k}$.
On peut v\'erifier que, partant de conditions TI1 \ref{OneCondEff}
pour les spins effectifs $\Je_i$,
le 6-j du membre de droite de \ref{sixjeff}
est bien du type \ref{sixjgen},
les quatre entiers de \ref{sixjgen} \'etant respectivement
$p_{1,2}$, $p_{12,3}$, $p_{1,23}$ et $p_{2,3}$.

Il est remarquable que nous ayons justement pu
g\'en\'eraliser les 6-j \`a des conditions \ref{sixjgen}
du type TI1, tout comme les op\'erateurs \ref{OneCondEff}.
Il n'est alors pas tr\`es \'etonnant que l'alg\`ebre
de nos op\'erateurs TI1 s'exprime justement gr\^ace \`a ces
6-j g\'en\'eralis\'es.
La m\'ethode de d\'emonstration est expos\'ee
dans la sous-partie suivante.
Le coefficient de fusion est tout simplement
\beq
F_{{\Je_{23}},{\Je_{12}}}\!\!\left[^{\Je_1}_{\Je_{123}}
\,^{\Je_2}_{\Je_3}\right]
=
{g_{\Je_1\Je_2}^{\Je_{12}}\
g_{\Je_{12}\Je_3}^{\Je_{123}}
\over
g _{\Je_2\Je_3}^{\Je_{23}}\
g_{{\Je_1}\Je_{23}}^{\Je_{123}}
}
\left\{\left\{ ^{\Je_1}_{\Je_3}\,^{\Je_2}_{\Je_{123}}
\left. \right |^{\Je_{12}}_{\Je_{23}}\right\}\right\}
\gaghat\gaghat\, ^{\Jehat_1}_{\Jehat_3}\,^{\Jehat_2}_{\Jehat_{123}}
\bverthat\bigr. \, ^{\Jehat_{12}}_{\Jehat_{23}}\gadhat\gadhat,
\label{Fgen}.
\eeq
Les coefficients $g$ g\'en\'eralis\'es \`a des spins continus
sont
\beq
g_{\Je_1,\Je_2}^{\Je_{12}}= (-1)^{p\phat}
(i/2)^{p+\phat}{H_{p\phat}(\varpi_{\Je_1})H_{p\phat}(\varpi_{\Je_2})
H_{p\phat}(-\varpi_{\Je_{12}})
\over{H_{p\phat}(\varpi_{p/2,\phat /2})}}
\label{ggen}
\eeq
o\`u la fonction $H$ est celle donn\'ee pr\'ec\'edemment en \ref{H}.
On peut v\'erifier que la restriction de ce coefficient \`a
des spins demi-entiers redonne bien le coefficient \ref{Fjj},
la phase $f_V$ sortant du lien entre le 6-j effectif
et le 6-j standard.

De la m\^eme mani\`ere, l'\'echange se g\'en\'eralise \`a
\beq
\epsffile{f4peff.eps}
=
\sum_{\Je_{13}}
B^{(\pm)}_{{\Je_{23}},{\Je_{13}}}\!\!\left[^{\Je_1}_{\Je_{123}}
\,^{\Je_2}_{\Je_3}\right]
\epsffile{f4pbrdef.eps}
\label{brdggen}.
\eeq
avec
\beq
B^{(\pm)}_{{\Je_{23}},{\Je_{13}}}\!\!\left[^{\Je_1}_{\Je_{123}}
\,^{\Je_2}_{\Je_3}\right]
=
e^{\pm i\pi (\Delta_{\Je_{123}}+\Delta_{\Je_3}
-\Delta_{\Je_{23}}-\Delta_{\Je_{13}})}
{g_{\Je_1 \Je_3}^{\Je_{13}} g_{\Je_{13} \Je_2}^{\Je_{123}} \over
g_{\Je_2 \Je_3}^{\Je_{23}} g_{\Je_1 \Je_{23}}^{\Je_{123}}}
\left\{\left\{
^{\Je_1}_{\Je_2}\,^{\Je_3}_{\Je_{123}}
\right. \left |^{\Je_{13}}_{\Je_{23}}\right\}\right\}
\gaghat\gaghat
\,^{\Jehat_1}_{\Jehat_2}\,^{\Jehat_3}_{\Jehat_{123}}
\bigr. \bverthat\, ^{\Jehat_{13}}_{\Jehat_{23}}\gadhat\gadhat
\label{Bgen}.
\eeq

\vskip 5mm

Nous terminons cette partie en remarquant qu'il est maintenant
facile de v\'erifier que l'exponentielle du champ
de Liouville esquiss\'e
en \ref{expphi} est bien local.
En effet, \`a condition d'\'eliminer les facteurs $g$ en les
introduisant dans les normalisations de \ref{expphi}
(\`a la place des pointill\'es),
la commutation de deux op\'erateurs de ce genre
est donn\'ee par le produit de quatre 6-j,
pour les deux param\`etres de d\'eformation diff\'erents $h$ et $\hhat$
et pour les deux chiralit\'es.
Les 6-j avec m\^eme d\'eformation provenant des deux chiralit\'es
se combinent par la relation d'orthogonalit\'e pour donner
finalement une commutation triviale.
On v\'erifie \'egalement que cet op\'erateur local poss\`ede
une alg\`ebre ferm\'ee pour la fusion.

L'exponentielle de Liouville \ref{expphi}
ne vaut cependant que pour le couplage faible, alors que notre
but essentiel ici est de traiter le couplage fort dans le prochain chapitre.
C'est pourquoi nous renvoyons le lecteur \`a la ref.\cite{G4} pour les
d\'etails
sur le r\'egime de couplage faible.
Dans le r\'egime de couplage fort l'op\'erateur \ref{expphi}
est toujours local mais a des poids conformes complexes.
Nous devrons donc en introduire un autre
dans le chapitre \ref{p5}.

\subsection{La m\'ethode de d\'emonstration}

\subsubsection{Les 6-j et leurs identit\'es}

Avant de pouvoir obtenir par r\'ecurrence l'alg\`ebre
des op\'erateurs g\'en\'eralis\'es \`a des spins continus,
il est n\'ecessaire de d\'emontrer les identit\'es
v\'erifi\'ees par les 6-j g\'en\'eralis\'es.
Ce sont elles qui permettront d'effectuer la r\'ecurrence.

La notation de spins effectifs $\Je$ continus provenait de la
physique ($\Je=J+(\pi/h)\Jhat$ dans le cas de spins demi-entiers...).
Nous pr\'ef\'erons dans cette partie plus math\'ematique
noter les spins continus tout simplement $j$ minuscule,
comme en \ref{sixjgen} ainsi qu'en ref.[P3].
Pour les 6-j g\'en\'eralis\'es \ref{sixjgen}, l'\'equation de Bidenharn-Elliot,
s'\'ecrit
\beq
\sum_{j_{23}}
\left\{ ^{j_2}_{j_4}
\,^{j_3}_{j_{234}}\right. \left |^{
j_{23}}_{j_{34}}\right\}
\left\{ ^{j_1}_{j_4}
\,^{j_{23}}_{j}\right. \left |^{j_{123}}_{j_{234}}\right\}
\left\{ ^{j_1}_{j_3}\,^{j_2}_{j_{123}}
\right. \left |^{j_{12}}_{j_{23}}\right\}
=
\left\{ ^{j_1}_{j_{34}}
\,^{j_2}_{j}\right. \left |^{j_{12}}_{j_{234}}\right\}
\left\{ ^{j_{12}}_{j_4}
\,^{j_3}_{j}\right. \left |^{j_{123}}_{j_{34}}\right\}
\label{penta6j}
\eeq
o\`u l'intervalle de sommation sur $j_{23}$ est maintenat control\'e
par les conditions TI1 (\ref{sixjgen}) des cinq 6-j:
il est ainsi \'etendu du domaine habituel (TI3)
$j_{23}\in$
$[|j_2-j_3|,j_2+j_3]\ \cap$
$[|j_{123}-j_1|,j_{123}+j_1]\ \cap$
$[|j_{234}-j_4|,j_{234}+j_4]$
au nouveau domaine (TI1)
$j_{123}-j_1,j_{234}-j_4\le$
$j_{23}\le$
$j_2+j_3$
parcouru par pas entiers, bien que ni $j_{23}$
ni les bornes ne le soient (la longueur de l'intervalle,
elle, est bien enti\`ere).
L'\'equation de Racah est la suivante
\beq
\sum_{j_{12}}
\left\{ ^{j_1}_{j_3}
\,^{j_2}_{j}\right. \left |^{
j_{12}}_{j_{23}}\right\}
\left\{ ^{j_3}_{j_2}
\,^{j_1}_{j}\right. \left |^{j_{13}}_{j_{12}}\right\}
e^{i\pi\epsilon\Delta_{j_{12}}}
=
e^{i\pi\epsilon(\Delta_{j_1}+\Delta_{j_2}+\Delta_{j_3}
+\Delta_{j}-\Delta_{j_{13}}-\Delta_{j_{23}})}
\left\{ ^{j_1}_{j_2}
\,^{j_3}_{j}\right. \left |^{
j_{13}}_{j_{23}}\right\}
\label{Racah6j}
\eeq
o\`u le domaine TI3 habituel
$j_{12}\in$
$[|j_1-j_2|,j_1+j_2]\ \cap$
$[|j-j_3|,j+j_3]$
est remplac\'e par le domaine TI1:
$j_{12}\in$
$[j-j_3,j_1+j_2]$.
Les poids $\Delta_j=-j-j(j+1)h/\pi$
sont les m\^emes que ceux de la th\'eorie conforme
que nous discutons.
Cette \'equation math\'ematique
existe bien \'evidemment ind\'ependemment de toute th\'eorie
conforme et les poids $\Delta_j$ sont alors de simples
fonctions des spins et du param\`etre de d\'eformation $h$.
Finalement, l'orthogonalit\'e s'\'ecrit
\beq
\sum_{j_{13}}
\left\{ ^{j_1}_{j_2}
\,^{j_3}_{j}\right. \left |^{j_{13}}_{j_{23}}\right\}
\left\{ ^{j_2}_{j_1}\,^{j_3}_{j}
\right. \left |^{j'_{23}}_{j_{13}}\right\}
=
\delta_{j_{23},j'_{23}}
\label{orth6j}
\eeq
o\`u le domaine TI3 habituel
$j_{13}\in$
$[|j_1-j_3|,j_1+j_3]\ \cap$
$[|j-j_2|,j+j_2]$
est remplac\'e par le domaine TI1:
$j_{13}\in$
$[j-j_2,j_1+j_3]$.
On peut \'egalement citer l'\'equation de Yang-Baxter
$$
\sum_{j_{134}}
\left\{ ^{j_1}_{j_2}
\,^{j_{34}}_{j}\right. \left |^{j_{134}}_{j_{234}}\right\}
\left\{ ^{j_1}_{j_3}
\,^{j_4}_{j_{134}}\right. \left |^{j_{14}}_{j_{34}}\right\}
\left\{ ^{j_2}_{j_3}
\,^{j_{14}}_{j}\right. \left |^{j_{124}}_{j_{134}}\right\}
e^{i\pi\epsilon(\Delta_{j}-
\Delta_{j_{124}}-\Delta_{j_{234}}-\Delta_{j_{134}})}
=
$$
\beq
\sum_{j_{24}}
\left\{ ^{j_2}_{j_3}
\,^{j_{4}}_{j_{234}}\right. \left |^{j_{24}}_{j_{34}}\right\}
\left\{ ^{j_1}_{j_3}
\,^{j_{24}}_{j}\right. \left |^{j_{124}}_{j_{234}}\right\}
\left\{ ^{j_1}_{j_2}
\,^{j_{4}}_{j_{124}}\right. \left |^{j_{14}}_{j_{24}}\right\}
e^{i\pi\epsilon(\Delta_{j_{4}}-
\Delta_{j_{14}}-\Delta_{j_{24}}-\Delta_{j_{34}})}
\label{YB6j}
\eeq
pour des intervalles de sommation d\'eduits ici encore
des conditions TI1 pour les 6-j.
Elle peut \^etre d\'eduite des pr\'ec\'edentes.

La m\'ethode de d\'emonstration de ces relations pr\'esent\'ee
en ref.[P3] est la suivante.
On exhibe tout d'abord une expression des 6-j avec spins continus
qui est une fraction rationnelle des variables
$e^{ihj_i}$.
En effet l'extension des q-factorielles des 6-j standards
\`a des arguments non entiers
donne des q-fonctions gamma, mais il est en fait possible de les
combiner deux \`a deux pour ne garder que des produits de combinaisons
lin\'eaires de spins q-d\'eform\'ees.
On peut ainsi les \'ecrire en termes d'une fonction
hyperg\'eom\'etrique $\,_4F_3$ [P3]
$$
\left\{ ^{j_1}_{j_3}\,^{j_2}_{j_{123}}
\right. \left |^{j_{12}}_{j_{23}}\right\}
=
{
\Xi _{j_2j_3}^{j_{23}}\
\Xi_{{j_1}j_{23}}^{j_{123}}
\over
\Xi_{j_1j_2}^{j_{12}}\
\Xi_{j_{12}j_3}^{j_{123}}
}
{
(-1)^{p_{1,2}}
\lfloor 2j_{23}+1 \rfloor
\lfloor p_{1,23} \rfloor \! !
\lfloor j_{123}+j_{23}-j_{2}-j_{12}+1 \rfloor_{p_{12,3}}
\over
\lfloor p_{12,3} \rfloor \! !
\lfloor p_{1,23}-p_{12,3} \rfloor \! !
\lfloor j_{23}+j_{123}-j_1+1 \rfloor_{p_{1,23}+p_{2,3}+1}
}
$$
\beq
{
\lfloor j_{123}+j_{1}-j_{2}-j_{3}+1 \rfloor_{p_{2,3}}
\over
\lfloor j_1-j_2-j_{12} \rfloor_{p_{12,3}-p_{1,23}}
}
\,_4F_3\left(
^{j_1+j_{23}+j_{123}+2,j_1-j_2-j_{12},-p_{2,3},-p_{12,3};}
_{p_{1,23}-p_{12,3}+1,j_{123}+j_{23}-j_{2}-j_{12}+1,
j_{123}+j_{1}-j_{2}-j_{3}+1;}
1\right),
\label{6j}
\eeq
avec
$p_{1,2}\equiv j_1+j_2-j_{12}$,
$p_{2,3}\equiv j_2+j_3-j_{23}$,
$p_{1,23}\equiv j_1+j_{23}-j_{123}$,
$p_{12,3}\equiv j_{12}+j_3-j_{123}$.
Tous les nombres not\'es $\lfloor x \rfloor$ sont q-d\'eform\'es:
\beq
\lfloor x \rfloor \equiv {\sin (hx)\over \sin h},
\quad
\lfloor x \rfloor_n\equiv
\prod_{k=0}^{n-1}
\lfloor x+k \rfloor,
\quad
\lfloor n \rfloor \! ! \equiv
\prod_{r=1}^n \lfloor r \rfloor
\label{nbdef}
\eeq
pour les nombres, produits et factorielles q-d\'eform\'es
respectivement\footnote{
Les math\'ematiciens \'ecrivent plus souvent
$(1-q^x)/(1-q)$,
avec $q=e^{-2ih}$.}.
La fonction $\Xi$ dans \ref{6j} est donn\'ee par
\beq
\Xi_{j_1j_2}^{j_{12}}\equiv
\sqrt{
\lfloor j_1-j_2+j_{12}+1 \rfloor_{p_{1,2}}
\lfloor -j_1+j_2+j_{12}+1 \rfloor_{p_{1,2}}
\lfloor 2j_{12}+2 \rfloor_{p_{1,2}}
\over
\lfloor p_{1,2} \rfloor \! !
}
\label{Xi}
\eeq
et nous rappelons la d\'efinition des q-fonctions hyperg\'eom\'etriques:
\beq
\,_{m}F_n\left(
^{(a_i)_{i=1..m};}
_{(b_j)_{j=1..n};}
z\right)
\equiv
\sum_{\nu=0}^\infty
{\prod_{i=1}^m
\lfloor a_i \rfloor_\nu
\over
\prod_{j=1}^n
\lfloor b_j \rfloor_\nu
}
{z^\nu
\over
\lfloor \nu \rfloor\! !}
\label{Fdef}.
\eeq
Pour les 6-j, il s'agira toujours d'une somme tronqu\'ee.
En effet si l'un des $a_i$ est un entier n\'egatif $-N$,
les termes de la somme avec $\nu$ sup\'erieur \`a $N$ sont nuls.

Consid\'erons ensuite la diff\'erence entre les membres de gauche et de droite
des
\'equations \ref{penta6j}, \ref{Racah6j}, \ref{orth6j}
ou \ref{YB6j}.
C'est \'egalement une fraction rationnelle des m\^emes
variables, avec cependant une somme dont les bornes
d\'ependent des spins.
Il est cependant possible de param\'etriser les spins par
des param\`etres bien choisis dont ces bornes
ne d\'ependent plus\footnote{
C'est en fait d\'ej\`a la m\^eme chose pour les 6-j qui contiennent
une somme et des produits dont les bornes d\'ependent des spins.
Mais on peut les \'ecrire sous une forme o\`u ces bornes ne d\'ependent
que des quantit\'es forc\'ees \`a \^etre enti\`eres par \ref{sixjgen}.
Si on fixe ces entiers, les 6-j sont alors de vraies fractions rationnelles
de certaines des variables $e^{ihj_i}$ (elles ne sont plus toutes
ind\'ependantes
quand les entiers sont fix\'es), ce qui suffit pour cette d\'emonstration.
}.
Pour des valeurs discr\`etes de ces param\`etres, les spins
sont demi-entiers.
Pour certaines de ces valeurs discr\`etes,
en nombre infini,
les spins v\'erifient m\^eme les in\'egalit\'es
triangulaires compl\`etes TI3.
Les 6-j g\'en\'eralis\'es se r\'eduisent dans ce cas
au 6-j normaux,
en particulier l'intervalle de la somme qui les d\'efinit
se r\'eduit \`a l'intervalle standard gr\^ace \`a des z\'eros
des arguments.
De plus, pour un sous-ensemble de ces valeurs discr\`etes,
de taille encore infinie,
l'intervalle de sommation des identit\'es \ref{penta6j},
\ref{Racah6j} et \ref{orth6j} pour des TI1 et TI3
sont confondus.
Les identit\'es de Bidenharn-Elliot, de Racah et l'orthogonalit\'e
se ram\`enent donc finalement
dans ces cas-l\`a aux identit\'es standard (TI3)
pour des spins entiers.
Elles sont d\'emontr\'ees en ref.\cite{KR} (cas q-d\'eform\'e bien s\^ur).
Les diff\'erences consid\'er\'ees entre membres de gauche et de droite
 valent donc zero pour toutes ces
valeurs discr\`etes\footnote{
Comme ici $h/\pi$ est irrationnel,
contrairement au cas particulier des mod\`eles minimaux,
ces valeurs discr\`etes obtenues par increments entiers des spins
donnent bien des valeurs discr\`etes toutes diff\'erentes des variables
$e^{ihj_i}$.
}.
Or, une fraction rationnelle qui a un nombre infini de z\'eros
ne peut \^etre qu'identiquement nulle,
ce qui prouve les identit\'es ci-dessus.
Elles ont \'egalement \'et\'e v\'erifi\'ees num\'eriquement
pour ces spins continus,
dans les cas q-d\'eform\'e ou non.

\vskip 2mm

On peut se demander ce qu'il se passerait si certains spins,
alors qu'ils sont \`a priori suppos\'es quelconques,
\'etaient demi-entiers.
La question se pose en effet car,
\`a cause de l'extension du domaine de sommation
dans la d\'efinition des 6-j,
il appara\^\i t alors dans ce cas
des z\'eros mais aussi des p\^oles dans les 6-j.
De plus, ce cas, que l'on pourrait imaginer marginal
(les demi-entiers sont de mesure nulle dans le continu),
va se pr\'esenter souvent dans le cas du couplage fort
(prochain chapitre) car il est alors possible,
et n\'ecessaire pour des raisons d'invariance modulaire,
d'introduire des spins qui sont des quarts ou des
sixi\`emes d'entiers.
Ceci nous oblige donc \`a prendre des r\`egles de s\'election TI1,
et cependant, ces spins fractionnaires
peuvent facilement se combiner pour donner des demi-entiers,
donc des p\^oles dans les 6-j.

Il reste n\'eanmoins vrai que la fraction rationnelle
constitu\'ee par la diff\'erence des deux membres de ces identit\'es
est identiquement nulle.
Les p\^oles qui peuvent appara\^\i tre seront donc automatiquement
annul\'es par multiplication par un 6-j nul ou par
soustraction du m\^eme p\^ole.
Ce m\'ecanisme est absolument n\'ecessaire,
en particulier pour l'orthogonalit\'e car c'est elle qui garantit
la fermeture de la sous-alg\`ebre physique du couplage fort.

En clair, il faut, une fois pour toutes, faire un choix entre
des conditions TI1, TI2 ou TI3.
Cela nous donne une alg\`ebre
plus ou moins \'etendue.
Une fois que ce choix est fait, il reste coh\'erent,
car garanti par le caract\`ere rationnel de ces identit\'es.
On peut ainsi par exemple avoir
choisi d'appliquer des TI1,
mais consid\'erer l'orthogonalit\'e (\ref{orth6j}) avec
des spins ``externes'' (i.e. $j_1,j_2,j_3,j$)
compatibles ``par hasard'' avec des TI3.
On sommera donc sur un intervalle pour $j_{13}$ donn\'e
par des TI1, ce qui donnera pour certains 6-j des p\^oles ou
de z\'eros.
Mais on peut \^etre s\^ur qu'ils vont s'annuler pour donner au
total une fonction delta,
c'est-\`a dire l'identit\'e,
sur l'espace des valeurs autoris\'ees pour $j_{23}$ par les TI1.
Ceci n'est pas du tout incompatible avec l'existence d'une relation
d'orthogonalit\'e pour les m\^emes spins externes mais un choix
g\'en\'eral de conditions TI3.
Dans ce cas-l\`a la sommation sur $j_{13}$ est en effet r\'eduite
au domaine impos\'e par ces TI3, cela fait toute la diff\'erence.
Le r\'esultat est alors encore l'identit\'e mais sur le domaine
de $j_{23}$ autoris\'e par des TI3, et si on calcule ceci
pour $j_{23}$ dans le domaine plus large autoris\'e par des TI1,
on obtient le projecteur sur le domaine TI3.
Dans la construction d'op\'erateurs physiques \`a partir de
ces op\'erateurs de base (\ref{expphi} pour le couplage faible
ou \ref{chidef-}, \ref{chidef+} pour le couplage fort)
ceci aura pour effet de d\'ecoupler ceux qui ne
r\'epondent pas aux conditions TI1, TI2 ou TI3 choisies.

\vskip 2mm

L'expression \ref{6j} des 6-j am\`ene une remarque:
sous cette forme, les sym\'etries des 6-j ne sont pas manifestes.
Mais on peut les faire appara\^\i tre par la transformation suivante
\beq
\,_{m+1}F_m\left(
^{(a_i)_{i=1..m},-N;}
_{(b_j)_{j=1..m};}
z\right)
=
(-1)^{N+1}
z^N
{\prod_{i=1}^m
\lfloor a_i \rfloor_N
\over
\prod_{j=1}^m
\lfloor b_j \rfloor_N
}
\,_{m+1}F_m\left(
^{(-N-b_i+1)_{i=1..m},-N;}
_{(-N-a_j+1)_{j=1..m};}
1/z\right)
\label{transfF}
\eeq
o\`u $N$ est un entier positif.
Il s'agit donc de sommes finies et \ref{transfF} se d\'emontre par
simple changement de variable\footnote{
Cette transformation triviale ne doit pas \^etre confondue avec
une transformation entre fonctions $\,_4F_3$ que nous avons
d\'emontr\'ee en ref.[P3] (Eq.5.1) pour effectuer une continuation
\`a des nombres n\'egatifs de charges d'\'ecran (elle n'\'etait connue
que dans le cas non d\'eform\'e).
Cette derni\`ere transformation n'est pas du tout triviale.
}
$\nu\to N-\nu$.
Dans le cas de spins demi-entiers tous les arguments de la
fonction $\,_4F_3$ dans \ref{6j} sont entiers.
On peut donc lui appliquer \ref{transfF} de nombreuses fa\c cons
diff\'erentes et prouver toutes le sym\'etries habituelles des 6-j:
elles sont engendr\'ees par
\beqa
\left\{ ^{J_1}_{J_3}\,^{J_2}_{J_{123}}
\right. \left |^{J_{12}}_{J_{23}}\right\}
&=&
\left\{ ^{J_2}_{J_{123}}\,^{J_1}_{J_{3}}
\right. \left |^{J_{12}}_{J_{23}}\right\}\nnn
\left\{ ^{J_1}_{J_3}\,^{J_2}_{J_{123}}
\right. \left |^{J_{12}}_{J_{23}}\right\}
&=&
\left\{ ^{J_3}_{J_1}\,^{J_{123}}_{J_{2}}
\right. \left |^{J_{12}}_{J_{23}}\right\}
\quad\hbox{   (sym\'etries TI3).}\nnn
\left\{ ^{J_1}_{J_3}\,^{J_2}_{J_{123}}
\right. \left |^{J_{12}}_{J_{23}}\right\}
&=&
\left\{ ^{J_3}_{J_1}\,^{J_2}_{J_{123}}
\right. \left |^{J_{23}}_{J_{12}}\right\}
\label{sym6jTI3}
\eeqa
On a donc les \'echanges de deux spins sup\'erieurs et des deux spins
en-dessous d'eux ainsi que des deux colonnes de gauche.
S'y ajoutent les \'echanges de deux colonnes quelconques pour des
6-j sym\'etriques dits de Racah-Wigner (normalisation l\'eg\`erement
diff\'erente).

Dans le cas de spins continus, en revanche,
le choix des quantit\'es qui devaient rester enti\`eres
($j_1+j_2-j_{12}$ entier mais pas $j_1-j_2+j_{12}$, par exemple)
a bris\'e un certain nombre de sym\'etries.
On ne peut plus appliquer que certaines des transformations
pr\'ec\'edentes (tous les arguments de la fonction $\,_4F_3$
ne sont plus entiers)
et finalement,
on ne trouve plus qu'une sym\'etrie r\'esiduelle
(les derni\`ere des trois \'ecrites plus haut)
\beq
\left\{ ^{j_1}_{j_3}\,^{j_2}_{j_{123}}
\right. \left |^{j_{12}}_{j_{23}}\right\}
=
\left\{ ^{j_3}_{j_1}\,^{j_2}_{j_{123}}
\right. \left |^{j_{23}}_{j_{12}}\right\}
\hbox{   (sym\'etrie TI1)}
\label{sym6jTI1}
\eeq
qui est d'ailleurs la seule qui respecte le choix
des quantit\'es $p_{i,j}$ enti\`eres.

\subsubsection{L'alg\`ebre des op\'erateurs}

Passons maintenant \`a la d\'emonstration
des formules de fusion et d'\'echange des op\'erateurs $V$.

La m\'ethode consiste \`a calculer les blocs conformes
\`a quatre points dont un des op\'erateurs est de spin 1/2.
L'\'equation de bootstrap est en effet dans ce cas une simple
\'equation diff\'erentielle lin\'eaire du deuxi\`eme
ordre (Eq.\ref{bootstrap}) dont la solution est une
fonction hyperg\'eom\'etrique.
Ceci permet de montrer que le changement de base
entre ces blocs conformes est bien de la forme g\'en\'erale
avanc\'ee par Moore et Seiberg (chapitre pr\'ec\'edent).
Demandant ensuite la coh\'erence de cette alg\`ebre
(associativit\'e de la fusion... etc),
ce qui s'exprime par les \'equations polynomiales,
on prouve que l'\'echange et la fusion des op\'erateurs
de spins sup\'erieurs \`a 1/2 est bien de la forme donn\'ee
dans le chapitre pr\'ec\'edent (Eqs.\ref{brd}, \ref{fus})
et on calcule les coefficients de fusion et d'\'echange.

Cette m\'ethode g\'en\'erale a \'et\'e appliqu\'ee une
premi\`ere fois en [P1] \`a des op\'erateurs
et des \'etats de spins demi-entiers,
donc \`a des r\`egles de s\'election TI3.
Nous l'avons ensuite appliqu\'ee en [P3] au cas le plus
g\'en\'eral possible.
Nous pr\'esenterons ici directement ce cas g\'en\'eral.
Elle ne peut pas \^etre g\'en\'eralis\'ee aux r\`egles TI1
pures puisqu'il est n\'ecessaire d'avoir un vecteur nul
dans une fonction de corr\'elation que l'on souhaite obtenir
par \'equation de bootstrap.
Le fonctions de corr\'elation les plus g\'en\'erales
dont on peut ainsi calculer les matrices d'\'echange et de fusion
sont celles du type
\beq
\epsffile{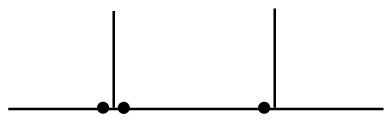}
\label{f4p1,5Cond}
\eeq
c'est-\`a-dire dont un vertex est de type TI2 et l'autre
de type TI1, ce qui donne pour la fusion\footnote{
Nous indiquons en \ref{fusion} de mani\`ere redondante
les projecteurs ${\cal P}_{\Je_i}$ et les nombres de shifts effectifs
$\me_i=\Je_j-\Je_{ij}$ pour plus de clart\'e.
La somme de \ref{fusion} sur ${\Je_{12}}$ est pour les valeurs autoris\'ees
par les r\`egles de s\'election TI1 et TI2.}
$$
\epsffile{f4p1x5.eps}=
{\cal P}_{\Je_{123} }
V^{(\Je_1)}_{\Je_{23}-\Je_{123}} {\cal P}_{\Je_{23}}
V^{(\Je_2)}_{\Je_3-\Je_{23}}
{\cal P}_{\Je_3 }
=
$$
$$
\sum _{\Je_{12}}
F_{{\Je_{23}},{\Je_{12}}}\!\!\left[^{\Je_1}_{\Je_{123}}
\,^{\Je_2}_{\Je_3}\right]
\sum _{\{\nu_{12}\}}
{\cal P}_{\Je_{123} }
V ^{(\Je_{12},\{\nu_{12}\})}_{\Je_3-\Je_{123}}
{\cal P}_{\Je_3 }
<\!\varpi_{\Je_{12}},{\{\nu_{12}\}} |
V ^{(\Je_1)}_{\Je_2-\Je_{12}} | \varpi_{\Je_2} \! >
=
$$
\beq
\sum _{\Je_{12}}
F_{{\Je_{23}},{\Je_{12}}}\!\!\left[^{\Je_1}_{\Je_{123}}
\,^{\Je_2}_{\Je_3}\right]
\epsffile{f4pfus1x.eps}
\label{fusion}
\eeq
donc pour tous les spins continus sauf $\Je_1$
qui doit \^etre de la forme $\Je_1=J_1+(\pi/h)\Jhat_1$
pour des spins $J_1$ et $\Jhat_1$ demi-entiers.

\vskip 2mm

Ces fonctions \`a quatre points ont d\'ej\`a \'et\'e calcul\'ees
en \ref{f4p} sur le cylindre.
Nous redonnons ici ce r\'esultat \'ecrit sur le plan complexe
et avec les notations introduites dans ce chapitre\footnote{
La fonction hyperg\'eom\'etrique $\,_2 F_1$ est ici comme en
\ref{f4p} standard (non q-d\'eform\'ee).}:
$$<\varpi_{\Je_{123}} | V_{\pm 1/2}^{(1/2)}(z_1) V^{(\Je_2)}_{\me}(z_2)
|\varpi_{\Je_3} > =z_2^{-\Delta_{\Je_2}-\Delta_{\Je_3}} \>
z_1^{-\Delta_{1/2}+\Delta_{\Je_{123}}}
\times
$$
\beq
 \left ( {z_2 \over
z_1}\right )^{\Delta_{\Je_{123}\pm 1/2} }  \left ( 1-{z_2 \over
z_1}\right )^{-h\Je_2 /\pi}\,_2 F_1(a_\pm ,b_\pm ;c_\pm ; {z_2\over z_1});
\label{f4pboot}
\eeq
avec
$$
a_\pm ={1\over 2}+
{h\over 2\pi}\left [-\varpi_{\Je_2}  \mp (\varpi_{\Je_{123}}-\varpi_{\Je_3}
)\right
]
;\quad
b_\pm ={1\over 2}+ {h\over 2\pi}\left [-\varpi_{\Je_2}
 \mp (\varpi_{\Je_{123}}+\varpi_{\Je_3} )
\right ]
;
$$
\beq
 c_\pm =1\mp{h\varpi_{\Je_{123}} \over \pi}
\eeq
o\`u nous rappelons (Eq.\ref{varpieff}) le lien entre mode-z\'ero et spin
effectif
$\varpi_{\Je_i}=\varpi_0+2\Je_i$
que nous noterons parfois par la suite plus simplement $\varpi_i$.

On a sur les fonctions hyperg\'eom\'etriques les relations suivantes
$$
F(a,b;c;x)=
{ \Gamma(c)\Gamma(b-a) \over \Gamma(b) \Gamma (c-b) }
(-x)^{-a}
\> F(a,1-c+a;1-b+a;{1\over x})+
$$
\beq
 {\Gamma(c)\Gamma(a-b) \over \Gamma(a)
\Gamma (c-b) }\>
(-x)^{-b} F(b,1-c+b;1-a+b;{1\over x}),
\label{trF1}
\eeq
pour $x\not\in R_+$, et
$$
F(a,b;c;x)=
{ \Gamma(c)\Gamma(c-b-a) \over \Gamma(c-a) \Gamma (c-b) }
\> F(a,b;a+b-c+1;1-x)+
$$
\beq
 {\Gamma(c)\Gamma(a+b-c) \over \Gamma(a)
\Gamma (b) }\>
(1-x)^{c-a-b} F(c-a,c-b;c-a-b+1;1-x),
\label{trF2}
\eeq
pour $(1-x)\not\in R_+$.
Calculant \'egalement la fonction
\`a quatre points avec $V^{(1/2)}_{\pm1/2}$
\`a droite,
on voit que la relation \ref{trF1}
relie pr\'ecis\'ement ces deux types de fonctions hyperg\'eom\'etriques
ce qui donne l'expression de l'\'echange.
La relation \ref{trF2} relie le corr\'elateur \ref{f4pboot}
\`a une autre fonction \`a quatre points avec $V^{(1/2)}_{\pm1/2}$,
et la sym\'etrie des fonctions \`a trois points montre
que ceci n'est pas autre chose que la fusion.
Ceci prouve donc la forme g\'en\'erale de la fusion \ref{fusion}
dans le cas le plus simple $\Je_1=1/2$.
Les coefficients de fusion sont donn\'es par les rapports
de fonctions gamma de \ref{trF2}
et valent
$$
F_{\Je_{123}+\epsilon_1/2,\Je_2+\epsilon_2/2}\!
\left[^{1/2}_{\Je_{123}}\,^{\Je_2}_{\Je_3}\right]=
{
\Gamma(1-\epsilon_1\varpi_{123}h/\pi)
\over
\Gamma(1/2+(
-\epsilon_1\varpi_{123}+\epsilon_2\varpi_2+\varpi_3)h/2\pi)
}
$$
\beq
{
\Gamma(\epsilon_2\varpi_2h/\pi)
\over
\Gamma(1/2+(
-\epsilon_1\varpi_{123}+\epsilon_2\varpi_2-\varpi_3)h/2\pi)
}
\label{fus1/2}.
\eeq
On obtient de m\^eme les coefficients d'\'echange les plus simples
et on v\'erifie dans ce cas le lien g\'en\'eral \ref{fusbrd}
entre les matrices de fusion et d'\'echange.

Il existe une solution connue aux \'equations polynomiales:
ce sont
les 6-j (g\'en\'eralis\'es),
avec en outre des coefficients de normalisation $g$
qui sont une pure
jauge pour ces \'equations,
c'est-\`a-dire les matrices d'\'echange
et de fusion \'ecrites en \ref{Fgen} et \ref{Bgen}.
Bien que le coefficient de fusion \ref{fus1/2}
soit donn\'e par des fonctions gamma
(alors que les 6-j d\'eform\'es sont trigonom\'etriques),
on suppose qu'il peut s'\'ecrire sous la forme
g\'en\'erale \ref{Fgen}.
On v\'erifie que c'est effectivement le cas,
les $g$ donn\'es en \ref{ggen}
emportant tous les facteurs non trigonom\'etriques.

Nous cherchons maintenant \`a g\'en\'eraliser
ceci, comme annonc\'e en introduction
\`a cette sous-partie,
c'est-\`a-dire pour la fusion \ref{fusion}
\`a un spin effectif $\Je_1=J_1+(\pi/h)\Jhat_1$
toujours discret mais
sup\'erieur \`a 1/2.
Les fonctions de corr\'elation pourraient en principe \^etre
calcul\'ees:
pour des spins $(J_1,\Jhat_1)$ demi-entiers, le d\'ecouplage du verteur
nul au niveau $nm=(2\Jhat_1+1)(2J_1+1)$
donne une \'equation diff\'erentielle de degr\'e $nm$
(aux d\'eriv\'ees partielles pour plus de quatre points).
On a \'egalement une repr\'esentation int\'egrale pour des spins continus.
On ne sait cependant ni r\'esoudre ces \'equations diff\'erentielles
ni calculer ces int\'egrales.
Renon\c cant donc \`a calculer les fonctions de corr\'elation\footnote{
Il s'agit ici des fonctions de corr\'elation
des op\'erateurs chiraux, ou blocs conformes.
Ceci n'exclut pas que l'on puisse calculer les fonctions \`a $N$ points
d'op\'erateurs physiques (locaux, habill\'es...)
de th\'eories simplifi\'ees (topologiques),
voir ref.\cite{dFK} et chapitre suivant.}
pour des spins sup\'erieurs \`a $1/2$,
nous cherchons uniquement \`a calculer
les matrices $F$ et $B$ de fusion et d'\'echange,
c'est-\`a-dire les transformations de dualit\'e
de ces fonctions de corr\'elation.
Et il ne faut pas croire que ce soit sans int\'er\^et,
bien au contraire.
Certes, seule l'obtention des fonctions de corr\'elation
constituerait une r\'esolution au sens plein de la th\'eorie,
mais l'alg\`ebre d'\'echange et de fusion permet d\'ej\`a de v\'erifier
sa coh\'erence.
La fermeture de l'\'echange et de fusion sur les seuls op\'erateurs
introduits nous garantit en effet qu'il n'appara\^\i tra pas d'autres
op\'erateurs
sur les ``pattes internes'' des fonctions de corr\'elation,
autrement dit
que nous avons bien tous les op\'erateurs de la th\'eorie.
La connaissance de cette alg\`ebre permettra ensuite
de d\'eterminer celle des op\'erateurs physiques,
et de v\'erifier qu'ils sont locaux.

Partant d'une fonction \`a cinq points avec deux spins 1/2,
l'\'equation pentagonale (fig.\ref{pentag}, chapitre 3),
qui n'est autre que l'associativit\'e de la fusion,
permet de d\'emontrer la forme g\'en\'erale de la fusion \ref{fusion}
pour un op\'erateur de spin 1.
Nous pr\'ecisons bien qu'elle prouve effectivement
que la d\'ependance en les descendants
est bien celle donn\'ee par l'\'equation \ref{fusion},
pourvu que le coeficient de fusion $F$
ob\'eisse \`a l'\'equation pentagonale \ref{penta}.
Nous utilisons ensuite des \'equations pentagonales du type
\beq
\epsffile{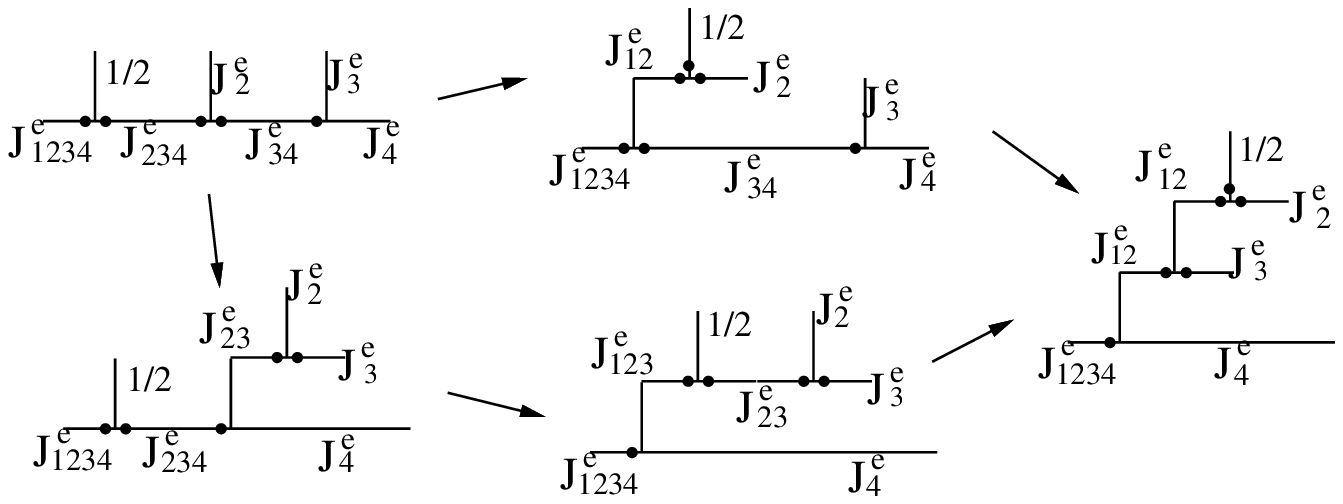}.
\eeq
Les r\`egles TI1 et TI2 indiquent que seuls les spins $J_2$ et $J_{12}$
sont des demi-entiers.
Consid\'erant le cas $J_{12}=J_2+1/2$,
on peut ains prouver  par r\'ecurrence la fusion \ref{fusion}
(avec la valeur \ref{Fgen} des coefficients de fusion)
pour $J_1$ demi entier.

Il faut ensuite sym\'etriquement
d\'eterminer la fusion des
op\'erateurs $\Vhat^{1/2}V^{\Je}$,
et par r\'ecurrence celle des op\'erateurs $\Vhat^{\Jhat}V^{\Je}$
pour $\Jhat$ demi-entier.
L'associativit\'e de la fusion de $V^{J}\Vhat^{\Jhat}V^{\Je}$
permet finalement d'obtenir \ref{fusion}
pour $\Je_1=J_1+(\pi/h)\Jhat_1$,
$J_1$ et $\Jhat_1$ demi-entiers.

On peut de m\^eme obtenir l'expression de l'\'echange
d'un op\'erateur de type TI2 avec un op\'erateur de type TI1.

\vskip 2mm

Pour les fonctions de corr\'elation les plus
g\'en\'erales TI1 (\ref{f4pggen})
dont tous les spins sont continus,
l'\'echange (\ref{brdggen})
a pu \^etre obtenu dans l'approche du gaz de Coulomb \cite{GS1,GS3}.
Il ne reste donc que la fusion de deux op\'erateurs tous les deux
de type TI1 qui n'ait pas \'et\'e d\'emontr\'ee directement,
mais elle peut \^etre d\'eduite de leur \'echange par sym\'etrie
de la fonction \`a trois points.

\section{La base du groupe quantique}
\label{p4.2}

Nous pr\'esentons ici dans une premi\`ere sous-partie (4.2.1)
l'autre base d'op\'erateurs, not\'es $\xi$.
Leur alg\`ebre est donn\'ee par la matrice $R$ universelle
du groupe quantique pour l'\'echange,
et par les coefficients de Clebsch-Gordan q-d\'eform\'es pour la fusion.
Elle met en \'evidence la sym\'etrie $U_q(sl(2))$
de la th\'eorie,
on l'appelle base du groupe quantique
(ou covariante).
Comme pour les op\'erateurs $V$,
nous donnons l'expression de leur alg\`ebre
d'abord pour des spins demi-entiers $(J,\Jhat)$
puis pour des spins continus $\Je$.
Nous faisons \'egalement le lien entre le changement de base
de Babelon\cite{B1,B2} et Gervais\cite{G1} pr\'esent\'e ici,
et un autre changement de base \cite{P,MR} dont la matrice
elle-m\^eme est donn\'ee
par des Clebsch-Gordan.
La matrice de Babelon et Gervais est essentiellement
la limite des Clebsch-Gordan dont un nombre
magn\'etique $M$ tend vers l'infini.

Nous \'etablissons dans
la deuxi\`eme sous-partie (4.2.2) des r\'esultats purement
math\'ematiques n\'ecessaires \`a
l'extension de l'alg\`ebre des $\xi$ \`a des spins continus.
Nous faisons ceci par deux limites successives.
On prouve d'abord que les 6-j ont pour limites,
lorsqu'on fait tendre certains
spins vers l'infini,
des Clebsch-Gordan ou des \'el\'ements de matrice $R$.
Les \'equations \ref{penta6j}, \ref{Racah6j}, \ref{orth6j} et \ref{YB6j}
donnent, dans cette limite,
des \'equations bien connues reliant les 6-j,
les Clebsch-Gordan et la matrice $R$.
La nouveaut\'e, outre cette limite infinie,
est que nous d\'efinissons par l\`a des Clebsch-Gordan et matrice $R$
pour des spins continus,
qui v\'erifient en outre les \'equations standard.
La deuxi\`eme limite infinie
est celle qui fait le lien entre le changement de base
de Pasquier \cite{P,MR} effectu\'e par des Clebsch-Gordan,
et la matrice de passage
des $V$ aux $\xi$ \'etudi\'es ici.
C'est ce qui nous amen\'es \`a \'etudier ces limites infinies.
Il faut pour cela faire tendre vers l'infini
un nombre magn\'etique $M$ des Clebsch-Gordan.
Ceci permet \`a partir des identit\'es pr\'ec\'edentes
sur la matrice $R$ les Clebsch-Gordan et les 6-j,
de prouver
toutes celles qui impliquent la matrice de passage,
et sont n\'ecessaires
\`a l'obtention de l'alg\`ebre des $\xi$ \`a partir de celle des $V$.

Tout ceci a \'et\'e fait dans l'article [P2]
dans le cas de spins demi-entiers.
La g\'en\'eralisation de ces limites \`a des spins continus pr\'esent\'ee
en partie 4.2.2 n'est pas publi\'ee et sera donc imprim\'ee en italique.
De plus, ces limites infinies exigeaient de sortir des in\'egalit\'es
triangulaires strictes des spins demi-entiers (TI3).
On avait donc d\^u, en ref.[P2], faire
des prolongements ad hoc.
Le probl\`eme a maintenant disparu
dans le cadre de spins continus (ref.[P3])
et d'in\'egalt\'es triangulaires relach\'ees
(TI1) o\`u les spins concern\'es peuvent tr\`es bien tendre
vers l'infini en respectant les TI1.

\subsection{Les op\'erateurs}

Les op\'erateurs $V$ sont ceux qui diagonalisent
la monodromie.
Ceci a pour cons\'equence la sym\'etrie (modulo une phase)
de la fonction \`a trois points.
On peut cependant changer de base.
Ceci revient pour les fonctions \`a quatre points \ref{ijklp}
\`a changer de base de l'espace vectoriel des solutions de
l'\'equation de bootstrap.
Les fonctions \ref{ijklp} \'etaient \'etiquet\'ees
par le z\'ero-mode ``interm\'ediaire'' ($p$ dans ${\cal F}_p^{ijkl}$),
alors que les nouvelles n'ont pas un z\'ero-mode interm\'ediaire
d\'etermin\'e.
On perd \'egalement la sym\'etrie de la fonction \`a trois points.
De mani\`ere \'equivalente, cela revient pour les op\'erateurs
$V^{(\Je)}_{\me}$ qui ont tous le m\^eme poids $\Delta_{\Je}$
\`a consid\'erer des combinaisons lin\'eaires d'eux-m\^emes.
Ils ne shiftent alors plus le z\'ero-mode entrant d'une quantit\'e
d\'etermin\'ee.

Ce changement de base pourrait tout \`a fait \^etre arbitraire.
Diagonaliser la matrice de monodromie avait cependant certains
avantages.
On voudrait quand m\^eme compenser leur perte
et on cherche par cons\'equent \`a effectuer un changement de base
qui donne des coefficients de fusion et d'\'echange
dans la nouvelle base
ind\'ependants du z\'ero-mode (ou du spin) des \'etats
entrants et sortants.
Ceci semble d'ailleurs pr\'ef\'erable car ces nouveaux op\'erateurs
shiftent ce z\'ero-mode de mani\`ere non contr\^ol\'ee.
Le changement de base s'\'ecrit avec la
matrice de passage\footnote{
Nous \'ecrivons \ref{xideV} dans le cas le plus simple
d'un seul spin demi-entier $J$,
car le but de cette formule est essentiellement d'introduire
la forme g\'en\'erique de ce changement de base pour le comparer
\`a celui de \cite{P,MR} que nous discutons ensuite
(Eq.\ref{xi2}...) dans ce cas simple pour plus de clart\'e.
Le changement de base des op\'erateurs de spins $(J,\Jhat)$
est donn\'e en [P2] Eq.3.46.} $A$:
\beq
{\cal P}_{J_{12}}
\
\xi^{(J_1)}_{M_1}
=
{\cal P}_{J_{12}}
\
\sum_{m_1}
A(J_1,J_{12})^{m_1}_{M_1}
\
\Vt^{(J_1)}_{m_1}
\label{xideV}
\eeq
o\`u les $\Vt$ sont les op\'erateurs de vertex renorm\'es
d\'efinis en \ref{Vtilde}.
Il s'av\`ere effectivement possible de trouver une matrice de passage
telle que les coefficients de fusion et d'\'echange des $\xi$
soient ind\'ependants du spin des \'etats entrants ou sortants.
Elle est unique \`a une normalisation pr\`es.
Elle a \'et\'e trouv\'ee par Babelon\cite{B1}
dans le cas du spin 1/2, et pour des spins sup\'erieurs
par Gervais en ref.\cite{G1} o\`u
elle est not\'ee
$|J_1,\varpi_{J_{12}})^{m_1}_{M_1}$.

Alors que nous avons uniquement exig\'e que leur alg\`ebre ne d\'epende
pas du mode-z\'ero $\varpi$,
le fait remarquable
est que l'\'echange des $\xi$ soit donn\'e par la matrice $R$
universelle de Drinfeld\cite{Dr}:
ceci s'\'ecrit dans le cas de spins demi-entiers
$$
\xi_{M_1,\Mhat_1}^{(J_1,\Jhat_1)}(z_1)
\,\xi_{M_2,\Mhat_2}^{(J_2,\Jhat_2)}(z_2)=
\sum_{ M'_1,\Mhat'_1,M'_2,\Mhat'_2}\>
e^{\pm i\pi b_\xi(J_i,M_i)}\times
$$
\beq
R^\pm(J_1,J_2)_{M_1\, M_2}^{M'_2\, M'_1}\,
\Rhat^\pm\pghat\Jhat_1,\Jhat_2\pdhat_{\Mhat_1\, \Mhat_2}^{\Mhat'_2\,
\Mhat'_1}\,
\xi_{M'_2,\Mhat'_2}^{(J_2,\Jhat_2)}(z_2)
 \,\xi_{M'_1,\Mhat'_1}^{(J_1,\Jhat_1)}(z_1).
\label{brdxi}
\eeq
o\`u la phase $b_\xi$ est donn\'ee en [P2] Eq.6.4 et
$R^\pm(..)$ et $\Rhat^\pm(..)$ sont des
\'el\'ements de matrice $R$ universelle du groupe quantique
pour les param\`etres de d\'eformation $h$ et $\hhat$ respectivement.
La valeur de $R$ ainsi que toutes les d\'emonstrations
seront donn\'es dans la sous-partie suivante,
nous nous contentons ici de pr\'esenter les id\'ees importantes.
La fusion s'\'ecrit \'egalement
$$
\xi ^{(J_1,\Jhat_1)}_{M_1,\Mhat_1}(z_1)
\,\xi^{(J_2,\Jhat_2)}_{M_2,\Mhat_2}(z_2) =\!\!\!\!
\sum _{J_{12}= | J_1 - J_2 |} ^{J_1+J_2}
\sum _{\Jhat_{12}= |\Jhat_1 - \Jhat_2 |} ^{\Jhat_1+\Jhat_2}\!\!\!\!
 e^{i\pi f_\xi(J_i,M_i)}\times
$$
$$
(J_1,M_1;J_2,M_2|J_{12})
\pghat\Jhat_1,\Mhat_1;\Jhat_2,\Mhat_2\verthat \Jhat_{12}\pdhat
\sum _{\{\nu\}}
\xi ^{(J_{12},\Jhat_{12},\{\nu\})} _{M_1+M_2,\Mhat_1+\Mhat_2}(z_2)
\times
$$
\beq
<\!\varpi _{J_{12},\Jhat_{12}},{\{\nu\}} |
\Vt^{(J_1,\Jhat_1)}_{J_2-J_{12},\Jhat_2-\Jhat_{12}}(z_1-z_2)
| \varpi_{J_2,\Jhat_2}\! >
\label{fusxi}
\eeq
avec un produit de deux Clebsch-Gordan des deux groupes quantiques
$U_q(sl(2))$ et $U_\qhat(sl(2))$,
la phase $f_\xi$ \'etant donn\'ee en [P2] Eq.6.2.
On remarque que l'\'el\'ement de matrice dans le membre de droite
est rest\'e un $\Vt$.
En effet le changement de base dans le membre de droite
de l'\'equation de fusion \ref{fusgjj} \'ecrite pour des $\Vt$
n'est effectu\'e que sur l'op\'erateur $\Vt^{(J_{12},\Jhat_{12})}$
et non sur l'\'el\'ement de matrice.
L'alg\`ebre des $\xi$ n'est donc ferm\'ee que pour l'\'echange,
pas pour la fusion \`a tous les ordres.

La g\'en\'eralisation \`a des spins continus
du passage des $V$ aux $\xi$
est possible gr\^ace aux identit\'es obtenues par limite infinie
dans la prochaine
sous-partie (4.2.2).
Les coefficients de fusion et d'\'echange des ces $\xi$ g\'en\'eralis\'es
s'expriment par des Clebsch-Gordan et des matrices $R$ effectifs qui,
comme les 6-j effectifs (Eq.\ref{sixjeff}),
sont simplement un moyen de faire le lien entre les nombres de charges
d'\'ecran de type $\alpha_-$ et $\alpha_+$,
et les deux 6-j matrices $R$ ou Clebsch-Gordan des deux
types diff\'erents.
Ceci a \'et\'e fait par J.-L. Gervais et J. Schnittger, en s'appuyant sur
les r\'esultats math\'ematiques de la partie 4.2.2, dans
des travaux non publi\'es.
Ils obtiennent ainsi comme pour les $V$ des op\'erateurs
$\xi_{\Me_i}^{(\Je_i)}$
de spins effectifs $\Je_i$ et de nombre magn\'etique
effectif $\Me_i$ continus soumis \`a la seule contrainte
\beq
\Je_i+\Me_i=p_i+\phat_i\pi/h
\quad,\quad
p_i,\phat_i\in {\cal N}
\label{condxi}
\eeq
dont l'\'echange est donn\'e par\footnote{
Nous rappelons que $\Jehat_i=\Je_i h/\pi$.}
$$
\xi_{\Me_1}^{(\Je_1)}(z_1)
\,\xi_{\Me_2}^{(\Je_2)}(z_2)=
\sum_{\Mep_1,\Mep_2}\>
e^{2ih\Je_1\Je_2}
$$
\beq
R^\pm((\Je_1,\Je_2))_{\Me_1\, \Me_2}^{\Mep_2\, \Mep_1}\,
\Rhat^\pm\pghat\pghat\Jehat_1,\Jehat_2
\pdhat\pdhat_{\Mehat_1\, \Mehat_2}^{\Mehatp_2\, \Mehatp_1}\,
\xi_{\Mep_2}^{(\Je_2)}(z_2)
 \,\xi_{\Mep_1}^{(\Je_1)}(z_1).
\label{brdxig}
\eeq
o\`u le domaine de sommation est d\'eduit des conditions
de type \ref{condxi} appliqu\'ees aux quatre op\'erateurs $\xi$.
Les doubles parenth\`eses $R((..))$ d\'esignent des matrices $R$ effectives
d\'efinies comme les 6-j effectifs (dont elles sont les limites) par:
\beq
R^\pm((\Je_1,\Je_2))_{\Me_1\, \Me_2}^{\Mep_2\, \Mep_1}\,
=
R^\pm(\Je_1,\Je_2)_{\Me_1-\phat_1\pi/h,\Me_2-\phat_2\pi/h}
^{\Mep_2i\phat'_2\pi/h, \Mep_1-\phat'_1\pi/h}\,
\label{Reff}
\eeq
o\`u
la matrice $R$ du membre de droite est valable pour des spins
continus\footnote{
On a comme en \ref{condxi} $\Je_i+\Mep_i=p'_i+\phat'_i\pi/h$.}.
Elle peut \^etre obtenue par l'application de son expression
universelle aux repr\'esentations infinies que constituent les spins
continus.
Nous l'obtenons dans la sous-partie suivante
comme limite de 6-j et son expression est donn\'ee en \ref{R}.
Alors que les matrice $R$ effectives (membre de gauche de \ref{Reff})
$R^\pm((\Je_1,\Je_2))_{\Me_1\, \Me_2}^{\Mep_2\, \Mep_1}$
sont soumises aux conditions du type \ref{condxi},
on peut v\'erifier sur \ref{Reff} que
les matrices $R$ normales avec spins continus (membre de droite)
$R^\pm(j_1,j_2)_{M_1,M_2}^{M'_2,M'_1}$
sont soumises aux conditions qui d\'ecouleront par
limite de celles des 6-j avec spins continus:
$j_i+M_i=p_i\in{\cal N}$.

La fusion est donn\'ee par
$$
\xi_{\Me_1}^{(\Je_1)}(z_1)
\,\xi_{\Me_2}^{(\Je_2)}(z_2)=
\sum _{\Je_{12}}
((\Je_1,\Me_1;\Je_2,\Me_2|\Je_{12}))
\pghat\pghat\Jehat_1,\Mehat_1;\Jehat_2,\Mehat_2\verthat \Jehat_{12}\pdhat\pdhat
\sum _{\{\nu\}}
\xi^{(\Je_{12},\{\nu\})} _{\Me_1+\Me_2}(z_2)
$$
\beq
<\!\varpi _{\Je_{12}},{\{\nu\}} |
\Vt^{(\Je_1)}_{\Je_2-\Je_{12}}(z_1-z_2)
| \varpi_{\Je_2}\! >
\label{fusxig}
\eeq
avec trois conditions du type \ref{condxi} issues des $\xi$ et
la condition $\Je_1+\Je_2-\Je_{12}=p_{1,2}+\phat_{1,2}\pi/h$,
$p_{1,2},\phat_{1,2}\in {\cal N}$ issues de l'op\'erateur $\Vt$.
Les Clebsch-Gordan effectifs peuvent \'egalement
\^etre obtenus comme limites de 6-j effectifs et sont donn\'es par
\beq
((\Je_1,\Me_1;\Je_2,\Me_2|\Je_{12}))
=
(\Je_1,\Me_1-\phat_1\pi/h;\Je_2,\Me_2-\phat_2\pi/h |
\Je_{12} +\phat_{1,2}\pi/h).
\eeq

\vskip 5mm

Outre celui d'avoir r\'ev\'el\'e la structure de groupe quantique
de la th\'eorie,
un des grands m\'erites de cette base est pratique:
c'est elle qui a permis
de d\'emarrer la r\'esolution de toute l'alg\`ebre,
c'est-\`a-dire des $\xi$ et des $V$.
En effet, chronologiquement c'est d'abord
l'\'echange des $\xi$ qui a \'et\'e d\'ecouvert
en ref.\cite{G1} puis leur fusion en ref.\cite{G3}
(c'\'etait plut\^ot un ansatz) par Gervais.
C'\'etait s\^urement plus facile car la structure
covariante du groupe quantique ($R$ et Clebsch-Gordan)
\'etait mieux connue et plus simple:
l'expression des coefficients de la matrice $R$
est un simple produit, contrairement aux Clebsch-Gordan
et 6-j qui contiennent une somme de produits.
Lorsque nous avons ensuite calcul\'e l'alg\`ebre
des $V$ en ref.[P1] puis des $\xi$ en ref.[P2],
nous avons pr\'esent\'e les $V$ comme venant avant
les $\xi$,
comme dans ce m\'emoire de th\`ese.
Il est en effet possible de ``remarquer'' que les
coefficients de l'alg\`ebre des $V^{(1/2)}$ sont des 6-j modulo
des normalisations $g$, et ensuite de prouver inductivement
qu'il en est de m\^eme pour tous les spins.
Cependant, dans la pratique,
les $\xi$ ont pr\'ec\'ed\'e les $V$:
connaissant la fusion des $\xi$ Eq.\ref{fusxi},
on peut en \'ecrire l'associativit\'e.
Elle s'\'ecrit en termes de quatre Clebsch-Gordan
mais aussi d'un coefficient de fusion des $V$
(cf ref.[P2] Eqs.2.22, 2.23).
C'est pr\'ecis\'ement l'\'equation de d\'efinition
des 6-j: cela prouve que la fusion des $V$ est essentiellement
donn\'ee par des 6-j.

Cependant, la fusion des $\xi$ n'\'etait alors qu'un ansatz.
C'est pourquoi nous nous sommes content\'es de
remarquer que la fusion
des $V^{(1/2)}$ \'etait donn\'ee par des 6-j modulo des
coefficients de normalisation $g$.
Comme illustr\'e dans la partie pr\'ec\'edente,
il \'etait alors facile de passer aux spins sup\'erieurs,
gr\^ace \`a l'\'equation
de Biedenharn-Elliot.
Pour en d\'eduire l'alg\`ebre des $\xi$,
nous n'avions plus besoin alors que
d'un certain nombre d'\'equations reliant les 6-j
et la matrice $R$ et les Clebsch-Gordan
par la matrice de passage.
Nous rel\'eguons leur d\'emonstration par limite infinie
\`a la sous-partie suivante.

\vskip 5mm

{\sl
Nous abordons maintenant une approche concurrente de ce
changement de base \cite{P,MR}.
Le rapport entre les deux approches est longtemps rest\'e myst\'erieux.
Ces travaux ont \'et\'e men\'es
pour un groupe quelconque,
mais nous ne pr\'esentons ici que le cas qui nous int\'eresse,
c'est-\`a-dire $U_q(sl(2))$,
et dans des notations compatibles
avec celles d\'ej\`a introduites.
On suppose donn\'es
des op\'erateurs de vertex
du type des $V$ (i.e. de type IRF, voir sous-partie suivante),
dont l'\'echange et la fusion sont donn\'es par des 6-j.
Cela s'applique donc aux n\^otres.
Ils agissent sur l'espace de Hilbert standard
$$
{\cal H}^{(V)}
=
\oplus_J
{\cal H}_J,
$$
somme directe de modules de Verma ${\cal H}_J$.
Nous nous contentons par souci de simplicit\'e
de traiter le cas le plus simple d'un seul spin
$J$ demi entier,
mais la g\'en\'eralisation \`a des repr\'esentations infinies
ne pose pas de difficult\'e.
Les \'etats (de plus haut poids) sont not\'es $|J\!>$.
On note
$$
{\cal P}_J
=
|J\!>
<\!J|
$$
le projecteur sur le module de Verma ${\cal H}_J$
la somme sur les descendants (cf \ref{ferm}) \'etant sous entendue.
Ceci est de toute mani\`ere valable pour de la simple
th\'eorie des groupes, auquel cas il n'y a pas de descendants.

On d\'efinit ensuite un autre espace de Hilbert pour
les op\'erateurs $\xi$:
$$
{\cal H}^{(\xi)}
=
\oplus_J
{\cal H}_J
\otimes
{\cal W}^J
$$
dont les \'etats sont $|J\!>\otimes|M\!>$.
Le projecteur sur $|J\!>\otimes|M\!>$ est not\'e
${\cal P}_J\otimes{\cal P}_M$.
On d\'efinit alors les op\'erateurs
\beq
\xi^{(J_1)}_{M_1}
\!\equiv
\!\!\!\!\!\!\!\sum_{J_2,J_{12},M_2}\!\!\!\!\!\!
\left(
|J_{12}\!>
\!\!
\otimes
|M_1+M_2\!>\!
\right)
<\!J_{12}|
V^{(J_1)}_{J_2-J_{12}}
|J_2\!>
(J_1,M_1;J_2,M_2|J_{12})
\left(\!
<\!J_2|\otimes\!\!
<\!M_2|
\right)
\label{xi2}
\eeq
ou, en en fixant l'\'etat de gauche
et en cherchant \`a se rapprocher au maximum de \ref{xideV}:
$$
\left(
{\cal P}_{J_{12}}\otimes{\cal P}_{M_1+M_2}
\right)
\xi^{(J_1)}_{M_1}
$$
\beq
\equiv
\left(
{\cal P}_{J_{12}}\otimes{\cal P}_{M_1+M_2}
\right)
\sum_{m_1}
V^{(J_1)}_{m_1}
(J_1,M_1;J_{12}+m_1,M_2|J_{12})
|M_1+M_2\!>
<\!M_2|.
\label{xi2'}
\eeq
Cette transformation diff\`ere peu de \ref{xideV}.
La matrice de passage $A(J_1,J_{12})^{m_1}_{M_1}$ de \ref{xideV}
est ici un Clebsch-Gordan
et d\'epend d'un cinqui\`eme param\`etre $M_2$,
nombre quantique suppl\'ementaire (magn\'etique) des \'etats.
On voit tr\`es facilement, gr\^ace aux relations
entre les 6-j, les Clebsch-Gordan et la matrice $R$
(Eqs.\ref{orthCG} \`a \ref{RCGCG=CGCG6j}), que ces op\'erateurs
$\xi^{(J_1)}_{M_1}$
ont eux aussi un \'echange donn\'e par la matrice $R$ et une fusion
donn\'ee par un Clebsch-Gordan.
Cette construction \ref{xi2}, \ref{xi2'}
appara\^\i t donc comme \'evidente \`a partir
d'op\'erateurs dont l'alg\`ebre est donn\'ee par des 6-j,
mais elle est \'egalement artificielle de par l'introduction
de ces \'etats $|M\!>$.
Ils ont en effet un int\'er\^et math\'ematique,
ce sont des repr\'esentations du groupe quantique,
mais ils n'ont aucun sens physique.

Jusqu'\`a r\'ecemment, on avait donc deux constructions diff\'erentes.
Soit la transformation \ref{xideV} donn\'ee par une matrice
$A(J_1,J_{12})^{m_1}_{M_1}$ ,
soit la transformation \ref{xi2}, \ref{xi2'} donn\'ee
par les Clebsch-Gordan de $U_q(sl(2))$
mais qui avait l'inconv\'enient d'agir sur un espace
de Hilbert ${\cal H}^{(\xi)}$ plus grand,
artificiel, sans interpr\'etation physique.
Nous nous sommes alors aper\c cus [P2] que la
matrice de passage $A$ de la premi\`ere construction \ref{xideV}
n'\'etait pas autre chose (\`a une jauge pr\`es) que la limite
du Clebsch-Gordan de \ref{xi2'} lorsque $M_2$,
le param\`etre suppl\'ementaire,
tendait vers l'infini.
Prenant la limite de \ref{xi2'} pour $M_2$
tendant vers l'infini,
le coefficient devient donc le m\^eme,
mais en plus les \'etats
$|M_1+M_2\!>$ et $<\!M_2|$ deviennent trivialement
$|\infty\!>$ et $<\!\infty|$.
Dans la limite asymptotique de $M_2$ grand,
l'espace $W^{J_2}$,
qui nous embarrassait, devient trivial.
Identifiant les \'etats $|J\!>$ pour \ref{xideV}
et les \'etats $|J\!>\otimes|\infty\!>$ pour \ref{xi2'},
la transformation \ref{xideV} est donc la limite de
\ref{xi2'} pour $M_2$ tendant vers l'infini.
}

\subsection{Les identit\'es par limite infinie}

{\sl
Nous d\'emontrons ici toutes les identit\'es
n\'ecessaires \`a l'obtention de l'alg\`ebre
des $\xi$ \`a partir de celle des $V$
pour des spins continus\footnote{
Comme pour la partie 4.1.2 traitant des 6-j g\'en\'eralis\'es
\`a des spins continus,
nous notons $j_i$ les spins continus dans cette partie
plus math\'ematique.
Les $j_i$ sont donc continus alors que les $J_i$
\'etaient des demi-entiers.
La convention pour les $M_i$ et $m_i$ est diff\'erente.
Nous continuons en effet \`a noter $M_i$ les nombres magn\'etiques
des Clebsch-Gordan,
tandis que la notation $m_i$
est r\'eserv\'ee \`a des diff\'erences de spins
comme pour les 6-j de l'alg\`ebre des $V$.
Le nombre magn\'etique $M_i$ sera donc demi-entier ou non
selon que le spin correspondant $J_i$ ou $j_i$ l'est ou non.}.
Elles sont toutes obtenues \`a partir
des identit\'es v\'erifi\'ees par les 6-j
(\ref{penta6j}, \ref{Racah6j}, \ref{orth6j} et \ref{YB6j}),
en faisant tendre certains spins vers l'infini.
Il y a deux limites successives.
Dans la premi\`ere nous d\'emontrons que les 6-j
ont pour limite des Clebsch-Gordan ou
des \'el\'ements de matrice $R$,
selon les spins qu'on fait tendre vers l'infini.
Ceci nous permettra d'obtenir toutes les identit\'es
concernant la matrice $R$, les Clebsch-Gordan et les 6-j,
pour des spins continus.
Dans la deuxi\`eme limite,
les Clebsch-Gordan tendent vers des \'el\'ements de la matrice
de passage $A$ des $V$ aux $\xi$.
Ceci nous donnera toutes les identit\'es n\'ecessaires \`a l'obtention
de l'alg\`ebre des $\xi$ \`a partir de celle des $V$,
pour des spins continus.

\vskip 2mm

Pour que les nombres q-d\'eform\'es aient un comportement asymptotique
d\'etermin\'e quand leur argument tend vers l'infini,
il faut introduire une partie imaginaire quelque part.
On obtient ainsi
\beq
\lfloor a+x \rfloor
\sim
{
-\epsilon
e^{-ih\epsilon(a+x)}
\over
2i\sin h
}
\hbox{  quand  Im}
(hx)\to\epsilon\infty,
\quad \epsilon=\pm 1.
\label{limqnb}
\eeq
On peut faire tendre la partie imaginaire de $hx$ vers l'infini
de deux mani\`eres:
soit pour h r\'eel, en faisant tendre Im($x$) vers l'infini,
soit pour h complexe (couplage fort, ou dans le cas du
couplage faible en lui donnant une petite
partie imaginaire)
en gardant $x$ r\'eel tendant vers l'infini.
Nous utiliserons les deux m\'ethodes.
On a \'egalement
\beq
\lfloor a+x \rfloor_\nu
\sim
\left(
{
-\epsilon
\over
2i\sin h
}
\right)^\nu
e^{-ih\epsilon\nu(a+x+{\nu-1\over 2})}
\hbox{  quand  Im}
(hx)\to\epsilon\infty.
\label{limqprd}
\eeq
Les 6-j (\ref{6j}) s'expriment en termes d'une fonction
hyperg\'eom\'etrique.
Nous utiliserons donc
la limite suivante:
\beq
\lim_{Im(hx)\to\epsilon\infty}
\,_{m+1}F_{n+1}\left(
^{(a_i)_{i=1..m},a+x;}
_{(b_j)_{j=1..n},b+x;}
z\right)
=
\,_{m}F_n\left(
^{(a_i)_{i=1..m};}
_{(b_j)_{j=1..n};}
ze^{ih\epsilon(b-a)}\right)
\label{limF}
\eeq
qui est valable au moins pour des fonctions $_{m+1}F_{n+1}$
et $_{m}F_{n}$ d\'efinies par des sommes tronqu\'ees,
c'est-\`a-dire \`a condition que l'un des $a_i$
soit un entier n\'egatif,
ce qui est le cas des 6-j.
Elle se d\'emontre facilement \`a partir de \ref{limqprd}.

\vskip 2mm

Gr\^ace \`a cela nous pouvons maintenant calculer
des limites de 6-j (avec spins continus).
Nous avons prouv\'e en ref.[P2]
la limite suivante pour des spins demi-entiers\footnote{
La bonne m\'ethode consiste ici \`a faire tendre la partie imaginaire
de $j$ vers l'infini puisque les r\`egles de s\'election permettent
de le faire tout en maintenant r\'eels les autres spins.}:
\beq
\lim_{\hbox{\scriptsize Im}(hj)\to-\infty}
\left\{ ^{\quad j_1}_{j+M_1+M_2}
\,^{j_2}_{j}\right. \left |^{\ j_{12}}_{j
+M_1}\right\}
=
(j_1,M_1;j_2,M_2|j_{12})
\label{6j->CG}
\eeq
qui peut \'egalement \^etre \'etendue \`a
\beq
\lim_{\hbox{\scriptsize Im}(hj)\to+\infty}
\left\{ ^{\quad j_1}_{j+M_1+M_2}
\,^{j_2}_{j}\right. \left |^{\ j_{12}}_{j
+M_1}\right\}
=
(j_2,M_2;j_1,M_1|j_{12}).
\label{6j->CG'}
\eeq
Nous avons maintenant tous les \'el\'ements pour
calculer la m\^eme limite pour les 6-j avec spins continus.
On obtient pour Im($hj$)
tendant vers $\epsilon\infty$:
$$
\sqrt{
\lfloor j_1+M_1 \rfloor \! !
\lfloor j_1+j_2-j_{12} \rfloor \! !
\lfloor j_1-M_1+1 \rfloor_{j_1+M_1}
\lfloor -j_1+j_2+j_{12}+1 \rfloor_{j_1+j_2-j_{12}}
\over
\lfloor j_2+M_2 \rfloor \! !
\lfloor j_{12}+M_{12} \rfloor \! !
\lfloor j_{2}-M_{2}+1 \rfloor_{j_{2}+M_{2}}
\lfloor j_{12}-M_{12}+1 \rfloor_{j_{12}+M_{12}}
}
$$
$$
(-1)^{j_2+M_2}
{e^{
ih\epsilon\left(
(-j_1+j_2+j_{12})(j_1+j_2+j_{12}+1)/2
+j_{12}M_2+j_2M_{12}+M_2
\right)
}
\over
\sqrt{
\lfloor j_1-j_2+j_{12}+1 \rfloor_{j_1+j_2-j_{12}}
\lfloor 2j_{12}+2 \rfloor_{j_1+j_2-j_{12}}}
}
{
\lfloor j_1-j_2-M_{12}+1 \rfloor_{j_2+M_2}
\over
\lfloor j_1-j_{12}-M_2 \rfloor \! !
}
$$
\beq
\,_3F_2\left(
^{j_1-j_2-j_{12},-M_{12}-j_{12},-M_2-j_2;}
_{j_{1}-j_{12}-M_{2}+1,
j_{1}-j_{2}-M_{12}+1;}
e^{-ih\epsilon(j_1+j_2+j_{12}+1)}
\right).
\label{CG1}
\eeq
Comme les Clebsch-Gordan q-d\'eform\'es ne sont
pas connus pour des spins continus,
tout au moins \`a ma connaissance,
on peut les d\'efinir par cette expression.
Il faut \'evidemment v\'erifier que leur restriction \`a des
spins demi-entiers et \`a des in\'egalit\'es triangulaires
compl\`etes\footnote{
Les in\'egalit\'es triangul\`eres TI1 pour
ces Clebsch-Gordan g\'en\'eralis\'es se d\'eduisent \'evidemment de celles
des 6-j gr\^ace \`a \ref{6j->CG}.
C'est donc: $j_1+j_2-j_{12}\in {\cal N}$ et $j_i+M_i\in {\cal N}$, $i=1,2,12$.
}
redonne bien l'expression standard des q-Clebsch-Gordan.
Cela ne pose pas de probl\`eme.
On peut trouver l'expression standard des q-Clebsch-Gordan
en ref.\cite{G3} (annexe C) o\`u
Gervais l'a d\'eriv\'ee par r\'ecurrence gr\^ace
\`a l'action du coproduit de $U_q(sl(2))$\footnote{
Il y a \'egalement en ref.\cite{G3} (Eqs.C.6, C.10),
une expression des Clebsch-Gordan en termes de la fonction
$\,_3F_2$ mais elle n'est pas ad\'equate pour une g\'en\'eralisation
\`a des spins continus (il faudrait d'abord la transformer
par \ref{transfF})
}.
Ces Clebsch-Gordan g\'en\'eralis\'es v\'erifient d'ailleurs
la m\^eme \'equation de r\'ecurrence issue du coproduit
$U_q(sl(2))$ g\'en\'eralis\'e.

Par cette limite \ref{6j->CG},
la premi\`ere des sym\'etries \ref{sym6jTI3} des 6-j
induit la sym\'etrie suivante des Clebsch-Gordan
\beq
(j_1,M_1;j_2,M_2| j_{12})
=
(j_2,-M_2;j_1,-M_1| j_{12})
\label{symCGTI3}
\eeq
pour des Clebsch-Gordan standards,
alors que les deux derni\`eres sym\'etries \ref{sym6jTI3} des 6-j
ne sont pas respect\'ees par la limite et ne donnent donc rien
pour les Clebsch-Gordan.
La sym\'etrie r\'esiduelle \ref{sym6jTI1} n'est pas
une sym\'etrie de la limite \ref{6j->CG},
et nos Clebsch-Gordan g\'en\'eralis\'es n'on donc aucune sym\'etrie,
ce que l'on peut v\'erifier explicitement.
La relation
\beq
\left[ (j_1,M_1;j_2,M_2| j_{12}) \right] ^*
=
(j_2,M_2;j_1,M_1| j_{12})
\label{CGcc}
\eeq
de conjugaison complexe reste n\'eanmoins valable pour les
Clebsch-Gordan g\'en\'eralis\'es.
Elle permet de passer de \ref{6j->CG} \`a \ref{6j->CG'}.

\vskip 2mm

Nous avons examin\'e en ref.[P2] une autre limite des 6-j:
\beq
\lim_{\im(jh)\to-\infty}
\left\{ ^{j_1}_{j_2}
\,^{j+M}_{j}\right.
\left |^{j+M'_2}_{j+M_1}\right\}
e^{i\pi\epsilon (\Delta_{j+M}+\Delta_{j}
-\Delta_{j+M'_2}-\Delta_{j+M_1})}
=
R^{\epsilon}\!(j_1,j_2)^{M'_2M'_1}_{M_1M_2}
\label{6j->R}
\eeq
extensible \`a
\beq
\lim_{\im(jh)\to+\infty}
\left\{ ^{j_1}_{j_2}
\,^{j+M}_{j}\right.
\left |^{j+M'_2}_{j+M_1}\right\}
e^{i\pi\epsilon (\Delta_{j+M}+\Delta_{j}
-\Delta_{j+M'_2}-\Delta_{j+M_1})}
=
R^{\epsilon}\!(j_2,j_1)^{M'_1M'_2}_{M_2M_1}
\label{6j->R'}
\eeq
o\`u $\epsilon=\pm1$, et $R^{+1}\!(j_1,j_2)$ et $R^{-1}\!(j_1,j_2)$
sont respectivement des \'el\'ements des matrices universelles
$R$ et $\Rb$.
Comme pour les Clebsch-Gordan nous g\'en\'eralisons ici
ce r\'esultat \`a des spins continus.

Le principe de la d\'emonstration est identique \`a celui du
cas discret d\'etaill\'e en ref.[P2], partie 5.
Contrairement \`a la limite \ref{6j->CG},
ici un des termes de la somme d\'efinissant la fonction
hyperg\'eom\'etrique domine les autres.
Ce terme n'est pas le m\^eme
selon que
$M_1$ est sup\'erieur ou inf\'erieur \`a $M'_1$.
On obtient une limite nulle dans un des cas et finie dans l'autre.
Le r\'esultat est donc une matrice triangulaire
(inf\'erieure ou sup\'erieure selon le signe de
$\epsilon$).
On v\'erifie qu'elle est bien identique \`a la matrice $R$ universelle.
Ses \'el\'ements de matrice valent pour des spins continus
$$
R^+\!(j_1,j_2)^{M'_2M'_1}_{M_1M_2}
=
(1-e^{2ih})^n
e^{-ihM_1M_2}
e^{-ihM'_1M'_2}
e^{-ihn(n+1)/2}
$$
\beq
{\sqrt{
\lfloor j_1-M_1+1 \rfloor_n
\lfloor j_1+M_1'+1 \rfloor_n
\lfloor j_2+M_2+1 \rfloor_n
\lfloor j_2-M_2'+1 \rfloor_n}
\over
\lfloor n \rfloor \! !}
\label{R}
\eeq
lorsque l'entier $n\equiv M'_2-M_2=M_1-M'_1$ est positif et $0$ sinon.

Comme pour les Clebsch-Gordan, les sym\'etries \ref{sym6jTI3}
qui sont respect\'ees par la limite \ref{6j->R}
(en l'occurence les deuxi\`eme et troisi\`eme)
donnent les sym\'etries suivantes
\beqa
R^\pm\!(J_1,J_2)^{M'_2M'_1}_{M_1M_2}
&=&
R^\pm\!(J_2,J_1)^{-M'_1\>-M'_2}_{-M_2\>-M_1}
\nnn
R^\pm\!(J_1,J_2)^{M'_2M'_1}_{M_1M_2}
&=&
R^\pm\!(J_2,J_1)^{M_1M_2}_{M'_2M'_1}
\ \hbox{   (sym\'etries TI3)}
\label{symRTI3}
\eeqa
pour la matrice $R$ avec spins demi-entiers
(la combinaison des deux en engendre une troisi\`eme).
Pour la matrice $R$ g\'en\'eralis\'ee \`a des spins continus, la
sym\'etrie r\'esiduelle \ref{sym6jTI1} des 6-j
ne laisse subsister que la sym\'etrie suivante
\beq
R^\pm\!(j_1,j_2)^{M'_2M'_1}_{M_1M_2}
=
R^\pm\!(j_2,j_1)^{M_1M_2}_{M'_2M'_1}
\hbox{   (sym\'etrie TI1)}.
\label{symRTI1}
\eeq
La transformation par conjugaison complexe
\beq
\left[ R^\pm\!(j_1,j_2)^{M'_2M'_1}_{M_1M_2} \right] ^*
=
R^\mp\!(j_2,j_1)^{M'_1M'_2}_{M_2M_1}
\label{Rcc}
\eeq
reste valable pour une matrice $R$ g\'en\'eralis\'ee.
Elle permet de relier les limites \ref{6j->R} et \ref{6j->R'}.

\vskip 3mm

L'int\'er\^et de ces limites est qu'elles s'appliquent
directement aux \'equations \ref{penta6j}, \ref{Racah6j},
\ref{orth6j}, \ref{YB6j}...\footnote{
Ces quatre relations ne sont ni exhaustives ni ind\'ependantes.}
v\'erifi\'ees par les 6-j.
Il faut n\'eanmoins v\'erifier sur chaque cas que l'intervalle
de sommation de ces relations reste
de longueur fixe (et donc finie),
c'est pourquoi nous l'expliciterons.

L'orthogonalit\'e des 6-j \ref{orth6j} donne
les deux relations d'orthogonalit\'e des Clebsch-Gordan:
en faisant tendre $j_2,j_{23},j'_{23},j$ avec diff\'erences
fix\'ees dans \ref{orth6j} (puis $j_3\to j_2$) on obtient
\beq
\sum_{j_{12}=-M_1-M_2}^{j_1+j_2}
(j_1,M_1;j_2,M_2| j_{12},M_{12})
(j_1,M'_1;j_2,M'_2| j_{12},M_{12})
=
\delta_{M_1,M'_1}\delta_{M_2,M'_2}
\label{orthCG}
\eeq
o\`u nous avons fait appara\^\i tre $M_{12}$ dans les Clebsch-Gordan
pour bien indiquer la contrainte $M_1+M_2=M'_1+M'_2=M_{12}$,
ce qui relie d'ailleurs les deux $\delta$ du membre de droite
pour en faire un seul.
Faire tendre $j_1,j_{13}$ et $j$ vers l'infini donne l'autre
relation d'orthogonalit\'e des Clebsch-Gordan:
\beq
\sum_{M'_1,M'_2\,/\,M'_1+M'_2=M_{12}}
(j_1,M_1;j_2,M_2| j_{12},M_{12})
(j_1,M'_1;j_2,M'_2| j'_{12},M_{12})
=
\delta_{M_1,M'_1}\delta_{M_2,M'_2}
\label{orthCG'}
\eeq
o\`u les bornes sont pour $M'_1$ par exemple
$-j_1\le M'_1\le j_2+M_{12}$.
On obtient \'egalement l'orthogonalit\'e des matrices $R$
en intercalant les bons facteurs $e^{\Delta...}$ (cf \ref{6j->R})
dans \ref{orth6j} o\`u on
fait tendre $j_3,j_{13},j_{23},j'_{23}$ et $j$
vers l'infini \`a diff\'erences fix\'ees:
\beq
\sum_{M'_1,M'_2\,;\,M'_1+M'_2=M_1+M_2}
R^{\pm}\!(j_1,j_2)^{M'_2M'_1}_{M_1M_2}
R^{\mp}\!(j_2,j_1)_{M'_2M'_1}^{M''_1M''_2}
=
\delta_{M_1,M''_1}\delta_{M_2,M''_2}
\label{orthR}
\eeq
avec une somme sur $M'_1\in [-j_1,j_2+M_1+M_2]$.

De la m\^eme mani\`ere, on d\'eduit des autres relations
v\'erifi\'ees par les 6-j de nombreuses relation
entre 6-j, Clebsch-Gordan et matrice $R$
d\'ej\`a connues dans le cas de spins demi-entiers
et d'in\'egalit\'es triangulaires TI3\footnote{
Les \'equations \ref{orthCG} \`a \ref{RCGCG=CGCG6j}
sont donn\'ees pour des TI3 en ref.\cite{KR},
sauf \ref{CGCG=R} qui n'est pas donn\'ee mais peut
se d\'eduire des r\`egles de n\oe uds donn\'ees dans
cette r\'ef\'erence.
}.
De la relation de Racah \ref{Racah6j}
pour $j_3,j_{13},j_{23},j$ tendant vers l'infini
on d\'eduit
\beq
\sum_{j_{12}=-M_1-M_2=-M'_1-M'_2}^{j_1+j_2}
\!\!\!\!\!\!\!\!\!\!
\!\!\!\!\!\!\!
(j_1,M_1;j_2,M_2| j_{12})
(j_2,M'_2;j_1,M'_1| j_{12})
e^{i\pi\epsilon(\Delta_{j_{12}}-\Delta_{j_1}-\Delta_{j_2})}
=
R^{\epsilon}\!(j_1,j_2)^{M'_2M'_1}_{M_1M_2}
\label{CGCG=R}
\eeq
et
pour $j_2,j_{12},j_{23},j$ tendant vers l'infini
(puis $j_3\to j_1$...):
\beq
\sum_{M'_1,M'_2}
R^{\epsilon}\!(j_1,j_2)^{M'_2M'_1}_{M_1M_2}
(j_2,M'_2;j_1,M'_1| j_{12})
=
e^{i\pi\epsilon(\Delta_{j_1}+\Delta_{j_2}-\Delta_{j_{12}})}
(j_1,M_1;j_2,M_2| j_{12})
\label{RCG=CG}.
\eeq

La relation pentagonale \ref{penta6j} donne pour
$j_1,j_{12},j_{123},j$ tendant vers l'infini
(puis changement d'indices):
$$
\sum_{j_{23}=max(j_{123}-j_1,-M_2-M_3)}^{j_2+j_3}
\left\{ ^{j_1}_{j_3}\,^{j_2}_{j_{123}}
\right. \left |^{j_{12}}_{j_{23}}\right\}
(j_2,M_2;j_3,M_3| j_{23})
(j_{1},M_{1};j_{23},M_{23}| j_{123})
=
$$
\beq
(j_1,M_1;j_{2},M_{2}| j_{12})
(j_{12},M_{12};j_3,M_3| j_{123})
\label{CG->6j}
\eeq
qui n'est autre que la relation de d\'efinition des 6-j
\`a partir des Clebsch-Gordan.
On faisant tendre $j_2,j_{12},j_{23},j_{123},j_{234},j$
vers l'infini on a
$$
\sum_{M'_1=-j_1}^{min(j_2+M_1+M_2,j_3+M_1+M_3)}
R^{\pm}\!(j_1,j_2)^{M'_2M'_1}_{M_1M_2}
R^{\pm}\!(j_1,j_3)^{M'_3M''_1}_{M'_1M_3}
(j_2,M'_2;j_3,M'_3| j_{23})
=
$$
\beq
(j_2,M_2;j_3,M_3| j_{23})
R^{\pm}\!(j_1,j_{23})^{M'_2+M'_3\>M''_1}_{M_1\>M_2+M_3}
\label{CGRR=CGR}.
\eeq

Finalement l'\'equation de Yang-Baxter pour les 6-j \ref{YB6j}
donne pour $j_4,j_{34},j_{234},j$ tendant vers l'infini
$$
\sum_{M'_1,M'_2,M'_3}
R^{\pm}\!(j_1,j_2)^{M'_2M'_1}_{M_1M_2}
R^{\pm}\!(j_1,j_3)^{M'_3M''_1}_{M'_1M_3}
R^{\pm}\!(j_2,j_3)^{M''_3M''_2}_{M'_2M'_3}
=
$$
\beq
\sum_{M'_1,M'_2,M'_3}
R^{\pm}\!(j_2,j_3)^{M'_3M'_2}_{M_2M_3}
R^{\pm}\!(j_1,j_3)^{M''_3M'_1}_{M_1M'_3}
R^{\pm}\!(j_1,j_2)^{M''_2M''_1}_{M'_1M'_2}
\label{YBR}
\eeq
qui est l'\'equation de Yang-baxter pour la matrice $R$.
Dans \ref{YBR}, les sommes, qui paraissent triples,
sont en fait simples \`a cause des contraintes issues des matrices $R$.
Si on prend $M'_2$ comme indice de sommation l'intervalle
est $[-j_2,\hbox{min}[j_1+M_1+M_2,j_3+M''_3+M''_2)]$
pour la somme du membre de gauche et
$M'_2\in[-j_2,\hbox{min}[j_1+M''_1+M''_2,j_3+M_3+M_2)]$
pour la somme du membre de droite.
Si dans l'\'equation de Yang-Baxter pour les 6-j \ref{YB6j}
on fait au contraire tendre $j_3,j_{34},j_{23},j_{234},j_{123},j$
vers l'infini on obtient
$$
\sum_{M'_1,M'_2}
R^{\pm}\!(j_1,j_2)^{M'_2M'_1}_{M_1M_2}
(j_1,M'_1;j_3,M_3|j_{13})
(j_2,M'_2;j_{13},M'_1+M_3|j_{123})
=
$$
\beq
\!\!\!\!\!\!\!\!\!\!\!\!\!
\!\!\!\!\!\!\!\!\!\!\!\!
\!\!\!\!\!\!\!\!\!\!\!\!
\sum_{\qquad\qquad\qquad j_{23}=max(j_{123}-j_1,-M_2-M_3)}^{j_2+j_3}
\!\!\!\!\!\!\!\!\!\!\!\!
\!\!\!\!\!\!\!\!\!\!\!\!
\!\!\!\!\!\!\!\!\!\!\!\!
\!\!\!\!\!\!
(j_1,M_1;j_{23},M_2+M_3|j_{123})
(j_2,M_2;j_3,M_3|j_{23})
\left\{ ^{j_1}_{j_2}
\,^{j_{3}}_{j_{123}}\right. \left |^{j_{13}}_{j_{23}}\right\}
e^{i\pi\epsilon(\Delta_{j_{3}}+\Delta_{j_{124}}-
\Delta_{j_{13}}-\Delta_{j_{23}})}
\label{RCGCG=CGCG6j}
\eeq
la somme du membre de gauche \'etant pour
$M'_1\in[\hbox{max}(-j_1,-j_{13}-M_3),j_2+M_2+M_1]$.

Il est clair que toutes ces relations permettent,
\`a partir de l'alg\`ebre des op\'erateurs $V$
de la partie \ref{p4.1},
d'obtenir l'alg\`ebre des op\'erateurs $\xi$ des
refs.\cite{P,MR} \'ecrits
en \ref{xi2}.
En effet dans cette construction, la matrice de passage
(avec changement d'espace)
est un Clebsch-Gordan.
Il est donc facile de voir que la relation \ref{RCGCG=CGCG6j}
montre que la fusion de ces $\xi$ est donn\'ee par la matrice $R$,
et que la relation \ref{CG->6j} montre que leur fusion
est donn\'ee par un Clebsch-Gordan, l'\'el\'ement de matrice
de $V$ dans la fusion restant inchang\'e.
Toutes les autres relations montrent la coh\'rence de l'alg\`ebre
des $\xi$.
}

\vskip 5mm

La deuxi\`eme \'etape de cette sous-partie consiste maintenant \`a
d\'emontrer l'alg\`ebre des $\xi$ de Babelon et Gervais
\'ecrits en Eq.\ref{xideV}.
Comme annonc\'e dans la sous-partie pr\'ec\'edente,
la matrice de changement de base est la limite du
Clebsch-Gordan de changement de base $A$ de l'autre
construction (Eq.\ref{xi2} lorsque le nombre $M$ suppl\'ementaire
introduit tend vers l'infini\footnote{
Il faut ici donner une petite partie imaginaire \`a $h$
(pour le couplage faible) et conserver $M_2$ r\'eel lorsqu'il tend
vers l'infini.
Sinon, la condition TI1 $j_2+M_2=p_2+\phat_2\pi/h$
imposerait que $j_2$ soit complexe.}:
\beq
\lim_{\hbox{\scriptsize Im}(hM_2)\to+\infty}
(j_1,M_1;j_2,M_2|j_{12})
=
A(j_1,j_{12})^{j_2-j_1}_{M_1}
\eeq
La matrice de passage $A$ est une notation propre \`a ce m\'emoire.
Elle exprime le passage des $\Vt=g_{...}^{..}V$ aux $\xi$.
Les $\Vt$ ont \'et\'e renorm\'es en \ref{Vtilde} par rapport aux $V$ pour que
leur
alg\`ebre soit purement donn\'ee par des 6-j,
sans normalisation suppl\'ementaire.
Les coefficients de normalisation $g$ contiennent des fonctions gamma
alors que les 6-j sont purement trigonom\'etriques.
C'est le guide que Gervais avait d\'ej\`a choisi en ref.\cite{G1}
pour d\'efinir des op\'erateurs $\psi=E_{..}^{..}(..)\,V$ dont
l'alg\`ebre devait \^etre purement trigonom\'etrique.
Cependant, \`a cette \'epoque,
la fusion n'\'etait trait\'ee correctement qu'\`a
l'ordre dominant.
Ainsi dans cette renormalisation par les coefficients
$E$, diff\'erents des $g$,
des fonctions gamma r\'eapparaissent aux ordres suivants.
Gervais avait cependant alors d\'efini
une matrice de passage $|j_1,\varpi_j)^{m_1}_{M_1}$ des $\psi=E.V$ aux $\xi$
de telle fa\c con que les $\xi$ soient covariants
pour le groupe quantique.
Il n'est donc pas \'etonnant que les deux matrices de passage
soient reli\'ees par
\beq
A(j_1,j)^{m_1}_{M_1}
=
{f(j+m_1)
\over
f(j)
}
{
E^{(j_1)}_{m_1}(\varpi_j)
\over
g_{j_1,j+m_1}^j
}
|j_1,\varpi_j)^{m_1}_{M_1}
\label{Ade|)}
\eeq
o\`u $f$ est une pure jauge.
On peut en effet v\'erifier que, quels que soient les facteurs $f$
introduits en \ref{Ade|)}, ils disparaissent
de l'alg\`ebre des $\xi$ ainsi d\'efinis.

Le principe de la d\'emonstration est identique
\`a celui de la limite des 6-j aux Clebsch-Gordan (Eq.\ref{6j->CG}).
Au lieu de passer d'une fonction hyperg\'eom\'etrique
$\,_4F_3$ \`a une $\,_3F_2$,
on passe d'une $\,_3F_2$ \`a une $\,_2F_1$.
Pour cela if faut d'abord transformer
l'expression \ref{CG1} des Clebsch-Gordan
car la fonction $\,_3F_2$ de l'Eq.\ref{CG1} a quatre
arguments qui tendent vers l'infini dans cette limite.
On la transforme
par \ref{transfF} avec $m=2$ et $-N=-M_2-j_2$,
pour obtenir
$$
(j_1,M_1;j_2,M_2|j_{12})
=
\sqrt{
\lfloor j_1+M_1 \rfloor \! !
\lfloor j_1-M_1+1 \rfloor_{j_1+M_1}
\lfloor -j_1+j_2+j_{12}+1 \rfloor_{j_1+j_2-j_{12}}
\over
\lfloor j_2+M_2 \rfloor \! !
\lfloor j_{12}+M_{12} \rfloor \! !
\lfloor j_1+j_2-j_{12} \rfloor \! !
\lfloor j_{2}-M_{2}+1 \rfloor_{j_{2}+M_{2}}
}
$$
$$
{
e^{
-ih\epsilon\left(
(j_1+j_2-j_{12})(j_1+j_2+j_{12}+1)/2
+j_{1}M_2+j_2M_{1}
\right)
}
\lfloor j_{12}-j_1-M_{2}+1 \rfloor_{j_{2}+M_{2}}
\over
\sqrt{
\lfloor j_{12}-M_{12}+1 \rfloor_{j_{12}+M_{12}}
\lfloor j_1-j_2+j_{12}+1 \rfloor_{j_1+j_2-j_{12}}
\lfloor 2j_{12}+2 \rfloor_{j_1+j_2-j_{12}}}
}
$$
\beq
{
\lfloor j_{12}-j_2+M_1+1 \rfloor_{j_2+M_2}
}
\,_3F_2\left(
^{-j_1-j_2+j_{12},M_{1}-j_{1},-M_2-j_2;}
_{j_{12}-j_{1}-M_{2}+1,
j_{12}-j_{2}+M_{1}+1;}
e^{ih\epsilon(j_1+j_2+j_{12}+1)}
\right)
\label{CG2}.
\eeq
Il est imm\'ediat de voir que la limite de la fonction
$\,_3F_2$ est
$$
\,_2F_1\left(
^{-j_1-j_2+j_{12},M_{1}-j_{1};}
_{
j_{12}-j_{2}+M_{1}+1;}
e^{-2ih\epsilon(j_1+j_{12})}
\right)
$$
qui est bien la fonction $\,_2F_1$
d\'efinissant la matrice de passage
$|j_1,\varpi_j)^{m_1}_{M_1}$
en ref.\cite{G1} Eq.2.24\footnote{
En ref.\cite{G1} Eq.2.24, deux cas sont consid\'er\'es
pour \'eviter tout entier n\'egatif $-N$ comme argument
inf\'erieur de la fonction hyperg\'eom\'etrique.
En effet pour un indice de sommation $\nu$ (cf Eq.\ref{Fdef})
sup\'erieur \`a $N$, il appara\^\i t alors un p\^ole.
Ceci peu tr\`es bien arriver dans les $\,_4F_3$,
$\,_3F_2$, $\,_2F_1$ consid\'er\'ees ici,
mais ce p\^ole sera toujours compens\'e par le z\'ero
d'un pr\'efacteur, ce qui aura pour effet de faire
commencer la sommation \`a $\nu=N+1$ et sera correct.
C'est en particulier la raison pour laquelle
l'expression \ref{CG1} des Clebsch-Gordan \'etait
inapropri\'ee dans la limite $M_2\to\infty$:
l'intervalle de sommation effectif obtenu par ce m\'ecanisme
\'etait pour $\nu$ de l'ordre de $M_2$,
tout en restant bien de taille fixe et finie.
}.
Il n'y a plus ensuite qu'\`a calculer les pr\'efacteurs.

Ceci nous permet ensuite de d\'eterminer l'alg\`ebre
des $\xi$ obtenus par la matrice A (Eq.\ref{xideV}).
On prend les \'equations \ref{CG->6j} et \ref{RCGCG=CGCG6j}
qui donnait la fusion et l'\'echange des $\xi$ obtenus par
changement de base par les Clebsch-Gordan et on applique la limite.
Lorsque $M_3$ tend vers l'infini,
l'\'equation \ref{CG->6j} donne
$$
\sum_{j_{23}=j_{123}-j_1}^{j_2+j_3}
\left\{ ^{j_1}_{j_3}\,^{j_2}_{j_{123}}
\right. \left |^{j_{12}}_{j_{23}}\right\}
A(j_1,j_{123})^{j_{3}-j_{123}}_{M_1}
A(j_2,j_{23})^{j_{23}-j_{3}}_{M_2}
=
$$
\beq
(j_1,M_1;j_{2},M_{2}| j_{12})
A(j_{12},j_{123})^{j_3-j_{123}}_{M_{12}}
\label{CG->6j'}
\eeq
et permet de prouver la fusion des $\xi$ \ref{fusxig},
alors que \ref{RCGCG=CGCG6j} donne
$$
\sum_{M'_1,M'_2\ ;\ M'_1=-j_1}^{j_2+M_2+M_1}
R^{\pm}\!(j_1,j_2)^{M'_2M'_1}_{M_1M_2}
A(j_1,j_{13})^{j_3-j_{13}}_{M'_1}
A(j_2,j_{123})^{j_{13}-j_{123}}_{M'_2}
=
$$
\beq
\sum_{j_{23}=j_{123}-j_1}^{j_2+j_3}
A(j_1,j_{123})^{j_{23}-j_{123}}_{M_1}
A(j_2,j_{23})^{j_3-j_{23}}_{M_2}
\left\{ ^{j_1}_{j_2}
\,^{j_{3}}_{j_{123}}\right. \left |^{j_{13}}_{j_{23}}\right\}
e^{i\pi\epsilon(\Delta_{j_{3}}+\Delta_{j_{124}}-
\Delta_{j_{13}}-\Delta_{j_{23}})}
\label{RCGCG=CGCG6j'}
\eeq
dans la m\^eme limite,
et permet de prouver leur \'echange \ref{brdxi}.

Les \'equations d'orthogonalit\'e des Clebsch-Gordan
\ref{orthCG} et \ref{orthCG'} permettent
\'egalement de prouver que la matrice de passage
$A$ est sa propre inverse.
Ceci est \'ecrit pour la matrice $|j_1,j)^{m_1}_{M_1}$
en ref.[P2] dans le cas de spins demi-entiers
et d'in\'egalit\'es TI3.
En revanche,
dans le cas de spins continus et d'in\'egalit\'es TI1,
les $\xi$ sont dans des repr\'esentations de dimension infinie.
Ainsi $M$ (et $m$ pour $V^{(j)}_m$)
ne sont pas born\'es inf\'erieurement,
la matrice $A$ est donc de dimension infinie
et la limite ``na\"\i ve'' de \ref{orthCG} et \ref{orthCG'}
donne une somme infinie.
Il faudrait donc pouvoir prouver une limite uniforme
mais cela para\^\i t \^etre faux car
cette somme infinie semble divergente.
Quant aux relations
\ref{CG->6j'} et \ref{RCGCG=CGCG6j'},
nous avons bien pris soin d'expliciter leurs intervalles
de sommation
pour v\'erifier qu'ils \'etaient fixes et finis.
Dans ces cas-l\`a, la limite est donc bien valide et ces relations justes.
Dans le cas de spins continus on connait donc
l'alg\`ebre des $\xi$ mais on ne sait
pas inverser le passage des $V$ aux $\xi$ (Eq.\ref{xideV}).

\vskip 2mm

Nous avons exhib\'e en ref.[P2] une troisi\`eme
et derni\`ere limite.
La premi\`ere limite d\'ecrite ici montre que
les 6-j tendent vers les Clebsch-Gordan ou la
matrice $R$ dans un certain r\'egime.
La deuxi\`eme est compl\`etement diff\'erente
et montre que le changement de base de Babelon et Gervais est un cas
limite du changement de base par les Clebsch-Gordan.

La premi\`ere limite montre en fait que la limite
des coefficients de fusion et d'\'echange des $V$
sont ceux des $\xi$.
En terme de mod\`eles de physique statisque
on a un mod\`ele \`a vertex qui est la limite
d'un mod\`ele IRF.
On peut alors se demander si la limite ne pourrait
pas s'\'etendre des coefficients de fusion et d'\'echange
aux op\'erateurs: les op\'erateurs $V$
eux-m\^emes ne pourraient-ils pas avoir les $\xi$ comme limite?
Et nous avons effectivement pu prouver
(pour des spins demi-entiers)
en ref.[P2] que dans la limite
o\`u ils agissent sur des \'etats de moment infini
les deux types d'op\'erateurs sont confondus:
la limite
$$
\lim_{\hbox{\scriptsize Im}(hj_{12})\to\infty}
A(J_1,j_{12})^{m_1}_{M_1}
=\delta_{m_1,M_1}
$$
montre que
$$
V^{(J)}_m
|\infty>
\sim
\xi^{(J)}_m
|\infty>.
$$
Ceci ne permet cependant de construire que des $\xi$
agissant sur l'espace trivial $|\infty>$
et non les $\xi$ g\'en\'eraux.
De plus ceci semble difficile \`a g\'en\'eraliser
au cas de spins continus \`a cause de probl\`emes
d'inversion de $A$ et d'interversion de limites.

\chapter{THEORIE DE LIOUVILLE: COUPLAGE FORT}
\label{p5}

Apr\`es avoir d\'etermin\'e l'alg\`ebre des op\'erateurs
de la th\'eorie de Liouville pour une charge centrale g\'en\'erique,
nous pouvons enfin \'etudier la physique du r\'egime de couplage fort
$d>1$ ou $C_{\hbox{\scriptsize Liou}}<25$.
Nous d\'eterminons dans une premi\`ere partie
une alg\`ebre coh\'erente \`a tous les ordres d'op\'erateurs
physiques de poids r\'eels
ce qui nous permet de construire dans une deuxi\`eme partie
un mod\`ele topologique fortement coupl\'e.

\vskip 2mm

Comme nous nous en sommes d\'ej\`a rendu compte
\`a plusieurs reprises,
le r\'egime de couplage fort conna\^\i t
des probl\`emes dus \`a l'apparition de quantit\'es complexes.
Dans le chapitre 2,
nous avons vu que la formule de KPZ \ref{gstring}
pour $\gamma_{\hbox{\scriptsize string}}$ \'etait complexe pour $d>1$.
De la m\^eme mani\`ere dans le chapitre 1,
la formule de Kac \ref{Kac}
donnait dans ce cas des poids complexes.
Par ailleurs, les masses carr\'ees des excitations d'une corde bosonique
non-critique vivant dans un espace-temps de dimension $d>1$,
pour un op\'erateur de vertex
dont l'habillage par la gravit\'e est
un op\'erateur de Liouville de poids $\Delta$,
valent $2(\Delta-1+N)$
et sont alors complexes.
Il en est de m\^eme pour les supercordes.
Tout ceci ne saurait \^etre acceptable physiquement.

Il existe certes dans ce r\'egime
des op\'erateurs et des \'etats de poids r\'eels,
mais ils engendrent par fusion et \'echange
des op\'erateurs et des \'etats de poids complexes,
\`a cause du caract\`ere complexe des charges d'\'ecran $\alpha_\pm$
(ou des param\`etres de d\'eformation $h$ et $\hhat$).

\vskip 3mm

Ceci conduit \`a l'id\'ee de la troncature expos\'ee dans la premi\`ere partie:
on cherche \`a d\'eterminer des combinaisons lin\'eaires
d'op\'erateurs de poids r\'eels
qui se d\'ecouplent des op\'erateurs
de poids complexes
et laissent stable un espace de Hilbert d'\'etats de poids
\'egalement r\'eels.
La fin de la preuve\footnote{En admettant que la continuation
des coefficients de l'alg\`ebre d'op\'erateurs par la sym\'etrie
$J\to-J-1$ donne bien l'alg\`ebre des op\'erateurs de
spins n\'egatifs (et nombre n\'egatif de charges d'\'ecran),
puisque c'est ainsi que nous avons obtenu l'alg\`ebre des op\'erateurs
de poids positifs.}
de cette troncature d\'ej\`a \'etudi\'ee
en refs\cite{GN5,GR,G2,G3} est obtenue en [P3]\footnote{
Nous travaillerons ici comme en [P3] dans la base des ondes de Bloch,
alors que les travaux \cite{GR,G2,G3} ont plut\^ot \'et\'e
r\'ealis\'es dans celle du groupe quantique,
puisque l'alg\`ebre des op\'erateurs n'\'etait alors pas connue
dans la premi\`ere base.}.
Elle n'est possible que pour les dimensions sp\'eciales
$C=7,13,19$
(et $C=3,5,7$ dans le cas supersym\'etrique),
et pour des spins fractionnaires (quarts ou sixi\`emes d'entiers).
L'inclusion de ces spins fractionnaires dans le spectre
est exactement ce qui n\'ecessaire \`a l'obtention d'une fonction
de partition invariante modulaire \cite{GR}
(c'est la m\^eme que celle d'un boson libre
compactifi\'e sur un cercle de rayon $\sqrt{(25-C)/3}$).
Il y a en fait deux alg\`ebres coh\'erentes d'op\'erateurs physiques,
les $\chi_-$ de poids r\'eels n\'egatifs et les $\chi_+$
de poids r\'eels sup\'erieurs \`a 1.

Comme d\'ej\`a indiqu\'e,
cette troncature de l'alg\`ebre d'op\'erateurs
pour ces dimensions sp\'eciales n'est pas
sans rappeler celle des mod\`eles minimaux.
Elles apparaissent toutes les deux pour les dimensions
o\`u le spectre des poids conformes donne lieu
\`a une d\'eg\'en\'erescence des valeurs propres de
la matrice de monodromie.
Dans une deuxi\`eme partie, nous montrons par une discussion
non publi\'ee jusqu'\`a pr\'esent (et donc imprim\'ee en italique)
qui para\^\i tra en [P5],
que la d\'eg\'en\'erescence de la matrice de monodromie est essentielle
pour obtenir une troncature de l'alg\`ebre d'op\'erateurs.
Ceci sera confirm\'e par le fait que
les dimensions sp\'eciales ainsi que les
spins fractionnaires peuvent \^etre obtenus
en demandant une d\'eg\'en\'erescence maximale
de ces valeurs propres.
Nous en profitons pour bien montrer
ce que signifie cette troncature de l'alg\`ebre d'op\'erateurs
en termes de fonctions de corr\'elations physiques.

L'\'etude des v\'eritables cordes non critiques,
c'est-\`a-dire ayant 19, 13 ou 7 degr\'es de libert\'e de
mati\`ere et vivant dans un espace-temps
de dimension 19, 13 ou 7,
a \'et\'e abord\'e en ref.\cite{BG}
(ainsi que le cas supersym\'etrique).
Cependant, le calcul des amplitudes
semble encore hors de port\'ee.
C'est pourquoi nous \'etudions dans la troisi\`eme partie
de ce chapitre un mod\`ele simplifi\'e\footnote{
Comme ce mod\`ele n'a fait l'objet que d'une simple
lettre [P4], nous faisons ici une \'etude
relativement d\'etaill\'ee qui sera reprise en ref.[P5],
publi\'ee tr\`es prochainement.}.
Ce mod\`ele est construit \`a partir de deux copies
diff\'erentes des th\'eories
construites dans la partie 5.1.
Comme les mod\`eles minimaux habill\'es par la gravit\'e,
il comporte un degr\'e de libert\'e de mati\`ere
habill\'e par la gravit\'e.
La mati\`ere est constitu\'ee d'op\'erateurs $\chi'_-$,
et la gravit\'e d'op\'erateurs $\chi_+$.
Les dimensions sp\'eciales 7, 13, 19 sont en effet
leurs propres compl\'ementaires \`a 26.
Il est en outre possible de construire des op\'erateurs
locaux de poids (1,1) \`a partir de $\chi'_-$ et de $\chi_+$.
La cohomologie BRST doit ensuite enlever les excitations
transverses
de deux degr\'es de libert\'e,
n'en laissant plus aucun dans le cas pr\'esent.
Ce mod\`ele doit donc \^etre topologique.
Nous calculons l'exposant critique\footnote{
Nous obtenons \'egalement
dans la deuxi\`eme partie, en \ref{gstrh}, une formule pour les genres
sup\'erieurs
qui n'est pas en ref.[P4] en annexe, mais figurera en [P5].}
\beq
\gamma_{\hbox{\scriptsize string}}
=
{d-1\over 12}
\qquad\qquad
d=7,13,19
\label{gstr}
\eeq
qui ne vaut pas que pour ce mod\`ele topologique mais pour tout
mod\`ele de cordes dans une des dimensions sp\'eciales
pour des champs de mati\`ere quelconques habill\'es par les $\chi_+$.
Cette valeur est la partie r\'eelle de la formule de KPZ \ref{gstring},
ce qui pourrait \^etre en accord avec les simulations
num\'eriques effectu\'ees dans le r\'egime de couplage fort\cite{BH}.
Nous calculons aussi
les fonctions \`a trois points
de cette th\'eorie.
Nous \'ecrivons ensuite une action effective
avec le nouveau terme cosmologique,
gr\^ace \`a laquelle,
nous inspirant de ce qui a \'et\'e fait dans le r\'egime
de couplage faible par Di Francesco et Kutasov \cite{dFK},
nous \'ecrivons ensuite une action effective et d\'eterminons
les fonctions \`a $N+1$ points avec un moment externe nul,
en d\'erivant les fonctions \`a $N$ points
par rapport \`a la constante cosmologique.
Les fonctions de Green avec moments externes quelconques
sont ensuite obtenus par sym\'etrie.
Soustrayant des fonctions de corr\'elation totales
les fonctions \`a une particule r\'eductible,
nous obtenons
les fonctions \`a $N$ points \`a une particule irr\'eductibles
(jusqu'\`a $N=6$ dans la pratique).
Elles sont analytiques et sym\'etriques,
ce qui confirme la pertinence
de l'action effective employ\'ee.

\section{Troncature de l'alg\`ebre d'op\'erateurs}
\label{p5.1}

Nous commen\c cons par une petite mise au point.
Nous consid\'erons ici des spins g\'en\'eralis\'es,
c'est-\`a-dire continus ou plus tard fractionnaires.
Nous avons dit dans le chapitre pr\'ec\'edent
que dans le cas de spins continus
la notion de spins $J$ et $\Jhat$ disparaissait,
dans le sens qu'un spin effectif $\Je$ continu
ne se d\'ecompose pas de mani\`ere unique sous la forme
$\Je=J+(\pi/h)\Jhat$.
Dans le cas pr\'esent $(\pi/h)$ est complexe,
et nous consid\'ererons toujours des spins
$J$ et $\Jhat$ r\'eels.
La d\'ecomposition $\Je=J+(\pi/h)\Jhat$ du spin effectif continu
complexe $\Je$ en termes de deux spins continus r\'eels
$J$ et $\Jhat$ est alors unique.
C'est pourquoi nous utiliserons dans ce chapitre
parfois $\Je$ et d'autres fois $J$ et $\Jhat$.
L'alg\`ebre des op\'erateurs s'\'ecrira n\'eanmoins
toujours en termes de spins effectifs et de 6-j effectifs.

\vskip 2mm

R\'e\'ecrivons le poids conforme d'un op\'erateur $V^{(J,\Jhat)}_{m,\mhat}$
donn\'e en \ref{deltajj}:
\beq
\Delta_{J,\Jhat}=
{C-1\over 24}-{1\over24}
\left(
(J+\Jhat+1)\sqrt{C-1}
-(J-\Jhat)\sqrt{C-25}
\right)^2
\label{deltajj2}
\eeq
qui est valable \'egalement pour $J$ et $\Jhat$ continus.
Dans le r\'egime de couplage fort,
la charge centrale de Liouville, $C$,
est comprise entre 1 et 25,
et ces poids sont complexes,
en g\'en\'eral.
Il y a cependant quelques cas particuliers pour lesquels
les poids sont r\'eels.
En effet,
$\sqrt{C-25}$ est imaginaire pur,
$\sqrt{C-1}$ est r\'eel,
et, dans les cas $J-\Jhat=0$ ou $J+\Jhat+1=0$
le poids est r\'eel ($J$ et $\Jhat$ sont r\'eels).
Si on les d\'esigne par les quantit\'es
$(n,m)=(2\Jhat+1,2J+1)$ (comme BPZ dans le cas de spins demi-entiers
ou $n$ et $m$ entiers),
ils s'agit des cas diagonaux.
On a des poids n\'egatifs\footnote{
Plus pr\'ecis\'ement,
ils sont strictement n\'egatifs
pour $J=\Jhat>0$,
nul pour $J=\Jhat=0$ qui est l'op\'erateur identit\'e,
et positif pour $J=\Jhat=-1/2$ i.e. $(n,m)=(0,0)$
qui est un cas pathologique.}
\beq
\Delta_{J,J}
=
-
{C-1\over 6}
J(J+1),
\label{w-}
\eeq
pour le cas de la digonale principale $(n,m)=(n,n)=(2J+1,2J+1)$
et des poids positifs
\beq
\Delta_{-J-1,J}=
1+{25-C\over 6}J(J+1)
\label{w+}
\eeq
pour les repr\'esentations
de l'autre diagonale $(n,m)=(n,-n)=(-2J-1,2J+1)$
pour $J$ et $n$ r\'eels quelconques.

Ceci n'est pas une grande d\'ecouverte et ne r\'esoud pas
le probl\`eme.
En effet, comme nous l'avons soulign\'e en
introduction de ce chapitre, par fusion ou \'echange,
ces op\'erateurs de poids r\'eels donnent des op\'erateurs
et des \'etats de poids complexes.
On ne peut donc pas se restreindre \`a ce sous ensemble d'op\'erateurs
de mani\`ere coh\'erente.
C'est le but de cette partie de pr\'esenter
la construction d'une sous alg\`ebre ferm\'ee\footnote{
Nous parlons souvent de sous-alg\`ebre ferm\'ee par
souci de clart\'e,  bien que
ce soit un pl\'eonasme. Une sous-alg\`ebre est par d\'efinition
un sous-espace vectoriel d'une alg\`ebre,
ferm\'e (ou stable) pour les op\'erations de l'alg\`ebre
(fusion et \'echange ici).}
(pour les op\'erations d'\'echange et de fusion)
d'op\'erateurs de poids r\'eels
agissant sur un espace de Hilbert d'\'etats de poids r\'eels.
Ceci garantira le d\'ecouplage des op\'erateurs et \'etats
de poids complexes.

Tous les d\'etails sont donn\'es en ref.[P3]
et nous nous contenterons ici d'esquisser
la d\'emonstration.
Ce travail fait suite aux refs.\cite{GN5,GR,G2,G3} o\`u
la fermeture de la fusion n'avait pas \'et\'e v\'erifi\'ee
pour les descendants.
De plus, ceci n'avait \'et\'e fait que pour
des op\'erateurs de spins demi-entiers
alors qu'il a \'et\'e d\'emontr\'e\cite{GR}  que l'inclusion
dans le spectre de spins fractionnaires
(cf Eqs.\ref{spfrac}, \ref{spfrac+} plus bas\footnote{
Nous utilisons la d\'enomination de spins fractionnaires
pour les spins \ref{spfrac}, \ref{spfrac+} qui peuvent \^etre des quarts
ou des sixi\`emes d'entiers,
et non pour les spins standards demi-entiers.
})
\'etait n\'ecesssaire
pour avoir l'invariance modulaire\footnote{
Ces spins fractionnaires avaient pu \^etre introduits pour les
\'etats, et la fermeture de l'alg\`ebre
(hors descendants pour la fusion)
avait pu \^etre prouv\'ee pour des op\'erateurs de spins
demi-entiers.
Mais la coh\'erence de la th\'eorie exige une correspondance
entre \'etats et op\'erateurs.
C'est pourquoi nous avonc d\^u \'etendre la construction
\`a des op\'erateurs de spins fractionnaires.
}.
C'est la raison essentielle qui nous a pouss\'es \`a
\'etudier l'alg\`ebre des op\'erateurs de spins continus
dans le chapitre pr\'ec\'edent.
L'alg\`ebre des spins fractionnaires n'est en effet pas
fondamentalement diff\'erente de celle des spins continus.
Nous verrons plus loin que ce sont
uniquement
les propri\'et\'es \ref{lien6j}, \ref{lien6j+}
qui distinguent les spins fractionnaires des spins continus.

\vskip 2mm

Consid\'erons pour le moment le cas le plus simple,
celui des op\'erateurs de spins effectifs $\Je_i=J_i+(\pi/h)J_i$
pour $J\in R$.
Nous notons ces spins effectifs particuliers
\beq
\Jme_i=J_i+(\pi/h)J_i.
\label{Jme}
\eeq
C'est la g\'en\'eralisation \`a des spins continus
de la diagonale de la table Kac et leurs poids r\'eels n\'egatifs sont
donn\'es par \ref{w-} avec $J=J_i$ continu.
Nous supposons donc que tous les op\'erateurs et \'etats initiaux
sont dans ce sous-ensemble, i.e. caract\'eris\'es par $\Je_i=\Jme_i$.
On a vu (cf Eqs.\ref{Fgen}, \ref{Bgen}) que
la fusion et l'\'echange de deux op\'erateurs $V^{(\Je)}$
est essentiellement donn\'ee par un produit de deux 6-j,
l'un avec param\`etre de d\'eformation $h$ et spins $\Je_i$,
l'autre avec param\`etre de d\'eformation $\hhat$ et spins $\Jehat_i$.
Pour les valeurs particuli\`eres $\Jme_i$ des spins,
on a $\Jehat_i\equiv(h/\pi)\Je_i=J_i+(h\pi)J_i=J_i+(\pi/\hhat)J_i$,
et $\Jehat_i$ est donc le transform\'e de $\Je_i$ par $h\to\hhat$
et est donc aussi not\'e $\Jmehat_i$.

Examinons plus en d\'etail ce qui se passe pour la fusion,
le m\'ecanisme est similaire pour l'\'echange.
Les deux 6-j du coefficient de fusion
\ref{Fgen} sont alors
\beq
\sixje \Jme_1,\Jme_3,\Jme_2,\Jme_{123},\Je_{12},\Jme_{23}
\sixjehat \Jmehat_1,\Jmehat_3,\Jmehat_2,\Jmehat_{123},\Jehat_{12},\Jmehat_{23}
\label{6j6j}
\eeq
Les cinq spins effectifs issus du membre de gauche
(point de d\'epart de la transformation)
sont, par choix, du type $\Je_i=\Jme_i$
alors que le spin $\Je_{12}$
de l'op\'erateur obtenu par fusion est (en g\'en\'eral)
quelconque:
$\Je_{12}=J_{12}+(\pi/h)\Jhat_{12}$.
La fusion engendre donc une combinaison lin\'eaire
d'op\'erateurs de poids
complexes.
Seuls certains d'entre eux, pour lesquels $J_{12}=\Jhat_{12}$,
ont un poids r\'eel.
Il faudrait donc que les coefficients des op\'erateurs
de poids complexes s'annulent.
Aucun des deux 6-j de \ref{6j6j} ne s'annule
dans ce cas et \ref{6j6j} ne peut donc s'annuler dans
les cas $J_{12}\ne \Jhat_{12}$ comme il le faudrait.
La solution consiste \`a ne plus consid\'erer aussi
simplement la fusion
d'op\'erateurs $V^{(\Jme)}_{\me}$
mais de combinaisons lin\'eaires de tels op\'erateurs.
La fusion de telles combinaisons lin\'eaires
engendrerait toujours tous les m\^emes op\'erateurs
$V^{(\Je_{12})}$ mais les coefficients de fusion
ne seraient plus donn\'es par \ref{6j6j}
mais par une combinaison lin\'eaire de produits du type \ref{6j6j}.
Et on voudrait que cette combinaison lin\'eaire soit
nulle pour $J_{12}\ne\Jhat_{12}$.
Ceci commence fort \`a ressembler \`a une relation
d'orthogonalit\'e entre 6-j,
sauf que nous avons des 6-j
de deux types diff\'erents, les uns d\'eform\'es
par $h$, les autres par $\hhat$.
Par cons\'equent,
\c ca ne marche pas,
en g\'en\'eral.
Mais il se trouve,
et c'est ce qui caract\'erise les dimensions sp\'eciales,
que lorsque $C=7,13$ ou 19,
les 6-j des deux types (pour des $\Je$ bien choisis) sont \'egaux,
\`a un simple signe pr\`es.
En introduisant les signes et normalisations ad\'equats dans la d\'efinition
d'op\'erateurs combinaisons lin\'eaires de $V^{(J,\Jhat)}_{m,\mhat}$,
le miracle se produit:
l'orthogonalit\'e des 6-j permet
le d\'ecouplage des op\'erateurs de spins $J_{12}\ne\Jhat_{12}$,
c'est-\`a-dire de poids complexes.
Et qui plus est,
le r\'esultat est encore une combinaison lin\'eaire
d'op\'erateurs du type consid\'er\'e initialement,
ce qui assure la stabilit\'e de l'alg\`ebre.

La propri\'et\'e fondamentale[P3] est
\beq
\sixjehat \Jmehat_1,\Jmehat_3,\Jmehat_2,\Jmehat_{123},\Jehat_{12},\Jmehat_{23}
=
(-1)^{(2+s)
\phi(J_1,J_2,J_3,J_{12},J_{23},J_{123})
}
\sixje \Jme_1,\Jme_3,\Jme_2,\Jme_{123},\Je_{12},\Jme_{23}
\label{lien6j}
\eeq
avec $p_{k,l}\equiv j_k+j_l-j_{kl}$ pour $k$ et $l$ dans $\{1,2,3,12,23\}$
et
$$
\phi(J_1,J_2,J_3,J_{12},J_{23},J_{123})=
p_{1,2}(2J_2+2J_3)+p_{2,3}2J_3+p_{1,23}2J_{23}+p_{12,3}2J_3
$$
\beq
+p_{1,2}(p_{1,2}+1)/2
+p_{2,3}(p_{2,3}+1)/2
+p_{1,23}(p_{1,23}+1)/2
+p_{12,3}(p_{12,3}+1)/2
\label{phi}.
\eeq
Comme on l'a annonc\'e plus haut,
elle peut \^etre prouv\'ee pour une charge centrale
\beq
C=1+6(s+2)\ ,\ s=-1,0,1
\hbox{ donc }
C=7,13,19
\eeq
et des spins
\beq
J_i={
n_i
\over
2(2+s)
}
\quad , \quad
n_i\in Z
\label{spfrac}
\eeq
qui sont donc des demi-entiers pour $C=7$
mais des quarts d'entiers pour $C=13$
et des sixi\`emes d'entiers pour $C=19$.

Nous avons \'etendu dans le chapitre pr\'ec\'edent la d\'efinition
des 6-j standards (d\'efinis pour des spins
demi-entiers) \`a des spins continus,
v\'erifiant \'egalement leur orthogonalit\'e,
l'\'equation de
Bidenharn-Elliot...etc.
En revanche, le lien \ref{lien6j} entre les 6-j d\'efinis
avec les param\`etres de d\'eformation $h$ et $\hhat$
ne peut \^etre \'etendu qu'au cas interm\'ediaire
des spins fractionnaires \ref{spfrac}.
On pourrait se demander s'il ne s'agit pas d'une simple
curiosit\'e math\'ematique et si ces spins fractionnaires
ne doivent pas \^etre exclus,
se restreignant aux spins demi-entiers.
On pourrait \'egalement se demander
s'il n'existerait pas une g\'en\'eralisation de \ref{lien6j}
\`a des spins continus,
que nous n'aurions pas su d\'emontrer.
Les r\'eponses \`a ces questions semblent \^etre n\'egatives
et les spins fractionnaires semblent avoir un sens plus profond.
En effet l'adjonction au spectre de ces spins fractionnaires
est exactement ce qui est n\'ecessaire \`a l'obtention
d'une fonction de partition invariante modulaire,
comme cela a \'et\'e prouv\'e en \cite{GR}.
La fonction de partition obtenue est alors identique
\`a celle d'un boson compactifi\'e sur un cercle
de rayon $\sqrt{2(2\pm s)}$,
bien que la physique d\'ecrite soit diff\'erente.

Nous d\'efinissons donc les op\'erateurs $\chi_-$
de poids \ref{w-} n\'egatifs
\`a partir des op\'erateurs $V^{(\Jme)}_{\me}$ pr\'ec\'edents\footnote{
${\cal N}$ d\'esigne l'ensemble des entiers naturels (i.e.
positifs ou nuls)}:
\beq
\chi_-^{(J_1)}
{\cal P}_{\Jme_2}
\equiv
\sum_{J_{12}\atop p_{1,2}\equiv J_1+J_2+J_{12}\in {\cal N}}
(-1)^{(2+s)(2J_2p_{1,2}
+{p_{1,2}(p_{1,2}+1)\over 2})}
g_{\Jme_1\Jme_2}^{\Jme_{12}}
\>
V^{(\Jme_1)}
_{\Jme_2-\Jme_{12}}
\>
{\cal P}_{\Jme_2}
\label{chidef-}
\eeq
pour un \'etat entrant de spin effectif $\Jme_2$ fix\'e
(${\cal P}_{\Jme_2}$ est le projecteur sur le module
de Verma correspondant).
La quantit\'e $p_{1,2}\equiv J_1+J_2+J_{12}$ doit
\^etre un entier positif ou nul, ce
qui constitue la seule contrainte des r\`egles de s\'election
pour les spins continus,
comme on l'a vu dans le chapitre pr\'ec\'edent.
Les coefficients de normalisation $g$ sont ceux qui apparaisent
dans les matrices de fusion et d'\'echange des $V$ en Eqs.\ref{Fgen},
\ref{Bgen}.
Ils sont d\'efinis pour des spins continus et donc pour
des spins fractionnaires du type \ref{spfrac}.
Les poids des op\'erateurs $\chi_-$ sont
\beq
\Delta(\chi_-^{(J)})
=
\Delta_{J,J}
=
-(2+s)J(J+1)
\label{wchi-}.
\eeq
Les ${\cal P}_{\Jme_i}$ projettent sur
des \'etats du m\^eme type et
la troncature a lieu dans l'espace de Hilbert suivant
\beq
{\cal H}^-_{s\,\hbox{\scriptsize phys}}
=
\bigoplus_{J=n/2(2+s),\> n\in Z}
{\cal H}_{J,J}
\label{H-}
\eeq
qui est stable par action des $\chi_-$.
Les poids des \'etats de cet espace de Hilbert
sont \'egalement donn\'es par \ref{wchi-}
et sont donc r\'eels n\'egatifs.

Lors d'une fusion ou d'un \'echange d'op\'erateurs $\chi_-$,
les normalisations $g$ s'\'eliminent,
tous les signes ont \'et\'e calcul\'es pour se combiner correctement,
et les 6-j
redonnent exactement la combinaison lin\'eaire
qui d\'efinit des op\'erateurs $\chi_-$
gr\^ace \`a la relation \ref{lien6j}
et \`a l'orthogonalit\'e des 6-j \ref{orth6j}.
Finalement l'\'echange et la fusion des $\chi_-$ ont des expressions
tr\`es simples:
\beq
\chi_-^{(J_1)}
\chi_-^{(J_2)}
=
e^{-2i\pi\epsilon(2+s)J_1J_2}
\chi_-^{(J_2)}
\chi_-^{(J_1)},
\label{brdchi-}
\eeq
$$
\chi_-^{(J_1)}
\chi_-^{(J_2)}
{\cal P}_{\Jme_3}
=
\sum_{J_{12}<J_1+J_2}
(-1)^{(2+s)2J_3(J_1+J_2-J_{12})}
$$
\beq
\sum_{\{\nu\}}
\chi_-^{(J_{12},\{\nu\})}
{\cal P}_{\Jme_3}
<\!\Jme_{12},{\{\nu_{12}\}} |
\chi_-^{(J_1)} | \Jme_2 \! >
\label{fuschi-}.
\eeq
On peut \'evidemment v\'erifier que les matrice de fusion
et d'\'echange des op\'erateurs
$\chi_-$ satisfont aux \'equations
polynomiales,
c'est-\`a-dire que les $\chi_-$ ont une alg\`ebre coh\'erente.

Examinons maintenant
la monodromie de ces op\'erateurs autour de 0.
Ils sont construits comme combinaisons lin\'eaires
d'op\'erateurs $V$ qui diagonalisent
cette monodromie.
Chacun de ces op\'erateurs $V$ a cependant en g\'en\'eral une
monodromie diff\'erente et la combinaison lin\'eaire
qui d\'efinit les $\chi$ peut ne plus redonner
un $\chi$ apr\`es monodromie.
D'apr\`es l'\'equation \ref{monodr3}, la monodromie des op\'erateurs $V$
impliqu\'es dans leur d\'efinition \ref{chidef-}
est donn\'ee par le facteur
\beq
e^{2i\pi\epsilon[\Delta(\Jme_1)+\Delta(\Jme_2)-\Delta(\Jme_{12})]}
\label{fact}
\eeq
o\`u $\Jme_{12}=\Jme_1+\Jme_2-p_{1,2}(1+\pi/h)$.
Si ce facteur
diff\`ere selon les valeurs de $p_{1,2}$, chaque op\'erateur de la somme
acquiert un facteur diff\'erent lors de la monodromie
et le r\'esultat n'est plus un $\chi_-$.
Fort heureusement, ce n'est pas le cas puisqu'on peut v\'erifier
que les poids $\Delta(\Jme_{12})$ pour des valeurs de l'entier
$p_{1,2}$ diff\'erentes ne diff\`erent que par des entiers
qui ne contribuent pas au facteur \ref{fact}.
La propri\'et\'e remarquable est
\beq
\Delta_{{n\over 2(2+s)}+p,{n\over 2(2+s)}+p}
=
\Delta_{{n\over 2(2+s)},{n\over 2(2+s)}}
+
p[n+(p+1)(2+s)]
\label{shiftdelta}
\eeq
o\`u $p[n+(p+1)(2+s)]$ est un entier.
Comme annonc\'e dans la partie 3.2,
nous sommes donc pr\'ecis\'ement dans le cas (exceptionnel)
o\`u la matrice de monodromie est d\'eg\'en\'er\'ee.
Cette propri\'et\'e est fondamentale et semble
caract\'eriser (presque) compl\`etement
les charges sp\'eciales, ainsi que les spins fractionnaires.
Nous y reviendrons dans la prochaine partie.

\vskip 5mm

On d\'efinit de mani\`ere similaire des op\'erateurs $\chi_+$
de poids \ref{w+} positifs.
Ils sont construits \`a partir d'op\'erateurs
de la diagonale secondaire de la table de Kac
g\'en\'eralis\'es \`a des spins continus
c'est-\`a-dire
\`a des spins effectifs
\beq
\Jpe_i=-J_i-1+(\pi/h)J_i
=-1+J_i(1-\pi/h)
\label{Jpe}.
\eeq
On a alors $\Jpehat_i=J_i+(-J_i-1)(\pi/h)$.

Il y a cependant dans ce cas une difficult\'e suppl\'ementaire.
Elle n'est pas directement li\'ee aux spins n\'egatifs ($-J-1$ ou $J$)
car les 6-j (et leurs \'equations) sont d\'efinis
pour des spins continus et en particulier n\'egatifs.
Les spins des $\chi_-$, par exemple, peuvent
tr\`es bien \^etre n\'egatifs.
La difficult\'e provient de nombres de charges d'\'ecran n\'egatifs.
En effet, la r\`egle TI1 pour le vertex \ref{OneCondEff}
implique que $p_{1,2}\equiv J_1+J_2-J_{12}$,
le nombre de charges d'\'ecran\footnote{
Nous parlons de nombre de charges d'\'ecran
pour employer des termes familiers au plus grand
nombre de lecteurs, bien qu'il
n'y en ait pas dans la construction op\'eratorielle
d\'ecrite ici.
Il s'agit simplement des nombres entiers des r\`egles de
s\'election \ref{OneCondEff},
qui correspondent \`a des nombres de charges d'\'ecran
dans une approche int\'egrale.}
de type $\alpha_-$,
soit un entier positif.
Mais dans le cas de spins \ref{Jpe} i.e. $(-J_i-1,J_i)$,
le nombre de charges d'\'ecran de type $\alpha_-$
est $\phat_{1,2}=(-J_1-1)+(-J_2-1)-(-J_{12}-1)=-p_{1,2}-1$
et les deux nombres de charges d'\'ecran ne peuvent pas \^etre simultan\'ement
positifs.
L'expression du coefficient de fusion \ref{Fgen} d'op\'erateurs
de spins $\Jpe_+$ contient les 6-j
\beq
\sixje \Jpe_1,\Jpe_3,\Jpe_2,\Jpe_{123},\Je_{12},\Jpe_{23}
\sixjehat \Jpehat_1,\Jpehat_3,\Jpehat_2,\Jpehat_{123},\Jehat_{12},\Jpehat_{23}
\label{6j6j+}
\eeq
et on peut v\'erifier que le deuxi\`eme
a des nombres de charges d'\'ecran
(c'est-\`a-dire les entiers dans \ref{sixjgen}) n\'egatifs.

L'alg\`ebre des op\'erateurs n'a \'et\'e d\'etermin\'ee
que pour des nombres positifs de charges d'\'ecran,
de m\^eme on ne conna\^\i t que des 6-j \`a nombres positifs
de charges d'\'ecran.
Comme d'autres auteurs l'ont fait pour la fonction \`a trois points
\cite{GL,D,G4}
nous effectuons donc une continuation
\`a des nombres n\'egatifs de charges d'\'ecran,
pour les coefficients de fusion et d'\'echange.
Pour la continuation de la fonction \`a trois points \`a des
nombres entiers
n\'egatifs de charges d'\'ecran \cite{D,G4}
il avait suffit de d\'efinir un produit de $a$ \`a $b$
pour $b$ inf\'erieur \`a $a$.
Il en est de m\^eme ici pour les coefficients $g$
(qui ne sont pas autre chose que les normalisations des
fonctions \`a trois points).
En revanche pour les 6-j contenus dans $F$ et $B$,
nous avons d\^u faire intervenir en [P3] une identit\'e
non triviale entre fonctions hyperg\'eom\'etriques\footnote{
Elle fait intervenir des r\'earrangements de termes
non triviaux et n'\'etait pas connue,
semble-t-il, dans le cas
q-d\'eform\'e.
J.-L. Gervais avait d\'ej\`a exprim\'e en \cite{G3} la sym\'etrie des
Clebsh-Gordan
sous $J\to-J-1$ gr\^ace \`a une identit\'e entre
fonctions hyperg\'eom\'etriques $_3F_2$.
Elle n'\'etait connue que dans le cas non-d\'eform\'e.
L'identit\'e  sur les fonctions $_4F_3$ q-d\'eform\'ees permet d'obtenir
celle sur les fonction $_3F_2$ par limite.}
$_4F_3$ (Eq.5.1 dans [P3]).
Elle permet de d\'emontrer que, via cette continuation,
les 6-j ont la sym\'etrie $j\to -j-1$:
\beq
\left\{ ^{-j_1-1}_{-j_3-1}\,^{-j_2-1}_{-j_{123}-1}
\right. \left |^{-j_{12}-1}_{-j_{23}-1}\right\}
=
\left\{ ^{j_1}_{j_3}\,^{j_2}_{j_{123}}
\right. \left |^{j_{12}}_{j_{23}}\right\}
\label{sixj-j-1}
\eeq
qui vaut pour des spins continus quelconques.
Ceci se traduit sur les 6-j effectifs par
\beq
\sixje \Je_1,\Je_3,\Je_2,\Je_{123},\Je_{12},\Je_{23}
=
\sixje -\Je_1-(1+\pi/h),-\Je_3-(1+\pi/h),-\Je_2-(1+\pi/h),
-\Je_{123}-(1+\pi/h),-\Je_{12}-(1+\pi/h),-\Je_{23}-(1+\pi/h),
\label{sixje-j-1}.
\eeq
et permet de remplacer le deuxi\`eme 6-j de \ref{6j6j+}
par un 6-j avec nombre positif de charges d'\'ecran.

Nous d\'eterminons ensuite l'\'equivalent de la propri\'et\'e \ref{lien6j}
qui est dans ce cas
$$
\sixjehat -\Jpehat_1-1-\pi/\hhat,-\Jpehat_3-1-\pi/\hhat,-\Jpehat_2-1-\pi/\hhat,
-\Jpehat_{123}-1-\pi/\hhat,-\Jpehat_{12}-1-\pi/\hhat,-\Jpehat_{23}-1-\pi/\hhat
=
$$
\beq
\!
(-1)^{(2-s)
\phi(J_1,J_2,J_3,J_{12},J_{23},J_{123})}
\sixje -\Jpe_1-1-\pi/\hhat,-\Jpe_3-1-\pi/\hhat,-\Jpe_2-1-\pi/\hhat,
-\Jpe_{123}-1-\pi/\hhat,-\Jpe_{12}-1-\pi/\hhat,-\Jpe_{23}-1-\pi/\hhat
\label{lien6j+}
\eeq
qui peut cette fois \^etre d\'emontr\'ee pour
\beq
J_i={
n_i
\over
2(2-s)
}
\quad , \quad
n_i\in Z
\label{spfrac+}
\eeq
et toujours $C=1+6(s+2)$, $s=-1,0,1$.
On voit que \ref{sixje-j-1} et \ref{spfrac+}
permettent de ramener les deux 6-j \ref{6j6j+}
\`a deux 6-j du m\^eme type (m\^eme param\`etre de d\'eformation
et m\^emes spins, sauf $\Je_{12}$).

Une fois ceci prouv\'e, le cas des $\chi_+$ est semblable \`a celui
des $\chi_-$.
On les d\'efinit par
\beq
\chi_+^{(J_1)}
{\cal P}_{\Jpe_2}
\equiv
\sum_{J_{12},p_{1,2}\equiv J_1+J_2-J_{12}\in {\cal N}}
(-1)^{(2-s)(2J_2p_{1,2}
+{p_{1,2}(p_{1,2}+1)\over 2})}
g_{\Jpe_1\Jpe_2}^{\Jpe_{12}}
{\cal P}_{\Jpe_{12}}
V^{(\Jpe_1)}
{\cal P}_{\Jpe_2}
\label{chidef+}
\eeq
pour des spins fractionnaires donn\'es
par \ref{spfrac+}.
L'espace de Hilbert correspondant est donc
\beq
{\cal H}^+_{s\,\hbox {\scriptsize phys}}
=
\bigoplus_{J=n/2(2-s),\> n\in Z}
{\cal H}_{-J-1,J}.
\label{H+}
\eeq
pour des poids
\beq
\Delta(\chi_+^{(J)})
=
\Delta_{-J-1,J}
=
1+(2-s)J(J+1)
\label{wchi+}.
\eeq

On obtient finalement l'expression de leur
\'echange et de leur fusion qui sont donn\'es par
\beq
\chi_+^{(J_1)}
\chi_+^{(J_2)}
=
e^{2i\pi\epsilon(2-s)J_1J_2}
\chi_+^{(J_2)}
\chi_+^{(J_1)},
\label{brdchi+}
\eeq
$$
\chi_+^{(J_1)}
\chi_+^{(J_2)}
{\cal P}_{\Jpe_3}
=
\sum_{J_{12}\le J_1+J_2}
(-1)^{(2-s)2J_3(J_1+J_2-J_{12})}
$$
\beq
\sum_{\{\nu\}}
\chi_+^{(J_{12},\{\nu\})}
{\cal P}_{\Jpe_3}
<\!\Jpe_{12},{\{\nu\}} |
\chi_+^{(J_1)} | \Jpe_2 \! >
\label{fuschi+}.
\eeq

De la m\^eme mani\`ere que les $\chi_-$,
ces op\'erateurs sont conserv\'es par monodromie autour de 0
gr\^ace \`a l'identit\'e
\beq
\Delta_{-{n\over 2(2+s)}-p-1,{n\over 2(2+s)}+p}
=
\Delta_{-{n\over 2(2+s)}-1,{n\over 2(2+s)}}
+
p[n+(p+1)(2-s)]
\label{shiftdelta+}.
\eeq

\vskip 3mm

Nous avons donc construit deux alg\`ebres d'op\'erateurs de poids r\'eels,
celle des $\chi_-$ d\'efinis en \ref{chidef-},
d'espace de Hilbert \ref{H-}, de poids n\'egatifs \ref{w-},
et celle des $\chi_+$ d\'efinis en \ref{chidef+},
d'espace de Hilbert \ref{H+}, de poids \ref{w+}.
Elles sont stables par fusion et \'echange ind\'ependamment
l'une de l'autre.

\section{Qu'est-ce qui caract\'erise les dimensions sp\'eciales ?}
\label{p5.1b}

{\sl
Nous montrons en premier lieu dans cette partie
que la situation o\`u la matrice de monodromie
est d\'eg\'en\'er\'ee est un cas tr\`es favorable
\`a une possible troncature de l'alg\`ebre d'op\'erateurs.
Nous montrons ensuite que les dimensions
sp\'eciales $C=7,13,19$
et les spins fractionnaires $J_i=k_i/2(2+s)$
sont pr\'ecis\'ement les cas o\`u
la matrice de monodromie est d\'eg\'en\'er\'ee.

Nous ne discutons que le cas des op\'erateurs $\chi_-$
et des spins fractionnaires \ref{spfrac},
le cas des op\'erateurs $\chi_+$
et des spins fractionnaires \ref{spfrac+}
pouvant \^etre trait\'e de mani\`ere similaire.

\vskip 5mm

Pour all\'eger la discussion,
introduisons les op\'erateurs renorm\'es
\beq
W^{(\Je_1)}
_{\Je_2-\Je_{12}}
\>
{\cal P}_{\Je_2}
=
(-1)^{(2+s)(2J_2p_{1,2}
+{p_{1,2}(p_{1,2}+1)\over 2})}
g_{\Je_1\Je_2}^{\Je_{12}}
\>
V^{(\Je_1)}
_{\Je_2-\Je_{12}}
\>
{\cal P}_{\Je_2}
\label{WdeV}
\eeq
en fonction desquels les op\'erateurs $\chi_-$
s'\'ecrivent simplement
\beq
\chi_-^{(J_1)}
{\cal P}_{\Jme_2}
\equiv
\sum_{J_{12}\ /\ p_{1,2}\equiv J_1+J_2+J_{12}\in {\cal N}}
W^{(\Jme_1)}
_{\Jme_2-\Jme_{12}}
\>
{\cal P}_{\Jme_2}
\label{chidefW}.
\eeq
Notons ici
\beq
{\cal F}^{\Jme_{123}\Jme_1\Jme_2\Jme_3}_{\Je_{23}}
=
\epsffile{f4pw.eps}
\label{f4pW}
\eeq
les blocs conformes des op\'erateurs $W^{\Jme}$.
Vu la nature des op\'erateurs $W$,
le spin interm\'ediaire effectif $\Je_{23}$ n'est pas contraint
\`a \^etre du type $\Jme$ (Eq.\ref{Jme}).
Il peut prendre les valeurs $\Je_{23}=\Jme_2+\Jme_3-n-\nhat\pi/h$
pour des entiers $n$ et $\nhat$ ind\'ependants entre 0 et
$p\equiv J_1+J_2+J_3-J_{123}$.
L'espace de ces blocs conformes est donc de dimension $(p+1)^2$.

Au contraire,
l'espace des blocs conformes des $\chi_-$ est de dimension 1
et leur expression \ref{chidefW} montre
que leur unique bloc conforme not\'e ${\cal F}_{(\chi)}$
s'exprime en termes de ceux des op\'erateurs $W$ par
\beq
{\cal F}^{\Jme_{123}\Jme_1\Jme_2\Jme_3}_{(\chi_-)}
=
\sum_{\Jme_{23}}
{\cal F}^{\Jme_{123}\Jme_1\Jme_2\Jme_3}_{\Jme_{23}}
\label{blocchi}
\eeq
pour $\Jme_{23}=\Jme_2+\Jme_3-n(1+\pi/h)$, $n=0... p$.
Nous consid\'erons maintenant les amplitudes physiques
nous limitant pour simplifier \`a celles qui ont des moments
gauches et droits identiques sur leur pattes externes.
Ceci n'est cependant pas n\'ecessaire puisque cette contrainte
n'est pas impos\'ee sur les pattes internes
(voir par exemple la partie suivante).
Nous verrons dans la prochaine partie que l'on peut
obtenir des op\'erateurs locaux par simple produit
d'op\'erateurs des deux chiralit\'es ($\chi_-\chib_-$).
Les amplitudes physiques s'\'ecrivent
alors tout simplement
\beq
A_{\hbox{\scriptsize Phys}}^{\Jme_{123}\Jme_1\Jme_2\Jme_3}
=
\left|{\cal F}^{\Jme_{123}\Jme_1\Jme_2\Jme_3}_{(\chi_-)}
\right|^2
\label{ampl0}
\eeq
o\`u la somme sur les blocs conformes n'appara\^\i t pas
puisqu'il y en a un seul.
Nous avons de plus suppos\'e en \ref{ampl0} que les nombres
quantiques sont choisis de mani\`ere \`a ce que ${\cal \Fb}={\cal F}^*$,
ce qui impose que les nombres gauches et droits soient
sym\'etriques sous l'\'echange de $\alpha_-$
et $\alpha_+$ qui sont complexes conjugu\'es
(voir Eqs.\ref{vertex} et \ref{chibdef+}
dans la prochaine partie).
Les propri\'et\'es de monodromie (autour de 0) des $\chi_-$,
leur \'echange et leur fusion,
montre que cette amplitude physique est locale.

\vskip 2mm

Faisons le lien avec la discussion de la partie 3.2
(page \pageref{f4pphysnd}
et suivantes)
sur les th\'eories diagonales et les th\'eories \`a
matrice de monodromie d\'eg\'en\'er\'ee (ou non diagonales).
Il est facile de voir que l'amplitude diagonale
\beq
\sum_{\Je_{23}}
\left|{\cal F}^{\Jme_{123}\Jme_1\Jme_2\Jme_3}_{\Je_{23}}\right|^2
\label{ampl1}
\eeq
est locale.
Les op\'erateur $W$ ont en effet \'et\'e renorm\'es
(par les coefficients $g$)
de mani\`ere \`a ce que leurs matrices d'\'echange et de fusion soient
donn\'es par des 6-j (et une simple phase)
et soient donc des matrices unitaires ce qui garantit
la localit\'e de \ref{ampl1} pour des
coefficients diagonaux $d_{\Je_{23}}=1$ (cf Eq.\ref{f4pdiag}),
donc omis dans \ref{ampl1}.
Cette amplitude, qui est celle d'exponentielles
de Liouville du couplage
faible (Eq.\ref{expphi}),
bien que ne comportant que des pattes externes de poids r\'eels,
contient des \'etats interm\'ediaires ($\Je_{23}$)
de poids complexes,
qui peuvent \'egalement appara\^\i tre sous forme d'op\'erateurs de poids
complexes par fusion.
Elle n'est donc pas acceptable physiquement.

Ce que nous avons donc fait dans la partie pr\'ec\'edente
en termes d'op\'erateurs,
consiste en termes de corr\'elateurs,
\`a chercher \`a \'ecrire une amplitude
physique locale, \'eventuellement non diagonale (cf Eq.\ref{f4pphysnd})
\beq
A_{\hbox{\scriptsize Phys}}^{\Jme_{123}\Jme_1\Jme_2\Jme_3}
=
\sum_{\Je_{23},\Jeb_{23}}
{\cal \Fb}^{\Jme_{123}\Jme_1\Jme_2\Jme_3}_{\Jeb_{23}}
X_{\Jeb_{23},\Je_{23}}
{\cal F}^{\Jme_{123}\Jme_1\Jme_2\Jme_3}_{\Je_{23}}
\label{ampl2}
\eeq
qui soit en fait restreinte \`a des poids $\Delta_{\Je_{23}}$
et $\Delta_{\Jeb_{23}}$ r\'eels.
Les seuls coefficients non nuls de la matrice $X$
doivent donc \^etre ceux des spins $\Je_{23}$ et $\Jeb_{23}$
de type $\Jme$ (Eq.\ref{Jme}).
On peut donc l'\'ecrire
\beq
A_{\hbox{\scriptsize Phys}}^{\Jme_{123}\Jme_1\Jme_2\Jme_3}
=
\sum_{\Jme_{23},\Jmeb_{23}}
{\cal \Fb}^{\Jme_{123}\Jme_1\Jme_2\Jme_3}_{\Jmeb_{23}}
X_{\Jmeb_{23},\Jme_{23}}
{\cal F}^{\Jme_{123}\Jme_1\Jme_2\Jme_3}_{\Jme_{23}}
\label{ampl3}
\eeq
Autrement dit,
la matrice $X$ doit avoir beaucoup de coefficients nuls
et ne contenir qu'une sous-matrice non nulle, celle
de l'espace
des blocs conformes de poids r\'eels ($\Je_{23}$ de type $\Jme$
dans \ref{f4pW}) dans lui-m\^eme.
Cet espace est de dimension $(p+1)$,
alors que l'espace total des blocs conformes
est de dimension $(p+1)^2$.
Le rang maximal de $X$ est donc $p+1$,
mais on voit par exemple dans notre cas que la matrice $X$
de \ref{ampl0} est de rang 1.
Ceci est d\^u au fait que comme nous l'avons dit
plus haut nous consid\'erons
des op\'erateurs locaux qui sont de simples produits
d'op\'erateurs des deux chiralit\'es.
De mani\`ere plus g\'en\'erale,
si cette matrice est de rang $k$,
on voit en la diagonalisant dans une base orthonorm\'ee
qu'elle projette sur un sous-espace $E^0$ de dimension $k$
de l'espace des blocs conformes de poids r\'eel
(i.e. Im($X$)=$E^0$ et Ker($X$)=$(E^0)^\bot$).

La localit\'e de l'amplitude physique \ref{ampl2}, \ref{ampl3}
impose ensuite de nombreuses conditions \`a cette matrice $X$.
Nous allons les discuter de mani\`ere g\'en\'erale,
mais nous donnons d'abord la solution obtenue dans la partie
pr\'ec\'edente:
la matrice $X_{\Jmeb_{23},\Jme_{23}}$
dont tous les \'el\'ements sont des 1
\beq
X_{\Jmeb_{23},\Jme_{23}}
=
\left[
\begin{array}{c}
1\ 1\ ...\ 1\\
1\ 1\ ...\ 1\\
.\ .\ ...\ .\\
1\ 1\ ...\ 1\\
\end{array}
\right]
\label{X}
\eeq
donne une amplitude physique \ref{ampl3}
locale\footnote{
Attention il s'agit de la matrice $X_{\Jmeb_{23},\Jme_{23}}$
de l'\'equation \ref{ampl3}
r\'eduite \`a des spins de type $\Jme$,
c'est-\`a-dire \`a des poids r\'eels.
Les \'el\'ements de la matrice totale $X_{\Jeb_{23},\Je_{23}}$
de l'\'equation \ref{ampl2}
dont un spin $\Jeb_{23}$ ou $\Je_{23}$ n'est pas
de type $\Jme$ sont nuls,
comme nous l'avons dit pr\'ec\'edemment.}.
Comme nous l'avons signal\'e dans la discussion g\'en\'erale
de la partie 3.2,
la matrice est hermitienne (pour que l'amplitude \ref{ampl3} soit r\'eelle)
et donc diagonalisable.
C'est bien le cas ici.
La matrice $X$ (\ref{X}) est de rang 1
et comporte donc une seule valeur propre non nulle.
Le vecteur propre correspondant est \'evidemment
le bloc conforme \ref{blocchi} des $\chi_-$.
On peut alors compl\'eter ce vecteur propre en une base
orthogonale arbitraire
dans laquelle tous les \'el\'ements de la matrice $X$
sont nuls sauf un,
ce qui donne l'amplitude \ref{ampl0}.

\vskip 5mm

Nous n'avons fait ici que r\'e\'ecrire
dans le langage des blocs conformes,
plus familier \`a un certain nombre de lecteurs,
les r\'esultats de la partie pr\'ec\'edente
exprim\'es dans un langage op\'eratoriel.
Mais ceci peut n\'eanmoins apporter
des r\'esultats tr\`es int\'eressants.

Imaginons donc que nous ne connaissons pas la solution
pr\'ec\'edente et que nous essayons
de d\'eterminer une matrice $X$
restreinte \`a des poids r\'eels
telle que l'amplitude \ref{ampl3} soit locale.
C'est ce qu'on fait Dotsenko et Fateev en ref.\cite{DF}
dans le cas du couplage faible.
Nous avons ici la contrainte suppl\'ementaire
que la matrice $X$ doit \'eliminer les blocs
conformes \ref{f4pW} de poids $\Delta_{\Je_{23}}$ complexes.

Comme nous l'avons dit, ceci signifie
que la matrice $X$ projette\footnote{
La matrice $X$ n'est pas \`a v\'eritablement parler un projecteur:
on n'a pas $X^2=X$, mais simplemement la matrice $X$
a des coefficients nuls en dehors du bloc correspondant
aux transformations de $E^0$ dans $E^0$,
c'est-\`a-dire Im($X$)=$E^0$ et Ker($X$)=$(E^0)^\bot$.
En termes de valeurs propres,
celles de l'orthogonal de $E^0$ sont bien nulles,
mais celle de $E^0$ peuvent \^etre diff\'erentes de 1.}
sur un sous-espace $E^0$ de l'espace
des \'etats de poids r\'eels.
L'amplitude \ref{ampl3} est exprim\'ee
en termes des blocs \ref{f4pW}
qui diagonalisent la monodromie (de $W^{(\Jme_2)}$) autour de 0.
Nous avons vu dans la partie 3.2
que l'invariance de cette amplitude
sous la monodromie autour de 0
exigeait que la matrice $X$ commute
avec la matrice de monodromie (Eq.\ref{f4pmon} et suite)
c'est-\`a-dire qu'elle soit diagonale
par bloc, chaque bloc correspondant \`a un espace propre
de la monodromie.
L'espace $E^0$ sur lequel la matrice $X$
effectue une projection est donc une somme directe de sous-espaces
des espaces propres de la matrice de monodromie.

Examinons ensuite la monodromie autour de l'infini.
La matrice de passage \`a la base qui diagonalise
cette monodromie est la matrice d'\'echange $B$.
Effectuant ce changement de base, le corr\'elateur physique \ref{ampl3}
s'\'ecrit
\beq
A_{\hbox{\scriptsize Phys}}^{\Jme_{123}\Jme_1\Jme_2\Jme_3}
=
\!\!\!\!\!\!
\sum_{\Jeb_{13},\Jme_{23},\Jmeb_{23},\Je_{13}}
\!\!\!\!\!\!
{\cal \Fb}^{\Jme_{123}\Jme_2\Jme_1\Jme_3}_{\Jeb_{13}}
\>
B^\dagger_{\Jeb_{13},\Jmeb_{23}}
\>
X_{\Jmeb_{23},\Jme_{23}}
\>
B_{\Jme_{23},\Je_{13}}
\>
{\cal F}^{\Jme_{123}\Jme_2\Jme_1\Jme_3}_{\Je_{13}}
\label{ampl4}
\eeq
o\`u nous n'avons pas \'ecrit tous les indices de la matrice
$B$.
L'expression dans cette base est donc identique \`a \ref{ampl2}
avec une nouvelle matrice
\beq
\Xt_{\Jeb_{13},
\Je_{13}}
=
\sum_{\Jme_{23},\Jmeb_{23}}
B^\dagger_{\Jeb_{13},\Jmeb_{23}}
X_{\Jmeb_{23},\Jme_{23}}
B_{\Jme_{23},\Je_{13}}
\eeq
qui n'est pas restreinte aux poids r\'eels,
car \'evidemment l'\'echange a fait r\'eappara\^\i tre
les poids complexes.
La matrice $X$ projette sur l'espace $E^0$,
la matrice $\Xt$ projette donc sur l'espace
$E^\infty$ transform\'e par la matrice $B$.
Exigeant que les \'etats de poids complexes
se d\'ecouplent de la th\'eorie,
l'espace $E^\infty$,
transform\'e de $E^0$ par $B$,
doit comme $E^0$ \^etre un sous-espace de
l'espace des \'etats de poids r\'eels.
La localit\'e du corr\'elateur physique
pour la monodromie autour de l'infini
exige ensuite,
comme pour celle autour de 0,
que la matrice $\Xt$ soit diagonale
par blocs,
chaque bloc correspondant \`a un espace
propre de la monodromie.
Et $E^\infty$ doit donc, comme $E^0$,
\^etre une somme directe de sous-espaces
des espaces propres de la matrice de monodromie.

On peut appliquer le m\^eme argument
\`a la monodromie autour de 1,
en effectuant le changement de base de blocs
conformes par la matrice de fusion,
mais ces trois monodromies sont li\'ees
et la localit\'e pour les deux premi\`eres (autour de 0 et l'inifini)
assure la localit\'e pour la troisi\`eme.

\vskip 2mm

Ce que nous venons de voir engendre
beaucoup de contraintes qui semblent
difficiles \`a toutes satisfaire simultan\'ement
pour obtenir une matrice $X$ admissible.
Mais ces contraintes sont loin
d'\^etre aussi importantes selon
la d\'eg\'en\'erescence de la matrice de monodromie.
Examinons les deux cas extr\^emes.

S'il n'y a aucune d\'eg\'en\'erescence,
c'est-\`a-dire que toutes les valeurs propres de la matrice de monodromie
sont diff\'erentes,
ces contraintes laissent peu de libert\'es.
En effet dans ce cas,
la matrice $X$ doit \^etre purement diagonale
et on a un nombre fini de choix
pour l'espace $E^0$ sur lequel elle
projette:
si on choisit $E^0$ de dimension $k$,
on a simplement un nombre fini de
mani\`eres (exactement $\left(\,^{p+1}_k\right)$)
de choisir $E^0$ comme somme directe
de $k$ espace propres de la monodromie
(dans l'espace des blocs de poids r\'eels
de dimension $(p+1)$).
L'\'echange doit donner pour l'autre base de blocs
conformes une matrice $\Xt$ et un espace $E^\infty$
soumis aux m\^emes contraintes.

Dans le cas inverse,
o\`u la matrice de monodromie est compl\`etement
d\'eg\'en\'er\'ee c'est-\`a-dire
proportionnelle \`a l'identit\'e (tout au moins pour les $(p+1)$
blocs de poids r\'eels qui nous int\'eressent),
nous avons en revanche beaucoup plus de libert\'es.
La matrice $X$ peut \^etre quelconque
et le choix de l'espace $E^0$
sur lequel elle projette
peut \^etre n'importe quel sous-espace
de l'espace des blocs de poids r\'eels
(de dimension $(p+1)$).
Pour un espace $E^0$ de dimension $k$,
ceci laisse donc le choix de $k(p+1-k)$
param\`etres r\'eels,
alors que pr\'ec\'edemment,
nous n'avions qu'un nombre fini de choix pour $E^0$ !
Les libert\'es sont les m\^emes pour la matrice
$\Xt$ et l'espace $E^\infty$ transform\'es.
Nous sommes donc dans ce cas
dans des circonstances beaucoup plus
favorables \`a l'existence d'une matrice $X$
donnant des corr\'elateurs physiques \ref{ampl3} locaux.
La seule contrainte qui subsiste effectivement
dans ce cas est que l'espace $E^\infty$
doit \^etre un sous-espace de l'espace des blocs
conformes de poids r\'eels.
Cette condition assure le d\'ecouplage des op\'erateurs
et des \'etats de poids complexes,
alors que la localit\'e du corr\'elateur
physique est assur\'ee automatiquement
par la d\'eg\'en\'erescence de la matrice de monodromie.

Cet argument demeure heuristique.
Bien que le deuxi\`eme cas (monodromie d\'eg\'en\'er\'ee)
semble infiniment plus favorable, il ne prouve
certes pas qu'il soit possible de d\'eterminer
une matrice $X$ qui rende local le corr\'elateur \ref{ampl3}.
Mais nous avons justement prouv\'e dans la partie pr\'ec\'edente
que pour les dimensions sp\'eciales $C=7,13,19$
et pour les spins fractionnaires (\ref{spfrac}, \ref{spfrac+})
o\`u nous avions une matrice de monodromie d\'eg\'en\'er\'ee
(\ref{shiftdelta}, \ref{shiftdelta}),
il \'etait possible d'obtenir une troncature de l'alg\`ebre
\`a des poids r\'eels
et des fonctions de corr\'elation physiques locales.
Ceci confirme donc le bien-fond\'e de l'argument d\'evelopp\'e ici.

\vskip 5mm

Il para\^\i t alors l\'egitime de chercher \`a d\'eterminer
tous les cas donnant lieu \`a une matrice de monodromie d\'eg\'en\'er\'ee.
C'est ce que nous allons faire maintenant.
Nous montrerons ainsi que le seul cas (ou presque)
qui donne lieu \`a une matrice de monodromie
compl\`etement d\'eg\'en\'er\'ee pour les blocs
conformes de poids r\'eels est celui
des dimensions sp\'eciales $C=7,13,19$
pour les blocs conformes dont les
spins sont les spins fractionnaires \ref{spfrac} ou \ref{spfrac+}.

Les blocs conformes du type
${\cal F}_p^{ijkl}(z_2,z_3)$ (Eq.\ref{ijklp})
ont une monodromie diagonale
(de $z_3$) autour de 0 avec valeurs propres
\beq
e^{2i\pi\epsilon(\Delta_i+\Delta_j-\Delta_p)}
\eeq
et celle-ci sera donc d\'eg\'en\'er\'ee (ind\'ependante de $p$)
si les poids $\Delta_p$ pour les divers blocs conformes
ne diff\`erent que d'un entier.
Examinons donc dans quel cas ceci peut se produire.

Dans le but d'utiliser des notations famili\`eres
au plus grand nombre,
d\'esignons comme BPZ un op\'erateur par son
pseudo-moment\footnote{
La relation avec les notations employ\'ees ailleurs
dans cette th\`ese est $\alpha=\alpha_-\varpi/2=
\alpha_-\Je+(\alpha_-+\alpha_+)/2$.}
$\alpha$
ce qui donne\footnote{
La notation $\Delta(\alpha)$ pourrait amener une confusion
avec la notation $\delta(\varpi)$ en Eq.\ref{deltaw}
par exemple.
Il n'en sera rien puisque nous n'utiliserons pas la notation
$\Delta(\alpha)$ en dehors de cette partie.}
\beq
\Delta(\alpha)
=
{C-1\over24}
+{1\over2}
\alpha^2.
\label{deltaalpha}
\eeq
On retrouve les poids de Kac $\Delta(n,m)$ \ref{Kac} pour les valeurs
discr\`etes $\alpha=
\alpha_+n/2
+
\alpha_-m/2$.
Les blocs conformes \ref{ijklp} que nous redonnons
\beq
{\cal F}_p^{ijkl}(z_2,z_3)
\equiv
<\!i|\phi_{ip}^j(z_2)\phi_{pl}^k(z_3)|l\!>
=
\epsffile{ijklp.eps}
\eeq
sont \'etiquet\'es
par leur \'etat interm\'ediaire d\'esign\'e g\'en\'eriquement par $p$.
Les r\`egles de s\'election montrent que cet \'etat $p$
aura un moment
\beq
\alpha_p=\alpha-n\alpha_+-m\alpha_-
\label{alphap}
\eeq
pour des entiers $n$ et $m$ (nombre de charges
d'\'ecran pour l'op\'erateur $\phi^j(z_3)$).
Dans \ref{alphap},
$\alpha$ vaut $\alpha_j+\alpha_k-(\alpha_++\alpha_-)$
o\`u $\alpha_j$ et $\alpha_k$ d\'esignent les moments
des op\'erateurs $\phi^j$ et $\phi^k$.
On consid\`ere des blocs conformes entre op\'erateurs de poids
r\'eels de moments $\alpha_j$ et $\alpha_k$ r\'eels.
Le moment $\alpha$ est alors r\'eel.
On a donc des blocs conformes de poids $\Delta(\alpha_p)$ r\'eels
pour $n=m$ dans \ref{alphap}.
La matrice de monodromie sur ces blocs conformes de poids r\'eels
sera donc compl\`etement d\'eg\'en\'er\'ee si
\beq
\hbox{ pour tout entier }n,
\qquad
\Delta(\alpha-n\alpha_+-n\alpha_-)
-
\Delta(\alpha)
\qquad\hbox{ est un entier}
\label{cond}.
\eeq
Utilisant l'expression \ref{alpha} de $\alpha_\pm$ en fonction de $C$,
ceci \'equivaut \`a
\beq
n
\sqrt{C-1\over3}
\left(
{n\over 2}
\sqrt{C-1\over3}
+
\alpha
\right)
\hbox{  est un entier.}
\eeq
Ceci est \'evidemment vrai pour $n=0$ et sera donc
vrai pour tout $n$ si la diff\'erence entre deux de ces
entiers successifs (obtenus pour $n$ et $n+1$)
\beq
\sqrt{C-1\over3}
\left(
(n+{1\over2})
\sqrt{C-1\over3}
+\alpha
\right)
\hbox{  est un entier.}
\eeq
Param\'etrisant comme pr\'ec\'edemment
la charge centrale $C=1+6(2+s)$
par le r\'eel $s$, ceci est finalement vrai pour tout $n$
si
\beq
{C-1\over3}
=
2(2+s)
\hbox{ est un entier}
\label{condC}
\eeq
et
\beq
\alpha=
(\alpha_-+\alpha_+)
\left({k\over2(2+s)}+{1\over2}\right)
\label{conda}
\eeq
pour un entier $k$.

On obtient donc \`a tr\`es peu de choses pr\`es
les conditions de la partie pr\'ec\'edente:
\ref{condC} force $s$ \`a \^etre un demi-entier,
ce qui constitue une g\'en\'eralisation
des conditions pr\'ec\'edentes qui exigaient que $s$ soit entier,
et la condition \ref{conda} correspond exactement
aux conditions pr\'ec\'edentes $\Je=\Jme=
J+(\pi/h)J$ (Eq.\ref{Jme}) pour
des spins fractionnaires $J=k/2(2+s)$ (Eq.\ref{spfrac}).

Les dimensions sp\'eciales $C=7,13,19$
et les spins fractionnaires $J_i=k_i/2(2+s)$,
pour lesquels nous avons pu effectuer
une troncature \'eliminant les poids complexes,
apparaissent donc comme (presque)
les seuls cas o\`u les blocs conformes
de poids r\'eels ont une monodromie
d\'eg\'en\'er\'ee.
Ceci ne para\^\i t pas tr\`es \'etonnant suite
\`a l'argument d\'evelopp\'e pr\'ec\'edemment
qui montrait que cette d\'eg\'en\'erescence
donnait beaucoup plus de libert\'e
pour une telle troncature.

\vskip 2mm

La condition \ref{condC}
constitue quand m\^eme un g\'en\'eralisation
\`a $s$ demi-entier donc \`a $C=4,10,16,22$
de la condition $s$ entier.
On peut donc se demander s'il ne serait pas possible
d'\'etendre la troncature pr\'ec\'edente \`a ces dimension,
ou \`a d\'efaut de comprendre pourquoi c'est impossible.

D'un point de vue ``technique'', c'est la propri\'et\'e \ref{lien6j}
des 6-j qui a permis la troncature.
Elle peut elle-m\^eme \^etre obtenue \`a partir
des propri\'et\'es 6.8 \`a 6.10 dans [P3] des nombres q-d\'eform\'es.
Grosso modo,
ces propri\'et\'es permettent de relier par un simple
signe les quantit\'es\footnote{$J$ est un spin
fractionnaire \ref{spfrac} et $k$ est un entier}
$2J=k/(2+s)$ d\'eform\'ees
par $h$ aux m\^emes quantit\'es d\'eform\'ees par $\hhat$.
Mais pour $s$ demi-entier ces quantit\'es $k/(2+s)$
n'incluent pas les entiers,
alors que des entiers q-d\'eform\'es
interviennent dans la d\'efinition des 6-j.
Ceci rend donc impossible la d\'emonstration de \ref{lien6j}.
Peut-\^etre serait-il possible de passer outre cette difficult\'e
technique en factorisant les entiers dans les 6-j,
mais cela ne para\^\i t pas \'evident.

D'un point de vue plus physique,
pour les dimensions $4,10,16,22$
obtenues pour $C=1+6(2+s)$ et $s$ demi-entier,
les spins fractionnaires $J_i=k_i/2(2+s)$
sont respectivement des entiers, des tiers d'entiers,
des cinqui\`emes d'entiers et des septi\`emes d'entiers.
Ils ne contiennent donc pas les demi-entiers,
contrairement aux dimensions $7,13,19$ ($s$ entier).
Ce n'est pas diff\'erent de ce que nous remarquions
pr\'ec\'edemment:
les quantit\'es $2J=k/(2+s)$ n'incluent pas les entiers dans ce cas.
Mais physiquement, ceci signifie
que le spectre n'inclut pas toutes les familles
d\'eg\'en\'er\'ees possibles,
en particulier la plus simple de poids r\'eels:
celle de spins (1/2,1/2), ou (2,2) dans la classification de BPZ.
Il n'est donc pas impossible que la m\^eme troncature
soit r\'ealisable dans ces dimensions.
Elle pourrait en effet \^etre masqu\'ee par une mauvaise
normalisation des 6-j faisant appara\^\i tre des entiers
artificiellement,
et elle n'a de toute fa\c con pas pu \^etre mise en \'evidence
dans l'approche initiale plus \'el\'ementaire \cite{GN5},
puisque cette troncature avait alors \'et\'e
recherch\'ee sur les spins $(1/2,1/2)$.
}

\section{Mod\`eles topologiques fortement coupl\'es}
\label{p5.2}

La troncature pr\'ec\'edente permet de construire
un mod\`ele de v\'eritable corde non critique,
constitu\'e dans le cas bosonique envisag\'e ici
de 19, 13 ou 7
degr\'es de libert\'e\footnote{
Evoluant donc un espace-temps de dimension 19, 13 ou 7,
\`a moins que certains de ces degr\'es de libert\'e
ne soient compactifi\'es.}
qui peuvent donc \^etre habill\'es par les op\'erateurs
pr\'ec\'edents, annulant ainsi la charge centrale totale.
Un op\'erateur de vertex habill\'e par un op\'erateur $\chi_\pm$
de poids $\Delta$ engendre des particules
de masses carr\'ees $2(\Delta-1+N)$,
o\`u $N$ est le nombre d'excitations.
Ceci exclut donc un habillage par les op\'erateurs $\chi_-$
de poids n\'egatifs qui engendrerait un nombre infini de tachyons.
Le calcul des amplitudes d'une telle th\'eorie
semble cependant pour le moment hors de port\'ee.

\subsection{Le mod\`ele - Exposants critiques}

C'est ce qui nous a amen\'es \`a l'id\'ee d'un mod\`ele
simplifi\'e, comportant uniquement deux degr\'es
de libert\'e,
constitu\'es de deux copies des op\'erateurs $\chi_\pm$ pr\'ec\'edents.
C'est la m\^eme chose que ce qui a d\'ej\`a \'et\'e
fait pour le couplage faible \cite{D,G4},
(ou les mod\`eles
minimaux coupl\'es \`a la gravit\'e
pour les valeurs de la charge centrale correspondantes).

En effet, les dimensions sp\'eciales $C=1+6(2+s)=7,13,19$
mises en \'evidence dans la partie pr\'ec\'edente
sont leurs propres compl\'ementaires \`a 26.
Il est donc possible de construire un mod\`ele
invariant conforme
\`a partir de deux degr\'es de libert\'e,
l'un donnant une charge centrale $C=1+6(2+s)$
et l'autre $C'=1+6(2+s')$,
si $s'=-s$.
Tous les op\'erateurs et les quantit\'es relatives
au degr\'e de libert\'e de charge centrale $C'$
seront \'egalement d\'enot\'es par des primes.
On remarque ensuite que les poids des op\'erateurs
$\chi'_-$ et $\chi_+$ permettent de construire
des op\'erateurs de poids 1:
\beq
\Delta(\chi_+^{(J)})
+
\Delta(\chi'\,_-^{(J)})
=
1+(2+s)J(J+1)
-(2-s')J(J+1)
=
1
\label{w1}
\eeq
o\`u $\chi'_-$ d\'esigne un op\'erateur $\chi_-$
introduit pr\'ec\'edemment en \ref{chidef-}
mais construit \`a partir de l'autre degr\'e de libert\'e
(de charge centrale $C'$).

\vskip 1mm

Quelle est l'interpr\'etation d'un tel mod\`ele ?
Comme on l'a vu sur le cas d'une v\'eritable corde
en introduction \`a cette partie,
les op\'erateurs $\chi_+$ sont les op\'erateurs
de Liouville qui doivent habiller la mati\`ere.
Le degr\'e de libert\'e
de charge centrale $C=1+6(2+s)$ pour lequel
on inclut des op\'erateurs $\chi_+$ dans la th\'eorie,
sera donc ici aussi interpr\'et\'e comme la gravit\'e.
L'autre degr\'e de libert\'e,
de charge centrale $C'=1+6(2+s')$
pour des op\'erateurs $\chi'_-$
est un degr\'e de libert\'e de mati\`ere
et nous notons sa charge centrale $c=C'=26-C$,
conform\'ement \`a la convention g\'en\'erale.

C'est la m\^eme chose que ce qui a \'et\'e fait
en refs.\cite{D,G4} dans le r\'egime de couplage faible.
Une copie du champ local de Liouville repr\'esente la mati\`ere
alors qu'une deuxi\`eme copie
constitue la gravit\'e
et habille la premi\`ere.
En effet, bien que tous les
op\'erateurs $V$ et $\chi$
aient \'et\'e construits \`a partir de solutions de la th\'eorie de Liouville,
ce n'en sont pas moins des champs conformes
qui peuvent aussi \^etre utilis\'es comme des champs de mati\`ere.
A la diff\'erence d'une v\'eritable corde
qui aurait dans le cas bosonique 19, 13 ou 7
degr\'es de libert\'e, il n'y en a ici qu'un seul.
Il contribue \`a lui tout seul pour 19, 13 ou 7 \`a
la charge centrale\footnote{
Dans une repr\'esentation de gaz de Coulomb de la th\'eorie,
il aurait une charge de ``background'' $Q=\sqrt{(C-1)/3}$
r\'eelle (elle est imaginaire pure pour le couplage faible)
correspondant \`a une charge centrale $C=1+3Q^2$
sup\'erieure \`a 1.}.
La cohomologie BRST supprime ensuite
deux degr\'es
de libert\'e et n'en laisse donc plus aucun
dans ce cas.
Le mod\`ele doit alors \^etre topologique.
C'est ce qui nous permettra dans la sous-partie
suivante de le d\'ecrire par une action effective
sans degr\'es de libert\'e transverses.
Les seuls degr\'es de libert\'e pris en compte
seront les modes-z\'ero.

\vskip 1mm

Les constructions pour la mati\`ere et la gravit\'e ne sont donc pas tr\`es
diff\'erentes.
Pour avoir des notations \'egalement sym\'etriques,
mais
aussi plus proches des notations les plus courantes,
nous introduisons les charges de background
\beq
Q_L=\sqrt{(C-1)/3}\quad \hbox{et} \quad Q_M=\sqrt{(c-1)/3}=\sqrt{(25-C)/3}
\label{Qdef}
\eeq
sym\'etriquement pour la mati\`ere et la gravit\'e.
Les charges d'\'ecran pour la gravit\'e peuvent alors s'\'ecrire
\beq
\alpha_\pm={Q_L/ 2}\pm
 i {Q_M/ 2}.
\label{alphadef}
\eeq
Les param\`etres de d\'eformation $h$ et $\hhat$
sont reli\'es \`a $\alpha_\pm$
par
\beq
h/\pi=\alpha_-^2/2=\alpha_-/\alpha_+
\hbox{ et }
\hhat/\pi=\alpha_+^2/2=\alpha_+/\alpha_-.
\eeq
Les charges d'\'ecran pour le degr\'e de libert\'e qui
repr\'esente la mati\`ere sont d\'esign\'ees
par des primes et s'\'ecrivent
\beq
-\alpha'_\pm={Q_M/ 2}\pm
 i {Q_L/ 2}.
\label{alpha'def}
\eeq
Elles sont reli\'ees \`a celles de la gravit\'e par
\beq
\alpha'_\pm=\mp i\alpha_\mp.
\label{a'a}
\eeq

\vskip 4mm

On peut imaginer pour cette th\'eorie
des op\'erateurs de vertex locaux habill\'es qui soient de simples
produits du type:
\beq
\vertex J_i, \Jb_i,
=
\chi'\, ^{(J_i)}_-
\chi_+^{(J_i)}
\chib'\, ^{(\Jb_i)}_-
\chib_+^{(\Jb_i)}\>
\label{vertex}
\eeq
qui sont de poids $(1,1)$ gr\^ace \`a \ref{w1}.
Il faut ensuite que ces op\'erateurs soient locaux.
Il a \'et\'e montr\'e en partie \ref{p5.1}
que les op\'erateurs $\chi$ avaient une monodromie autour de 0
bien d\'efinie
et que leurs coefficients de fusion et d'\'echange
\'etaient de simples phases ou signes.
Les op\'erateurs chiraux sont donc
d\'ej\`a ``presque'' locaux,
c'est pourquoi nous pouvons envisager qu'un simple produit
de $\chi$ soit local.
Examinant les phases issues des deux composants chirales
lors de la monodromie autour de 0, la fusion et l'\'echange
on voit que l'op\'erateur \ref{vertex} est local
si et seulement si les spins
gauches et droits diff\`erent d'un entier:
\beq
J_i=\Jb_i
+N
\label{J-Jb}
\eeq
pour n'importe quel entier $N$.
Il est remarquable que la m\^eme condition ($J-\Jb$ entier) ait \'et\'e
obtenue en \cite{GR} en demandant
l'invariance modulaire de la th\'eorie.
Ceci est en opposition compl\`ete
avec le cas du couplage faible o\`u la construction
des op\'erateurs locaux impose toujours $J=\Jb$.
Ici, au contraire,
non seulement il est possible de consid\'erer
des op\'erateurs de vertex de spins $J_i\ne\Jb_i$,
mais c'est n\'ecessaire:
ils seront de toute fa\c con engendr\'es par fusion
puisque la fusion \ref{fuschi-}, \ref{fuschi+}
donnera lieu \`a des sommations ind\'ependantes sur $J_{12}$
et $\Jb_{12}$.
De plus, les sommes \ref{chidef-}, \ref{chidef+}
qui d\'efinissent les $\chi$ doivent pouvoir
\^etre faites ind\'ependamment pour les spins $J_i$
et les spins $\Jb_i$,
sinon il ne pourrait y avoir une premi\`ere sommation sur les 6-j
pour assurer la fermeture de l'alg\`ebre des $\chi$
et une autre ind\'ependante sur les 6-j de l'autre chiralit\'e
pour assurer la fermeture de celle des $\chib$.
En outre, les sommations \ref{chidef-}, \ref{chidef+}
ainsi que la fusion
se font par pas entiers pour les spins,
ce qui est parfaitement coh\'erent avec la condition \ref{J-Jb}.
On en conclut que
l'inclusion dans l'alg\`ebre de
tous les op\'erateurs de vertex \ref{vertex}
avec des spins $J$ et $\Jb$ diff\'erant d'un entier (\ref{J-Jb})
est une une condition n\'ecessaire et suffisante \`a
l'obtention d'une alg\`ebre coh\'erente.
Il y a donc un d\'econfinement de la chiralit\'e
dans la phase de couplage fort.

La d\'efinition \ref{chidef+} de $\chi_+^{(J)}$
n'est pas sym\'etrique entre $\alpha_-$
et $\alpha_+$,
puisqu'elle correspond en fait \`a des op\'erateurs
de spins $(-J-1,J)$.
Dans le r\'egime de couplage faible,
un choix d'habillage s'imposait
pour obtenir le terme cosmologique
$e^{\alpha_-\Phi}$
qui redonne la bonne limite classique\footnote
{Ce choix est \'egalement confirm\'e (au moins
pour le terme cosmologique) par la limite de Seiberg
(voir \cite{GM,Seib,Polch})
qui dans nos notations affirme
que le mode-z\'ero $p_0=i((J+1/2)\alpha_-+(\Jhat+1/2)\alpha_+)$,
s'il est imaginaire pur (cas du couplage faible), doit \^etre de
partie imaginaire positive.
En revanche, s'il est r\'eel, il n'y a pas de choix de signe
impos\'e et donc pas de choix d'habillage.
C'est le cas pour le couplage fort puisque alors
$p_0=\pm i(J+1/2)(\alpha_--\alpha_+)$ est r\'eel.
Le signe $\pm$ d\'epend du choix d'habillage par les $\chi_+$:
les $\chi_+^{(J)}$ peuvent au choix \^etre d\'efinis par des spins $(-J-1,J)$
comme en \ref{chidef+}
ou par des spins $(J,-J-1)$ comme en \ref{chibdef+}.}.
Dans le r\'egime de couplage fort,
au contraire,
la sym\'etrie entre $\alpha_-$
et $\alpha_+$ est un point cl\'e.
C'est ce qui nous permettra d'obtenir un exposant $\gamma_{\hbox{\scriptsize
string}}$
r\'eel.
Nous exigeons donc que l'op\'erateur \ref{vertex}
soit sym\'etrique par conjugaison complexe (donc
\'echange de $\alpha_-$ et $\alpha_+$)
\`a condition d'\'echanger les deux chiralit\'es.
C'est pourquoi,
sym\'etriquement par rapport \`a la d\'efinition \ref{chidef+}
des $\chi_+$, on d\'efinit
les op\'erateurs $\chib_+$ par
\beq
\chib_+^{(\Jb_1)}
{\cal P}_{\Jpeb_2}
\equiv
\sum_{\Jb_{12},\pb_{1,2}\equiv \Jb_1+\Jb_2-\Jb_{12}\in {\cal N}}
(-1)^{(2-s)(2\Jb_2\pb_{1,2}
+{\pb_{1,2}(\pb_{1,2}+1)\over 2})}
g_{\Jpeb_1\Jpeb_2}^{\Jpeb_{12}}
{\cal P}_{\Jpeb_{12}}
V^{(\Jpeb_1)}
{\cal P}_{\Jpeb_2}
\label{chibdef+}
\eeq
mais avec
\beq
\Jpeb_i\equiv \Jb_i+(-\Jb_i-1)(\pi/h)
\eeq
toujours pour des spins $\Jb_i$ fractionnaires (\ref{spfrac+}).
Ceci correspond \`a l'autre choix d'habillage.
Les poids et l'alg\`ebre de ces op\'erateurs sont identiques
\`a ceux des $\chi_+$ d\'efinis pr\'ec\'edemment.

Ceci donne des blocs conformes gauches et droits complexes conjugu\'es
l'un de l'autre
$$
{\cal \Fb}^{\Jme_{123}\Jme_1\Jme_2\Jme_3}_{(\chi_-)}
=
\left(
{\cal F}^{\Jme_{123}\Jme_1\Jme_2\Jme_3}_{(\chi_-)}
\right)^*
$$
comme nous l'avions demand\'e en \ref{ampl0}.
Cela est vrai pour des spins gauches et droits \'egaux pour les pattes
externes,
c'est-\`a-dire le cas particulier auquel nous nous \'etions
restreints pour simplifier la discussion de la partie 5.2.
Dans le cas g\'en\'eral consid\'er\'e ici,
les blocs gauches et droits sont sym\'etriques par conjugaison
complexe \`a condition d'\'echanger $J_i$ et $\Jb_i$.
La conjugaison
complexe aura donc pour effet d'\'echanger les spins gauches
et droits ($J_i\leftrightarrow\Jb_i$) des op\'erateurs
physiques \ref{vertex} ou des fonctions de corr\'elation.

\vskip 2mm

Nous pouvons d'ores et d\'ej\`a calculer l'exposant $\gamma_{\hbox{\scriptsize
string}}$
pour ce mod\`ele par la m\^eme m\'ethode que
pour le couplage faible.
Le terme cosmologique est l'op\'erateur
\beq
\vertex 0, 0, =\chi_+^{(0)}(z) \chib_+^{(0)}(\zb)
\label{cosmodef}
\eeq
qui est de poids (1,1) et n'inclut pas de mati\`ere.
Comme pour le couplage faible, les op\'erateurs
consid\'er\'es jusqu'\`a maintenant
sont des op\'erateurs pour une valeur particuli\`ere
de la constante cosmologique,
mettons $\mu=1$.
La loi de changement d'\'echelle \ref{scalV1}
s'\'ecrit en termes de spins effectifs
\beqa
\left.
V^{(\Je)}_{\me}\right | _{(\mu)}
&=&
\mu^{\Je}
\mu^{-\varpi/2} V_{\me}^{(\Je )}
\>\mu^{\varpi/2} \nnn
\left.
\Vb^{(\Jeb)}_{\meb}\right | _{(\mub)}
&=&
\mub^{\Jeb}
\mub^{-\varpib/2} \Vb_{\meb}^{(\Jeb)}
\>\mub^{\varpib/2}
\label{scalV}
\eeqa
o\`u nous rappelons que $\varpi$ est le mode-z\'ero
des champs libre $P^{(i)}$ sous-jacents
(renorm\'e en \ref{wdep}).
Comme les op\'erateurs physiques \ref{vertex}
sont des produits d'op\'erateurs des deux chiralit\'es
de spins $J$ et $\Jb$ diff\'erents,
leur loi d'\'echelle ne sera pas fonction uniquement
du produit $\mu\mub$ comme dans le cas du couplage faible
o\`u les spins (et mode-z\'ero) droits et gauches \'etaient \'egaux.
On d\'etermine alors le lien entre $\mu$ et $\mub$
par la sym\'etrie d\'ej\`a choisie pour \ref{vertex}:
les op\'erateurs de vertex \ref{vertex}
sont sym\'etriques par conjugaison complexe
($\alpha_-\leftrightarrow\alpha_+$) \`a condition
d'\'echanger les $\Jb_i$ et les $J_i$,
il doit donc en \^etre de m\^eme pour les facteurs dans \ref{scalV}.
Pour $\mu$ r\'eel, ceci conduit \`a $\mub=\mu^{h/\pi}=\mu^{\alpha_-/\alpha_+}$.
On en d\'eduit la loi d'\'echelle pour les op\'erateurs de Liouville
habillant l'op\'erateur de vertex \ref{vertex}:
\beq
\left. \chi_{+}^{(J)}\right|_{(\mu)}
\left.\chib_{+}^{(\Jb)}\right|_{(\mub)} =
\mu^{-2+Q_M(\alpha'_-J+\alpha'_+\Jb)/2}
\mu^{-(\varpi+\varpib\alpha_-/\alpha_+)/2}
\chi_{+}^{(J)} \chib_{+}^{(\Jb)}
\mu^{(\varpi+\varpib\alpha_-/\alpha_+)/2}
\label{chichibmu}
\eeq
Le facteur $\mu^{-2+Q_M(\alpha'_-J+\alpha'_+\Jb)/2}$
vaut $\mu^{-2}$ pour le terme cosmologique
($J=\Jb=0$).
La constante cosmologique physique est donc $\mu_c=\mu^2$.
Les fonctions de corr\'elation
des op\'erateurs de vertex \ref{vertex}
se calculent sur les vides $Sl(2,C)$
invariants complets $|\varpi'_0,\varpi_0,\varpib'_0,\varpib_0\!>$
et $<\!-\varpi'_0,-\varpi_0,-\varpib'_0,-\varpib_0|$
o\`u les moments $\varpi',\varpib'$ sont ceux des champs libres
sous-jacents aux op\'erateurs $\chi'_-,\chib'_-$ de mati\`ere.
On obtient ainsi
\beq
\left< \prod_{\ell=1}^N \chi_{+}^{(J_\ell)}
\chib_{+}^{(\Jb_\ell)} \right > _{\mu_c}
=
\left < \prod_{\ell=1}^N \chi_{+}^{(J_\ell)}
\chib_{+}^{(\Jb_\ell)} \right >_1
\mu_c^{-\sum_\ell\left [1-{Q_M\over 4}(\alpha'_-J_\ell+ \alpha'_+
\Jb_\ell)\right ]
+{s+2\over 2}}
\label{mudep}
\eeq
o\`u le dernier terme
(i.e. $(s+2)/2$) est obtenu par action
des op\'erateurs $\mu^{\pm\varpi/2}$ de \ref{chichibmu}
sur les vides entrants et sortants.
On peut transformer cette \'equation en
\beq
\left < \prod_{\ell=1}^N \chi_{+}^{(J_\ell)}
\chib_{+}^{(\Jb_\ell)} \right >_{\mu_c}
=
\left < \prod_{\ell=1}^N \chi_{+}^{(J_\ell)}
\chib_{+}^{(\Jb_\ell)} \right >_1
\mu_c^{\sum_\ell {Q_M\over 4}\left [ \alpha'_-(J_\ell+{1\over 2})+ \alpha'_+
(\Jb_\ell+{1\over2})\right ]
-\left({N\over2}-1\right)
\left({s+2\over2}\right)}
\label{mudep2}
\eeq
en utilisant les liens entre $Q_M$ et $\alpha'_\pm$
qui donnent en particulier
\beq
-Q_M(\alpha'_-+\alpha'_+)=2(2-s).
\eeq

L'\'el\'ement d'aire \ref{cosmodef}
permet de d\'efinir
la fonction de partition \`a aire
fix\'ee:
\beq
{\cal Z}_{\mu_c}(A)\equiv \left <\> \delta\!\left (\int dz d\zb
\left. \chi_{+}^{(0)}\right |_{(\mu_c)}
\left. \chib_{+}^{(0)}\right |_{(\mu_c)} -A\right )
\right >.
\label{Zmu(A)def}
\eeq
La loi de changement d'\'echelle \ref{chichibmu}
nous permet ensuite de
transformer \ref{Zmu(A)def} en
\beq
{\cal Z}_{\mu_c}(A)
=
\left <\>
\mu_c^{-(\varpi+\varpib\alpha_-/\alpha_+)/4}
\delta\!\left (\int dz d\zb
\mu_c^{-1}
\left. \chi_{+}^{(0)}\right |_{(1)}
\left. \chib_{+}^{(0)}\right |_{(1)} -A\right )
\mu_c^{(\varpi+\varpib\alpha_-/\alpha_+)/4}
\right >
\eeq
puis, calculant l'action des \'el\'ements extr\^emes
sur le vide et sortant $\mu_c^{-1}$ de la fonction delta:
\beq
{\cal Z}_{\mu_c}(A)
=
\mu_c^{(s+2)/2}
\left <\>
\mu_c
\delta\!\left (\int dz d\zb
\left. \chi_{+}^{(0)}\right |_{(1)}
\left. \chib_{+}^{(0)}\right |_{(1)} -\mu_c A\right )
\right >
\eeq
c'est-\`a-dire
\beq
{\cal Z}_{\mu_c}(A)
=
\mu_c^{1+(s+2)/2}
{\cal Z}_{1}(\mu_cA).
\eeq
Ceci traduit un comportement de la fonction de partition
\`a aire fix\'ee en
\beq
{\cal Z}(A)
\propto
A^{-1-(2+s)/2}.
\label{Z(A)}
\eeq

Ce r\'esultat peut \'egalement \^etre obtenu en remarquant
que \ref{mudep} sans op\'erateurs donne la d\'ependance
de la fonction de partition en $\mu_c$:
\beq
{\cal Z}(\mu_c)\propto \mu_c^{(2+s)/2}
\eeq
et que donc la fonction de partition \`a aire fix\'ee
qui est sa transform\'ee de Laplace
\beq
{\cal Z}(A)=\int d\mu_c {\cal Z}(\mu_c) e^{-\mu_c A}
\eeq
a le comportement \ref{Z(A)}.

La loi d'\'echelle \ref{Z(A)} montre donc que pour ce mod\`ele
\beq
\gamma_{\hbox{\scriptsize string}}
=
{2-s
\over
2}
\label{gs}
\eeq
qui est r\'eelle.
Exprim\'ee en fonction de la dimension d'espace $d=1+6(2-s)$,
ceci donne la valeur \ref{gstr} annonc\'ee en introduction.
Cette expression permet de voir que la valeur
obtenue pour $\gamma_{\hbox{\scriptsize string}}$
est la partie r\'eelle de la formule de KPZ \ref{gstring}.

Les simulations num\'eriques effectu\'es en ref.\cite{BH}
sur $n$ mod\`eles d'Ising coupl\'es \`a la gravit\'e
(jusqu'\`a $n=8$ donc $c=4$)
ont donn\'e des r\'eultats fort proches de la partie r\'eelle
de la formule de KPZ
(sans que l'ad\'equation soit cependant parfaite).
Ceci confirme donc partiellement notre r\'esultat
qui est n\'eanmoins dans une zone ($c=7$ ou plus)
non atteinte par ces simulations num\'eriques.

Les valeurs de $\gamma_{\hbox{\scriptsize string}}$
pour les trois valeurs sp\'eciales de la charge centrale
sont les suivantes
\beq
\left \{
\begin{array}{rccc}
s& c & C & \gamma_{\hbox{\scriptsize string}} \nnn
1& 7  & 19 & 1/2 \nnn
0& 13 & 13 & 1 \nnn
-1 & 19 & 7 & 3/2 \nnn
\end{array} \right.
\eeq
La valeur de $1/2$ pour $c=7$ est la m\^eme
que celle de surfaces d\'eg\'en\'er\'ees
en polym\`eres branch\'es.
Peut-\^etre ceci peut-il \^etre mis en relation
avec le fait que notre
terme cosmologique (ainsi que tous les op\'erateurs)
est un simple produit d`op\'erateurs des deux chiralit\'es.
Une interpr\'etation satisfaisante de ces valeurs de $\gamma_{\hbox{\scriptsize
string}}$
reste \`a trouver.

\subsection{Fonctions \`a $N$ points}

Passons maintenant au calcul effectif des fonctions
de corr\'elation des op\'erateurs habill\'es \ref{vertex}.
La fonction \`a trois points peut \^etre obtenue directement
par son expression op\'eratorielle (cf [P3] partie 7.2).
Elle s'\'ecrit
\beq
\left <
\vertex J_1, \Jb_1,
(z_1)
\vertex J_2, \Jb_2,
(z_2)
\vertex J_3, \Jb_3,
(z_3)
\right >
=
{\cal C}_{1,2,3} /
\Bigl ( \prod_{k<l} | z_k-z_l |^2 \Bigr  )
\label{f3p1}.
\eeq
Comme dans le cas du couplage faible,
pour le calcul de fonctions de corr\'elation
dans l'approche op\'eratorielle
les op\'erateurs des deux extr\'emit\'es du corr\'elateur,
ceux qui engendrent les \'etats \`a partir du vide,
Ils doivent contenir des op\'erateurs du type $V^{(J)}_{-J}$
correspondants (cf \ref{etatsL}),
ce qui n'est pas le cas des $\chi_+$.
Cette diff\'erence est due au fait que le vide $Sl(2,C)$
invariant ne fait pas partie de l'espace de Hilbert
de la th\'eorie.
Deux des op\'erateurs de \ref{f3p1},
bien que not\'es $\vertex J_i, \Jb_i, $
pour ne pas alourdir les notations,
ne contiennent donc pas de vrais $\chi_+$.
Ils sont n\'eanmoins d\'efinis \`a partir de $V^{\Jpe_i}$
ce qui ne remet pas en question la loi d'\'echelle \ref{mudep}
puisqu'elle ne d\'epend que du spins des op\'erateurs et du vide.
Ceci n'est pas sp\'ecifique au couplage fort et on pourra
voir une discution d\'etaill\'ee
dans le cas du couplage faible en ref.\cite{G4} par exemple.
Nous donnons \`a titre d'exemple\footnote{
Les pointill\'es dans \ref{bouts}
d\'esignent des termes qui s'annulent sur le vide.}
$$
\vertex J_3, \Jb_3,
(z_3)
\sim
g'\,_{0,\Jme_3}^{\Jme_3}
V'^{(\Jme_3)}_{-\Jme_3}\
g_{0,\Jpe_3}^{\Jpe_3}
V^{(\Jpe_3)}_{-\Jpe_3}\
\gb'\,_{0,\Jmeb_3}^{\Jmeb_3}
\Vb'^{(\Jmeb_3)}_{-\Jmeb_3}\
\gb_{0,\Jpeb_3}^{\Jpeb_3}
\Vb^{(\Jpeb_3)}_{-\Jpeb_3}
+...
$$
$$
\sim
g'\,_{0,\Jme_3}^{\Jme_3}
\> :\!
e^{2\Jme_3\alpha'_-P'^{(1)}(z_3)}
\!:\
g_{0,\Jpe_3}^{\Jpe_3}
\> :\!
e^{2\Jpe_3\alpha_-P^{(1)}(z_3)}
\!:
\>\times
$$
\beq
\gb'\,_{0,\Jmeb_3}^{\Jmeb_3}
\> :\!
e^{2\Jmeb_3\alpha'_-\Pb'\,^{(1)}(z_3)}
\!:\
\gb_{0,\Jpeb_3}^{\Jpeb_3}
\> :\!
e^{2\Jpeb_3\alpha_-\Pb^{(1)}(z_3)}
\!:
\>+...
\label{bouts}
\eeq
qui engendre l'\'etat entrant
\beq
|\varpi'_{J_3,J_3},\varpi_{-J_3-1,J_3},
\varpib'_{\Jb_3,\Jb_3},\varpib_{\Jb_3,-\Jb_3-1}\!>
\label{etatin}
\eeq
alors que $\vertex J_1, \Jb_1,$ engendre
l'\'etat sortant
$$
<\!-\varpi'_{J_1,J_1},-\varpi_{-J_1-1,J_1},
-\varpib'_{\Jb_1,\Jb_1},-\varpib_{\Jb_1,-\Jb_1-1}|
=
$$
\beq
<\!\varpi'_{-J_1-1,-J_1-1},\varpi_{J_1,-J_1-1},
\varpib'_{-\Jb_1-1,-\Jb_1-1},\varpib_{-\Jb_1-1,\Jb_1}|
\label{etatout}
\eeq
\`a partir du vide $<\!-\varpi'_0,-\varpi_0,-\varpib'_0,-\varpib_0|$.
Il n'y a de facteur de normalisation dans aucun des deux cas
puisque les coefficients $g$ donn\'es
en \ref{ggen} (cf \ref{bouts} pour le premier cas)
sont tous \'egaux \`a un.
L'op\'erateur $\vertex J_2, \Jb_2,$ est donc appliqu\'e aux \'etats
\ref{etatout}, \ref{etatin}, et sa d\'efinition \ref{vertex} montre alors que
la fonction \`a trois points
s'exprime simplement en termes des coefficients
de normalisation $g$
\beq
{\cal C}_{1,2,3}=
\gp_{\Jme_{2} , \Jme_{3} }^{-\Jme_{1} -1-\pi/h'}\>\>
g_{\Jpe_{2} , \Jpe_{3} }^{-\Jpe_{1} -1-\pi/h}\>\>
\gbp_{\Jmeb_{2} , \Jmeb_{3} }^{-\Jmeb_{1} -1-\pi/h'}\>
\gb_{\Jpeb_{2} , \Jpeb_{3} }^{-\Jpeb_{1} -1-\pi/h}\> \>
\label{C}.
\eeq
On peut noter que cette fonction
est automatiquement sym\'etrique dans l'\'echange de 1, 2 et 3
puisque c'est une sym\'etrie des $g$ contenus dans \ref{C} (cf \ref{ggen}).
Comme dans le cas du couplage faible,
les coefficients issus de la mati\`ere ($g'$)
se simplifient \'enorm\'ement avec ceux
qui proviennent de la gravit\'e ($g$).
L'expression des coefficients $g$ en terme de chemins
permet assez facilement de montrer
que la fonction \`a trois points totale se factorise
(comme dans le cas du couplage faible):
\beq
{\cal C}_{1,2,3}
=
\lf J_1, \Jb_1,
\lf J_2, \Jb_2,
\lf J_3, \Jb_3,
\eeq
avec
\beq
\lf J, \Jb, =
\sqrt{
F
\left(
-Q_M\alpha'_+(J+1/2)
\right)}
\sqrt{ F
\left(
-Q_M\alpha'_-(\Jb+1/2)
\right)}
\eeq
o\`u la fonction $F$ est d\'efinie en \ref{H}.

Nous avons \'ecrit en [P3] la fonction \`a trois points
sous la forme \ref{f3p1}, essentiellement parce qu'elle donnait
un r\'esultat automatiquement sym\'etrique,
comme cela a \'et\'e fait en ref.\cite{G4} pour le couplage faible.
Il est cependant apparu dans l'\'etude en ref.\cite{dFK}
du m\^eme genre
de mod\`ele pour le couplage faible dans une repr\'esentation
int\'egrale,
que les fonctions de corr\'elation \`a consid\'erer\footnote{
Plus pr\'ecis\'ement de valeur non nulle par repr\'esentation int\'egrale.}
devaient contenir un op\'erateur de moment de signe oppos\'e
et de choix d'habillage diff\'erent.
Sans toutefois avoir la justification de la repr\'esentation
int\'egrale,
nous devons ici \'egalement introduire un op\'erateur
d'habillage diff\'erent,
faute de quoi nous ne pourrions pas obtenir les fonctions
\`a $N+1$ points par d\'erivation des fonctions
\`a $N$ points.
Ceci nous am\`ene donc \`a d\'efinir l'op\'erateur que
nous app\`elerons conjugu\'e par
\beq
\vertexc J_i, \Jb_i,
=
\chi'\, ^{(-J_i-1)}_-
\chi_+^{(J_i)}
\chib'\, ^{(-\Jb_i-1)}_-
\chib_+^{(\Jb_i)}\>
\label{vertexc}.
\eeq
Nous avons pris les op\'erateurs conjugu\'es de type
$\chi'\, ^{(-J_i-1)}_-$ pour la mati\`ere en transformant les
spins par $J_i\to -J_i-1$ par rapport \`a \ref{vertex}.
Nous avons fait pour ces op\'erateurs l'autre choix d'habillage,
ce qui revient \`a faire une autre sym\'etrie $J_i\to -J_i-1$
pour la gravit\'e et donc \`a employer les m\^emes
op\'erateurs de type $\chi_+^{(J_i)}$.
La loi d'\'echelle \ref{mudep2} sera
par cons\'equent inchang\'ee pour ces op\'erateurs \ref{vertexc}.
On peut par ailleurs v\'erifier que ces op\'erateurs sont \'egalement locaux
(avec le lien \ref{J-Jb} entre $J_i$ et $\Jb_i$).

Nous ne consid\'ererons donc que des fonctions de corr\'elation
contenant un op\'erateur conjugu\'e \ref{vertexc}
et $N-1$ op\'erateurs de type standard \ref{vertex}.
Ceci nous permettra d'obtenir des r\`egles de Feynman
coh\'erentes.
Ces fonctions de corr\'elation auront donc toujours une loi
d'\'echelle donn\'ee par \ref{mudep2}:
\beq
\left <
\vertexc J_1, \Jb_1,
\vertex J_2, \Jb_2,
...
\vertex J_N, \Jb_N,
\right >_{\mu_c}
=
\left <
\vertexc J_1, \Jb_1,
\vertex J_2, \Jb_2,
...
\vertex J_N, \Jb_N,
\right >_1
\>
\mu_c^{\sum_{i=1}^NP_i
-\left({N\over2}-1\right)
\left({s+2\over2}\right)}
\label{mudep3}
\eeq
avec
\beq
P_i=P(J_i,\Jb_i)=
{Q_M\over 4}
[\alpha'_-(J_i+{1\over 2})
+\alpha'_+(\Jb_i+{1\over 2})]
\label{Pi}
\eeq
qui d\'esigne un pseudo-moment
qui intervient dans la d\'ependance en $\mu_c$
des fonctions de corr\'elation.
C'est lui qui donnera la valeur du propagateur (\ref{prop} plus loin)
et il semble \^etre le ``moment effectif'' de cette th\'eorie
o\`u les deux chiralit\'es sont d\'ecoupl\'ees.

Utilisant un op\'erateur de ce type pour
engendrer les \'etats sortants,
on voit que la fonction \`a trois points
est alors donn\'ee par
\beq
\left <
\vertexc J_1, \Jb_1,
(z_1)
\vertex J_2, \Jb_2,
(z_2)
\vertex J_3, \Jb_3,
(z_3)
\right >
\sim
\gp_{\Jme_{2} , \Jme_{3} }^{\Jme_{1} }\>\>
g_{\Jpe_{2} , \Jpe_{3} }^{-\Jpe_{1}\!-1-\pi/h}\>\>
\gbp_{\Jmeb_{2} , \Jmeb_{3} }^{\Jmeb_{1} }\>
\gb_{\Jpeb_{2} , \Jpeb_{3} }^{-\Jpeb_{1}\!-1-\pi/h}\> \>
\label{f3p2}
\eeq
o\`u nous avons omis la d\'ependance en $z_i$
qui est la m\^eme qu'en \ref{f3p1}
puisqu'on a toujours des op\'erateurs de poids (1,1).
Ce r\'esultat n'est pas manifestement sym\'etrique
et la simplification entre mati\`ere et gravit\'e ne peut
avoir lieu exactement de la m\^eme fa\c con que pr\'ec\'edemment
\`a cause de l'habillage diff\'erent
de l'op\'erateur $\vertexc J_1, \Jb_1,$ (les
nombres de charges d'\'ecran pour la mati\`ere et la
gravit\'e ne sont plus \'egaux
ou reli\'es de la m\^eme fa\c con).
Il est remarquable que le r\'esultat se simplifie quand m\^eme
(le calcul qui utilise \`a nouveau l'expression des $g$
en termes de chemins sera d\'etaill\'e en [P5])
pour donner\footnote{
Il ne faut pas s'\'etonner d'avoir ici
(comme en \ref{mudep3} et plus loin en \ref{f4p1}...)
une puissance de $\mu_c$ sym\'etrique dans les trois pattes,
alors que le premier vertex ($\vertexc J_1, \Jb_1, $)
a un moment oppos\'e pour la mati\`ere
(cf \ref{vertexc} qui a \'et\'e transform\'e
par $J_i\to-J_i-1$),
et qu'il est donc logiquement not\'e comme sortant
sur le graphe de Feynman \ref{f3p2b}.
Ceci est simplement d\^u au fait que les spins $J_i$
(par lesquels on d\'efinit le moment $P_i$
en Eq.\ref{Pi})
sont en fait ceux de la gravit\'e pour les
op\'erateurs \ref{vertex}
et \'egalement pour les op\'erateurs conjugu\'es \ref{vertexc}
dont l'habillage est diff\'erent.}
\beq
\left <
\vertexc J_1, \Jb_1,
(z_1)
\vertex J_2, \Jb_2,
(z_2)
\vertex J_3, \Jb_3,
(z_3)
\right >_{\mu_c}
\!\!\!\!\!
\sim
\lfc J_1, \Jb_1,
\lf J_2, \Jb_2,
\lf J_3, \Jb_3,
\>
\mu_c^{P_1+P_2 + P_3
-{s+2\over 4}}
\!\!\!\!
=
\!\!
\epsffile{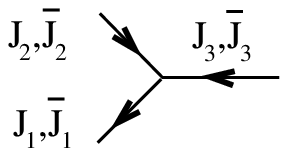}
\label{f3p2b}
\eeq
avec
\beq
\lfc J,\Jb, =
\sqrt{
F
\left(
Q_M\alpha'_-(J+1/2)
\right)}
\sqrt{F
\left(
Q_M\alpha'_+(\Jb+1/2)
\right)}
{}.
\eeq

Nous faisons ensuite l'hypoth\`ese
que cette th\'eorie peut \^etre
d\'ecrite par un action effective
contenant le terme cosmologique introduit plus haut:
\beq
S_{\mu_c}=S_0-\mu_c \int \vertex 0, 0,
\eeq
o\`u $S_0$ est une action qui ne d\'epend pas de $\mu_c$
et dont nous ne connaissons pas la forme.
La d\'erivation par rapport \`a $\mu_c$
d'une fonction \`a $N$ points calcul\'ee
avec cette action effective donne une fonction \`a $N+1$ points\footnote{
Nous parlons donc maintenant de fonctions de corr\'elation
avec points d'insertion int\'egr\'es sur toute la surface.
Les signes d'int\'egration dans \ref{der1}, \ref{f4p1}
d\'esignent donc cette int\'egrale sur les points de la surface.}:
\beq
{\partial\over\partial{\mu_c}}
\left<
e^{-S_{\mu_c}}
\int
\vertexc J_1,\Jb_1,
\vertex J_2,\Jb_2,
...
\vertex J_N,\Jb_N,
\right>
=
\left<
e^{-S_{\mu_c}}
\int
\vertexc J_1,\Jb_1,
\vertex J_2,\Jb_2,
...
\vertex J_N,\Jb_N,
\vertex 0,0,
\right>
\label{der1}
\eeq
Partant de la fonction \`a trois points
particuli\`ere
\beq
\left <
\vertexc J_i, \Jb_i,
\vertex 0, 0,
\vertex J_i,  \Jb_i,
\right >
_{\mu_c}
=
\mu_c
^{2P_i
-1
}
\lfc J_i,\Jb_i,
\lf 0,0,
\lf J_i,\Jb_i,
\label{f3p0}
\eeq
et prenant sa primitive,
nous
obtenons
\beq
\left <
\vertexc J_i,  \Jb_i,
\vertex J_i, \Jb_i,
\right >_{\mu_c}
\!\!\!\!
=
{
1
\over 2P_i}
\>
\mu_c
^{2P_i}
\>
\lfc J_i,\Jb_i,
\lf J_i,\Jb_i,
\eeq
o\`u il a \'et\'e utilis\'e que $\lf 0,0,=1$,
ce qui est une propri\'et\'e tr\`es particuli\`ere
aux dimensions sp\'eciales.
La fonction \`a deux points avec pattes externes coup\'ees
est l'inverse du propagateur\footnote{\`a un facteur (1/2) pr\`es,
que nous ajoutons, car il
s'av\`ere n\'ecessaire par la suite} qui vaut donc
\beq
P_{\mu_c}(J_i,\Jb_i)=
P_i
\>
{
\mu_c
^{-2P_i}
\over(
\lfc J_i,\Jb_i,
\lf J_i,\Jb_i,
)}
=
\epsffile{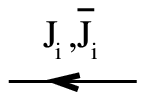}
\label{prop}.
\eeq
On peut aussi d\'eriver la fonction \`a trois points
\ref{f3p2b} et obtenir une fonction \`a quatre points
dont la quatri\`eme patte ($\vertex 0, 0,$) a un moment particulier.
Comme l'expression obtenue est sym\'etrique pour les deux autres pattes
engendr\'ees par des op\'erateurs de type \ref{vertex}
(i.e.
$\vertex {J_2}, {\Jb_2}, $
et
$\vertex {J_3}, {\Jb_3}, $)
on peut, sans trop de risques de se tromper, \'etendre
ceci \`a la quatri\`eme patte.
On obtient ainsi la fonction \`a quatre points tous de
moments quelconques:
$$
\left <
e^{-S_{\mu_c}}
\int
\vertexc J_1, \Jb_1,
\vertex J_2, \Jb_2,
\vertex J_3, \Jb_3,
\vertex J_4, \Jb_4,
\right >_{\mu_c}
=
\lfc J_1,\Jb_1,
\lf J_2,\Jb_2,
\lf J_3,\Jb_3,
\lf J_4,\Jb_4,
\mu_c^{P_1+P_2+P_3+P_4
-{s+2\over 2}}
$$
\beq
\left(
1+{Q_M\over 4}\Bigl[ \alpha'_-
(J_1+J_2+J_3+J_4+1)
+\alpha'_+(\Jb_1+\Jb_2+\Jb_3+\Jb_4+1)\Bigr]
\right)
\label{f4p1}
\eeq
Si cette description par une action effective
est valide, cette fonction \`a quatre points,
qui est une fonction de corr\'elation totale,
doit inclure les diagrammes construits \`a partir
de vertex \`a trois points:
\beq
\epsffile{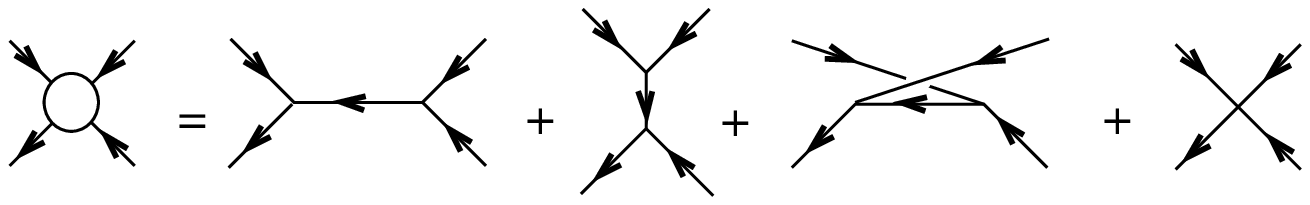}
\label{f4ps}
\eeq
ainsi que la fonction \`a quatre points \`a une particule
irr\'eductible figur\'ee \`a droite.
Nous pouvons calculer les graphes de Feynman
du membre de droite construits \`a partir
du vertex \`a trois points que nous avons obtenu en \ref{f3p2b}
et du propagateur obtenu en \ref{prop}.
On v\'erifie que pour un graphe du type
\beq
\epsffile{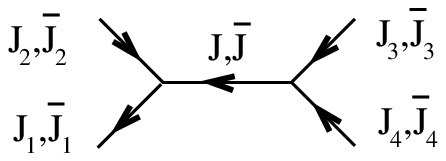}
\label{f4p1d}
\eeq
les facteurs $\lf J, \Jb,
\lfc J, \Jb,$ ainsi que la d\'ependance en $\mu_c$
dus \`a l'\'etat interm\'ediaire $(J,\Jb)$
issus des deux fonctions \`a trois points
(\ref{f3p2b})
et du propagateur (\ref{prop})
s'annulent bien.
Nous omettrons donc d\'esormais les facteurs $L,L^{conj}$
qui sont une pure jauge
(ou normalisation des op\'erateurs)
et nous placerons \`a $\mu_c=1$ pour simplifier
les expressions.

Il faut \'egalement sommer sur les \'etats
qui peuvent se propager sur les pattes internes.
Pour le diagramme
\ref{f4p1d} par exemple
les r\`egles de s\'election
se traduisent par
$J=J_3+J_4-\ns_2$ et $\Jb=\Jb_3+\Jb_4-\nsb_2$
o\`u $\ns_2$ est le nombre de charges d'\'ecran
attach\'ees au vertex de droite,
et de m\^eme pour l'autre chiralit\'e avec $\nsb_2$.
Ces nombres de charges d'\'ecran doivent \^etre inf\'erieurs au nombre
total de charges d'\'ecran
$\ns=J_2+J_3+J_4-J_1,\nsb=\Jb_2+\Jb_3+\Jb_4-\Jb_1$
(ceci est issu des r\`egles de s\'elections du premier vertex de \ref{f4p1d}).
Ces r\`egles de s\'election peuvent para\^\i tre \'etonnantes
(surtout par comparaison au cas du couplage faible),
mais elles refl\`etent simplement celles des op\'erateurs $\chi$.
Les deux chiralit\'es sont d\'ecoupl\'ees,
et en revanche pour une chiralit\'e donn\'ee
les spins $J$ et $\Jhat$ li\'es \`a $\alpha_-$ et $\alpha_+$ sont reli\'es
(pour avoir des poids r\'eels)
ainsi que ceux de la mati\`ere et de la gravit\'e
(pour avoir des op\'erateurs $(1,1)$).
Les nombres de charges d'\'ecran
li\'es \`a $\alpha_-,\alpha_+,\alpha'_-,\alpha'_+$
pour la mati\`ere et la gravit\'e sont donc reli\'es
de la m\^eme mani\`ere\footnote{
Plus pr\'ecis\'ement, notant $\nu_-,\nu_+,\nu'_-,\nu'_+$,
les nombres de charges d'\'ecran,
pour la gravit\'e et pour la mati\`ere,
li\'es respectivement \`a $\alpha_-,\alpha_+,\alpha'_-,\alpha'_+$,
on a $\ns=\sum_{i=2}^N J_i\>-J_1=\nu'_-=\nu'_+=\nu_+-2J_1-1=-\nu_--2J_1-(N-1)$,
pour
les fonctions \`a $N$ points,
alors que pour l'autre chiralit\'e
les r\^oles de $\nu_-$ et $\nu_+$ sont \'echang\'es.
Les nombres $\ns$ et $\nsb$ d\'esignent donc les nombres de charges
d'\'ecran pour la mati\`ere.}.

Imaginons ce que va donner la sommation sur les moments internes
$(J,\Jb)$ d'un diagramme du type \ref{f4p1d}.
Chaque terme de la sommation s'exprimera par un propagateur
du genre de \ref{prop} (avec moment shift\'e).
Mais il y aura autant de termes de
que de fa\c cons de r\'epartir $\ns$ charges d'\'ecran sur deux vertex
et ind\'ependamment $\nsb$ charges de l'autre chiralit\'e,
c'est-\`a-dire $(\ns+1)(\nsb+1)$ possibilit\'es.
Vu que les moments admissibles sont en progression arithm\'etique
et que le propagateur est lin\'eaire dans les moments,
ce nombre $(\ns+1)(\nsb+1)$ sera m\^eme toujours factoris\'e.
La somme des trois voies de ce type (Eq.\ref{f4ps})
donnera donc une fonction \`a quatre points incluant un tel
facteur (multipli\'e par une combinaison lin\'eaire des moments,
qui provient des propagateurs).
Mais ce n'est pas le cas de la fonction \`a quatre
points \ref{f4p1} obtenue par d\'erivation.
Il faut donc supposer que l'amplitude physique \`a $N$ points
doit inclure un facteur combinatoire suppl\'ementaire
exprimant le nombre de fa\c cons de r\'epartir $\ns$
charges d'\'ecran sur $N-2$ vertex\footnote{
C'est normal, puisque elles ne peuvent pas \^etre plac\'ees
sur les 2 op\'erateurs extr\^emes de genre diff\'erent.}.
On d\'efinit donc les amplitudes physiques \`a $N$ points par\footnote{
Le nombre de fa\c cons de r\'epartir $\ns$
objets indiff\'erentiables en $N-2$ points peut \^etre vu comme le
coefficient d'ordre $\ns$ du d\'eveloppement en $x$
de $(1+x+x^2...)^{N-2}=1/(1-x)^{N-2}$.
Ceci donne $(N-2)(N-1)...(N+\ns-3)/\ns!$ qui est \'ecrit
en \ref{Aphys} sous forme de nombre de combinaisons.
Il ne semble pas en revanche y avoir d'interpr\'etation
de ce nombre en termes de choix de $N-3$ (ou $\ns$) objets parmi $\ns+N-3$.}
$$
\>^{\hbox{\scriptsize Phys}}\!
A^{(N)}_{(\ns,\nsb)}
\left(
(J_1,\Jb_1),
(J_2,\Jb_2)...
(J_N,\Jb_N)
\right)
$$
\beq
=
\left(
\,^{\ns+N-3}_{N-3}
\right)
\left(
\,^{\nsb+N-3}_{N-3}
\right)
\left<
e^{-S_{\mu_c}}
\int
\vertexc J_1,\Jb_1,
\vertex J_2,\Jb_2,
...
\vertex J_N,\Jb_N,
\right>.
\label{Aphys}
\eeq
Nous rappelons que les nombres de charges d'\'ecran sont donn\'es par
$\ns=J_2+...+J_N-J_1$ et $\nsb=\Jb_2+...+\Jb_N-\Jb_1$.

Ceci donne alors la r\`egle de d\'erivation
pour les amplitudes physiques (utilisant \ref{der1}):
$$
\>^{\hbox{\scriptsize Phys}}\!
A^{(N+1)}_{(\ns,\nsb)}
\left(
(J_1,\Jb_1),
(J_2,\Jb_2)...
(J_N,\Jb_N),
(0,0)
\right)
$$
\beq
=
-
\left({\ns+N-2\over N-2}\right)
\left({\nsb+N-2\over N-2}\right)
\>{\partial\over\partial{\mu_c}}
\>^{\hbox{\scriptsize Phys}}\!
A^{(N)}_{(\ns,\nsb)}
\left(
(J_1,\Jb_1),
(J_2,\Jb_2)...
(J_N,\Jb_N)
\right)
\label{der2}.
\eeq
On peut donc maintenant calculer la vraie amplitude
physique \`a quatre points
$$
\epsffile{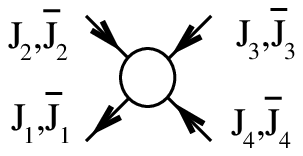} =
\>^{\hbox{\scriptsize Phys}}\!
A^{(4)}_{(\ns,\nsb)}
\left(
(J_1,\Jb_1),
(J_2,\Jb_2),
(J_3,\Jb_3),
(J_4,\Jb_4)
\right)
=
(\ns+1)(\nsb+1)\times
$$
\beq
\left(
1+{Q_M\over 4}\Bigl[ \alpha'_-
(J_1+J_2+J_3+J_4+1)
+\alpha'_+(\Jb_1+\Jb_2+\Jb_3+\Jb_4+1)\Bigr]
\right)
\label{f4p2}
\eeq
qui comprend un simple facteur suppl\'ementaire
par rapport \`a \ref{f4p1}
(nous nous sommes plac\'es \`a $\mu_c=1$ et avons omis les facteurs
$L,L^{conj}$ conform\'ement \`a ce que nous avions annonc\'e).
Il est ensuite facile de calculer les sommes correspondant
aux trois premiers graphes de Feynman de \ref{f4ps}
et par soustraction d'obtenir
la fonction \`a quatre points \`a une particule irr\'eductible:
$$
\epsffile{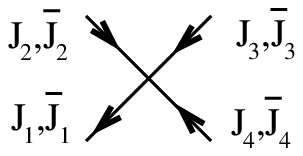} =
\>^{1PI}\!
A^{(4)}_{(\ns,\nsb)}
\left(
(J_1,\Jb_1),
(J_2,\Jb_2),
(J_3,\Jb_3),
(J_4,\Jb_4)
\right)
$$
\beq
=
(\ns+1)(\nsb+1)
{1\over 4}
\left(
(2+s)
-{Q_M\over 2}(\alpha'_-\ns
+\alpha'_+\nsb)
\right)
\label{f4pirr}
\eeq
qui ne d\'epend pas des moments.

On peut ensuite d\'eterminer de la m\^eme mani\`ere les fonctions d'ordre
sup\'erieur.
La fonction \`a cinq points est obtenue par d\'erivation
et se d\'ecompose en graphes de Feynman du type:
\beq
\epsffile{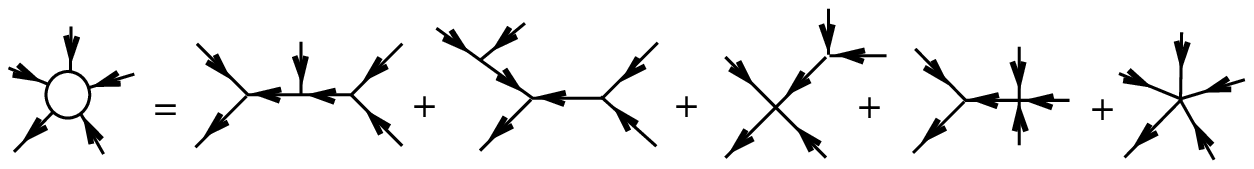}
\eeq
o\`u nous n'avons figur\'e que les cinq graphes
de topologies (orient\'ees) diff\'erentes.
Ecrivant tous les graphes obtenus par permutation
des quatre op\'erateurs ``entrants'',
on en obtient 26.
Ceux par exemple qui contiennent deux fonctions
\`a trois points s'expriment par une quadruple
somme sur les deux moments internes
et les r\'esultats n'ont pu \^etre obtenus
qu'avec l'aide de programmes de calcul symbolique.
Par soustraction, on obtient la fonction \`a cinq points
irr\'eductible
$$
\>^{1PI}\!
A^{(5)}_{(\ns,\nsb)}
\left(
(J_1,\Jb_1),
(J_2,\Jb_2),
(J_3,\Jb_3),
(J_4,\Jb_4),
(J_5,\Jb_5)
\right)
$$
\beq
=
\left(
\,^{\ns+2}_{2}
\right)
\left(
\,^{\nsb+2}_{2}
\right)
\left(
B^{(5)}(\ns,\nsb)
-
\sum_{i=1}^5
(P_i
)^2
\right)
\label{f5pirr}
\eeq
o\`u $B^{(5)}(\ns,\nsb)$ est un polyn\^ome
en $\ns,\nsb$ d\'ependant de la charge centrale
donn\'e en [P4] Eq.5.15.
De la m\^eme mani\`ere,
la fonction \`a six points n\'ecessite le calcul
de 236 graphes de Feynman de 12 topologies diff\'erentes.
Il y en 2752 de 34 topologies diff\'erentes pour la fonction
\`a sept point et nous nous sommes donc arr\^et\'es
\`a la fonction \`a six points irr\'eductible qui vaut
$$
\>^{1PI}\!
A^{(6)}_{(\ns,\nsb)}
\left(
(J_1,\Jb_1)
...
(J_6,\Jb_6)
\right)
=
$$
\beq
\left(
\,^{\ns+3}_{3}
\right)
\left(
\,^{\nsb+3}_{3}
\right)
\left[
B^{(6)}(\ns,\nsb)
-
{3\over 2}
\left(
(2+s)
-{Q_M\over8}
(\alpha'_-\ns
+\alpha'_+\nsb)
\right)
\left(
\sum_{i=1}^6
(
P_i )^2
\right)
\right]
\label{f6pirr}
\eeq
o\`u le polyn\^ome $B^{(6)}(\ns,\nsb)$ est donn\'e en [P4] Eq.5.17.

Il est normal que la somme des graphes de Feynman calcul\'es
soient sym\'etriques dans l'\'echange de $(J_2,\Jb_2)$...$(J_N,\Jb_N)$,
puisque tous les graphes ainsi sym\'etris\'es ont \'et\'e inclus.
En revanche, la sym\'etrie pour tous les op\'erateurs,
de 1 \`a $N$ n'est pas \'evidente a priori.
Le moment
$P_1$ est un moment sortant, et,
aux charges d'\'ecran pr\`es,
il est la somme des autres qui sont entrants.
Il est donc tout \`a fait remarquable les fonctions irr\'eductibles
obtenues (\ref{f4pirr},\ref{f5pirr},\ref{f6pirr})
soient sym\'etriques dans les $N$ moments et en particulier
pour le changement de signe $P_i\to-P_i$.
Ceci semble \^etre une bonne confirmation
de la pertinence de cette formulation par action effective
utilis\'ee ici.

Ce r\'esultat est similaire \`a celui
obtenu en ref.\cite{dFK} dans le r\'egime de couplage faible.
Les fonctions \`a une particule irr\'eductibles
apparaissent l\`a-bas comme des fonctions analytiques et
sym\'etriques des moments.
Les fonctions d'ordre sup\'erieur \`a 6 (sans charges d'\'ecran)
ont \'et\'e calcul\'ees dans ce cas en\cite{dFK}
et sont toujours des sommes de puissances paires
des moments:
la puissance est de fa\c con g\'en\'erale\footnote{
$E(x)$ d\'esigne la partie enti\`ere de $x$.}
$2 E((N-3)/2)$ pour une
fonction \`a $N$ points.
Nous avons donc obtenu le m\^eme type de comportement
jusqu'\`a $N=6$ (avec charges d'\'ecran)
pour le couplage fort.

\vskip 2mm

Nous concluons en remarquant que
les r\'esultats obtenus dans cette deuxi\`eme sous-partie
permettent de d\'eterminer l'exposant critique
$\gamma_{\hbox{\scriptsize string}}$ pour des genres sup\'erieurs \`a z\'ero.
On peut \`a partir des fonctions \`a $N$ points
et du propagateur d\'eterminer la d\'ependance
en $\mu_c$ des fonctions de corr\'elation avec boucle\footnote{
Il faudrait s\^urement calculer aussi des corr\'elateurs
avec deux op\'erateurs conjugu\'es \ref{vertexc},
mais de toute fa\c con, la d\'ependance en $\mu_c$,
qui seule nous importe ici,
serait inchang\'e}:
consid\'erons une fonction \`a $N+2$ points
dont deux moments externes sont \'egaux ($=P_{N+2}$ par exemple)
et refermons la en une boucle par un propagateur.
La deuxi\`eme ligne de \ref{mudep} montre que la
fonction \`a $N+2$ points
comporte un facteur $\mu_c^{2P_{N+2}-(s+2)/2}$ suppl\'ementaire
par rapport \`a la fonction \`a $N$ points.
Le propagateur contribue pour $\mu_c^{-2P_{N+2}}$
et donc la boucle pour $\mu_c^{(2+s)/2}$ au total.
Chaque boucle suppl\'ementaire en fait autant.
Malgr\'e les fondements assez fragiles de cet argument,
imaginons que cette d\'ependance en $\mu_c$
survive \`a la renormalisation n\'ecessaire
\`a ces fonctions avec boucles.
Ceci donne donc en genre $h$ un comportement
de la fonction \`a $N$ points
$$
\sim
\mu_c^{\sum_{i=1}^NP_i-(N/2+g-1)(2+s)/2}.
$$
Prenant le cas $N=0$ et r\'ep\'etant l'argument de la
sous-partie pr\'ec\'edente on obtient donc finalement
\beq
\gamma_{\hbox{\scriptsize string}}=
2+(h-1)\left({2+s\over2}\right)
\label{gstrh}
\eeq
en genre $h$.

Ce raisonnement, bien qu'assez sp\'eculatif,
permet pour le couplage faible
de retrouver la formule \ref{gstring}
pour $\gamma_{\hbox{\scriptsize string}}$ en tout genre,
ce qui semble confirmer sa validit\'e.

\newpage

\centerline{\bf\large CONCLUSION ET PERSPECTIVES}

\vskip 3cm

La r\'esolution
de la th\'eorie de Liouville dans le r\'egime de
couplage fort vient donc de conna\^\i tre des progr\`es importants.
C'est une des r\'ealisations majeures de cette th\`ese.
La physique du couplage fort semble tr\`es diff\'erente de celle du
couplage faible.
Outre le traitement sym\'etrique des deux charges d'\'ecran,
une diff\'erence essentielle r\'eside dans le d\'econfinement de la
chiralit\'e:
les moments (ou les spins) des composantes chirales droites
et gauches sont diff\'erents,
contrairement au cas du couplage faible o\`u les op\'erateurs
locaux sont construits \`a partir des m\^emes spins
gauches et droits\footnote{
Ce sont les spins du groupe quantique
qui sont diff\'erents pour les deux chiralit\'es.
Les poids droits et gauches des op\'erateurs de mati\`ere ou de gravit\'e seuls
sont \'egalement diff\'erents.
En revanche, on a \'evidemment toujours des op\'erateurs
habill\'es de poids (1,1).}
Ceci est tr\`es certainement valable aussi bien pour le mod\`ele
topologie \'etudi\'e dans la partie 5.3
que pour un mod\`ele de v\'eritable corde dont les champs de mati\`ere
seraient habill\'es par les op\'erateurs $\chi_+^{(J)}\chib_+^{(\Jb)}$.
Ce dernier op\'erateur est en effet (quasiment)
local pour $J$ et $\Jb$ diff\'erents.
Comme pour le mod\`ele topologique,
il est fort probable qu'on
puisse construire des op\'erateurs de vertex
de mati\`ere locaux de modes gauches et droits diff\'erents
de mani\`ere \`a obtenir apr\`es habillage des op\'erateurs de poids (1,1).
Certes, ceci n'a pas encore \'etait fait explicitement,
mais la fonction de partition de tels mod\`eles a
n\'eanmoins d\'ej\`a \'et\'e calcul\'ee en ref.\cite{BG},
et son invariance modulaire a paru n\'ecessiter
l'introduction dans le spectre de spins gauches et droits
diff\'erant d'un entier arbitraire.
Au del\`a de notre mod\`ele topologique,
ceci confirme donc le d\'econfinement de la chiralit\'e
pour de v\'eritables mod\`eles de cordes.
On ne peut cependant pas compl\`etement exclure que la
localit\'e des op\'erateurs de vertex
de mati\`ere puisse imposer l'\'egalit\'e des moments droits et gauches
(pour un spectre particulier qui assurerait l'invariance modulaire),
ce qui donnerait alors \'egalement des spins $J$ et $\Jb$ \'egaux pour la
gravit\'e.
Dans cette hypoth\`ese, au demeurant peu probable,
le d\'econfinement de la chiralit\'e serait
particulier au mod\`ele topologique.

\vskip 3mm

Examinons comment on peut dans cette approche construire
des mod\`eles topologiques,
c'est-\`a-dire incluant deux degr\'es de libert\'e,
un pour la mati\`ere, l'autre pour la gravit\'e.
Nous essayons les constructions les plus g\'en\'erales
possibles,
nous permettant m\^eme de m\'elanger de mani\`ere
inhabituelle gravit\'e et mati\`ere.
Ce cadre plus g\'en\'eral
nous permettra de bien distinguer ce qui a \'et\'e
fait dans le cas du couplage faible (cf ref.\cite{G4} et
la conclusion de le chapitre 2) et du couplage fort (chapitre 5),
ainsi que d'ouvrir quelques perspectives.

Nous cherchons donc \`a construire un op\'erateur local \`a partir
des champs de mati\`ere d\'enot\'es par des primes (spins $J',\Jhat'...$)
et de ceux de gravit\'e, sans primes.
Chacun d'entre eux a deux composantes chirales not\'ees avec et sans barres.
Chaque composante chirale a \`a son tour deux spins $J$ et $\Jhat$
correspondant aux deux charges d'\'ecran $\alpha_\pm$.
Notre op\'erateur local peut donc finalement \^etre
obtenu comme combinaison lin\'eaire du produit\footnote{
Combinaison sur des $m,\mhat,m'...$ diff\'erents,
\`a $J,\Jhat,J'...$ fix\'es.}
$$
V^{J,\Jhat}_{m,\mhat}\ V'\,^{J',\Jhat'}_{m',\mhat'}\
\Vb^{\Jb,\Jhatb}_{\mb,\mhatb}\ \Vb'\,^{\Jb',\Jhatb'}_{\mb',\mhatb'}.
$$
Comme on l'a vu, la localit\'e pour les op\'erateurs
comporte deux volets:
il faut d'abord v\'erifier l'invariance pour la monodromie autour
de l'origine,
ce qui donne des conditions tr\`es simples dans cette base o\`u
elle est diagonale,
et ensuite que l'\'echange et la fusion donne bien des op\'erateurs
du m\^eme type (ce qui garantit l'invariance des corr\'elateurs physiques
\`a quatre points sous les monodromies
autour de 1 et l'infini).
On a au total $2^3=8$ spins diff\'erents.
Dans une bonne normalisation\footnote{
C'est-\`a-dire o\`u on a renorm\'e par les coefficients $g$
de mani\`ere \`a ce qu'ils disparaissent de la fusion,
comme pour les op\'erateurs $\Vt$ en Eq.\ref{Vtilde}.}
l'\'echange et la fusion de ces op\'erateurs sont essentiellement
donn\'es par les huit 6-j correspondants.
La m\'ethode g\'en\'erale pour obtenir un op\'erateur local
consiste \`a consid\'erer des combinaisons lin\'eaires (sur $m$)
de ces op\'erateurs pour des spins \'egaux deux \`a deux.
Ceci permet d'avoir aussi des 6-j \'egaux deux \`a deux,
et ensuite,
si on a choisi une combinaison lin\'eaire appropri\'ee,
leur orthogonalit\'e prouve la localit\'e
des op\'erateurs physiques.

Dans le cas du couplage faible
on consid\`ere ainsi des spins gauches et droits \'egaux
($J=\Jb$, $\Jhat=\Jhatb$, $J'=\Jb'$, $\Jhat'=\Jhatb'$),
ce qui garantit l'invariance sous la monodromie autour de 0.
L'orthogonalit\'e des 6-j droits et gauches montre que
l'\'echange
et la fusion de ces op\'erateurs sont triviaux:
ces op\'erateurs sont locaux et en particulier les conditions $J=\Jb$
sont stables.
Ce n'est pas diff\'erent de ce qui a \'et\'e fait dans l'approche
par repr\'esentation int\'egrale\cite{DF}, o\`u le corr\'elateur physique local
est obtenu comme combinaison lin\'eaire de produits de blocs conformes
dont les coefficients sont l'\'equivalent de nos $g$ (ils sont quasiment
\'egaux).

Dans le cas du couplage fort,
nous avons au contraire \'et\'e amen\'es \`a choisir des spins
$J$ et $\Jhat$ \'egaux de mani\`ere a obtenir des poids r\'eels
(ou plus pr\'ecis\'ement $J'=\Jhat'$, $\Jb'=\Jhatb'$
pour la mati\`ere et $J=-\Jhat-1$, $\Jb=-\Jhatb-1$ pour la gravit\'e).
La fermeture de l'alg\`ebre de ces op\'erateurs
et donc la stabilit\'e de cette condition
n\'ecessite alors l'\'egalit\'e des 6-j issus des spins $J$
et des spins $\Jhat$ ($=J$ ou $=-J-1$).
Ceci n'est cependant pas automatique comme pour les 6-j des
deux chiralit\'es dans le cas du couplage faible,
puisqu'ici ces deux 6-j sont calcul\'es \`a partir des param\`etres
de d\'eformation $h$ et $\hhat$ respectivement.
Ce n'est donc qu'exceptionnellement, pour les dimensions sp\'eciales
$C=1+6(2+s)$, $s=-1,0,1$ et des spins fractionnaires $J=n/2(2\pm s)$,
que les 6-j des deux types ont pu \^etre reli\'es
pour donner une relation d'orthogonalit\'e
(en utilisant en outre la sym\'etrie des 6-j sous $J\to-J-1$).
C'est aussi pour ces dimensions et spins sp\'eciaux
que la matrice de monodromie est d\'eg\'en\'er\'ee,
ce qui assure l'invariance de ces op\'erateurs sous
la monodromie autour de l'origine.

Poussant plus loin la logique de ces constructions,
il vient tout de suite \`a l'esprit une troisi\`eme fa\c con
d'obtenir des op\'erateurs locaux:
puisqu'il y a $2^3=8$ types de spins,
on peut choisir une contrainte non entre spins gauches et droits
($J=\Jb$, couplage faible),
non entre $J$ et $\Jhat$ (couplage fort),
mais entre spins de la mati\`ere $J'$ et de la gravit\'e $J$.
Le choix $J=\Jhat'$, $\Jhat=-J'-1$, $\Jb=-\Jhatb'-1$ et $\Jhatb=J'$,
par exemple, donne des op\'erateurs de poids $(1,1)$
que l'on d\'efinirait donc par
$$
\sum_{m,\mhat,\mb,\mhatb}
(g^{.}_{..}...)
V^{J,\Jhat}_{m,\mhat}\ V'\,^{-\Jhat-1,J}_{-\mhat,m}\
\Vb^{\Jb,\Jhatb}_{\mb,\mhatb}\ \Vb'\,^{\Jhatb,-\Jb-1}_{\mhatb,-\mb}.
$$
Comme les param\`etres de d\'eformation pour la mati\`ere
et la gravit\'e sont oppos\'es ($h+\hhat'=0=\hhat+h'$),
les 6-j correspondants seront \'egaux et il sera facile
d'obtenir ainsi
des op\'erateurs locaux
pour une charge centrale quelconque.
Ces op\'erateurs physiques ne sont donc pas comme
d'habitude le simple produit
d'un vertex de mati\`ere par un vertex de gravit\'e.
Au contraire mati\`ere et gravit\'e sont intimement m\'el\'ees.
Il ne semble d'ailleurs pas y avoir de terme cosmologique
c'est-\`a-dire de terme de poids (1,1) issu uniquement
du degr\'e de libert\'e qu'on souhaiterait interpr\'eter
comme la gravit\'e.
Peut-\^etre cette construction est-elle en fait \'equivalente
\`a un mod\`ele $c=1$
par transformation lin\'eaire des champs libres sous-jacents.
Une diff\'erence essentielle avec les deux  cas pr\'ec\'edemment envisag\'es
est que ces op\'erateurs sont de poids (1,1),
et qu'ils donnent encore uniquement des op\'erateurs (1,1) par fusion et
\'echange.
Dans les constructions du couplage faible ou fort,
ce n'est que la cohomologie BRST qui permet de s\'electionner
les op\'erateurs de vertex de poids (1,1), alors qu'ici cela
d\'ecoule automatiquement des r\`egles de s\'election des op\'erateurs.
Ce mod\`ele serait certainement tr\`es int\'eressant \`a \'etudier plus avant.

\vskip 3mm

Je souhaite conclure
par un examen un peu plus approfondi du parall\`ele d\'ej\`a \'evoqu\'e
entre les mod\`eles minimaux et les mod\`eles
topologiques
fortement coupl\'es de la partie 5.3.
Ces deux cas pr\'esentent en effet des caract\'eristiques
similaires.
On a dans les deux cas un ensemble discret de dimensions particuli\`eres,
d'un c\^ot\'e $c=1-6(p-q)^2/pq$
et de l'autre $c=1+6(2-s)$.
On a dans les deux cas une troncature d'une sous-alg\`ebre ferm\'ee
correspondant \`a un ensemble discret de moments.
En termes de spins on a $J=(m-1)/2$, $\Jhat=(n-1)/2$
pour $n,m$ entiers respectivement
inf\'erieurs \`a $p$ et $q$ dans le cas des mod\`eles
minimaux,
et $J=\Jhat=n/2(2\pm s)$ dans le deuxi\`eme cas.
On a ensuite,
comme on l'a soulign\'e dans la partie 5.2,
une d\'eg\'en\'erescence de la matrice de monodromie
dans les deux cas.
On peut d'ailleurs se demander s'il ne s'agirait pas des cas
de d\'eg\'en\'erescence maximale.
Il serait certainement tr\`es int\'eressant d'examiner syst\'ematiquement
(en fonction du param\`etre qu'est $c$)
la d\'eg\'en\'erescence des valeurs propres\footnote{
Voir Eqs.\ref{deltaalpha}, \ref{alphap}... pour les notations
$\alpha$, $\Delta(\alpha)$.}
$\propto e^{2i\pi\Delta(\alpha)}$
pour les valeurs $\alpha\to \alpha+p\alpha_-+\phat\alpha_+$
donn\'ees par les r\`egles de fusion ($p,\phat$ entiers).
Peut-\^etre que les dimensions $c=1+6(2-s)$
avec $s$ fractionnaire, qui donnent \'egalement une d\'eg\'en\'erescence
de la monodromie, mais moins importante,
pourraient \^etre reli\'es aux autres mod\`eles
minimaux.
Les nombres q-d\'eform\'es \`a partir des param\`etres $h$ et $\hhat$
correspondants ont quelques propri\'et\'es remarquables
qui permettent d'esp\'erer une troncature (partielle ou diff\'erente ?)
\'egalement dans ces cas-l\`a.

Il ne se passe certes pas exactement la m\^eme chose dans le cas des mod\`eles
minimaux et dans le n\^otre.
Le nombre de valeurs sp\'eciales pour $c$ est infini ou fini
pour le couplage faible ou fort respectivement,
au contraire le nombre de familles conformes dans une th\'eorie
est soit fini soit infini,
et la d\'eg\'en\'erescence de la matrice de monodromie
ne donne pas lieu \`a une troncature exactement de la m\^eme mani\`ere:
dans le cas des mod\`eles minimaux
les \'etats $(n,m)$ et $(n+p,m+q)$ qui sont dans le m\^eme
espace propre pour la monodromie se d\'ecouplent l'un de l'autre,
et dans le cas du couplage fort c'est au contraire
tout l'espace propre des \'etats de spins
$J=\Jhat=n/2(2\pm s)+$entier qui se d\'ecouple des autres espaces
propres de la monodromie.

Il y a n\'eanmoins encore un autre lien
entre ces deux types de mod\`eles,
ce qui pourrait r\'ev\'eler quelque chose de plus profond.
Cela a d\'ej\`a \'et\'e not\'e en conclusion de ref.[P4].
Par la formule de KPZ (Eq.\ref{gstring} en genre 0:
$\gamma_{\hbox {\scriptsize
str}}=(d-1-\sqrt{(d-1)(d-25)})/12$),
un mod\`ele $(p,q)$ en dimension $d=1-6(p-q)^2/pq$ donne une susceptibilit\'e
de corde n\'egative $\gamma_{\hbox {\scriptsize
str}}=(p-q)/p$.
Cependant,
dans les mod\`eles de matrices ou les flots de KdV,
on obtient souvent d'abord $\gamma_{\hbox {\scriptsize str}}$
d'o\`u l'on d\'eduit la dimension $d$ par la formule de KPZ.
On obtiendrait donc la m\^eme dimension \`a partir
de l'autre branche de la formule de KPZ:
$\gamma_{\hbox {\scriptsize
str}}=(d-1{\bf +}\sqrt{(d-1)(d-25)})/12$
ou $\gamma_{\hbox {\scriptsize str}}=(q-p)/q>0$
pour un mod\`ele $(p,q)$.
On s'aper\c coit alors qu'un mod\`ele $(s,2)$
donne dans l'autre branche,
une susceptibilit\'e
$\gamma_{\hbox {\scriptsize str}}=(2-s)/2$
identique \`a celle que nous avons obtenu
dans le couplage fort (Eq.\ref{gs}) !
Notre mod\`ele topologique pourrait donc
n'\^etre qu'une autre branche des mod\`eles
minimaux $(s,2)$.
Il correspondent \`a des dimensions -2 pour un mod\`ele $(1,2)$
et $-\infty$ pour un mod\`ele $(0,2)$.
Des mod\`eles de matrices
de susceptibilit\'e
$\gamma_{\hbox {\scriptsize str}}=(q-p)/q$ positive
ont d\'ej\`a \'et\'e obtenus en refs.\cite{Durh,KH}.
Celle-ci a \'et\'e interpr\'et\'ee comme \'etant
donn\'ee par l'autre branche
de la formule de KPZ.
Peut-\^etre pourrait-on au contraire
l'interpr\'eter par notre formule $\gamma_{\hbox {\scriptsize str}}=
(2-s)/2=(c-1)/12$,
ce qui donnerait des dimensions fractionnaires dans le r\'egime de
couplage fort,
et effectuer le lien entre ces mod\`eles de matrices
(ou d'autres)
et notre mod\`ele topologique,
au moins pour nos op\'erateurs de spins $J=\Jb$
puisqu'il n'y a de toute fa\c con pas de d\'ecomposition chirale
dans les mod\`eles de matrices.

\newpage

\centerline{\large\bf LISTE DES PUBLICATIONS}

\vskip 3cm

\noindent
[P1] ``The genus-zero bootstrap of chiral vertex operators
in Liouville
theory'',

E. Cremmer, J.-L. Gervais, J.-F. Roussel,
{\sl Nucl. Phys.} {\bf B413}
(1994) 244.

\vskip 5mm
\noindent
[P2] ``The Quantum Group structure of 2D Gravity and minimal models:

the genus-zero chiral bootstrap'',
E. Cremmer, J.-L. Gervais, J.-F. Roussel,

\cmp 161, 1994, 597.

\vskip 5mm
\noindent
[P3] ``Solving the strongly coupled 2D Gravity II:
Fractional-spin operators

and topological
three-point functions'',
J.-L. Gervais, J.-F. Roussel, {\sl Nucl.}
{\sl Phys.}

{\bf B426} (1994) 140.

\vskip 5mm
\noindent
[P4] ``From weak to strong coupling regime
in two-dimensional gravity'',
J.-L.

Gervais, J.-F. Roussel,
\pl B338, 1994, 437.

\vskip 5mm
\noindent
[P5] ``Solving the strongly coupled 2D Gravity III:
string susceptibility and

topological N-point function.'',
J.-L. Gervais, J.-F. Roussel,
\`a para\^\i tre.

\vskip 5mm
\noindent
[P6] ``Approche op\'eratorielle de l'\'equation de Liouville'',

J.-F. Roussel,
compte-rendu des Journ\'ees Jeunes Chercheurs 93.

\end{document}